\documentclass[reqno,centertags, 12pt]{amsart}
\usepackage{amsmath,amsthm,amscd,amssymb,siunitx}
\usepackage{latexsym}
\usepackage{graphicx,wrapfig}
\usepackage{enumitem}
\usepackage{hyperref}
\usepackage[mathscr]{eucal}
\usepackage{tikz}
\usepackage{url}
\usepackage{float}

\usepackage{totcount}
\newtotcounter{citnum} 
\def\oldbibitem{} \let\oldbibitem=\bibitem
\def\bibitem{\stepcounter{citnum}\oldbibitem}

\newcommand{\bbA}{{\mathbb{A}}}
\newcommand{\bbC}{{\mathbb{C}}}
\newcommand{\bbD}{{\mathbb{D}}}
\newcommand{\bbE}{{\mathbb{E}}}

\newcommand{\bbL}{{\mathbb{L}}}
\newcommand{\bbP}{{\mathbb{P}}}

\newcommand{\bbR}{{\mathbb{R}}}
\newcommand{\bbS}{{\mathbb{S}}}
\newcommand{\bbT}{{\mathbb{T}}}

\newcommand{\bbZ}{{\mathbb{Z}}}

\newcommand{\fre}{{\frak{e}}}
\newcommand{\frg}{{\frak{g}}}

\newcommand{\x}{{\mathbf{x}}}

\newcommand{\calA}{{\mathcal{A}}}
\newcommand{\calB}{{\mathcal{B}}}
\newcommand{\calC}{{\mathcal{C}}}
\newcommand{\calD}{{\mathcal{D}}}
\newcommand{\calE}{{\mathcal{E}}}

\newcommand{\calH}{{\mathcal H}}
\newcommand{\calI}{{\mathcal I}}

\newcommand{\calK}{{\mathcal K}}
\newcommand{\calL}{{\mathcal L}}

\newcommand{\calN}{{\mathcal N}}
\newcommand{\calP}{{\mathcal P}}

\newcommand{\calU}{{\mathcal U}}

\newcommand{\bdone}{{\boldsymbol{1}}}

\newcommand{\lb}{\label}

\newcommand{\wti}{\widetilde  }

\newcommand{\tr}{\text{\rm{Tr}}}

\newcommand{\spec}{\text{\rm{spec}}}

\newcommand{\rank}{\text{\rm{rank}}}
\newcommand{\ran}{\text{\rm{ran}}}

\newcommand{\bi}{\bibitem}
\newcommand{\hatt}{\widehat}
\newcommand{\beq}{\begin{equation}}
\newcommand{\eeq}{\end{equation}}
\newcommand{\ba}{\begin{align}}
\newcommand{\ea}{\end{align}}

\let\det=\undefined\DeclareMathOperator{\det}{det}

\newcounter{smalllist}
\newenvironment{SL}{\begin{list}{{\rm\roman{smalllist})}}{%
\setlength{\topsep}{0mm}\setlength{\parsep}{0mm}\setlength{\itemsep}{0mm}%
\setlength{\labelwidth}{2em}\setlength{\leftmargin}{2em}\usecounter{smalllist}%
}}{\end{list}}

\newcommand{\bigtimes}{\mathop{\mathchoice%
{\smash{\vcenter{\hbox{\LARGE$\times$}}}\vphantom{\prod}}%
{\smash{\vcenter{\hbox{\Large$\times$}}}\vphantom{\prod}}%
{\times}%
{\times}%
}\displaylimits}

\newcommand{\comm}[1]{}

\DeclareMathOperator{\Real}{Re}
\DeclareMathOperator{\Ima}{Im}

\DeclareMathOperator{\arccosh}{Arc\,cosh}

\DeclareMathOperator{\sgn}{sgn}
\allowdisplaybreaks
\numberwithin{equation}{section}

\newtheorem{theorem}{Theorem}[section]

\newtheorem*{p2.1}{Proposition 2.1}

\theoremstyle{definition}
\newtheorem{definition}[theorem]{Definition}

\newtheorem*{remark}{Remark}
\newtheorem*{remarks}{Remarks}

\newcommand{\jap}[1]{\langle #1 \rangle}
\newcommand{\bigjap}[1]{\left\langle #1 \right\rangle}
\newcommand{\norm}[1]{\lVert#1\rVert}

\AtBeginDocument{%
  \LetLtxMacro{\TheRealLabel}{\label}%
  \LetLtxMacro{\TheRealRef}{\ref}%
  \LetLtxMacro{\TheRealPageRef}{\pageref}%
}

\begin{document}

\title[Twelve Tales]{Twelve Tales in Mathematical Physics: An Expanded Heineman Prize Lecture}
\author[B.~Simon]{Barry Simon$^{1,2}$}

\thanks{$^1$ Departments of Mathematics and Physics, Mathematics 253-37, California Institute of Technology, Pasadena, CA 91125.
E-mail: bsimon@caltech.edu}

\thanks{$^2$ Research supported in part by NSF grants DMS-1265592 and DMS-1665526 and in part by Israeli BSF Grant No. 2014337.}

\

\date{\today}
\keywords{Simon, Schr\"{o}dinger operators, quantum mechanics, quantum field theory, statistical mechanics}
\subjclass[2010]{Primary: 81Q10, 81T08, 82B24; Secondary: 47A55, 81Q15, 81Q20, 81Q70, 81T25, 81U24}

\begin{abstract}  This is an extended version of my 2018 Heineman prize lecture describing the work for which I got the prize.  The citation is very broad, so this describes virtually all my work prior to 1995 and some afterwards.  It discusses work in non-relativistic quantum mechanics, constructive quantum field theory and statistical mechanics.
\end{abstract}

\maketitle

\tableofcontents

\section{Introduction} \lb{s0}

The citation for my 2018 Dannie Heineman prize for Mathematical Physics reads: \emph{for his fundamental contributions to the mathematical physics of quantum mechanics, quantum field theory, and statistical mechanics, including spectral theory, phase transitions, and geometric phases, and his many books and monographs that have deeply influenced generations of researchers}.  This is very broad so I decided to respond to the invitation to speak at the March 2018 APS which says the talk should be \emph{preferably on the work for which the Prize is being awarded}, by discussing the areas of my most important contributions to mathematical physics.  I couldn't say much in the 30 minutes allotted to the talk so it seemed to make sense to prepare this expanded Prize Lecture.

I will discuss 12 areas in theoretical and mathematical physics.  The first seven involve areas where my work was largely done during my Princeton years, 1969--1980 (a kind of golden era in mathematical physics \cite{IAMPPrinceton}) and the last four during my Caltech years, 1980--1995 (I've remained at Caltech since 1995 but my interests shifted towards the spectral theory of long range potentials and of orthogonal polynomials whose connection to physics is more remote).  The eighth area is one where I had work both before and after I moved to Caltech.  It's a pleasure to thank Michael Aizenman, Michael Cwikel, David Damanik, Jan Derezinski, Rupert Frank, J\"{u}rg Fr\"{o}hlich, Fritz Gesztesy, Leonard Gross, George Hagedorn, Bernard Helffer, Svetlana Jitomirskaya, Martin Klaus, Elliott Lieb, John Morgan, Derek Robinson, Israel Sigal, Alan Sokal and Maxim Zinchenko for feedback on drafts of this article.

Many of the topics I'll discuss have spawned industries (as shown by my current Google scholar h-index of 113); any attempt to quote all the related literature would stretch the number of references far beyond the \total{citnum} so I'll mainly settle for quoting relevant review articles or books where they exist or perhaps limit to on a very small number of later papers that shed light on my earlier work.  In particular, I focus very much on my own work and make no pretense of doing comprehensive reviews or a serious history even of all the ideas floating around at the time of my work and certainly not all the work after I essentially left a subject.

\section{Summability of Divergent Eigenvalue Perturbation Series} \lb{s1}

Eigenvalue perturbation theory depends on formal perturbation series (aka RSPT (or just RS) for Rayleigh-Schr\"{o}dinger Perturbation Theory) introduced by Rayleigh \cite{RayPT} and Schr\"{o}dinger \cite{SchrPT}.  The core of the rigorous theory about 1970 when I began my research in this area were results of Rellich \cite{RellichPT}, extended by Nagy \cite{NagyPT} and Kato \cite{KatoPT} (see my review \cite{SimonKato} of Kato's work written on the centenary of his birth) and summarized in Kato's magnificent 1966 book \cite{KatoBk}.

The Kato--Rellich theory in its simplest form considers operator families
\begin{equation}\label{1.1}
  A(\beta)=A_0+\beta B
\end{equation}
where $A_0$ and $B$ are typically unbounded self--adjoint operators ($B$ need only be symmetric; see \cite[Chap. 7]{OT} for a presentation of the language of unbounded self--adjoint operators) on a Hilbert space, $\calH$.  One demands that there are $a$ and $b$ so that
\begin{equation}\label{1.2}
  D(B)\supset D(A_0); \qquad \forall \varphi \in D(A_0): \, \norm{B\varphi}\le a\norm{A_0\varphi}+b\norm{\varphi}
\end{equation}
In Kato's language, one says that $A(\beta)$ is a \emph{type A perturbation} of $A_0$.  The big result of this theory is

\begin{theorem} \lb{T1.1} If $A(\beta)$ is a family of type A and $E_0$ is an isolated eigenvalue of $A_0$ of finite multiplicity, $\ell$, then there exist $\ell$ analytic functions, $\{E_j(\beta)\}_{j=1}^\ell$, near $\beta=0$ which are all the eigenvalues of $A(\beta)$ near $E_0$ when $\beta$ is small.  Moreover, there exists an analytic choice $\{\varphi_j(\beta)\}_{j=1}^\ell$ of eigenvectors, orthonormal when $\beta$ is real and small.  The Taylor coefficients of $E_j$ and $\varphi_j$ are given by the Rayleigh--Schr\"{o}dinger perturbation theory.
\end{theorem}

\begin{remarks} 1.  For textbook presentations of this theorem, see Kato \cite{KatoBk}, Reed--Simon \cite{RS4} or Simon \cite[Sections 1.4 and 2.3]{OT}.

2. The \emph{Kato---Rellich theorem} assets that for $\beta \in (-a^{-1},a^{-1})$, one has that $A(\beta)$ is self--adjoint on $D(A_0)$.

3.  The theory is more general than the self--adjoint case.  It suffices that $A_0$ is closed, that $B$ obey \eqref{1.2} and $E_0$ be a point of the discrete spectrum (isolated point of finite algebraic multiplicity).  In that case, one has analyticity in the non-degenerate case ($\ell=1$) but, in general, $E_j$ may be one or more convergent Puiseux series (fractional powers in $\beta$)).  The is also a generalization that only requires quadratic form estimates.

4.  It is also not necessary that the $\beta$ dependence be linear; a suitable kind of analyticity suffices.
\end{remarks}

While this is elegant mathematics, the striking thing is that it doesn't cover many cases of interest to physics.  Perhaps, the simplest example is
\begin{equation}\label{1.3}
  A(\beta) = -\frac{d^2}{dx^2}+x^2+\beta x^4
\end{equation}
the quantum anharmonic oscillator.  This is the usual textbook model of RSPT because the sum over intermediate states is finite and one can compute the first few terms in the RSPT by hand.

Moreover, it can be regarded as a toy model for a $\varphi^4$--field theory.  Indeed, if one specializes a $(\varphi^4)_{d+1}$ QFT in $d$ space dimensions to $d=0$, one gets a path integral for the $A(\beta)$ of \eqref{1.3} and the RSPT terms can also be written in terms of Feynman diagrams, at least for the ground state (see, for example, Simon \cite{SimonPphi2, SimonFI}).

In this regard, a celebrated argument of Dyson \cite{DysonQED} is relevant.  He noted that quantum electrodynamics (QED) isn't stable if $e^2 <0$ since electrons then attract.  Since there isn't a sensible theory for such $e$, he argued the Feynman perturbation series must diverge.  Similarly, \eqref{1.3} for $\beta<0$ isn't bounded below (indeed, even worse, the operator isn't self--adjoint and a boundary condition is needed at $\pm\infty$ - see, for example, \cite[Theorem 7.4.21]{OT}).  In fact, various estimates \cite{JaffePT, BW1, SimonAHO} show that the perturbation coefficients for the Rayleigh--Schr\"{o}dinger series of \eqref{1.3} grow like $n!$. (see the further discussion around \eqref{2.9} below).

Two other standard models to which RSPT is applied are the Stark effect in Hydrogen(indeed the title of Schr\"{o}dinger's paper \cite{SchrPT} where he introduced his version of RSPT is ``Quantization as an Eigenvalue Problem, IV.  Perturbation Theory with Application to the Stark Effect of Balmer Lines'')
\begin{equation}\label{1.4}
  A(\beta) = -\Delta -\frac{1}{r}+\beta z
\end{equation}
and the Zeeman effect in Hydrogen
\begin{equation}\label{1.5}
  A(\beta) = -\frac{1}{2}\Delta -\frac{1}{r}+\frac{\beta^2}{8}(x^2+y^2)+\beta L_z
\end{equation}
The $z$ in \eqref{1.4} and $x^2+y^2$ terms in \eqref{1.5} are clearly not bounded at infinity by $A(0)$ (i.e. \eqref{1.2} fails) and it is known that both problems have divergent RSPT.  The Stark effect is more singular than the other two examples in that, as first noted by Oppenheimer \cite{Opp}, its bound states turn into resonances, an issue that I will discuss in Section \ref{s2}.

One might think, on the basis of these three examples that convergent RSPT is irrelevant to physics but that is wrong.  First of all, the Kato--Rellich theorem implies that in the Born--Oppenheimer limit (i.e. infinite nuclear masses), the electronic energies are real analytic in the nuclear coordinates (at non-coincident points if the internuclear repulsion is included); see \cite{LiebSimonBO, MorganSimon} for some of my work on Born--Oppenheimer curves.

Moreover, we have the following interesting example on $L^2(\bbR^3\times\bbR^3)$
\begin{equation}\label{1.6}
  A(\beta) = -\Delta_1-\Delta_2-\frac{1}{r_1}-\frac{1}{r_2}+\beta\frac{1}{|\mathbf{r_1}-\mathbf{r_2}|}
\end{equation}
where the Kato--Rellich theory applies.  Up to a scale factor of $Z^{-2}$, when $\beta=Z^{-1}$, this describes a two electron system moving around a nucleus of charge $Z$.  This $A(0)$ is the sum of two independent hydrogen atoms so it has continuous spectrum $[-\tfrac{1}{4},\infty)$ and eigenvalues
\begin{equation}\label{1.6A}
  E_{n,m} =-\frac{1}{4n^2}-\frac{1}{4m^2};\qquad n,m=1,2\dots
\end{equation}
For $n$ or $m$ equals $1$, these are below $-\tfrac{1}{4}$, so discrete and Theorem \ref{T1.1} applies.

There is a huge literature on the discrete eigenvalues of this system, especially the ground state.  Some of it is summarized in
\cite[Example 2.1]{SimonKato}. I have a joint paper \cite{HO2Simon} on what happens at $\beta_c$, the coupling where the ground state hits the continuous spectrum.

The major theme of this section is that RSPT tells you something about the eigenvalues, even when the series diverges.  Before my work, the standard connection, where Kato \cite{KatoAsymPT} was the pioneer, concerned asymptotic series, a notion first formalized by Poincar\'{e} \cite{Poin} in 1886. Given a function, $f$, defined in a region $R$ with $0$ in its closure, we say that $f(\beta)$ has $\sum_{n=0}^{\infty} a_n\beta^n$ as an asymptotic series on $R$ if an only if, for any $N$, we have that
\begin{equation}\label{1.7}
   \lim_{|\beta|\to 0,\,\beta\in R} (f(\beta)-\sum_{n=0}^{N} a_n\beta^n)\beta^{-N} = 0
\end{equation}
Kato's method allows one to prove that RSPT is asymptotic when $R=(0,B)$ for any eigenvalue of the anharmonic oscillator, \eqref{1.3}, and for the Zeeman effect, \eqref{1.5}, and the method in his book \cite{KatoBk} allows one to take $R$ to be suitable sectors in the complex plane.

\eqref{1.7} shows that $f$ determines $\{a_n\}_{n=0}^\infty$ but since, for example, when $R=(0,B)$, if \eqref{1.7} holds for $f$, it also holds for $f_1(\beta)=f(\beta)+10^6 \exp(-1/10^6\beta)$, if we only know \eqref{1.7} and $\{a_n\}_{n=0}^\infty$, we can't say anything about the value of $f(\beta)$ for any particular fixed, non--zero $\beta$.  Over the years, mathematicians have developed a number of methods for recovering a unique function among the several associated to a given asymptotic series.  Hardy \cite{HardyBk} is a discussion of many of them.  Two of them -- Pad\'{e} and Borel summability are relevant to our discussion here.

Truncated Taylor series are polynomial approximations to a formal series $\sum_{n=0}^{\infty} a_n\beta^n$. Pad\'{e} approximation involves rational approximation (the name is after the thesis of Pad\'{e} \cite{Pade}; his advisor, Hermite, was a great expert on rational approximation). Given a formal series, $\sum_{n=0}^{\infty} a_n\beta^n$, we say that $f^{[N,M]}(\beta)$ is the $[N,M]$ Pad\'{e} approximant if
\begin{equation}\label{1.8}
  f^{[N,M]}(\beta)= \frac{P^{[N,M]}(\beta)}{Q^{[N,M]}(\beta)}; \quad \deg P^{[N,M]} = M, \quad \deg Q^{[N,M]} = N
\end{equation}
\begin{equation}\label{1.9}
  f^{[N,M]}(\beta) - \sum_{n=0}^{N+M} a_n \beta^n = \textrm{O}\left(\beta^{N+M+1}\right)
\end{equation}

A formal power series is called a \emph{series of Stieltjes} if it has the form
\begin{equation}\label{1.10}
  a_n = (-1)^n \int_{0}^{\infty} x^n d\mu (x)
\end{equation}
for some positive measure, $d\mu$ on $[0,\infty)$ with all moments finite.  This is related to the \emph{Stieltjes transform} of $\mu$ which is defined by
\begin{equation}\label{1.11}
  f(\beta) = \int_{0}^{\infty} \frac{d\mu (x)}{1+x\beta}
\end{equation}
since it is easy to see that $\sum_{n=0}^{\infty} a_n\beta^n$ is an asymptotic series for such an $f$ in any region of the form $\{\beta\,|\,|\arg\beta|<\pi-\epsilon\}$.  A basic result on convergence of Pad\'{e} approximants is

\begin{theorem} (Stieltjes Convergence Theorem) \lb{T1.2} If $\{a_n\}_{n=0}^\infty$ is a series of Stieltjes, then for each $j \in \bbZ$, the diagonal Pad\'{e} approximants, $f^{[N,N+j]}(z)$, converge as $N \to \infty$ for all $\beta \in \bbC\setminus [0,\infty)$ to a function $f_j(\beta)$ given by \eqref{1.11} with $\mu$ replaced by $\mu_j$ which obeys \eqref{1.10} (with $\mu=\mu_j$).  The $f_j$ are either all equal or all different depending on whether \eqref{1.10} has a unique solution, $\mu$, or not.
\end{theorem}

\begin{remarks}  1.  For proofs, see Baker \cite{Baker, BakerBk, BakerGamel} or Simon \cite[Section 7.7]{OT}

2.  Stieltjes \cite{Stie} didn't discuss Pad\'{e} approximates by name but instead studied continued fractions which lead to the result for $j=0,1$ from which one can deduce the general result.

3. There is a huge literature on convergence of Pad\'{e} approximants in cases where they are not series of Stieltjes (see Lubinsky \cite{Lub}).
\end{remarks}

In applying Theorem \ref{T1.1} it is useful to know when \eqref{1.10} has a unique solution.  A sufficient (but certainly not necessary) condition is that
\begin{equation}\label{1.12}
  |a_n| \le AC^n (kn)!
\end{equation}
for some $k \le 2$.

The second summability method relevant to us here may work if \eqref{1.12} holds for $k=1$.  One forms the Borel transform
\begin{equation}\label{1.13}
    g(w) = \sum_{n=0}^{\infty} \frac{a_n}{n!} w^n
\end{equation}
which defines an analytic function in $\{w \,|\, |w| < C^{-1}\}$.  Under the assumption that $g$ has an analytic continuation to a neighborhood of $[0,\infty)$, one defines for $\beta$ real and positive
\begin{equation}\label{1.14}
  f(\beta) = \int_{0}^{\infty} e^{-a} g(a\beta) da
\end{equation}
if the integral converges.  Since $\int_{0}^{\infty} e^{-a}a^n\,da=n!$, formally, $f(\beta)=\sum_{n=0}^{\infty} a_n \beta^n$.  Here, one has a theorem of Watson \cite{Watson}; see Hardy \cite{HardyBk} for a proof:

\begin{theorem} \lb{T1.3} Let $\Theta \in \left(\tfrac{\pi}{2},\tfrac{3\pi}{2}\right)$ and $B > 0$.  Define
\begin{align}
  \Omega &= \{z\,|\,0 < |z| < B, |\arg z| < \Theta\} \lb{1.15} \\
  \wti{\Omega} &= \{z\,|\,0 < |z| < B, |\arg z| < \Theta - \tfrac{\pi}{2} \} \lb{1.16} \\
  \Lambda &= \{w\,|\,w \ne 0, |\arg w| < \Theta - \tfrac{\pi}{2} \} \lb{1.17}
\end{align}
Suppose that $\{a_n\}_{n=0}^\infty$ is given and that $f$ is analytic in $\Omega$ and obeys
\begin{equation}\label{1.18}
  \left|f(z) - \sum_{n=0}^{N} a_n z^n\right| \le A C^{N+1} (N+1)!
\end{equation}
on $\Omega$ for all $N$.  Define
\begin{equation}\label{1.19}
  g(w) = \sum_{n=0}^{\infty} \frac{a_n}{n!} w^n; \qquad |w| < C^{-1}
\end{equation}
Then $g(w)$ has an analytic continuation to $\Lambda$ and for all $z \in \wti{\Omega}$, we have that
\begin{equation}\label{1.20}
  f(z) =  \int_{0}^{\infty} e^{-a} g(az)\, da
\end{equation}
\end{theorem}

\begin{remark}  By writing \eqref{1.20} as $f(z) = z^{-1}\int_{0}^{\infty} e^{-a/z}g(a)\,da$, one sees that the natural regions of analyticity aren't sectors but regions of the form $\Real(1/z)>1/R$ which are circles tangent to the imaginary axis.  A few years after Watson, Nevanlinna \cite{Nevan} proved a stronger version of Theorem \ref{T1.3} using such regions; see also Sokal \cite{Sokal}.
\end{remark}

My own work on the anharmonic oscillator was motivated by my thesis advisor, Arthur Wightman, who had the idea of exploring this as a way of understanding QFT perturbation theory.  He wanted to exploit an idea of Symanzik to use scaling.  One looks at
\begin{equation}\label{1.21}
  H(\alpha,\beta)=-\frac{d^2}{dx^2}+\alpha x^2+\beta x^4
\end{equation}
and notes that if $\lambda$ is positive, then
\begin{equation}\label{1.22}
  (U(\lambda)f){x} = \lambda^{1/4}f(\lambda^{1/2} x)
\end{equation}
is unitary and
\begin{equation}\label{1.23}
  U(\lambda)H(\alpha,\beta)U(\lambda)^{-1} = \lambda^{-1}H(\alpha\lambda^2,\beta\lambda^3)
\end{equation}
So for any $\alpha,\beta,\lambda$ real with $\beta,\lambda > 0$, one has that
\begin{equation}\label{1.24}
  E_n(\alpha,\beta) = \lambda^{-1}E_n(\alpha\lambda^2,\beta\lambda^3)
\end{equation}
where $E_n$ is the $n$th eigenvalue of $H(\alpha,\beta)$.

Wightman gave the problem to another graduate student, Arnie Dicke, but they came to me with a technical problem they ran into.  Then, in early 1968, I was a second year graduate student in physics but I had been charmed by Kato's book \cite{KatoBk} and was regarded as a local expert on some of the material.  The problem was that $U(\lambda)$ was only a bounded operator if $\lambda>0$ and so \eqref{1.23} only made sense for such $\lambda$ and they wanted \eqref{1.24} for complex $\lambda$.

I came up with the following argument.  In the region $R = \{\alpha,\beta \in \bbC, \, \beta\notin (-\infty,0]\}$, $H(\alpha,\beta)$ is an analytic family of type A (I proved estimates like \eqref{1.2} for $A=H(\alpha,\beta),\,\beta\notin (-\infty,0]$ and $B=x^2$ or $x^4$).  Thus, as long as the eigenvalue is simple at $(\alpha_0,\beta_0)\in R$, we have that $E_n(\alpha,\beta)$ is analytic near $(\alpha_0,\beta_0)$.  Since \eqref{1.24} holds for $\lambda$ real, it holds for small complex $\lambda$ by analyticity.  (There was an issue of eigenvalue labelling - there was no guarantee that if one went around a loop starting and ending on $\bbR\times (0,\infty)$, that $n$ couldn't change.)

Here, I missed a golden opportunity.  I had proven an invariance of discrete spectrum under complex scaling.  It didn't occur to me to ask about an operator like $-\tfrac{d^2}{dx^2}-\beta e^{-\mu x}$ or $-\Delta-\tfrac{1}{r}$ which like $H(\alpha,\beta)$ has an analytic continuation for $H(\lambda)=U(\lambda)HU(\lambda)^{-1}$ from real $\lambda$ to complex $\lambda$.  If I had, I might have found Combes great discovery of a year later (I'll discuss his work in Section 2).

After I found this, given that Dicke was bogged down in his construction of solution with the expected WKB asymptotics at infinity (which turned into his thesis and which he asked me to publish as an appendix to my long paper \cite{SimonAHO}), he and Wightman felt that I should explore aspects of this problem beyond the existence of solutions that Dicke was looking at.  I immediately noticed that \eqref{1.24} implies that
\begin{equation}\label{1.25}
  E_n(1,\beta)=\beta^{1/3}E_n(\beta^{-2/3},1)
\end{equation}
so since $E_n(\alpha,1)$ is analytic near $\alpha=0$, $E_n(1,\beta)$ has a convergent series near infinity, not in $\beta^{-1}$, but in $\beta^{-2/3}$, so that $E_n(1,\beta)$ has a kind of three sheeted structure.

In some of my work, I made an assumption that $E_n(\alpha,1)$ has no natural boundaries -- this was proven to be true many years later (Eremenko--Gabrielov \cite{Erem}) but for $|\arg\alpha|<2\pi/3$, as we'll see shortly, it was proven there were no singularities at all in the same time frame as my paper.

In 1968--69, Wightman was on leave in Europe and he thought about and talked to others about the anharmonic oscillator and wrote me letters.  Andre Martin pointed out to him that the large $\beta$ expansion couldn't converge for all $\beta\ne 0$.  For, if it were, $E_n(\alpha,1)$ would be an entire Herglotz function and so linear which one can easily see isn't true for $E_n(\alpha,1)=\alpha^{1/2}E_n(1,\alpha^{-3/2})$ shows that
\begin{equation*}
  \lim_{\alpha\to\infty\,\,\alpha\in\bbR} E_n(\alpha,1)/\sqrt{\alpha} = E_n(1,0)
\end{equation*}

I'd never seen the theorem about entire Herglotz functions which I'm sure Martin got by using the Herglotz representation theorem.  While I'd later often use that representation theorem heavily in my career and even find a useful extension for meromorphic Herglotz functions on the disk \cite{SimonMeroHerg}, I'd never heard of it at the time.  In those pre--Google days, I couldn't easily find much about Herglotz functions which was good because it forced me to find my own unconventional proof of the entire Herglotz theorem and that allowed me to prove that $E_n(\alpha,1)$ couldn't have an isolated singularity at infinity.  I'd already proven using Kato's methods that for any fixed $\omega=e^{i\theta},\,|\theta|<\pi$, one had that $p^2+\omega x^2+\beta x^4$ has a RS asymptotic perturbation series as $\beta\downarrow 0, \,\beta>0$ which, by scaling, implied that $E_n(1,\beta)$ has an asymptotic series in $\{\beta\,|\, |\arg\beta|<3\pi/2 - \epsilon,\,|\beta|<R_\epsilon\}$.  This in turn implied that on the three sheeted Riemann surface, there were an infinity of singularities with limit point 0 (on the natural three sheeted surface) and asymptotic phase  $\pm 3\pi/2$.

Around this time, I got a hold of a preprint of Bender and Wu \cite{BW1}.  (In those days, Xeroxing was expensive so preprints were mimeographed and of limited distribution.  While I had known Carl Bender when I was a senior at Harvard and he a graduate student  and we were in Schwinger's QFT course together, I certainly didn't know him well enough to get a preprint, but fortunately, he and Arnie Dicke were friends and I got it from Arnie). Bender and Wu computed the first 75 coefficients, $\{a_n\}_{n=1}^{75}$ for the anharmonic oscillator ground state RS series and they did a numerical analysis of the $a_n$ leading to a conjecture of the large $n$ asymptotics (I'll say more about this subject in the next section).  They also did a mathematically unjustified WKB analysis of the analytic behavior of $E_n(1,\beta)$ which was consistent with what I had found.  (I still remember that my first seminar outside Princeton was a physics talk at Chicago where I made reference to the ``notoriously unreliable WKB approximation''.  Afterwards, a kindly older gentleman came up to me and introduced himself: ``I'm the W of WKB''!).

Early in 1969, I got a letter from Arthur Wightman that began ``The specter of Pad\'{e} is haunting Europe...''.  Various theoretical physicists had the idea of using diagonal Pad\'{e} approximants on some field theoretic Feynman series and Wightman suggested that I try it on the anharmonic oscillator.  I'd never done any scientific computing (and haven't done any since!) but with the first 41 coefficients from the Bender--Wu preprint and explicit determinantal formulae from Baker's book \cite{BakerBk}, it was straightforward.

In those days, one did computer calculations by writing the program in Fortran on punch cards, submiting the deck and waiting a day to get back the results.  My initial output was nonsense, but I realized I'd left out a $(-1)^n$, fixed it, and the second time was golden!  I computed $f^{[N,N]}(\beta)$ for $N=1,2,\dots,20$ and $\beta=0.1,0.2,\dots,1$ and got rapid convergence to answers consistent with less accurate variational calculations already in the literature.

The approximants $f^{[N,N]}(\beta)$ were monotone in $N$ suggesting that the underlying series was a series of Stieltjes.  I realized that with my methods, to prove this, one needed to show that on $\{\beta\,|\,|\arg\beta|<\pi\}$, the $E_j(\beta)$ have no natural boundaries and no eigenvalue crossing, equivalently the same for $E_j(\alpha,1)$ within  $\{\alpha\,|\,|\arg\alpha|<2\pi/3\}$.  Nick Khuri, a physicist at Rockefeller, heard of my work and invited me to talk while Martin was visiting there and I explained the situation to him.  Loeffel--Martin \cite{LM}, using a clever argument tracking the zeros of eigenfunctions were able to show no eigenvalue crossing assuming one could make analytic continuation and I could show, using their results, that one could be sure one could analytically continue.

The four of us (Loeffel, Martin, Simon and Wightman \cite{LMSW}) then published an announcement putting everything together.  The analyticity results implies that for a positive measure on $(0,\infty)$, one has that
\begin{equation}\label{1.26}
  E_j(\beta) = E_0-\beta\int_{0}^{\infty} \frac{d\rho_j(x)}{1+x\beta}
\end{equation}
for all $\beta\in\bbC\setminus (-\infty,0]$.  My results on the RS series being asymptotic in the cut plane then implied that the RS series was a series of Stieltjes, so the diagonal Pad\'{e} approximants converge.  Moreover, I had shown that \eqref{1.12} holds for $k=1$, so the limits are the same and equal the eigenvalues.

While this Pad\'{e} result is nice, the known scope where one can prove Pad\'{e} summability is very limited.  Loeffel et al. \cite{LMSW} note that their methods imply that for $m=2,3,\dots$, the RS series for the eigenvalues of $p^2+x^2+\beta x^{2m}$ are series of Stieltjes so the diagonal Pad\'{e} approximants converge.  However, \eqref{1.12} holds for $k=(m-1)$ so they only knew uniqueness when $m=2,3$.  In fact, several years later, Graffi--Grecchi \cite{GGx8} proved that for the $x^8$ oscillator, the $f^{[N,N+j]}$ converge as $N\to\infty$ to $j$ dependent limits, none of which is the eigenvalue!!  Moreover, the Loeffel--Martin \cite{LM} method tracks zeros and so is limited to ODEs and there is no rigorous Pad\'{e} result known for anharmonic oscillators with more than one degree of freedom.

Borel summability turns out to be much more widely applicable.  Shortly after the four author announcement appeared, I got contacted by Sandro Graffi and Vincenzo Grecchi whom I hadn't previously known.  They enclosed a Xerox of the pages of Hardy's book dealing with Watson's Theorem and more importantly some numerical calculations of the Borel sum of the $x^4$ ground state (based on a not rigorously justified use of Pad\'{e} approximants of the Borel transform, $g$, of \eqref{1.13}) which not only converged but more  rapidly than ordinary Pad\'{e} approximants.  I quickly determined that my techniques showed the hypotheses of Watson's theorem held for $x^4$ oscillators in any dimension and that a higher order (i.e. $(kn)!$ instead of $n!$) Borel summability works for the $x^{2m}$ oscillator so we published a paper with these results \cite{GGS}.  Before leaving the issue of the perturbation series for the anharmonic oscillator, I note that using the first 60 terms in the series and the computer power available in 1978, Seznec and Zinn--Justin \cite{SZJ}, using modified Borel summability and large order expansions, claimed to be able to find the ground state for all values of $\beta$ to one part in $10^{23}$!

I wrote several papers on applying Borel summability in $\Phi_2^4$ cutoff field theory \cite{SimonPphiBorel1, SimonPphiBorel2} and other contexts \cite{SimonSAC1, SimonSAC2}.  Avron--Herbst--Simon \cite{AHS3} proved Borel summability of Zeeman Hamiltonians and there have been proofs by others of Borel summability of various quantum field theoretic perturbation series (Feynman diagram expansion of Schwinger functions): $P(\phi)_2$ \cite{EMS}, $\phi^4_3$ \cite{MSPhi43}, $Y_2$ \cite{Renou}, $Y_3$ \cite{MSY3}.  As we'll see in the next section, there is a sense in which the Stark series is Borel summable.

Before leaving asymptotic perturbation theory, I mention a striking example of Herbst--Simon \cite{HerbS2}
\begin{equation*}
  A(\beta) = -\frac{d^2}{dx^2}+x^2-1+\beta^2 x^4 + 2\beta x^3 -2 \beta x
\end{equation*}

If $E_0(\beta)$ is the lowest eigenvalue, we proved that for all small, non--zero positive $\beta$
\begin{equation*}
  0 < E_0(\beta) < C \exp(-D\beta^{-2})
\end{equation*}
Thus $E_0(\beta)$ has $\sum_{n=0}^{\infty} a_n \beta^n$ as asymptotic series with $a_n \equiv 0$.  The asymptotic series converges but, since $E_0$ is strictly positive, it converges to the wrong answer!

\section{Complex Scaling Theory of Resonances} \lb{s2}

Our second tale also concerns eigenvalue perturbation theory, but in situations where the eigenvalue turns into a resonances.  One of the simplest real physical examples where, at the time of my work, this was expected to occur involves the $1/Z$ expansion of \eqref{1.6}.  The eigenvalues $E_{m,n}$ of $A(0)$ given by \eqref{1.6A} when $m,n \ge 2$ are embedded in the continuous spectrum.  For example $E_{2,2} = -\tfrac{1}{8} > -\tfrac{1}{4}$.  For $\beta\ne 0$, one expects the bound state to dissolve into a resonance.

There is a standard physics textbook calculation called time--dependent perturbation theory (TDPT).  The lifetime, $\tau$, is by the Wigner--Weisskopf formula $\tau=\hbar/\Gamma$ with
$\Gamma= 2\mbox{Im}\,E(\beta)$.  The leading order for $\Gamma$ is called the \emph{Fermi golden rule} and is given by $\Gamma=\Gamma_2 \beta^2+\mbox{O}(\beta^3)$ where
\begin{equation}\label{2.1}
  \Gamma_2 = \left.\frac{d}{d\lambda}\jap{B\varphi_0,\tilde{P}_{(-\infty,\lambda)}(A_0)B\varphi_0}\right|_{\lambda=E_0}
\end{equation}
Here $\tilde{P}$ is a spectral projection for $A_0$ with the eigenspace at $E_0$ removed and $\varphi_0$ is the eigenvector with $A_0\varphi_0=E_0\varphi_0$ for the embedded eigenvalue with $\norm{\varphi_0}=1$ (this is only the correct form if the eigenvalue is simple).  Usually, the right side of \eqref{2.1} is written as $\jap{B\varphi_0,\delta(A_0-E_0)B\varphi_0}$.  This version is from Simon \cite{SimonTDPT}.  There is a subtlety here that we won't discuss in detail (but see \cite{RS4} or \cite[Example 3.2]{SimonKato}): $E_{2,2}$ is actually a degenerate eigenvalue and a subspace of the eigenspace at $-\tfrac{1}{8}$ has a symmetry with a continuum only beginning at $-\tfrac{1}{16}$ (put differently, the continuum it is imbedded in has a different symmetry from part of the eigenspace), so only part of the eigenspace dissolves into resonances.  These resonances are observed in nature and are called autoionizing states or Auger resonances.

A second important model is the Stark Hamiltonian, \eqref{1.4}. If $\beta\ne 0$, it is not hard to see that spec($A(\beta))$ is all of $\bbR$ (for $A_0(\beta))$. For the operator without Coulomb term, one can write down $\exp(itA_0(\beta))$ explicitly \cite{AvronHerbstStark} and show that $-1/r$ doesn't change the spectrum, indeed, wave operators exist and are complete \cite{SimonEtAlStark, HerbstStark1}.  Thus the discrete eigenvalues are swamped by continuous spectrum.  The theoretical physics literature based on a formal tunnelling calculation studied the leading asymptotics of the width, which is $\mbox{O}(\beta^n)$ for all $n$, and found that the leading order is
\begin{equation}\label{2.2}
  \Gamma(\beta) = (2\beta)^{-1} \exp\left(-\frac{1}{6\beta}\right) \left(1+\textrm{O}\left(\tfrac{1}{n}\right)\right)
\end{equation}
This formula, first found correctly by Lanczos \cite{Lanc}, is called the Oppenheimer formula after \cite{Opp}.

There were fundamental mathematical questions discussed by Friedrichs \cite{FriedCont}, who was the first person to look mathematically at issues of eigenvalues turning into resonances.  First, in cases like the Stark effect, where there are RSPT series but no eigenvalues, what is the meaning of the perturbation coefficients.  Second, what exactly is a resonance?  Third, in a case like autoionizing states, what exactly are the higher order terms of TDPT (the physics literature was unclear on this point) and is the series ever convergent?

Before the complex scaling approach, there was the idea of solving the first problem by connecting the series to the asymptotics of the spectral projections of the perturbed operators.  This notion, called spectral concentration was pioneered by Titchmarsh \cite{TitPT} and Kato \cite{KatoThesis, KatoBk} and later by others \cite{CRejto, Ridd}.  It works well for the Stark effect where the width is $\mbox{o}(\beta^n)$ for all $n$ so one can hope to prove spectral concentration to all orders \cite{Rejto1, Rejto2} although it does not seem possible to fit a result like \eqref{2.2} into this framework.  But for the case of autoionizing states, the widths go as $c\beta^2, c\ne 0$ and there is only spectral concentration to first order.

Howland wrote several papers \cite{Howland3, Howland5, Howland4, Howland1, Howland2} that addressed both kind of models, but they required either the perturbation or some other object be finite rank so they didn't cover either physical model mentioned above.

In two remarkable papers, Combes with his collaborators Aguilar and Balslev \cite{AC, BC} developed a framework to study the absence of singular spectrum that I realized was ideal to study autoionizing states.  They called it the theory of dilation analytic potentials but after the quantum chemists started using it in calculations, the name shifted to complex scaling.

Consider first a two body potential, $V(x),\,x\in\bbR^\nu$ and for $\theta\in\bbR$ define
\begin{equation}\label{2.2a}
  (U(\theta)f)(x) = e^{\nu\theta/2}f(e^\theta x)
\end{equation}
which is a one parameter semigroup of unitary operators.  Define, again for $\theta\in\bbR$:
\begin{equation}\label{2.3}
  V(\theta)=U(\theta)VU(\theta)^{-1}
\end{equation}
which is, of course, multiplication by $V(e^\theta x)$.  For general $V$, this doesn't make sense for $\mbox{Im}\,\theta\ne 0$, e.g. let $V$ be a square well. But for some $V$'s, one can analytically continue.  Particular examples are $V(x)=|x|^{-\beta}, 0<\beta<2$, in particular for $\beta=1$, where $V(\theta)$ continues to an entire function and $V(x)=e^{-\mu r}/r$ where $V(\theta)$ can be continued (as a relatively bounded operator) so long as $|\mbox{Im}\,\theta|<\pi/2$.  $V$ is called dilation analytic if $V(\theta)(-\Delta+1)^{-1}$ has an analytic continuation from $\bbR$ to all $\theta$ with $|\mbox{Im}\,\theta|<\Theta$ for some $\Theta>0$.

Let $H_0=-\Delta$ and $H=H_0+V$.  Then
\begin{equation}\label{2.4}
  H(\theta) = U(\theta)HU(\theta)^{-1}
\end{equation}
will have an analytic continuation to $\{\theta\,|\,|\mbox{Im} \, \theta| < \Theta\}$ as an analytic family of type (A).

The Kato--Rellich theory is applicable.  As in the last section, discrete eigenvalues are $\theta$--independent at least for $\mbox{Im}\,\theta$ small.  But since
\begin{equation}\label{2.5}
  H(\theta) = -e^{-2\theta}\Delta+V(\theta)
\end{equation}
we know that if $V(-\Delta+1)^{-1}$ is compact, then we have that $H(\theta)$ has continuous spectrum $e^{-2\mbox{Im}\theta}[0,\infty)$, i.e. the continuous spectrum rotates about the threshold $0$.

Balslev--Combes \cite{BC} analyzed the spectrum for $N$--body Hamiltonians and found a spectrum like that shown in Figure 1

\begin{figure}[H]
\includegraphics[scale=0.4,clip=true]{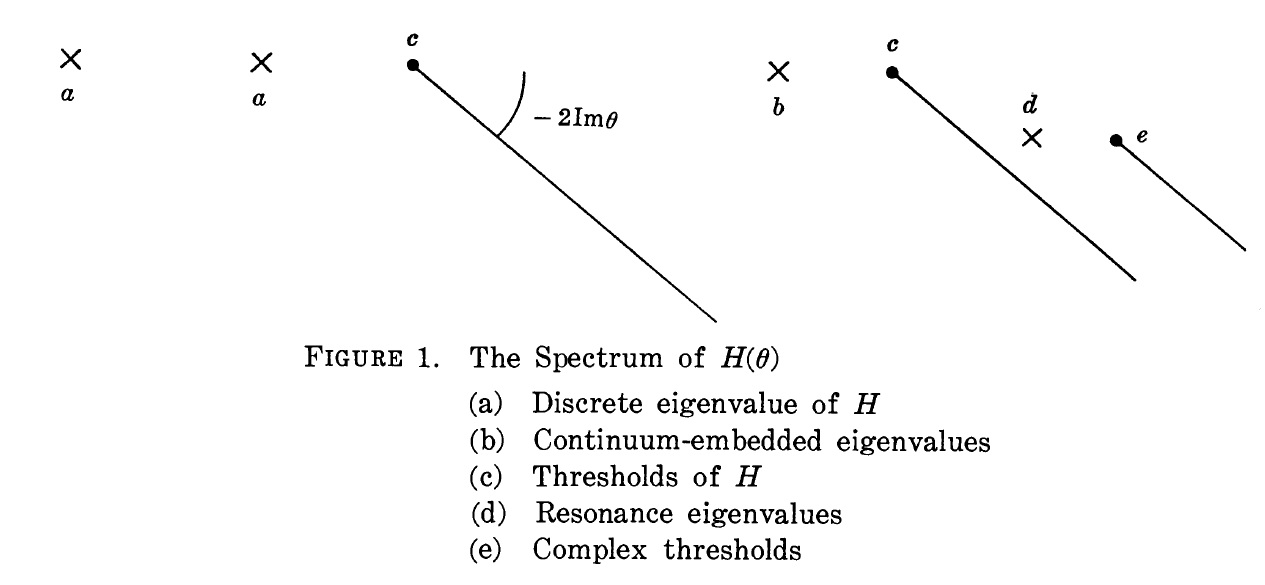}
\caption{Spectrum of a Complex Scaled Hamiltonian}
\end{figure}
Instead of continuous spectrum rotating about zero, it rotates about each scattering threshold.  By an induction argument, one can prove that the set of thresholds is a closed countable set.  An important point is that as the spectrum swings down, it can uncover eigenvalues which then persist until perhaps hit by another piece of continuous spectrum when they can disappear.  Combes and company interpreted these complex eigenvalues as resonances.

A key use Balslev--Combes made of their theory was to prove the absence of singular continuous spectrum (see Section \ref{s6} below).  One of my later results on complex scaling that I should mention is a quadratic form version \cite{SimonCSQF} which has some significant technical simplifications, some of them involving work with Mike Reed on the spectrum of tensor products \cite{SimonTP1, SimonTP2, SimonTP3}.

In \cite{SimonTDPT}, I realized that complex scaling was an ideal tool for understanding autoionizing states.  One can prove that embedded eigenvalues also don't move if $\mbox{Im}\,\theta$ is moved away from zero to positive values while continuous spectrum does move.  Thus in studying $H+\beta W$, one can look at $H(\theta)+\beta W(\theta)$.  While $E_0$ might be an embedded eigenvalue of $H$, so long as it is not at a scattering threshold of $H$, it is a discrete eigenvalue of $H(\theta)$ when $\theta=i\epsilon$ with $\epsilon$ small and positive. So $E_0$ will become a an eigenvalue, $E_0(\beta)$, of $H(\theta)+\beta W(\theta)$ given by a convergent power series in $\beta$ (if $E_0$ is degenerate, there are extra subtleties).  In general, $\mbox{Im}\,E_0(\beta) < 0$, i.e. embedded eigenvalues turn into resonances.  The Rayleigh--Schr\"{o}dinger series for $E_0(\beta)$ provide an unambiguous higher order TDPT which is convergent!  Moreover, one can manipulate the second order term to validate the Fermi golden rule and so get a rigorous proof of it.

For the Stark effect, the conventional wisdom among mathematical physicists was that complex scaling couldn't work.  Because it was known (see, e.g. \cite{HerbstStark1}) that $-\Delta+F\hat{e}\cdot x$ for $F \ne 0$ and $\hat{e}$ a unit vector has no scattering thresholds, there was no place for the continuous spectrum of
\begin{equation}\label{2.6}
  H_0(\theta) = -e^{-2\theta}\Delta+e^\theta F\hat{e}\cdot x
\end{equation}
to go when $\mbox{Im}\,\theta$ is small and non--zero.  But W. Reinhardt, a quantum chemist, was fearless and found \cite{Rein} calculations gave sensible answers.

This made I. Herbst reconsider the conventional wisdom \cite{HerbstStark2}.  In fact for $\mbox{Im}\,\theta \in (0,\pi/3)$, $H_0(\theta)$ defines a closed operator with empty spectrum!  So, since there is no place for the continuous spectrum to go, it disappears! It is, of course, known that a bounded operator cannot have empty spectrum (see, e.g. \cite[Theorem 2.2.9]{OT}) but $H_0(\theta)$ is not bounded and $H_0(\theta)^{-1}$ has a single point, namely $0$ in its spectrum; in some sense, $H_0(\theta)$ has $\infty$ as the only point in its spectrum.  Herbst was able to analyze \cite{HerbstStark2} the Hydrogen Stark Hamiltonian whose resonance energies he showed have width that are $\mbox{O}(F^k)$ for all $k$ and had RS series as asymptotic series.

Herbst and I \cite{HerbS} then extended this work to analyze the Stark Hamiltonian for general atoms.  We also proved a kind of Borel summability.  The Rayleigh--Schr\"{o}dinger perturbation series is Borel summable to a function defined about the positive imaginary axis in the $F$ plane whose analytic continuation back to real $F$ is the resonance.   We proved this for atoms.  About the same time, Graffi--Greechi \cite{GG1, GG2, GGStark} discovered this for the hydrogen Stark effect using the separability of that problem into 1D problems (see below). Sigal \cite{SigStark1, SigStark2, SigStark3, SigStark4} and Herbst--M{\o}ller--Skibsted \cite{HMSkib} have further studied Stark resonances in multi--electron atoms proving that the widths are strictly positive and exponentially small in $1/F$.

Harrell and I then wrote a paper \cite{HS} that was able to analyze the small coupling behavior of the imaginary part of some resonance energies that are exponentially small.  Essentially, this allowed a rigorous proof of some results obtained earlier by theoretical physicists using a formal WKB analysis.  First of all we proved the Lanczos--Oppenheimer formula \eqref{2.2}.  As noted earlier by Herbst--Simon \cite{HerbS} this implies asymptotics of the perturbation coefficients

\begin{equation}\label{2.7}
  E(F) \sim \sum_{n=0}^{\infty} A_{2n} F^{2n}
\end{equation}
\begin{equation}\label{2.8}
  A_{2n} = -6^{2n+1} (2\pi)^{-1} (2n)! \left(1+\textrm{O}\left(\frac{1}{n}\right)\right)
\end{equation}
since one can write $f(x) = \oint_\gamma (2\pi i)^{-1}\tfrac{f(z)}{z-x}\, dz$ where $\gamma$ is a contour that is a small circle with a loop around the negative axis (in the $-F^2$ variable) and a large circle.

In the context of the anharmonic oscillator, the same idea of precise asymptotics of RS coefficients occurred earlier than the work of Herbst--Simon and Harrell--Simon.  As noted in Section \ref{s1}, Bender--Wu \cite{BW1} had computed the first 75 coefficients for the ground state $E_0(1,\beta)$ and they did a numerical fit and conjectured that
\begin{equation}\label{2.9}
  a_n = 4 \pi^{-3/2} (-1)^{n+1} \left(\tfrac{3}{2}\right)^{n+1/2} \Gamma(n+\tfrac{1}{2}) \left(1+\textrm{O}\left(\tfrac{1}{n}\right)\right)
\end{equation}
They had the leading constant to 8 decimal places and guessed its analytic form.  In my anharmonic oscillator paper \cite{SimonAHO}, I noted that \eqref{2.9} was equivalent to leading asymptotics
\begin{equation}\label{2.10}
  \mbox{Im}\,E_0(1,\beta) \underset{\beta=-b+i0}\sim 4 \pi^{-1/2}b^{-1/2}e^{-2/3b}
\end{equation}

Without noticing my remark, Bender and Wu noted \cite{BW2} that \eqref{2.10}, and so \eqref{2.9}, follow from a formal WKB calculation of the tunnelling in a potential x$^2-bx^4$.  Harrell-Simon \cite{HS} have rigorous proofs of \eqref{2.10} and so \eqref{2.9}.  Helffer-Sj\"{o}strand \cite{HelfS2} proved Bender-Wu type formulae for higher dimensional oscillators.

Harrell--Simon uses ODE (i.e. 1D) techniques.  The Stark effect can be separated into 1D problems in elliptic coordinates (noted by Jacobi \cite{Jac} in classical mechanics and then Schwarzschild \cite{SchwSch} and Epstein \cite{Epst1} in old quantum theory and in parabolic coordinates by  Schr\"{o}dinger \cite{SchrPT} and Epstein \cite{Epst2}) and this was later used mathematically by Titchmarsh \cite{TitPT, TitBk}, Harrell--Simon \cite{HS} and by Graffi--Grecchi and collaborators \cite{GG1, GG2, GG3, GG4, GG5, GG6, GG7, GGStark}.

The Zeeman effect for Hydrogen can be reduced to a two dimensional problem.  Avron \cite{AvronZeeman} used this and an instanton calculation of tunnelling (see section \ref{s8}) to formally compute the asymptotics for RS coefficients for the ground state of the Zeeman Hamiltonian \eqref{1.5}.
\begin{equation}\label{2.11}
  E_k = \left(\frac{4}{\pi}\right)^{5/2} (-1)^{k+1} \pi^{-2k} \Gamma\left(2k+\frac{3}{2}\right) \left(1+\textrm{O}\left(\frac{1}{k}\right)\right)
\end{equation}
Helffer-Sj\"{o}strand \cite{HelfS2} then gave a rigorous proof of this using PDE techniques.

Quantum Chemists embraced the complex scaling method to do calculations of resonance energies in atoms and molecules.  I wrote a review of the mathematical theory \cite{SimonCSRev} for a joint conference.  The calculations for molecular resonance curves were done in a Born--Oppenheimer approximation with fixed nuclei which lead to potentials which are analytic outside a large ball.  I introduced exterior complex scaling Simon \cite{SimonExtCS} to justify what they did and wrote a paper with Morgan \cite{MorganSimon2} explaining why exterior scaling did indeed justify their calculations.  A more elegant approach (smooth exterior scaling) was subsequently developed by Hunziker \cite{HunCS} and G\'{e}rard \cite{GerCS}.

I should note that I have reason to believe that, at least at one time, Kato had severe doubts about the physical relevance of the complex scaling approach to resonances.  \cite{HS} was rejected by the Annals of Mathematics, the first journal it was submitted to.  The editor told me that the world's recognized greatest expert on perturbation theory had recommended rejection so he had no choice.  I had some of the report quoted to me.  The referee said that the complex scaling definition of resonance was arbitrary and physically unmotivated with limited significance.  My review of Kato's work on non-relativistic quantum mechanics (henceforth NRQM) \cite[Part 1, pg. 154-155]{SimonKato} has a long discussion of why I believe the complex scaling definition is physically relevant with many references to the literature.

I should mention that I used complex scaling \cite{SimonNoPosBS} to show $N$--body systems with local potentials that can be continued to the right half plane (in particular, with Coulomb potentials) can't have positive energy bound states or thresholds.

While I've focused on the complex scaling approach to resonances, there are other methods.  One, called distortion analyticity, works sometimes for potentials which are the sum of a dilation analytic potential and a potential with exponential decay (but not necessarily any $x$--space analyticity).  The basic papers include Jensen \cite{JenScatt}, Sigal \cite{SigalDistort}, Cycon \cite{CyconDistort}, and Nakamura \cite{NakaDistort1, NakaDistort2}.  Some approaches for non-analytic potentials include G\'{e}rard-Sigal \cite{GerSig}, Cattaneo--Graf--Hunziker \cite{CGHRes}, Cancelier--Martinez--Ramond \cite{CMRRes} and Martinez--Ramond--Sj\"{o}strand \cite{MRSRes}.  There is an enormous literature on the theory of resonances from many points of view. I should mention a beautiful set of ideas about counting asymptotics of resonances starting with Zworski \cite{ZworskiOrig}; see Sj\"{o}strand \cite{SjRes} for unpublished lectures that include lots of references, a recent review of Zworski \cite{ZworskiReview} and the book of Dyatlov--Zworski \cite{DyaZwor} (I have one paper related to these ideas \cite{SiFred}). The form of the Fermi Golden Rule at thresholds is discussed in Jensen--Nenciu \cite{JNFermi}.  A review of the occurrence of resonances in NR Quantum Electrodynamics and of the smooth Feshbach--Schur map is Sigal \cite{SigalNRQED} and a book on techniques relevant to some approaches to resonances is Martinez \cite{MarBk}.

Finally, I note that these two sections have dealt with eigenvalue perturbation theory.  I'll return in Section \ref{s8} to a different issue involving perturbations that give birth to eigenvalues from the edge of continuous spectrum and to eigenvalues at limiting values of coupling constant, namely $-\hbar^2\Delta+V(x)$ as $\hbar\downarrow 0$.

\section{Statistical Mechanical Methods in EQFT} \lb{s3}

The fifteen years following 1965 saw the development of a subject known as constructive quantum field theory (CQFT) which successfully constructed interacting quantum fields in $2$ and $3$ space-time dimensions obeying all the Wightman axioms \cite{WG,SW,Jost}.  Because of the failure to get to $4$ space-time dimensions (except for some negative results \cite{AizDQFT, FrohDQFT,AizDC}), the long lasting impact to rigorous quantum physics has been more limited than initially hoped (extending to the physically relevant $4$ dimensional case is a million dollar problem \cite{JaffeMill}).  Still, the spinoff to various areas of mathematics and theoretical physics has been substantial.

My main goal in this section is to focus on my work, much of it jointly with Francesco Guerra and Lon Rosen, on using methods from classical statistical physics to study Bose CQFT, but I'll begin with some of my other work motivated by CQFT that had important mathematical spinoffs.

CQFT was initially developed by many researchers including J. Fr\"{o}hlich, F. Guerra, K. Osterwalder, L. Rosen, R. Schrader, I. Segal, E. Seiler, T. Spencer, A. Wightman and especially J. Glimm, A. Jaffe and E. Nelson.  I refer the reader to the books of Simon \cite{SimonPphi2} and Glimm-Jaffe \cite{GJBook}.

The initial work mainly on $(\varphi^4)_2$ theories focused on the Hamiltonian viewpoint where controlling spatially cutoff theories is hard because the operators act on an infinite number of variables and the potential is not bounded from below (we use $(X)_d$ shorthand to describe theories where $d$ is the number of space-time dimensions and $X$ an abbreviation for the interaction term).  The first breakthrough was by Nelson \cite{NelHyper} who realized that the free Bose Hamiltonian, $H_0$, in a periodic box in one space dimension, viewed as an infinite sum of harmonic oscillators (with different frequencies), could be realized as a Gaussian process by shifting from $dx$ to $\varphi_0^2\,dx$, where $\varphi_0$ is the ground state, so that $H_0$ acted on $\bbR^\infty$ with a Gaussian probability measure, $d\mu_0$.  The operator $H_0$ was then realized as a pure Dirichlet form (i.e. $\jap{\psi,H_0\psi} = \int\,|\nabla\psi|^2\,d\mu_0$).  For differential operators, this shift to ground state measure and Dirichlet form goes back to Jacobi (!) and since, Nelson's representation has been used many times in mathematical analysis of quantum theories with finitely many degrees of freedom, e.g. \cite{FSW}.  In this representation, Nelson proved that
\begin{equation}\label{3.1}
  \norm{e^{-tH_0}f}_p \le \norm{f}_p
\end{equation}
for all $f\in L^p(\bbR^\infty)$ and all $t>0$ and he proved that for some $T>0$
\begin{equation}\label{3.2}
  \norm{e^{-TH_0}f}_4 \le C\norm{f}_2
\end{equation}
for some fixed $C$ and all $f\in L^2$.  He also showed that while the $(\varphi^4)_2$ spatially cutoff interaction, $V$, is not bounded from below, it obeys
\begin{equation}\label{3.3}
  \int e^{-sV} \, d\mu < \infty \qquad \text{all } s>0
\end{equation}
\begin{equation}\label{3.4}
  V \in \bigcap_{p<\infty} L^p(\bbR^\infty,d\mu)
\end{equation}
and most importantly that \eqref{3.1}-\eqref{3.4} imply that $H_0+V$ is bounded from below.

Two important followups were by Glimm \cite{GlimmHyper}, who proved that \eqref{3.2} plus a mass gap imply that by increasing $T$, \eqref{3.2} holds with $C=1$ (this yields dimension independence and allows removing the need for Nelson to restrict to periodic boundary conditions) and by Federbush \cite{FedHyper}, who used interpolation to prove that $\norm{e^{-sH_0}f}_{p_s} \le C_s \norm{f}_2$ with $p_s\downarrow 2$ as $s\downarrow 0$ and then took derivatives, implicitly getting the first Gaussian logarithmic Sobolev inequality but which was dimension dependent.

The next step is to prove essentially self-adjointness of $H_0+V$ on $D(H_0)\cap D(V)$ for spatially cutoff $\varphi^4_2$.  This was accomplished by Glimm-Jaffe \cite{GJSA} who proved it using additional estimates beyond those of Nelson and subsequently by Segal \cite{SegalSA1, SegalSA2, SegalSA3} who only needed the estimates \eqref{3.1}-\eqref{3.4}.

At this point, my work enters via a widely quoted joint paper with H{\o}egh-Krohn \cite{SiHK} entitled \emph{Hypercontractive semi-groups and two dimensional self-coupled Bose fields}.  We abstracted and simplified Segal's self-adjointness result.  One significant aspect was inventing the term ``hypercontractive'' for groups obeying \eqref{3.1} and \eqref{3.2} (Nelson complained to me that since \eqref{3.2} has a $C$ which might not be one, we should have used ``hyperbounded'' but I replied that hypercontractive sounded better).  Other terms that I've introduced that have caught on include Agmon metric, almost Mathieu equation, Berry's phase, Birman-Schwinger bound, CLR inequality, CMV matrix, coupling constant threshold, diamagnetic inequalities, HVZ theorem, Kato class, Kato's inequality, ten martini problem, Verblunsky coefficients and ultracontractivity.

Hypercontractivity and its differential version, logarithmic Sobolev inequalities (first completely explicated by Gross \cite{GrossLogSob}), have had an enormous number of applications outside quantum field theory; they are even used in Perelman's proof of the Poincar\'{e} conjecture.  See \cite[Section 6.6]{HA} for a discussion of the various sides of the mathematical theory with historical notes, additional references and presentation of some of the applications.  Several years later, in 1983, Brian Davies and I \cite{DavSi} found a variant of hypercontractivity called ultracontractivity which has evoked considerable mathematics.

Before turning to the discussion of statistical mechanical methods in QFT, I should mention another aspect of my work in CQFT with mathematical spinoff.  I wrote a series of papers with E. Seiler \cite{SeilerSi1, SeilerSi2, SeilerSi3, SeilerSi4} on the Yukawa QFT in two space-time dimensions, aka $Y_2$, that developed some mathematical tools in the theory of trace ideals that have had many applications including to quantum information theory.

The work on statistical mechanical methods depends on the second big breakthrough in CQFT, namely Euclidean Quantum Field Theory (EQFT).  The Wightman axioms show that the Wightman functions (vacuum expectation values of the product of quantum fields as tempered distributions on Minkowski space) of any QFT can be analytically continued in time to pure imaginary time differences and that these continued functions are invariant under the Euclidean group.  Schwinger \cite{Schw} first emphasized this, so the analytic continuation to imaginary times are sometimes called Schwinger functions.  Symanzik \cite{Syman1, Syman2} noted the analogy between classical statistical mechanics and EQFT focusing on the analog of the Kirkwood-Salzburg equations.

The central development was due to Nelson \cite{NelEQFT1, NelEQFT2}. He understood that for Bose QFT, EQFT is essentially an infinite dimensional path integral with the extra bonus of Euclidean invariance.  A key role was played by the extension of the Feynman-Kac formula that Guerra-Rosen-Simon called the Feynman-Kac-Nelson formula.  This immediately implied a symmetry, later dubbed Nelson's symmetry:
\begin{equation}\label{3.5}
  \jap{\Omega_0,e^{-tH_\ell}\Omega_0} = \jap{\Omega_0,e^{-\ell H_t}\Omega_0}
\end{equation}
where $H_\ell=H_0+\int_{0}^{\ell} :P(\varphi):(x)\,dx$ is the spatially cutoff Hamiltonian and $H_0\Omega_0=0$.  Nelson also realized the key multidimensional Markov property which allowed one to go from Euclidean fields to Minkowski fields (later Osterwalder-Schrader \cite{OstSch} found an alternate way to do this, which, because it extended to Fermi fields and provided necessary and sufficient conditions, supplanted this part of Nelson's approach).

Nelson gave a few lectures on this new approach in Princeton early in 1971 and lectured at a Berkeley summer school that summer attended by many experts on CQFT (but I was not there).  Even though this work eventually rapidly revolutionized the subject, initially, it had little impact.  I think part of the reason this happened was that the language, especially as presented by Nelson, was so foreign to the functional analysts working in the field, part was that Nelson's lectures seemed obscure and, most importantly, his original work provided no new technical results in conventional CQFT.  Indeed, the only CQFT technical result was a new proof of a lower bound for $(\varphi^4)_2$ theories
\begin{equation}\label{3.6}
  E_\ell \equiv \inf\spec(H_\ell) \ge -c\ell-d
\end{equation}
a result originally proven by Glimm-Jaffe \cite{GJLLB}.  Nelson's proof was much simpler than theirs but its impact was lessened by a simple proof that I found (Simon \cite{SimonLLB}) shortly before Nelson.

The work that made Nelson's theory take off was a remarkable note of Francesco Guerra \cite{Guerra}, then a postdoctoral visitor at Princeton.  Guerra was out of town when Nelson lectured but he got notes from Sergio Albeverio, then another fellow postdoc.  Guerra realized that \eqref{3.5} and
\begin{equation}\label{3.7}
  E_\ell = \lim_{t\to\infty} -\frac{1}{t} \log\jap{\Omega_0,e^{-tH_\ell}\Omega_0}
\end{equation}
provided tools to study $E_\ell$ and $\Omega_\ell$, the vector with $H_\ell\Omega_\ell=E_\ell\Omega_\ell$ (for example, these two equations immediately imply that $\ell\mapsto E_\ell$ is concave).  He proved that $E_\ell/\ell$ had a limit, $\alpha_\infty$, and that $|\jap{\Omega_\ell,\Omega_0}| = \text{O}(\ell^{-k})$ for all $k$.  This was way beyond anything obtained via the purely operator theory used previously in CQFT.

Indeed, I have a vivid memory of how I first learned of these results.  Guerra had been visiting Princeton at that point for about 18 months.  He was very quiet - I'd probably exchanged only a few words with him and he'd given no talks.  Wightman told me that Guerra had asked Wightman to set up a meeting with Lon Rosen (another postdoc and a student of Glimm with several significant CQFT results) and me and we met in Wightman's office in early January, 1972.  Guerra began by writing three facts that he was going to prove.  Lon and I later compared notes and we had the same thought ``yeah, sure, you're going to do that''.  These went so far beyond what was known that it was literally unbelievable.  He began by writing \eqref{3.5} on the blackboard which we'd seen since it was part of Nelson proof of \eqref{3.6} and ten minutes later, he'd proven the three facts.  We were shell shocked!

After Guerra told us of these results, Lon, Francesco and I began working together on exploiting these ideas (our work went through two phases - first we mainly exploited consequences of \eqref{3.5} and similar results but later we fully embraced the Euclidean viewpoint).  In short order, we found \cite{GRS1, GRS2} improvements on what Guerra had found: first $E_\ell = -\alpha_\infty\ell - \beta_\infty +\text{o}(1)$ as $\ell\to\infty$ and secondly, for some $c, d>0$ and $\ell\ge 1$, one has that $e^{-c\ell}\le|\jap{\Omega_\ell,\Omega_0}|\le e^{-d\ell}$ (see Lenard-Newman \cite{LenNew} for further developments on these subjects).  Moreover, we found a new and much simpler proof of some bounds of Glimm-Jaffe \cite{GJPT} that allow one to show that limit points of the cutoff Wightman functions (as the spatial cutoff in $H_\ell$ is removed) are tempered distributions.

The above mentioned work of Guerra and GRS got the attention of experts in CQFT and virtually all papers in the subject after early 1972 used the EQFT framework.  I recall that a few weeks after GRS started working together, Glimm came to Princeton to talk about the bounds in \cite{GJPT} and spent the hour seminar sketching their subtle proof.  Afterwards, Francesco, Lon and I waylaid him and explained in 10 minutes the short proof we had found using an extended version of Nelson's symmetry.  It was Glimm's chance to be shell shocked!

The further introduction of techniques from rigorous statistical mechanics and, in particular, the use of correlation inequalities, the major accomplishment highlighted in this section, were introduced in two papers, one by Guerra-Rosen-Simon \cite{GRS4} and one by Griffiths-Simon \cite{GriffSi}.  GRS \cite{GRS4} was a long paper, so long that the Annals of Mathematics broke it into two parts so it spread between two issues.  Among other things, it provided a detailed exposition of EQFT so that it and my book on the $P(\varphi)_2$ theory (Simon \cite{SimonPphi2}, based on lectures I gave at the ETH) served as the standard references on the subject for a time.

The most important set of ideas in GRS \cite{GRS4} involve the lattice approximation.  Our work was announced \cite{GRS3} a year earlier than Wilson's work \cite{Wilson} on lattice QCD, which of course went much further by allowing Fermion and Gauge fields albeit without mathematical rigor. (It appears from Wilson's historical note \cite{Wilson2} that he didn't start to think about lattice approximations to EQFT until early 1974 while we were already working on it in the spring of 1972; that said, there is no reason to think that Wilson knew of our work in 1974 or even in 2005!).  The free EQFT is a Gaussian random field with covariance $(-\Delta+m^2)^{-1}$.  One gets the lattice approximation by replacing $-\Delta$ by a finite difference operator.  Since it is the inverse of the covariance that appears in the exponent of the Gaussian field, the free lattice field is formally
\begin{equation}\label{3.8}
  Z^{-1} \exp\left(-\tfrac{1}{2}\sum_{|i-j|=1} (s_i-s_j)^2\right)\displaystyle\prod_{j\in\bbZ^2} e^{-m^2s_j^2}\,ds_j
\end{equation}
which is an Ising type ferromagnet with nearest neighbor interactions and spins lying in $\bbR$ (rather than just $\pm 1$ with single site distribution $e^{-as^2}\,ds$.  While \eqref{3.8} is a formal infinite product, if one puts it in a finite box, the spins lie in $\bbR^k$ and the product is a simple finite measure.  An analysis of the interaction just changes $e^{-as^2}$ to $e^{-Q(s)}$ for a suitable semibounded even polynomial.

One powerful tool in the statistical mechanics of spin systems is correlation inequalities, a method initiated by Griffiths \cite{G1onGKS, G2onGKS, G3onGKS} whose inequalities were extended by Kelly-Sherman \cite{KSonGKS} (hence GKS inequalities).  A different set of inequalities are due to Fortuin, Kasteleyn and Ginibre \cite{FKG} (hence FKG inequality).  Relevant to EQFT are versions tailor made for spins with continuous values due to Ginibre \cite{GinibreGKS} for GKS and Cartier \cite{Cartier} for FKG.  With these extensions, GRS \cite{GRS4} obtained GKS and FKG inequalities for Euclidean $P(\varphi)_2$ theories.

The most important application of these correlation inequalities  (namely of GKS) is to show monotonicity in volume of the so-called half-Dirichlet Schwinger functions, a suggestion of Nelson \cite{NelDirichlet}, exploited by GRS \cite{GRS4} to obtain $P(\varphi)_2$ quantum fields obeying all the Wightman axioms except perhaps uniqueness of the vacuum (for this last axiom, see below).  It should be mentioned that the earliest construction of $P(\varphi)_2$ theories (indeed the first construction of non-trivial examples of theories obeying all the Wightman axioms, albeit in 2 space-time dimensions), using cluster expansions, was by Glimm-Jaffe-Spencer \cite{GJS1} for $\lambda P(\varphi)_2$ theories with small $\lambda$ and, then, by Spencer \cite{SpencerExt} for $P(\varphi)_2 + \mu\varphi$ with $|\mu|$ large,  The Nelson-GRS work (for $P(X)=Q(X)+\mu X$ with $Q$ even) was the first results without restrictions on coupling constant.

The second application that I mention is results by Simon \cite{SimonCI1} who, following work of Lebowitz \cite{LebMass} on spin systems, used the FKG inequalities to show that decay of the truncated two point function dominates the decay of all the truncated vacuum expectation values.  This means to prove uniqueness of the vacuum (respectively, existence of a mass gap), it is enough to prove that as $x-y\to\infty$, one has that $\jap{\varphi(x)\varphi(y)}-\jap{\varphi(x)}\jap{\varphi(y)}$ goes to zero (respectively goes to zero exponentially).

While the work of Ginibre and Cartier nicely proves GKS and FKG inequalities for fairly general single spin distributions, there are other results for $\pm 1$ spins that don't extend so generally.  In this regard, Griffiths \cite{GriffTrick} introduced a simple but powerful tool.  Consider two $\pm 1$ spins, $s_1$ and $s_2$ and let $t=\tfrac{1}{2}(s_1+s_2)$.  Then $t$ takes values $0,\pm 1$ just like a spin $1$ spin but if $s_1$ and $s_2$ are uncoupled, the weights are $\tfrac{1}{4},\tfrac{1}{2},\tfrac{1}{4}$ rather than equal weights.  If we find a coupling with energy $H$ so that the Gibbs weight $e^{-H}$ has the values $2,1,2$, then the adjusted weights are all equal.  We thus pick $H=-(\log\, 2)t^2$ which is ferromagnetic.  Thus any correlation inequality that holds for ferromagnetically coupled spin $1/2$ spins extends to spin $1$ spins.  Using this idea and a second order deMoivre-Laplace limit theorem, Griffiths-Simon \cite{GriffSiAnon, GriffSi} realized a lattice system with real valued spins with a weight $\exp(-as^4-bs^2)\,ds\quad (a>0,b\in\bbR)$ as a limit of scaled spin $1/2$ ferromagnetic chains..  This allowed one to obtain GHS (after Griffiths-Hurst-Sherman \cite{GHS}) and Lebowitz \cite{LebIneq} inequalities and a Lee-Yang \cite{LeeYang} theorem for $P(\varphi)$ theories with $P(X)=aX^4+bX^2+\mu X,\,a,\mu\ge 0, b\in\bbR$.  This in turn can be used to prove that when $\mu>0$ such theories have a unique vacuum (Simon \cite{SimonCI2}) and even a mass gap (GRS \cite{GRS5}).  It is known (see Newman \cite{Newman}) that if the polynomial $P$ is of (even) degree larger than $4$ then $\exp(-P(x)dx)\,dx$ may not be approximated by ferromagnetic arrays of $\pm 1$ spins, so the Griffiths-Simon result is restricted to $\varphi^4$ theories.

I should remark that correlation inequalities are useful in the study of Schr\"{o}dinger operators on $\bbR^\nu$.  For example, it is known \cite{SimonVanc, EMN}, using GHS inequalities, that if $V(x)$ is an even function on $\bbR$ with $V'''(x) \ge 0$ for $x > 0$ and if $E_1 <E_2 < E_3$ are the first three eigenvalues of $-d^2/dx^2+V(x)$, then $E_3-E_2 \ge E_2-E_1$.  And, in Section \ref{s7}, we'll discuss applications of FKG inequalities to Schr\"{o}dinger operators in magnetic fields.

While I only worked on CQFT in two space-time dimensions, there is some deep work by others on the three dimensional case.  This, as well as work by others on two dimensions, is presented in the book by Glimm-Jaffe \cite{GJBook}.

I should close the discussion of my work in CQFT by mentioning a paper with Fr\"{o}hlich \cite{FrohSi} that, among other things, constructs $P(\varphi)_2$ theories obeying all the Wightman axioms for any semibounded polynomial $P$.  It relies on Spencer's large $\mu$ expansion \cite{SpencerExt} and FKG inequalities.

\section{Thomas--Fermi Theory} \lb{s4}

In 1972-73, Elliott Lieb and I found results on the Thomas-Fermi (TF) theory that we announced in 1973 \cite{TFAnon} with full details only published in 1977 \cite{TFfull} due, in part, to a long journal backlog.  We first of all established existence and uniqueness of solutions to the TF equations for neutral (and positive ionic) atoms and molecules and, more importantly, proved that TF theory was an exact approximation to quantum theory in suitable $Z\to\infty$ limits.  Since then, an entire industry has been spawned from this work.

TF theory goes back to Thomas \cite{ThomasTF} and Fermi \cite{FermiTF} in 1927 near the dawn of quantum mechanics as an approximation expected to be valid in regions of high electron density.  Interestingly enough, it approximated a linear equation in $3N$ variables as $N\to\infty$ by a non-linear equation in $3$ variables! They originally found their basic equation using a Fermi surface heuristic argument but we relied on the 1932 approach of Lenz \cite{LenzTF} who used energy functionals giving birth to the density functional method of atomic and molecular physics that has become such a standard that the 1998 Nobel Prize in Chemistry was awarded to Walter Kohn ``for his development of the density-functional theory''.

In units where
\begin{equation}\label{4.1}
  \hbar^2\left(\tfrac{3}{2}\right)^{2/3}(2m)^{-1} = 1
\end{equation}
(where $m$ is the electron mass and we assume the electron has $2$ spin states - state counting is important because one assumes Fermi statistics), the Lenz functional is
\begin{equation}\label{4.2}
  \calE(\rho;V) = \frac{3}{5}\int \rho^{5/3}(x)\,d^3x+\frac{1}{2}\int \frac{\rho(x)\rho(y)}{|x-y|}\,d^3xd^3y - \int \rho(x) V(x) \,d^3x
\end{equation}
Here $\rho(x)$ is the electron density, so, if there are $N$ electrons, we have that
\begin{equation}\label{4.3}
  \int \rho(x)\,d^3x=N
\end{equation}
and $V(x)$ is the one electron potential; for a molecule with nuclear charges $z_1,\dots,z_k$ at distinct points $R_1,\dots,R_k$, we have that
\begin{equation}\label{4.4}
  V(x) = \sum_{j=1}^{k} \frac{z_j}{|x-R_j|}
\end{equation}
We set
\begin{equation}\label{4.5}
  Z=\sum_{j=1}^{k} z_j
\end{equation}

The last term in \eqref{4.2} is just the interaction of the electrons with the nuclei and is exact, not an approximation.  The second term is an electron repulsion and assumes no electron correlation so that the two point density is
\begin{equation}\label{4.6}
  \rho_2(x,y) = \rho(x)\rho(y)
\end{equation}
which cannot be even approximately true unless $N$ is large, since $\int \rho_2(x,y)\,d^3xd^3y = N(N-1)$, while, by \eqref{4.3}, $\int \rho(x) \rho(y)\,d^3xd^3y = N^2$.  The first term relies on a quasi-classical calculation.  If one has $N$ particles in a box, $\Omega$, of size $|\Omega|$ with $\rho=N/|\Omega|$ and fills phase space by putting particles in $\{p\,|\,|p|\le p_F\}$, then $N=\tfrac{4\pi}{3}p_F^3|\Omega|/2h^3$ ($2$ in the denominator from $2$ spin states) by the rule that each particle takes volume $h^3$ in phase space.  The total energy of this is then $C|\Omega|\rho^{5/3}$ with an explicit $C$ which explains where the first term in \eqref{4.2} comes from.  The choice \eqref{4.1} leads to $C=3/5$.  Of course, the notion of states taking $h^3$ in phase space is an approximation justified in a large $N$ limit by Weyl's celebrated eigenvalue counting result (see later and \cite[Section 7.5]{OT} for exposition and references).

The Euler-Lagrange equation with Lagrange multiplier to take the condition \eqref{4.3} into account with the restriction $\rho(x)\ge 0$ is that there is $\varphi_0\ge 0$ so that with
\begin{align}
  \varphi(x) &= V(x) - \int\frac{\rho(x)}{|x-y|}\,d^3y \label{4.7} \\
  \rho^{2/3}(x) &= \left\{
                     \begin{array}{ll}
                       \varphi(x)-\varphi_0, & \hbox{ if }\varphi(x) \ge \varphi_0 \\
                       0, & \hbox{ if } \varphi(x) \le \varphi_0
                     \end{array}
                   \right. \lb{4.8}
\end{align}
This is the Thomas-Fermi integral equation which implies the Thomas Fermi PDE
\begin{equation}\label{4.8A}
  \Delta\varphi = \left[\max(\varphi-\varphi_0,0)\right]^{3/2}
\end{equation}

One result that Lieb and I proved \cite{TFAnon, TFfull} is the following:

\begin{theorem} \lb{T4.1} Let $V$ be given by \eqref{4.4} and $N, Z$ given by \eqref{4.3}/\eqref{4.5}.  Then $\calE(\rho;V)$ is well defined if $\rho\ge 0$ lies in $L^1\cap L^{5/3}$.  Moreover:

(a) If $N \le Z$, there is a unique minimizer of $\calE(\rho;V)$ among those $\rho$'s obeying \eqref{4.3}.

(b) If $N>Z$, there is no minimizer of $\calE(\rho;V)$ among $\rho$'s obeying \eqref{4.3}

(c) If $N<Z$, the minimizing $\rho$ has compact support and obeys the TF integral equation for some $\varphi_0 > 0$ and is real analytic on the open set $\{x\,|\,\varphi(x) > \varphi_0; x\notin\{R_j\}_{j=1}^k\}$.

(d) If $N=Z$, the minimizer minimizes $\calE(\rho;V)$ on all $\rho\in L^1\cap L^{5/3},\rho\ge 0$ without any condition \eqref{4.3}.  This minimizing $\rho$ obeys the TF integral equation with $\varphi_0=0$.  One has that, for all $x$, $\varphi(x)>0$ and so $\rho(x)>0$ on all of $\bbR^3$.  $\varphi$ is real analytic on $\bbR^3\setminus\{R_j\}_{j=1}^k$ and
\begin{equation}\label{4.9}
  \rho(x)\sim 1728|x|^{-6}
\end{equation}
as $x\to\infty$.
\end{theorem}

\begin{remarks} 1. The only prior results on existence were for the neutral atomic case ($k=1,R_1=0,\varphi_0=0$) where one looks for spherically symmetric solutions of \eqref{4.8A}.  Since, if $\varphi$ is spherically symmetric, one has that $\Delta\varphi=r^{-1}(r\varphi)''$, we see that if $Y(r)=r\varphi(r\omega)$, then \eqref{4.8A} when $r\ne 0$ is equivalent to
\begin{equation}\label{4.9A}
  Y''(r) = r^{-1/2}Y(r)^{3/2}
\end{equation}
which goes back to the work of Thomas and Fermi.  Thomas noticed that $Y(x) = 144x^{-3}$ solves \eqref{4.9A} (which leads to $\varphi=144x^{-4}$ and $\rho=\varphi^{3/2}=1728|x|^{-6}$). In 1929, already, Mambriani \cite{Mamb} proved existence and uniqueness of solutions of \eqref{4.9A} with $\lim_{r\downarrow 0} Y(x)=a$ and $\lim_{x\to\infty} Y(x) = 0$; see Hille \cite{Hille} for further work.  But Lieb-Simon had the first results on existence and uniqueness going beyond the spherically symmetric case.  We note that uniqueness of spherically symmetric solutions of the PDE doesn't prove that the minimizer for $\calE$ is spherically symmetric nor that the minimizer is unique.

2. Sommerfeld \cite{Sommer} suggested that the singular solution $144x^{-4}$ should control general asymptotics of the TF PDE and that was proven by Hille \cite{Hille} in the spherically symmetric case and by Lieb-Simon \cite{TFfull} in the non-central case.  We used subharmonic comparison ideas, a technique we learned from Teller \cite{Teller}, who used it in a different context.

3. Uniqueness of minima follows from strict convexity of $\calE$, i.e.
\begin{equation*}
  0<\theta<1,\,\rho_1\ne\rho_2 \Rightarrow \calE(\theta\rho_1+(1-\theta)\rho_2)<\theta\calE(\rho_1)+(1-\theta)\calE(\rho_2)
\end{equation*}
since one term in $\calE$ is linear and the other two are strictly convex.

4.  Existence uses what has come to be called the direct method of the calculus of variations (see, for example Dacorogna \cite{Dac}).  Namely, one looks at $\{\rho\,|\,\rho\in L^1\cap L^{5/3},\,\rho>0,\,\int \rho dx \le N\}$ which is compact in a suitable weak topology (if $\le N$ is replaced by $=N$, it is not weakly closed, so not compact) and one proves that $\rho\mapsto\calE(\rho)$ is weakly lower semicontinuous.  A potential theory argument shows that if $N\le Z$, the minimizer has $\int\rho = N$ but if $N>Z$, the minimizer obeys $\int\rho = Z$.  These weak compactness, lsc ideas are now standard analysis but, at the time, while they were used in some areas, they were not widely known in mathematical physics.  While Lieb eventually became a world expert in subtle extensions of this method, he learned the necessary functional analysis from me at the time of our work.
\end{remarks}

With the existence out of the way, we turned to figuring out the connection to atomic physics.  In this regard, there are several reasons that the TF theory might not have anything to do with true quantum systems.  As we saw, in the neutral case, $\rho_{TF}$ decays as $x\to\infty$ as $C|x|^{-6}$ but true atomic bound states decay exponentially (see Section \ref{s6} below; O'Connor's work was done before Lieb and I were working).  As we'll see, scaling shows that in the atomic case the TF density shrinks as $Z$ grows (at a $Z^{-1/3}$ rate), while true atoms expand in extent (although it might be that atomic radii, defined as where $Z-1$ electron live, might be bounded as $Z\to\infty$, they certainly don't shrink).  Finally, it is a result of Teller \cite{Teller} that molecules don't bind in TF theory while they do exist in nature.  For technical reasons, Teller had a short distance cutoff in the Coulomb potential in his argument leading some people to question whether his result held in TF theory without cutoffs, but, it does, as Lieb and I showed.  Interestingly enough, this apparent negative result in TF theory was a key, several years later, in the elegant Lieb-Thirring proof \cite{LTSM} of the stability of matter!

Early in our work, Lieb understood why none of these issues were problems in connecting TF theory to quantum mechanical atoms.  TF theory describes the cores of atoms while chemistry involves the outermost electrons so it isn't surprising that molecules don't bind in TF theory - it is an expression of the repulsion of the cores.  The $|x|^{-6}$ Sommerfeld asymptotics describes the mantle of the core while exponential decay describes the last few electrons.  I still remember the start of our collaboration while we were both visitors at IHES in the fall of 1972.  Lieb had the idea that Weyl type estimates should show that TF was a proper semiclassical limit of atoms.  At the end of a long day of discussing this idea, I told him of Teller's result which I'd learned about in a course taught by Wightman, so since TF theory didn't bind atoms, it couldn't describe physics.  The next morning Lieb walked in and said to me: ``Mr. Dalton's hooks are in the outer shell.''  In other words, chemistry had nothing to with region in which a leading quasi-classical limit is valid. (The notion behind density functional theory is that chemistry can be connected to non-leading terms).

One key to the large $Z$ results is scaling. The following is easy to check.  If $Z>0$ and
\begin{equation}\label{4.10}
  V_Z(x) = Z^{4/3}V(Z^{1/3}x);\qquad\rho_Z(x) = Z^2\rho(Z^{1/3}x)
\end{equation}
then
\begin{equation}\label{4.11}
  \calE(\rho_Z;V_Z) = Z^{7/3}\calE(\rho;V);\qquad\int \rho_Z(x)\,d^3x = Z\int \rho(x)\,d^3x
\end{equation}
In particular if
\begin{equation}\label{4.12}
  E_V(Z;N) = \inf \{\calE(\rho;V_Z)\,|\,\rho\in L^1\cap L^{5/3},\,\rho\ge 0\,\int \rho(x)\,d^3x=N\}
\end{equation}
then
\begin{equation}\label{4.13}
  E_V(Z;nZ) = Z^{7/3}E_V(1;n)
\end{equation}

Given $z_1,\dots,z_k,R_1,\dots,R_k$, we let $E_{TF}(N;z_1,\dots,z_k;R_1,\dots,R_k)$ be the TF energy (i.e. minimum of $\calE(\rho;V)$ with $V$ given by \eqref{4.4} over $\rho$'s obeying \eqref{4.3}).  Then \eqref{4.13} says that
\begin{align}
  E_{TF}(nZ;z_1Z,\dots,z_kZ;&Z^{-1/3}R_1,\dots,Z^{-1/3}R_k) \nonumber \\
                            &= Z^{7/3}E_{TF}(n;z_1,\dots,z_k;R_1,\dots,R_k) \label{4.14}
\end{align}

We next describe quantum atomic energies.  Let $\calH_{phys}$ be those elements in $L^2(\bbR^{3N};\bbC^{2N})$ which are functions of $N$ points $x_1,\dots,x_N$ in $\bbR^3$ and spins $\sigma_1,\dots,\sigma_N$ in $\bbC^2$ which are antisymmetric under permutations of $(x_j,\sigma_j)$ (see \cite[Section 7.9]{OT} for more on this formalism).  On $\calH_{phys}$ let
\begin{equation}\label{4.15}
  H = -\sum_{j=1}^{N}\frac{\hbar^2}{2m}\Delta_j +\sum_{i<j} \frac{1}{|x_i-x_j|}-\sum_{j=1}^{N}\sum_{\ell=1}^{k}\frac{z_\ell}{|x_j-R_\ell|}
\end{equation}
where $\hbar$ is given by \eqref{4.1}.  We set
\begin{equation}\label{4.16}
  E_{Q}(N;z_1,\dots,z_k;R_1,\dots,R_k) = \inf_{\varphi\in\calH_{phys}} \jap{\varphi,H\varphi}
\end{equation}

If $N\ge Z\equiv\sum_{j=1}^{k}z_j$, it is known (see \cite{ZhislinInfinite, SimonInfinite}) that $H$ has a discrete ground state, $\psi$; we set
\begin{align}
  \rho_Q(x;&z_1,\dots,z_k;R_1,\dots,R_k) \nonumber\\
           &= \sum_{\sigma_j=\pm 1; j=1,\dots,N}\int |\psi(x,x_2,\dots,x_N;\sigma_1,\dots,\sigma_N)|^2\,d^3x_2\dots d^3x_N \label{4.17}
\end{align}
(the ground state can be degenerate in which case in the theorem below one can take any ground state eigenfunction).

The main result of Lieb-Simon \cite{TFfull} is

\begin{theorem} \lb{T4.2} For any distinct $R_1,\dots,R_k$ and positive $z_1,\dots,z_k$, and $n>0$, we have that
\begin{align}
  \lim_{Z\to\infty}  Z^{-7/3}E_{Q}(nZ;&z_1Z,\dots,z_kZ;Z^{-1/3}R_1,\dots,Z^{-1/3}R_k) \nonumber\\
                              &=  E_{TF}(n;z_1,\dots,z_k;R_1,\dots,R_k) \label{4.18}
\end{align}
Moreover, if $n \le \sum_{j=1}^{k}z_j$, then
\begin{align}
  \lim_{Z\to\infty} nZ^{-2} \rho_{Q}(Z^{-1/3}x;&z_1Z,\dots,z_kZ;Z^{-1/3}R_1,\dots,Z^{-1/3}R_k) \nonumber\\
                                               &=  \rho_{TF}(x;n;z_1,\dots,z_k;R_1,\dots,R_k) \label{4.19}
\end{align}
in the sense of convergence of the integral over $x$ in any fixed open set.
\end{theorem}

Our proof of \eqref{4.18} uses the method of Dirichlet-Neumann bracketing.  This goes back to Weyl \cite{WeylQC} as formalized by Courant-Hilbert \cite{CH} (see \cite{AW} for the discrete analog and \cite[Section 7.5]{OT} for another textbook discussion).  They used it to count eigenvalues of the Laplacian in regions with smooth boundary.  It was later used to prove that when $V\in C_0^\infty(\bbR^\nu)$, then as $\lambda\to\infty$, one has that $N(\lambda V)$, the number of negative eigenvalues of $-\Delta+\lambda V$ on $L^2(\bbR^\nu)$ obeys
\begin{equation}\label{4.20}
  \lim_{\lambda\to\infty} \lambda^{-\nu/2} N(\lambda V) = (2\pi)^{-\nu}\tau_\nu \int \max(-V(x),0)^{\nu/2}\,d^\nu x
\end{equation}
(where $\tau_\nu$ is the volume of the unit ball in $\bbR^\nu$).  This was discovered independently about the same time by Birman-Borzov \cite{BirmanBorzov}, Martin \cite{Martin}, Robinson \cite{Robinson} and Tamura \cite{Tamura}.  (Lieb and I only knew of Martin's work when we looked at Thomas-Fermi, although all but Tamura existed at the time.)  In Section \ref{s8}, I'll discuss what happens if $V$ is not $C_0^\infty(\bbR^\nu)$.

Using these ideas of dividing space into small boxes, it wasn't hard to show that $E_Q(V_Z)/E_{TF}(V_Z)\to 1$ as $Z\to\infty$ if $V\in C_0^\infty(\bbR^3)$.  These methods don't deal with the boxes around the nuclei at $R_j$.  Basically, one needs to show that the system doesn't collapse on those points, i.e. most of the electrons wind up in the boxes containing those points. When I left IHES for Marseille at the end of 1972, Lieb and I were at this point and were left with the problem we called between ourselves ``pulling the poison Coulomb tooth''.  I spent a long weekend in Paris in March, 1973 and we figured out how to pull the tooth.  With current technology, one would use Lieb-Thirring inequalities (discussed in Section \ref{s8} below; see also \cite{LTSM, LTonLT} for the original papers, \cite{HundSi} for the discrete case and \cite[Section 6.7]{HA} for a textbook discussion) but they didn't exist, so we used an ad hoc argument exploiting the angular momentum barrier.

The proof of \eqref{4.19} isn't hard.  The $\rho$'s are functional derivatives of the energy under adding an infinitesimal $V_Z$ to the Coulomb attraction.  Normally convergence of functions doesn't imply convergence of derivatives but it does for concave functions and one can show the energies, as minima of a set of functions linear in $\lambda$, are concave under $\lambda\to \lambda V$.

I have one other result on large Z ions.  As noted above, it is known (see \cite{ZhislinInfinite, SimonInfinite}) that the Hamiltonian, $H(Z,N)$ for a charge $Z$ nucleus and $N$ electrons has infinitely many bound states if $N\le Z$.  What happens if $N>Z$?  It is a result of Ruskai \cite{RuskaiRS}  and Sigal \cite{IMSS, SigalRS} that there is a finite number $N(Z)$, so that $H(Z,N(Z))$ has no discrete spectrum (i.e. there is a negative ion with nuclear charge $Z$ and total charge $-(N(Z)-Z)$) and so that for all $N>N(Z)$, we have that $N(Z)$ has no bound states below the continuum.  In \cite{LSST}, Lieb, Sigal, Thirring and I showed that $N(Z)/Z \to 1$ as $Z\to\infty$.  That this is especially subtle is seen by the fact that if one replace fermionic electrons by bosons (with negative charge), then Benguria-Lieb \cite{BL} have shown that the analogous $\liminf$ is strictly bigger than $1$.  I note that there are no twice negatively charged ions known in nature so it is possible that $N(Z)$ is always either $Z$ or $Z+1$.  In fact, one of the fifteen open problems in my 2000 list (Simon \cite{Simon15} of which 11 remain open) is to prove that $N(Z)-Z$ remains bounded as $Z\to\infty$.

Since 1973, there has been a huge literature on large $Z$ atoms and molecules and on density functional theory.  I will not attempt a comprehensive review, but I should mention the work on non-leading asymptotics beyond \eqref{4.18} and \eqref{4.19}.  For the energy, Hughes \cite{Hughes} and Siedentop-Weikard \cite{SW1, SW2} obtained the O$(Z^2)$ term for atoms and Ivrii-Sigal \cite{IvrSig} for molecules.  Later, Fefferman-Seco \cite{FS} obtained the O$(Z^{5/3})$ term.  For the density, Iantchenko-Lieb-Siedentop \cite{ILS} found the O$(Z^2)$ term (recently, I was coauthor of a paper \cite{FMSS} that proved an analog for a relativistic Hamiltonian).  Since much of the work on higher order corrections to \eqref{4.2} was done by Lieb and collaborators,  I refer the reader to the relevant volume of Lieb's Selecta \cite{LiebSelecta} for references.  The reader can also look at two somewhat dated review articles by Lieb \cite{LiebTFReview} and Hundertmark \cite{HundTFReview}.

Before leaving this subject, I should mention that Lieb and I \cite{LiebSimonHF} used methods similar to those we used to prove existence of solutions to the TF equation to prove existence of solutions to the Hartree and Hartree-Fock equations for neutral (and positive) atoms and molecules. Later works on existence of solutions of Hartree-Fock include Lions \cite{Lions} and Lewin \cite{Lewin}.

\section{Infrared Bounds and Continuous Symmetry Breaking} \lb{s5}

A fundamental problem in statistical physics concerns the following. The Gibbs states of statistical mechanics are clearly analytic in all parameters, yet nature is full of discontinuities, for example the direction of a magnet as a magnetic field is slowly varied through zero field.  We now realize that the way to understand this is by looking at the thermodynamic limit, i.e. infinite volume, where states can become non-analytic in parameters.  That this is far from evident can be seen by a story told in Pais' book \cite[pp 432-433]{Pais} that as late as 1937, at the van der Waals Centenary conference, a vote of the physicists present was taken on whether this view was correct and the vote was close (although, given that Peierls' work mentioned below was in 1936, it shouldn't have been close; that means that Peierls' work was not well known, or at least not understood, at the time).

The simplest models on which this can be explored are the lattice gases whose formalism is described in the books of Ruelle \cite{Ruelle} and Simon \cite{SimonLG}.  Two of the simplest examples both have spins on a lattice, say $\bbZ^\nu$, $\boldsymbol{\sigma}_\alpha$, at points $\alpha\in\bbZ^\nu$.  In a finite box $\Lambda\subset\bbZ^\nu$, the Hamiltonian (energy functional) is
\begin{equation}\label{5.1}
  H_\Lambda = -J\sum_{\alpha,\gamma\in\Lambda,\,|\alpha-\gamma|=1} \boldsymbol{\sigma}_\alpha\cdot\boldsymbol{\sigma}_\gamma
\end{equation}
The sign is there so that when $J>0$, energies are lowest when spins are parallel, i.e. the model is ferromagnetic.  $J<0$ describes the antiferromagnet.  We will normally take $J=\pm 1$ which is no loss since we can vary the inverse temperature, $\beta$ in $e^{-\beta H}$ defining the Gibbs measure.  We have not been careful about boundary conditions; we will most often take periodic BC although sometimes free or plus BC.

I said two models because I haven't described the set of single spins and their distributions.  If each $\boldsymbol{\sigma}_\alpha=\pm 1$ with equal apriori weights, we have the nearest neighbor \emph{Ising model}. If instead each $\boldsymbol{\sigma}_\alpha\in S^2$, the unit sphere is $\bbR^3$ with the rotation invariant apriori weight, we have the \emph{classical Heisenberg model}.  More generally if each $\boldsymbol{\sigma}_\alpha\in S^{d-1}$, the unit sphere is $\bbR^d$, we have the \emph{$d$-rotor model}.  Significant here is the global symmetry of the system: discrete, $\boldsymbol{\sigma}_\alpha\to - \boldsymbol{\sigma}_\alpha$ for the Ising model and continuous, $\boldsymbol{\sigma}_\alpha\to R\boldsymbol{\sigma}_\alpha$ with $R\in SO(3)$, the rotation group for the Heisenberg model.

In 1936, Peierls \cite{Peierls} found a simple argument proving that when $\nu\ge 2$, the nearest neighbor Ising model on $\bbZ^\nu$ has a phase transition at low temperature. However, the argument depends on the sharp difference between spin up and spin down and fails for the classical Heisenberg model where the spins vary continuously.  Indeed, in 1966, Mermin-Wagner \cite{MW} (Hohenberg \cite{Hohen} had asimilar result on a related model in the same time frame) proved that in $2D$, the classical Heisenberg model has no broken symmetry states at positive temperature (see also \cite{KLSh}).  Their argument relies on the fact that if the spin wave energy (when $J=1$) is given by
\begin{equation} \lb{5.2}
  E_p= \frac{1}{2}\sum_{|\alpha|=1}(1-e^{ip\cdot\alpha}) = \sum_{j=1}^{\nu} (1-\cos(p_j))
\end{equation}
(the Fourier transform of the nearest neighbor coupling with a constant added so that $E_p\ge 0$ with minimum value $0$), then $E_p\sim p^2$ for $p$ small so that
\begin{equation}\label{5.3}
  \int_{|p_j|<\pi} \frac{d^\nu p}{E_p} = \infty \text{   if } \nu=2
\end{equation}

In 1976, J\"{u}rg Fr\"{o}hlich, Tom Spencer and I (henceforth FSS) \cite{FSS1, FSS2} proved that

\begin{theorem} [\cite{FSS1, FSS2}] \lb{T5.1} The classical $d$-vector model ($d\ge 1)$ with nearest neighbor interactions on $\bbZ^\nu$ with $\nu\ge 3$ has multiple phases (with broken symmetry) if $\beta\ge\beta_c$ where
\begin{equation}\label{5.4}
  \beta_c \le \beta_c^{FSS} \equiv \frac{d}{2}I(\nu)
\end{equation}
with
\begin{equation}\label{5.5}
  I(\nu)=\frac{1}{(2\pi)^\nu}\int_{|p_j|<\pi} \frac{d^\nu p}{E_p}
\end{equation}
so $I(\nu)<\infty$ when $\nu\ge 3$
\end{theorem}

\begin{remark} While this was the first result on the existence of phase transitions in the isotropic model, there were earlier results on the (in the quantum case, highly) anisotropic case for the classical (see Malyshev \cite{MalCH} and Kunz et al. \cite{KPVCH}) and quantum (see Ginibre \cite{GinQH} and Robinson \cite{RobQH}) Heisenberg models.  These works used variants of the Peierls method.  Of course, only the isotropic model has a continuous symmetry to break.
\end{remark}

$I(3)$ can be computed exactly in terms of elliptic integrals so one finds (with ``errors'' computed by comparing with high temperature expansions in parentheses)
\begin{equation}\label{5.6}
  T_c(\nu=3,d=3)\ge 1.3189\,(1.44;\,9\% \text{ error})
\end{equation}
and
\begin{equation}\label{5.7}
  T_c(\nu=3,d=1)\ge 3.9567\,(4.5108;\,14\% \text{ error})
\end{equation}

The method of FSS which I sketch below is basically the only method known for rigorously proving spontaneous continuous symmetry breaking with a nonabelian symmetry group.  We note that such continuous symmetry breaking is not only central to statistical mechanics but also to models of particle physics.

A basic notion is reflection positivity.  This is one of the Osterwalder Schrader axioms mentioned in Section \ref{s3}.  FSS realized it also played an important role in statistical mechanical models.

Consider spins in a box, $\Lambda$, with even sides with periodic boundary conditions and slice the box across bonds into two halves, $\Lambda_+$ and $\Lambda_-$. There is a natural reflection $\Theta$ of spins in $\Lambda_+$ onto spins in $\Lambda_-$ that extends to a map of polynomials in the spins.  A state $\jap{\cdot}$ is called \emph{reflection positive} (RP) if and only if for any polynomial, $A$, in the spins of $\Lambda_+$ we have that
\begin{equation} \lb{5.8}
  \jap{\Theta(A)A}   \, \ge 0
\end{equation}

First, suppose that we consider uncoupled spins (i.e. a product measure over sites). Then $\jap{\Theta(A)A}= |\jap{A}|^2 \ge 0$ so the measure is RP.  Now suppose we have a Hamiltonian of the form
\begin{equation} \lb{5.9}
  -H = A+\Theta(A) + \sum_j \Theta(B_j)B_j
\end{equation}
We claim that if $\jap{\cdot}_0$ is RP, so is $\jap{\cdot} = \jap{\cdot e^{-H}}_0/\jap{e^{-H}}_0$ for $\exp(A+\Theta(A)) = \exp(A)\Theta(\exp(A))$ and we can expand $\exp(\sum_j \Theta(B_j)B_j)$ into a Taylor series.

Consider a box, $\Lambda$, with periodic boundary conditions and state $\jap{\cdot}_\Lambda$.  Define the magnetization (here and below we notationally suppress a $\Lambda$ dependence).
\begin{equation} \lb{5.10}
  \mathbf{m} = \frac{1}{|\Lambda|}\sum_{\alpha\in\Lambda} \boldsymbol{\sigma}_\alpha
\end{equation}

Our sign that there is a phase transition will be that
\begin{equation} \lb{5.11}
  \liminf \jap{\mathbf{m}^2}_\Lambda \equiv M^2 >0
\end{equation}
This implies many other notions of phase transition.   For example, one can show that the derivative of the free energy per unit volume with respect to an external magnetic field has a discontinuity of at least $2M$.  Also, there are multiple equilibrium states in the sense of Dobrushin \cite{Dob1, Dob2} and Lanford-Ruelle \cite{LR} (see \cite[Section III.2]{SimonLG}).

We let $\Lambda^*$ be the dual lattice to $\Lambda$ so that
$\{\alpha\mapsto |\Lambda|^{-1/2} e^{i\mathbf{p}\cdot\boldsymbol{\alpha}}\}_{\mathbf{p}\in\Lambda^*}$ is an orthonormal basis for the $\ell^2(\Lambda)$.  Define the Fourier spin wave variables
\begin{equation} \lb{5.12}
  \hatt{\boldsymbol{\sigma}}_p = \frac{1}{\sqrt{|\Lambda|}}\sum_{\alpha\in\Lambda} e^{-ip\cdot\alpha}\boldsymbol{\sigma}_\alpha
\end{equation}
and define the spin wave expectation function
\begin{align}
  g_\Lambda(\mathbf{p}) &= \jap{\hatt{\boldsymbol{\sigma}}_p\cdot\hatt{\boldsymbol{\sigma}}_{-p}} \lb{5.13} \\
                        &= \frac{1}{|\Lambda|}\sum_{\alpha,\beta\in\Lambda} e^{-i\mathbf{p}\cdot(\boldsymbol{\alpha-\beta})}
                                  \jap{\boldsymbol{\sigma}_{\alpha}\cdot\boldsymbol{\sigma}_{\beta}} \nonumber \\
                        &= \sum_\alpha e^{-i\mathbf{p}\cdot\boldsymbol{\alpha}}\jap{\boldsymbol{\sigma}_{\alpha}\cdot\boldsymbol{\sigma}_{0}} \lb{5.14}
\end{align}

Note that $\mathbf{m}=|\Lambda|^{-1/2}\hatt{\boldsymbol{\sigma}}_{p=0}$ so that
\begin{equation} \lb{5.15}
   \jap{\mathbf{m}^2}_\Lambda = |\Lambda|^{-1} g_\Lambda(p=0)
\end{equation}
Since $\hatt{\boldsymbol{\sigma}}_p$ are components of the functions $\alpha\mapsto\boldsymbol{\sigma}_\alpha$ in an ON basis, the Plancherel theorem implies that
\begin{equation} \lb{5.16}
  \sum_{p\in\Lambda^*} g_\Lambda(p) = \sum_\alpha \jap{|\boldsymbol{\sigma}_\alpha|^2}=|\Lambda|
\end{equation}
Since there are $|\Lambda|$ values of $p$ in $\Lambda^*$, this says that normally each $g_\Lambda(p)$ should be of size $1$ while the condition of there being a phase transition is that $g_\Lambda(p=0)$ is of order $\Lambda$.  This allows one to interpret the phase transition as due to a Bose condensation of spin waves.

The key to the proof will be what FSS dubbed an \emph{infrared bound} (IRB), that for $p\ne 0$, one has that:
\begin{equation} \lb{5.17}
  g_\Lambda(p) \le \frac{d}{2\beta E_p}
\end{equation}
where $E_p$ is the spin wave energy.

By \eqref{5.15}-\eqref{5.17}
\begin{equation} \lb{5.18}
  \liminf_{|\Lambda|\to\infty} \jap{\mathbf{m}^2}_\Lambda \ge 1-\frac{dI(\nu)}{2\beta}
\end{equation}
where $I(\nu)$ is given by \eqref{5.5} and we use the fact that $\Lambda^*$ fills out the $\nu$-fold product of $[-\pi,\pi]$ as $|\Lambda|\to\infty$.  Thus infrared bounds imply Theorem \ref{T5.1}.

The first step in the proof of infrared bounds from reflection positivity is to use RP to prove something called Gaussian domination, namely if we define, for arbitrary $\{\mathbf{h}_\alpha\}\in\bbR^{d|\Lambda|}$
\begin{equation}\label{5.19}
  Z(\{\mathbf{h}_\alpha\}) = \int_{S^{(d-1)|\Lambda|}} \exp\left(-\beta\frac{1}{2}\sum_{|\alpha-\gamma|=1} (\boldsymbol{\sigma_\alpha-\sigma_\gamma}-\mathbf{h_\alpha-h_\beta})^2\right)
\end{equation}
then one has that
\begin{equation}\label{5.20}
  Z(\{\mathbf{h}_\alpha\})\le Z(\{\mathbf{h}_\alpha\}\equiv 0)
\end{equation}

One first proves that if $\Lambda$ is split into two halves $\Lambda_+$ and $\Lambda_-$ and, given $\mathbf{h}$, we let $\mathbf{h}_+$ be the $\mathbf{H}$ obtaining by restricting $\mathbf{h}$ to $\Lambda_+$ and reflecting it and similarly for $\mathbf{h}_-$, then
\begin{equation}\label{5.21}
  Z(\{\mathbf{h}\}) \le Z(\{\mathbf{h}_+\})^{1/2}\,Z(\{\mathbf{h}_-\})^{1/2}
\end{equation}
The details of the proof of \eqref{5.20} can be found in \cite{FSS2} or \cite{FILS2}.

Once one has Gaussian domination, fix $\mathbf{h}$ all real and use the fact that $Z(\{\lambda\mathbf{h}\})$ is maximized at $\lambda=0$ so the second derivative is negative.  This implies that
\begin{equation}\label{5.22}
  \bigjap{\sum_{|\alpha-\gamma|=1}|(\boldsymbol{\sigma}_\alpha-\boldsymbol{\sigma}_\gamma)\cdot(\mathbf{h}_\alpha-\mathbf{h}_\gamma)|^2}_\Lambda \le \frac{1}{2\beta} \sum_{|\alpha-\gamma|=1}|\mathbf{h}_\alpha-\mathbf{h}_\gamma|^2
\end{equation}
Adding the results for the real and imaginary part extends this inequality to complex $\mathbf{h}$.  Taking $\mathbf{h}$ to be a plane wave with a single component (and summing over possible components) proves the infrared bound and completes the proof that RP$\Rightarrow$Phase Transition.

A little about the history of this work with Fr\"{o}hlich and Spencer.  In October, 1975 I heard indirectly that J\"{u}rg and Tom had found that there was spontaneously symmetry breaking in the multi-component $\Phi^4_3$ EQFT.  The analog of infrared bounds for this case was easy.  There is a K\"{a}llen-Lehmann representation for the two point function
\begin{equation}\label{5.22a}
  \jap{\boldsymbol{\varphi(x)}\cdot\boldsymbol{\varphi(x)}} = \alpha+\int d\rho(m) \int \frac{d^3k}{(2\pi)^3}e^{ik\cdot(x-y)}(k^2+m^2)^{-1}
\end{equation}
where if $\boldsymbol{\varphi}$ has $N$ components, one has that
\begin{equation}\label{5.22b}
  \int d\rho(m) = N
\end{equation}
The infrared bound then just needs that $(k^2+m^2)^{-1}\le k^{-2}$.  Since I'd heard of this indirectly and they were looking at the field theory, I felt, perhaps unfairly, that I could think about the statistical mechanical analog.  I realized that the key was \eqref{5.22b} which followed from canonical commutation relations and I found a commutation inequality for the transfer matrix and could use that to push through a phase transition (this is the argument that appears in \cite{FSS1}).  The three of us met at the AMS meeting in San Antonio in Jan 1976 and agreed to publish our results jointly. We found the Gaussian domination approach during the writeup of the full paper.

\bigskip

It is not surprising that the work on infrared bounds generated considerable further work (479 Google scholar citations).  I will not try to describe all of it but will focus on two further developments in which I played a role. The first concerns  phase transitions in spin systems with long range interactions.  It was known for many years that finite range $1D$ systems could not have phase transitions (see, for example \cite[Theorem II.5.3]{SimonLG}).  Ruelle \cite{Ruelle1D} proved that this remains true for infinite range interactions with not too slow decay.  In particular, for the pair interacting ferromagnetic model with $J(n)=(1+|n|)^{-\alpha}$, he proved there were no phase transitions if $\alpha>2$.

On the other hand, Dyson \cite{Dyson1D}, exploiting correlation inequalities,  proved there are phase transitions if $1<\alpha<2$.  For $2D$, as we've seen Mermin-Wagner \cite{MW} proved that plane rotors had no symmetry breaking for nearest neighbor interactions but Kunz-Pfister \cite{KP2D} used Dyson's method to prove that 2D plane rotors with a similar pair interaction has a broken symmetry phase transition if $2<\alpha<4$.  Because these results depend on correlation inequalities which fail for classical Heisenberg models, the proofs do not extend to such Heisenberg models.  Indeed, Dyson conjectured but could not prove that his result was still true in the Heisenberg case.

With Fr\"{o}hlich, Israel and Lieb, I showed \cite{FILS1} infrared bounds could be proven for such long range models and, in particular, we proved Dyson's conjecture about long range order in the $1D$ classical Heisenberg model with slowly decaying pair interaction.  We called a function $J$ on $\bbZ_+$, the strictly positive integers, RP if and only if for all positive integers, $n$ and $\mathbf{z}\in\bbC^n$, one has that

\begin{equation} \lb{5.23}
  \sum_{i, j\ge 1} \bar{z}_iz_j J(i+j-1) \ge 0
\end{equation}

Then FILS show that 1D ferromagnets with $-H = {\sum_{i<j} J(j-i)\boldsymbol{\sigma}_j\cdot\boldsymbol{\sigma}_i}$ yield periodic BC RP states and infrared bounds.  \eqref{5.23} comes up already in the study of the Hamburger moment problem \cite[Section 4.17]{RA} and, using this, \cite{FILS1} easily proves that $J(n)=(1+|n|)^{-\alpha},\,\alpha>0$ is RP. The corresponding $E_p$ then has $\int E_p^{-1}\,dp<\infty$ precisely if $\alpha<2$ recovering Dyson's result and extending it to the $d$-vector model.  \cite{FILS1} is also able to recover the Kunz-Pfister result and extend it to the classical Heisenberg model.

\bigskip

The second extension concerns quantum lattice gases.  In that case the spins, $\boldsymbol{\sigma}_\alpha$ of \eqref{5.1} are non-commuting matrices, indeed quantum spins.  The basic formalism (see \cite[Section II.3]{SimonLG}) has a vector space $\bbC^{2s+1}$ for quantum spins with total spin $s$ (so $\boldsymbol{\sigma}\cdot\boldsymbol{\sigma}=s(s+1)\bdone$) for each site and the Hilbert space associated with a box $\Lambda$ is $\otimes_{\alpha\in\Lambda}\bbC^{2s+1}$ so a space of total dimension $(2s+1)|\Lambda|$.  Statistical mechanical states are defined in terms of traces.  For classical Heisenberg models on $\bbZ^\nu$ (or any bipartite lattice) the Heisenberg ferro- and antiferromagnet are equivalent under flipping every other spin.  It is critical to realize this is not true in the quantum case!   It is not even true for two spins.  If $\boldsymbol{\sigma}_j; \, j=1,2$ are two spin $\tfrac{1}{2}$ quantum spins, the lowest energy of $-\boldsymbol{\sigma}_1\cdot\boldsymbol{\sigma}_2$ is $-\tfrac{1}{4}$ (with multiplicity $3$) while the lowest energy of $\boldsymbol{\sigma}_1\cdot\boldsymbol{\sigma}_2$ is $-\tfrac{3}{4}$ (with multiplicity $1$)! In fact the ground state energy density of the ferromagnet is explicit while for the antiferromagnet, it is not, so before the work of Dyson, Lieb and Simon \cite{DLS1, DLS2} (henceforth DLS), to be discussed below, the quantum anti-ferromagnet was invariably regarded as harder than the quantum ferromagnet.  But DLS could prove

\begin{theorem} [\cite{DLS2}] \lb{T5.2}  The nearest neighbor quantum Heisenberg antiferromagnet for $s\ge 1$ and $\nu\ge 3$ and for spin $1/2$ and sufficiently large $\nu$ has a phase transition with N\'{e}el order.
\end{theorem}

\begin{remarks} 1. N\'{e}el order means that for any $\Lambda$ with even sides, one has that $\liminf\bigjap{\left(\frac{1}{|\Lambda|} \sum_{\alpha\in\Lambda}(-1)^{|\alpha|}\boldsymbol{\sigma}_\alpha\right)^2}_\Lambda>0$.

2. Kennedy, Lieb and Shastry \cite{KLS} in 1988 using a more subtle analysis filled in the missing spin $1/2$ cases when $\nu=3$.

3. While the bounds on $\beta_c$ are concrete, they involve an implicit equation which includes the ground state energy of the antiferromagnet which is not known in closed form.
\end{remarks}

There are two issues involving the quantum case viz a viz the classical case that should be mentioned.  First infrared bounds cannot hold in the form of \eqref{5.17} because they imply that as $\beta\to\infty$ that $\jap{\boldsymbol{\sigma}_\alpha\cdot\boldsymbol{\sigma}_\gamma}$ goes to a  constant (i.e. independent of $\alpha$ and $\gamma$). In the classical case, this quantity goes to $1$ uniformly in the sites (for $\Lambda$ fixed).  But in the quantum case with spin $S$, it goes to $S(S+1)$ for $\alpha=\gamma$ but only to $S^2$ for $\alpha=\gamma$ (for the maximum spin value of $\boldsymbol{\sigma}_\alpha+\boldsymbol{\sigma}_\gamma$ is $2S$ so  the maximum value of $\boldsymbol{\sigma}_\alpha\cdot\boldsymbol{\sigma}_\gamma= \tfrac{1}{2}[ (\boldsymbol{\sigma}_\alpha+\boldsymbol{\sigma}_\gamma)^2-\boldsymbol{\sigma}_\alpha^2-\boldsymbol{\sigma}_\gamma^2] = \tfrac{1}{2}[2S(2S+1)-2S(S+1)]=S^2$).  The solution is to get an initial inequality not on the thermal expectation $\jap{AB}=\tr(ABe^{-\beta H})/\tr(e^{-\beta H})$ but what DLS call the Duhamel two point function $(A,B)=\int_{0}^{1}\tr(e^{-x\beta H}Ae^{-(1-x)\beta H}B)/\tr(e^{-\beta H})$.  Since DLS prove that $(A^*,A)\ge g(A)f(c(A)/4g(A)$ where $g(A)=\tfrac{1}{2}\jap{A^*A+AA^*}$, $c(A)=\jap{[A^*,[H,A]]}$ and $f$ is the function given implicitly by $f(x\tanh x)=x^{-1}\tanh x$ and this implies a direct bound with $\coth$. Various formula involving $\coth$ occur, indeed, DLS conjecture (but do not prove) that the correct analog of \eqref{5.17} is $g_\Lambda(p)\le \sqrt{\tfrac{3}{2}}S\coth\left(\sqrt{\tfrac{2}{3}}S\beta E_p\right)$ where $g_\Lambda(p)$ is thermal expectation of $\hatt{\boldsymbol{\sigma}}_p\cdot\hatt{\boldsymbol{\sigma}}_{-p}$.

Secondly, to get infrared bounds on the Duhamel functions, one needs that the algebra of matrices on which the reflection acts in non-commutative RP to be real matrices.  Of course, the usual representation of Pauli spins is not real.  $\sigma_1$ and $\sigma_3$ are but $\sigma_2=
\bigl(\begin{smallmatrix}
  0 & i \\
  -i & 0
\end{smallmatrix}\bigr)$
is not!  Indeed, because of the commutation relations $[\sigma_1,\sigma_2]=2i\sigma_3$, there is no representation in which all spins can be real.  For the antiferromagnet, one can take $s_1=\sigma_1$, $s_2=i\sigma_2$, $s_3=\sigma_3$ and let the reflection be
\begin{equation}\label{5.24}
  \Theta((s_\alpha)_j) = \left\{
                           \begin{array}{ll}
                             -(s_{R\alpha})_j & \hbox{ , }j=1,3 \\
                             (s_{R\alpha})_j & \hbox{ , } j=2
                           \end{array}
                         \right.
\end{equation}
so that $-\boldsymbol{\sigma}_\alpha\cdot\boldsymbol{\sigma}_{R\alpha}= s_\alpha\cdot\Theta(s_\alpha)$ and so get positivity under a reflection on a real algebra for the antiferromagnet.  The corresponding infrared bound on the Duhamel two point function then reads
\begin{equation}\label{5.25}
  \left((\hatt{\boldsymbol{\sigma}}_p)^j,(\hatt{\boldsymbol{\sigma}}_{-p})^j\right) \le \frac{1}{2\widetilde{E}_p} \qquad \widetilde{E}_p=\nu+\sum_{1}^{\nu} \cos(p_j)
\end{equation}
(the sign in $\widetilde{E}_p$ is such that it vanishes at $p_j=\pi$ consistent with N\'{e}el order).  This bound leads to Theorem \ref{T5.2}.

Dyson, Lieb and I initially thought that we had a trick for getting real matrices for the ferromagnet.  One can double dimension and replace multiplication by $i$ by
$\bigl(\begin{smallmatrix}
  0 & 1 \\
  -1 & 0
\end{smallmatrix}\bigr)$
and thereby homomorphically map $n\times n$ complex matrices to a subset of all $2n\times 2n$ real matrices.  However it turns out when you do this at each site for Pauli matrices, the natural reflection no longer has reflection positivity.  Our announcement \cite{DLS1} focused on the ferromagnet and so did the preprint of \cite{DLS2}.  However, Fr\"{o}hlich was giving a course at Princeton on the work of FSS and DLS and didn't understand one step in our preprint.  He found this and came to us on the same day we'd finishing correcting the galley proofs for the longer article; indeed, after we placed the envelope with them in the outgoing departmental mailbox.  We immediately realized that there was a problem and retrieved and then fixed the galley proofs so that the published version of \cite{DLS2} is correct. Later Speer \cite{Speer} proved that reflection positivity must fail for this model. I note that now, almost 45 years later, there is no rigorous proof of the existence of a phase transition in the quantum Heisenberg ferromagnet!  It is fortunate that none of us was a young unknown when this work was done for while there is a correct very important result, the wrong result was embarrassing.  The paper \cite{DLS2} does have results on the quantum $xy$ ferromagnet where the coupling drops the $\boldsymbol{\sigma}_{\alpha,z}\boldsymbol{\sigma}_{\beta,z}$ (the $xy$ model has an abelian continuous symmetry); this is possible because there is a representation in which two of the Pauli matrices are both real.

There has been considerable literature on the quantum Heisenberg model since.  There is a lovely online bibliography on this subject posted by Kennedy-Nachtergale \cite{KenNach}.

\bigskip

Another application of RP methods involves what is called the \emph{Chessboard Peierls Method}.  Fr\"{o}hlich, Israel, Lieb and I wrote two papers \cite{FILS1, FILS2} that systematized both infrared estimates and this method and, in particular, applied the Chessboard Peierls method to a number of models.  The key is what is called \emph{Chessboard Estimates}. The name was introduced by Fr\"{o}hlich-Simon \cite{FrohSi} in a paper in the Annal of Mathematics on the structure of states in general $P(\varphi)_2$ quantum field theories.  They could not use the less fancy term ``checkerboard estimate'' because that had already been used by GRS \cite{GRS4} for a different bound.

While FS systematized the estimates and introduced the name, the idea had appeared earlier in works of Glimm-Jaffe Spencer \cite{GJS2}, Guerra\cite{Guerra2}, Seiler-Simon \cite{SeilerSi3} and Park \cite{Park1, Park2}.  Fr\"{o}hlich-Lieb (henceforth FL) \cite{FL} following up on their use in QFT by Glimm-Jaffe-Spencer and exploited these estimates with the Peierls argument to prove phase transitions in spin models and this was pushed further by FILS.

We consider a box $\Lambda$, typically with periodic BC, that can be partitioned by hyperplanes into boxes, $\{\Delta_\alpha\}_{\alpha\in Q}$, so that there are an even number of boxes in each direction and so that there is RP in each hyperplane.  Given a function $F$ of the spins in box $\Delta_\alpha$, we cover $\Lambda$ by continually reflecting $F$ in hyperplanes and let $\gamma(F,\Delta_\alpha)$, be the $|\Lambda|^{th}$ root of the $\jap{\cdot}_\Lambda$ expectation of the product of these reflected copies of $F$.

The chessboard estimate says that given functions, $F_\alpha$, of the spins in $\Delta_\alpha$, one has that
\begin{equation}\label{5.26}
   \left|\bigjap{\prod_{\alpha\in Q}\, F_\alpha}_\Lambda\right| \le \prod_{\alpha\in Q} \gamma(F_\alpha,\Delta_\alpha)
\end{equation}

If the number of edges in each direction is a power of $2$, it is easy to prove the estimate directly by multiple use of the Schwarz inequality.  In general, one uses an argument reminiscent of the proof of Gaussian domination (in fact, one can prove Gaussian domination from Chessboard Estimates).  One considers the ratio of the two sides of the Chessboard estimate as each $F_\alpha$  runs through the various $F$'s and their reflections, considers the one that maximizes it and then uses RP to prove among the maximizers is one where the ratio is $1$.

The Peierls strategy sums on contours that separate various states of the system.  A key part of the strategy is the estimation of the probability of large contours.  Those can typically be thought of expectations of products of bad events, typically one for each link in the contour.  One can use a checkerboard estimate to get upper bounds on these probabilities in terms of thermodynamic quantities and this is the Chessboard Peierls method.

In particular, FL showed this approach was effective in studying anisotropic classical Heisenberg models; they succeeded in proving a phase transition in $2D$ for arbitrarily small anisotropy.  FILS used this technique in a wide variety of models including ones with no symmetry.  In particular, FILS recovered results of the Piragov Sinai \cite{PS1, PS2} approach in different way.

\bigskip

Since this section is the only one on statistical mechanics, per se, I end it with a brief discussion of some of my other work in the subject.  First a paper \cite{SiCLQS} (and a brief report in \cite{SiCLQSA}) on the classical limit of quantum spin models.  This was motivated by a wonderful paper of Lieb \cite{LiebCLQS} who considered a classical Hamiltonian, $H_\Lambda(\{\boldsymbol{\sigma}_\alpha\}_{\alpha\in\Lambda})$, which is affine in the spins $\boldsymbol{\sigma}_\alpha\in S^3, \alpha\in\Lambda$.  The classical partition function is
\begin{equation}\label{5.29}
  Z_{cl}(\gamma) = \int \prod_{\alpha\in\Lambda}\left[d\Omega(\boldsymbol{\sigma}_\alpha)/4\pi\right] \,\exp(-H_\Lambda[\{\boldsymbol{\sigma}_\alpha\}])
\end{equation}
where $d\Omega$ is the usual unnormalized measure on the unit sphere, $S^2$, in $\bbR^3$.  For $\ell=\tfrac{1}{2},1.\tfrac{3}{2},\dots$, define
\begin{equation}\label{5.30}
  Z^\ell_Q(\gamma) = (2\ell+1)^{-|\Lambda|} \tr(\exp[-H(\gamma L_\alpha/\ell)])
\end{equation}
where $L_\alpha$ is an independent spin $\ell$ quantum spin at each site $\alpha\in\Lambda$.

Then Lieb proved that
\begin{equation}\label{5.31}
  Z_{cl}(\gamma) \le Z_Q^\ell(\gamma)\le Z_{cl}(\gamma+\ell^{-1}\gamma)
\end{equation}
Among other things, this immediately implies convergence of $Z_Q^\ell(\gamma)$ to $Z_{cl}(\gamma)$ as $\ell\to\infty$ because $Z_{cl}(\gamma)$ is continuous in $\gamma$ (indeed it is analytic).  Moreover, in situations where one knows that the infinite volume limit object (the pressure), $p_\cdot(\gamma) = \lim\tfrac{1}{|\Lambda|}Z_{\cdot,\Lambda}(\gamma)$ exists, it implies convergence of the limit objects, since while they might lose analyticity in the limit, they are convex and so continuous.

I had begun teaching a course on group representations which eventually turned into a book \cite{SimonGpRep} and it occurred to me to wonder what the analog of \eqref{5.31} was if the representations of $SU(2)$ or $SO(3)$ were replaced by a more general compact Lie group.  In particular, what classical limit space replaces $S^2$.

While it was mathematical elegance that attracted me, I had another motivation.  Dunlop-Newman \cite{DN} had proven a Lee-Yang zero theorem for $S^2$ spins by using the fact that Asano \cite{Asano} had proven one for spin $\tfrac{1}{2}$ quantum spins, the Griffiths trick \cite{GriffTrick} then gets it for spin $\ell$ quantum spins and the limit theorem implied by \eqref{5.31} then implies one for $S^2$ spins.  It was natural to worry about spins on $S^{d-1}$ (i.e. $d$-component rotors).

Since I eventually showed the classical limit spaces are symplectic manifolds, $S^N$ is never a classical limit if $N\ge 3$ but it turns out it is a quotient of one.  I reduced proving Lee-Yang for $S^{d-1}$ for all $d$ to proving a conjectured Asano type result for spin $\tfrac{1}{2}\,SO(2k)$ spinors.  To this day, not only is that conjecture still open but so is whether Lee-Yang holds for $S^{d-1}$ spins with $d\ge 4$!

To avoid going too far afield, I'll describe the main results of \cite{SiCLQS} assuming a knowledge of the theory of representations of compact Lie groups as described in Simon \cite{SimonGpRep}, Adams \cite{AdamsGpRep} or Fulton-Harris \cite{FHGpRep}.  One fixes a fundamental weight, $\lambda$, on a compact Lie group, $G$, and for each $L=1,2,\dots$, one considers the irreducible representation, $U_L$, on $\calH_L$, with maximal weight $L\lambda$.  By picking a basis in the Lie algebra, $\frg$, of $G$, one considers Hamiltonians multilinear in the basis vectors at the various sites in a box $\Lambda$.  If $d_L=\dim(\calH_L)$, then
\begin{equation}\label{5.32}
  Z_Q^L(\gamma) = d_L^{-|\Lambda|}\tr(\exp[-H(\gamma S_\alpha/L)])
\end{equation}

Extend $\lambda$ to $\frg^*$, the dual of $\frg$ by setting it to zero on the orthogonal complement of the Cartan subalgebra.  Under the dual of the adjoint action of $G$ on $\frg$, one gets a manifold by looking at the orbit of this extension of $\lambda$.  These coadjoint orbits, $\Gamma_\lambda$, which also play a role in the Kirilov \cite{Kir} theory of representations of nilpotent Lie groups and in the closely related Kostant \cite{Kost}-Souriau \cite{Sou} method of geometric quantization, are the classical limits.  Haar measure on $G$ induces a measure on $\Gamma_\lambda$ and one uses this to define a suitable classical limit partition function $Z_{cl}(\gamma)$.  By using coherent vectors based on the maximal weight vectors and the same Berezin-Lieb inequalities that Lieb did, I could extend his result to this case.

There is a magic weight, $\delta$, which is the sum of all the fundamental weights.  Let $a=2\jap{\lambda,\delta}/\jap{\lambda,\lambda}$ where $\jap{\cdot,\cdot}$ is the Killing inner product on the weight space.  Then I extend \eqref{5.31}to
\begin{equation}\label{5.33}
  Z_{cl}(\gamma) \le Z_Q^L(\gamma) \le Z_{cl}(\gamma+aL^{-1}\gamma)
\end{equation}

$SU(2)$ is rank 1, so there is a single fundamental weight and $\delta=\lambda$ so $a=2$.  Moreover, $L=1$ corresponds to $\ell=\tfrac{1}{2}$, so $\ell=\tfrac{1}{2}L$. Thus, \eqref{5.33} in this case is just \eqref{5.31}.  One surprise of this analysis is that there are several distinct classical limit spaces if the rank is $2$ or more.  For example, for $SO(4)$, the space for the limit of spherical harmonics is the $4$-d space $S^2\times S^2$ while for the spinor representations, it is the $2$-d space $S^2\cup S^2$.

\bigskip

At Princeton, I ran a ``brown bag seminar'' which included brief presentations about current research both on one's own work and work of others.  There were typically about 25 participants that often included all the senior math physics faculty at Princeton (Lieb, Nelson, Wightman and me) and Dyson from the Institute as well as our wonderful group of postdocs/junior faculty/grad students (see \cite{IAMPPrinceton} for a complete list but included were Aizenman, Avron, Fr\"{o}hlich, Deift and Sigal).  In the fall of 1979, Michael Aizenman came back from a conference in Hungary and, at a brown bag, reported on some work of Dobrushin-Pecherski (a small part of \cite{DP}) that showed sufficiently fast power decay of correlations in spin systems implied exponential decay.  In trying to understand why this might be, I proved the following:

\begin{theorem} [\cite{SimonLS,SimonLSAnon}] \lb{T5.3} Let $\jap{\sigma_\alpha \sigma_\gamma}$ denote the two point function of a spin $\tfrac{1}{2}$ nearest neighbor (infinite volume, free boundary condition) Ising ferromagnet at some fixed temperature. Fix $\alpha,\gamma$ and $B$, a set of spins whose removal breaks the lattice in such a way that $\alpha$ and $\gamma$ lie in distinct components. Then:
\begin{equation}\label{5.34}
  \jap{\sigma_\alpha \sigma_\gamma} \le \sum_{\delta\in B} \jap{\sigma_\alpha \sigma_\delta}\jap{\sigma_\delta \sigma_\gamma}
\end{equation}
\end{theorem}

\begin{remarks} 1. One consequence of this is that if the lattice is $\bbZ^\nu$, then if $\jap{\sigma_\alpha \sigma_\gamma}\le C|\alpha-\gamma|^{-\mu}$ with $\mu> \nu-1$, then for some $C_1$ and $m>0$, one has that $\jap{\sigma_\alpha \sigma_\gamma}\le C_1 e^{-m|\alpha-\gamma|}$.

2. I talked about this at a later brown bag which stimulated additional work: Lieb found an improvement and Aizenman and I found a version for multicomponent models.  We arranged for these three papers and one by Rivasseau \cite{Riva} to appear successively in CMP.  Lieb's improved result \cite{LiebLS} involved the component $\Lambda$ of $\bbZ^\nu\setminus B$ with $\alpha\in\Lambda$ and allowed $\jap{\sigma_\alpha \sigma_\delta}$ in \eqref{5.34} to be replaced by $\jap{\sigma_\alpha \sigma_\delta}_{B\cup\Lambda}$, the expectation with interactions outside $B\cup\Lambda$ dropped.  This included Griffiths third inequality \cite{G3onGKS}. These geometric correlation inequalities are sometimes called Lieb-Simon inequalities as a result.

3. As mentioned, Aizenman and I \cite{AizSiLS} proved a version for $d$-vector models. Rivasseau \cite{Riva} extended Lieb's improved inequality to $d=2$ models.

4. Related inequalities appeared earlier in work of Kasteleyn-Boel \cite{KB}.
\end{remarks}

\bigskip

Among some of my other results on lattice gases are

1. A work with Sokal \cite{SiSok} which made rigorous an argument of Thouless \cite{ThouEE} exploiting energy-entropy estimates, that, for example, provided another proof of the result of Ruelle \cite{Ruelle1D} that a pair of spin $\tfrac{1}{2}$ Ising ferromagnets whose coupling obeys $\sum n|J(n)|< \infty$ has zero spontaneous magnetization.

2. A paper \cite{SimonCritical1D} on the one dimensional $d$-rotor model with critical $J(n)=|n|^{-2}$ (for $n\ge 1$).  As mentioned above, for $J(n)=|n|^{-\alpha}$ Dyson showed phase transitions for Ising spins if $1<\alpha<2$, FILS proved phase transitions for $d$-rotor models with $1<\alpha<2$ and Ruelle proved no phase transition if $\alpha>2$.  The case $\alpha=2$ is borderline.  Fr\"{o}hlich-Spencer \cite{FSCritical1D} proved Ising models with this borderline $\alpha$ have discrete symmetry breaking.  In \cite{SimonCritical1D}, I proved $d$-rotor models with $d\ge 2$ and this critical coupling do \emph{not} have continuous symmetry breaking.

3. A note with Aizenman \cite{AizSiRotor} comparing Ising and plane rotors that gives a lower bound on the Berezinskii-Kosterlitz-Thouless transition temperature.

4.  A note \cite{SiHTDecay} showing that the rather complicated directional dependence of the high temperature decay in the Ising model is explained by leading order perturbation theory.

5. A note \cite{SimonUBMF} showing that for $d$-rotor models, mean field theory provides upper bounds on transition temperatures.

6. A book \cite{SimonLG} that discusses the lattice models that have been at the center of this section.  It focuses on formalism and does not discuss correlation inequalities, Lee-Yang, the Peierls argument and infrared bounds, some of the most fascinating aspects of the subject.  I have a book on those aspects of this subject in preparation \cite{SimonLG2}.

\section{$N$--Body quantum mechanics} \lb{s6}

The previous sections have focused on one or two problems, all (but the last section) within a limited area of mathematical physics.  This section is much more diffuse dealing with general $N$-body NRQM, so I will leave more background to references and only briefly discuss a lot of work.  In particular, by thinking of $2$ as a possible value of $N$, I'll throw in some subjects that are not usually considered $N$-body QM like some inverse potential scattering and even a little bit of general $1D$ Schr\"{o}dinger operators.

A full $N$-body Hamiltonian acts on $L^2(\bbR^{\nu N})$ where $\boldsymbol{x}\in\bbR^{\nu N}$ is written $\boldsymbol{x}=(\boldsymbol{r}_1,\dots,\boldsymbol{r}_N)$ with $\boldsymbol{r}_j\in\bbR^\nu$.  We write
\begin{equation}\label{6.1}
  \wti{H}_0 = -\sum_{j=1}^{N}(2m_j)^{-1}\Delta_{\boldsymbol{r}_j},\quad\wti{V}=\sum_{1\le i<j\le N}V_{ij}(\boldsymbol{r}_i-\boldsymbol{r}_j),\quad\wti{H}=\wti{H}_0+\wti{V}
\end{equation}
Then a basic preliminary is
\begin{theorem} \lb{T6.1} In any coordinate system, $\boldsymbol{\rho}_1,\dots,\boldsymbol{\rho}_N$ where $\boldsymbol{\rho}_j,\,j=1,\dots,N-1$ is a linear combination of $\boldsymbol{r}_k-\boldsymbol{r}_\ell$ and (with $M=\sum_{j=1}^{N}m_j$)
\begin{equation}\label{6.2}
  \boldsymbol{\rho}_N=\frac{1}{M}\sum_{j=1}^{N} m_j\boldsymbol{r}_j
\end{equation}
we have that, realizing $\wti{\calH} \equiv L^2(\bbR^{\nu N})=L^2(\bbR^\nu)\otimes L^2(\bbR^{\nu (N-1)})\equiv \calH_{CM}\otimes\calH$, where the first factor is functions of $\rho_N$ and the second functions of $\{\rho_j\}_{j=1}^{N-1}$,
\begin{equation}\label{6.3}
  \wti{H}_0 = h_0\otimes\bdone + \bdone\otimes H_0
\end{equation}
\begin{equation}\label{6.4}
  \wti{H}=h_0\otimes\bdone + \bdone\otimes H
\end{equation}
where $h_0=-(2M)^{-1}\Delta_{\boldsymbol{\rho}_N}$, $H_0$ is a positive quadratic form in $-i\boldsymbol{\nabla}_{\boldsymbol{\rho}_j}, \, j=1,\dots,N-1$ and $H=H_0+V$.
\end{theorem}

I refer the reader to \cite[Section 11]{SimonKato} for a discussion of various coordinate systems and the formalism of Sigalov--Sigal \cite{SigSig} (see also Hunziker--Sigal \cite{HunzSig}).  In that formalism, a major role is played by the inner product
\begin{equation}\label{6.5}
  \jap{r^{(1)},r^{(2)}} = \sum_{j=1}^{N} m_j \boldsymbol{r^{(1)}}_j \cdot \boldsymbol{r^{(2)}}_j
\end{equation}
In this inner product, $\wti{H}_0$ is the Laplace-Beltrami operator and the reason that \eqref{6.3} holds is that $\rho_N$ is orthogonal to the other $\rho_j$'s.  One coordinate system that we'll need soon is atomic coordinates where
\begin{equation}\label{6.6}
     \boldsymbol{\rho}_j = \boldsymbol{r}_j-\boldsymbol{r}_N,\, j=1,\dots,N-1; \qquad   \boldsymbol{\rho}_N=\frac{1}{M}\sum_{j=1}^{N} m_j\boldsymbol{r}_j
\end{equation}
In this coordinate system when $m_1=m_2=\dots=m_{N-1}=m$ and $\frac{1}{m}+\frac{1}{m_N} \equiv \frac{1}{\mu}$, one has that (see \cite[(11.48)]{SimonKato} for the calculation)
\begin{equation} \lb{6.7}
  H_0 = -\sum_{j=1}^{N-1} \frac{1}{2\mu}\Delta_j - \frac{1}{m_N} \sum_{j<k} \boldsymbol{\nabla}_j\cdot\boldsymbol{\nabla}_k
\end{equation}

One normally studies $H$, \emph{the $N$-body Hamiltonian with center of mass removed}, aka \emph{reduced $N$-body Hamiltonian}.  If one takes $m_N$ to infinity, the annoying extra last term in \eqref{6.7} known as the \emph{Hughes-Eckert} term is gone.  This extra term is present because in the inner product \eqref{6.5}, the $\rho_j$ of \eqref{6.6} are not mutually orthogonal.  For this reason, it is often convenient to use coordinate systems like Jacobi coordinates \cite[Example 11.6]{SimonKato}.

One last piece of kinematics we need is the notion of  a \emph{cluster decomposition} or clustering, $\calC = \{C_\ell\}_{\ell=1}^k$, which is a partition, i.e. a family of disjoint subsets whose union is $\{1,\dots,N\}$. We use $\bbP$ for the set of all non-trivial clusterings, i.e. those with $\ell \ge 2$.  We set $\#(C_\ell)$ to be the number of particles in $C_\ell$.  A coordinate, $\boldsymbol{\rho}$, is said to be internal to $C_\ell$ if it is a function only of $\{\boldsymbol{r}_m\}_{m \in C_\ell}$ which is invariant under $\boldsymbol{r}_m \to \boldsymbol{r}_m+\boldsymbol{a}$ (all $m\in C_\ell$), equivalently, it is a linear combination of $\{\boldsymbol{r}_m-\boldsymbol{r}_q\}_{m,q \in C_\ell}$.  A \emph{clustered Jacobi coordinate system} is a set of $\#(C_\ell)-1$ independent internal coordinates for each cluster together with $\boldsymbol{R}_\ell = (\sum_{q \in C_\ell} m_q\boldsymbol{r}_q)/(\sum_{q \in C_\ell} m_q)$,  If we write $\calH(C_\ell)$ to be $L^2$ of the internal coordinates  of cluster $C_\ell$ and $\calH^{(\calC)}$ to be $L^2$ of all the centers of mass of the clusters then
\begin{equation}\label{6.8}
  \wti{\calH} = \wti{\calH}^{(\calC)} \otimes \bigotimes_{\ell=1}^k \calH(C_\ell)
\end{equation}
\begin{equation}\label{6.9}
  \wti{H}_0 = \wti{H}_0^{(\calC)}\otimes\bdone\dots\otimes\bdone+ \sum_{\ell=1}^{k} \bdone\otimes\dots\otimes H_0(C_\ell) \otimes\dots\otimes\bdone
\end{equation}
where $\wti{H}_0^{(\calC)} = -\sum_{\ell=1}^{k} (2M(C_\ell))^{-1}\Delta_{\boldsymbol{R}_\ell}$ and $H_0(C_\ell)$ is a quadratic form in the derivatives of the internal coordinates.

In \eqref{6.9}, the operator $\wti{H}_0^{(\calC)}$ has a decomposition like \eqref{6.3} where $\calH$ is replaced by $\calH^{(\calC)}$, the functions of the differences of the centers of mass of the $C_j$.  We write
\begin{equation}\label{6.10}
  \wti{H}_0^{(\calC)} = h_0\otimes\bdone+\bdone\otimes H_0^{(\calC)}
\end{equation}

Given a cluster decomposition, $\calC=\{C_\ell\}_{\ell=1}^k$, we write $(jq) \subset \calC$ if $j$ and $q$ are in the same cluster of $\calC$ and $(jq) \not\subset \calC$ if they are in different clusters.  We define
\begin{align}
  V(C_\ell)  &= \sum_{\substack{j,q\in C_\ell \\ j<q }} V_{jq} \lb{6.11} \\
  V(\calC)   &= \sum_{\ell=1}^{k} V(C_\ell) = \sum_{\substack{ (jq) \subset \calC \\ j<q}} V_{jq} \lb{6.12} \\
  I(\calC)   &= \sum_{j<q} V_{jq} - V(\calC) = \sum_{\substack{ (jq) \not\subset \calC \\ j<q}} V_{jq} \lb{6.13}
\end{align}
$V(\calC)$ is the \emph{intracluster interaction} and $I(\calC)$ \emph{the intercluster interaction}.  We define on $\calH(C_\ell)$
\begin{equation}\label{6.14}
  H(C_\ell) = H_0(C_\ell) + V(C_\ell)
\end{equation}
\begin{align}
  H(\calC)  &= H_0^{(\calC)}\otimes\bdone\dots\otimes\bdone+\sum_{\ell=1}^{k} \bdone\otimes\dots\otimes H(C_\ell)\otimes\dots\otimes\bdone \lb{6.15} \\
            &= H-I(\calC) \nonumber
\end{align}

\begin{equation}\label{6.16}
   \Sigma(\calC) = \sum_{\ell=1}^{k} \inf\sigma(H(C_\ell))
\end{equation}

We note that
\begin{equation}\label{6.17}
    \calC \in \bbP \Rightarrow \sigma(H_0^{(\calC)}) = [0,\infty)
\end{equation}

By \eqref{6.15}, we have that (where $\sigma(\cdot)$ is the spectrum)  $\sigma(H(\calC))=\sigma(H_0^{(\calC)})+\sigma(H(C_1))+\dots+\sigma(H(C_k))$.  By \eqref{6.17}
\begin{equation}\label{6.18}
  \calC \in \bbP \Rightarrow\sigma(H(\calC)) = [\Sigma(\calC),\infty)
\end{equation}
fir some $\Sigma(\calC)$. When I discuss $N$--body spectral and scattering theory below, I'll be interested in thresholds.  A \emph{threshold}, $t$, is a decomposition $\calC=\{C_\ell\}_{\ell=1}^k \in\bbP$ and an eigenvalue, $E_\ell$ of $H(C_\ell)$ for each $\ell=1.\dots,k$.  The \emph{threshold energy} is $E(t)=\sum_{\ell=1}^{k} E_\ell$.  Of course, $E(t) \ge \Sigma(\calC)$.

With these preliminaries in hand, I can describe the central mathematical questions in the analysis of $N$-body NRQM.  I assume that the reader is familiar with the basic notions of self-adjointness of unbounded operators, the spectral theorem for them \cite[Chapters 5 and 7]{OT} and the spectral decompositions into discrete and essential spectrum and into absolutely continuous, singular continuous and point spectrum (see \cite[Theorem 5.1.12]{OT}).  One always supposes that the two body potentials, $V_{ij}$, go to zero at infinity, usually faster than $r^{-1-\varepsilon}$.

(1) \textbf{The self-adjointness of $H$}.

(2) \textbf{The determination of the essential spectrum of $H$}.  This is solved by the celebrated (see \cite[Section 3.3]{CFKS} for a proof and references to the original papers of Hunziker, van Winter and Zhislin whose proofs are very different from the proof in \cite{CFKS}).

\begin{theorem} [HVZ Theorem] \lb{T6.2}   For reduced $N$ body Hamiltonians with two body potentials vanishing at infinity, one has that
\begin{equation}\label{6.19}
   \sigma_{ess}(H) = [\Sigma,\infty) \qquad \Sigma = \inf_{\calC\in\bbP} \Sigma(\calC)
\end{equation}
\end{theorem}

(3) \textbf{Absence of singular continuous spectrum for $H$}. My advisor, Arthur Wightman had a colorful name for this: ``the no goo hypothesis''.  His point was that a.c. spectrum had an interpretation as scattering states and point spectrum as bound states.  If there were singular continuous spectrum, it would have to be goo.  Connected to this is that point spectrum should only have limit points at thresholds; one might expect no embedded point spectrum but the examples discussed at the start of Section 2 show that is too simple minded although one might like to prove the absence of positive energy eigenvalues and thresholds.

(4) \textbf{Asymptotic Completeness}.  To describe this, we need some additional preliminaries.  Let $t$ be a threshold and $\calC$ the associated cluster decomposition.  Under the decomposition \eqref{6.8}, we let $\calH_t$ be all states of the form $\varphi\otimes\eta$ where $\varphi$ is an arbitrary vector in $\wti{\calH}^{(\calC)}$ (i.e. function of the differences of centers of mass of the clusters) and $\eta$ a sum of products of eigenvectors of $H(C_\ell)$ with eigenvalue $E_\ell$.  Let $P_t$ be the projection onto $\calH_t$.  In 1959, Hack \cite{Hack} proved that, for each threshold the limits (the funny convention that has $\Omega^\pm$ associated to limits as $t\to\mp\infty$ comes from the physics literature where $\Omega^\pm$ defined this way is connected in time independent scattering to $\lim_{\pm\varepsilon\downarrow 0} (H-E-i\varepsilon)^{-1}$).
\begin{equation}\label{6.20}
  \Omega^\pm_t=s-\lim_{t\to\mp\infty} e^{itH}e^{-itH(\calC)}P_t
\end{equation}
exists so long as the two body potentials decay faster that $r^{-1-\varepsilon}$ (for longer range, including the physically important Coulomb case, following Dollard \cite{Dollard}, one needs to use modified wave operators - see, for example \cite[Section XI.9]{RS3} - we'll refer to this case below without further technical detail).  These are the \emph{cluster wave operators}.  If $\psi = \Omega^-_t\gamma$, then as $t\to\infty$, we have that $e^{-itH}\psi$ looks like bound clusters of $C_\ell$ in eigenstates with energy $E_\ell$ moving freely relative to each other, i.e. intuitively scattering states.  One can show that for distinct thresholds, $t\ne s$, one has that $\ran\,\Omega^-_t$ is orthogonal to $\ran\,\Omega^-_s$.  Asymptotic completeness is the assertion that
\begin{equation}\label{6.21}
  \bigoplus_{\text{all thresholds }t}\ran\,\Omega^+_t = \bigoplus_{\text{all thresholds }t}\ran\,\Omega^-_t = \calH_{ac}(H)
\end{equation}
where $\calH_{ac}(H)$ is the space of all vectors whose spectral measures for $H$ is purely absolutely continuous.

For each of these four, I made significant, albeit not the definitive, contributions as I'll describe soon.  Kato (see \cite[Sections 7-10]{SimonKato}) was both the pioneer and continuing master of the self-adjointness problem but I made a basic discovery on allowed local positive singularities and followed up on Kato's work on what I called Kato's Inequality.   I not only named the HVZ theorem (where Hunziker, van Winter and Zhislin were the initiators) but reworked and extended it twice, including my work with Last on the ultimate HVZ theorem.  Perry, Sigal and I were the first to prove the absence of singular continuous spectrum for fairly general $N$-body operators although we relied heavily on ideas of Mourre (and Balslev-Combes had earlier handled suitable analytic potentials including the important Coulomb case).  Sigal-Soffer were the first to establish $N$-body asymptotic completeness but they (and later, others) relied in part on my work with Deift which reduced the problem to the existence of what are now called Deift-Simon wave operators.

In the remainder of this section, I'll discuss in more detail these and other works on $N$-body and related problems.  My earliest paper on general N-body systems is Simon \cite{SimonInfinite} which proved that general atoms and positive ions have infinitely many eigenvalues (aka bound states) below their continuous spectrum.  This is really a remark on a paper of Kato \cite{KatoHe}, written 20 years earlier.  In that paper, Kato showed that Helium in the approximation of infinite nuclear mass had infinitely many eigenvalues and with its physical mass has at least 25,585 eigenvalues (counting multiplicity). Kato could not go beyond Helium and had the 25,585 limitation because he only used crude methods to estimate the bottom of the continuous spectrum.  The basic point of \cite{SimonInfinite} is that since Kato's work, Hunziker \cite{HunzHVZ} had proven Theorem \ref{T6.2} above and that by using that, it was not difficult to exploit the method of Kato in \cite{KatoHe} to get the very general result.  I should mention that ten years earlier, Zhislin \cite{ZhislinInfinite} using more involved methods had proven this general result, so my proof was new, but the result was not.

I knew about Hunziker's paper \cite{HunzHVZ} because he had done the work while a postdoc at Princeton and it was something that my advisor Arthur Wightman discussed in his course.  To understand the basis of Hunziker's proof, it pays to recall the essence of one argument for the reduced $2$-body case: if $H_0=-\Delta,\,H=H_0+V$ and $V$ goes to zero at infinity, then $\sigma_{ess}(H)=[0,\infty)$.  One writes down the second resolvent equation:
\begin{align}
  (H-z)^{-1} &= (H_0-z)^{-1} - (H-z)^{-1}V(H_0-z)^{-1}\Rightarrow \nonumber \\
             &(H-z)^{-1} = (H_0-z)^{-1}\left[1+V(H_0-z)^{-1}\right]^{-1} \label{6.22}
\end{align}
Since $V$ goes to zero, $z\mapsto V(H_0-z)^{-1}$ is a compact analytic function on $\bbC\setminus [0,\infty)$, so, by the analytic Fredholm Theorem (\cite[Theorem VI.4]{RS1},
\cite[Theorem 3.14.3]{OT}), $\left[1+V(H_0-z)^{-1}\right]^{-1}$ is meromorphic on $\bbC\setminus [0,\infty)$ with finite rank residues.  This implies the claimed result on $\sigma_{ess}(H)$.  For $N>2$, Hunziker instead used the fact that Weinberg \cite{Weinberg} and van Winter \cite{vW}, essentially by resumming perturbation theory, proved the Weinberg-van Winter equations
\begin{equation} \lb{6.23}
  (H-z)^{-1} = D(z) + (H-z)^{-1} I(z)
\end{equation}
where $D(z)$ and $I(z)$ are built out of the potentials and the resolvents of the $H(\calC)$ and so analytic in $\bbC\setminus [\Sigma,\infty)$.  Moreover, Hunziker \cite{HunzCpt} proved that $I(z)$ was compact (this was proven by Weinberg when $N=3$ and conjectured in general) so, as in the reduced $2$-body case, one gets the full $N$ body result.

Over the next few years, I became aware that in \cite{vW}, van Winter (who like Hunziker specifically looked at $\nu=3$ but further restricted to $V_{ij}\in L^2(\bbR^3)$ so she could use Hilbert Schmidt rather than just compact operators) implicitly had Theorem \ref{T6.2} when her conditions hold by a method close to Hunziker's.  Moreover, Zhislin \cite{ZhislinInfinite} had the result for atoms using very different, geometric methods, but his method, as explicated by J\"{o}rgens-Weidmann \cite{JW}, could also obtain Theorem \ref{T6.2}.  Thus, by the time of Reed-Simon, vol. 4 \cite{RS4}, I had decided to call the result the HVZ theorem, a name which stuck.

I provided two generations of improvements in this result and its proof.  In Simon \cite{SimonHVZ}, motivated in part by work on Deift-Simon wave operators (see below), I found a geometric way of understanding the theorem (independently, Enss \cite{EnssHVZ}, at about the same time, found a proof similar in spirit).  Given a partition, $\calC$, one defines
\begin{equation}\label{6.24}
  |\boldsymbol{r}|_\calC = \min_{(jq)\not\subset\calC}\{|\boldsymbol{r}_j-\boldsymbol{r}_q|\}\quad |\boldsymbol{r}| = \max_{j\ne q}\{|\boldsymbol{r}_j-\boldsymbol{r_q}|\}
\end{equation}
One constructs a $C^\infty$ partition of unity (i.e. $\sum_{\calC\in\bbP} j_\calC = 1,\,j_\calC \ge 0$) so that for some $d_N>0$, $j_\calC$ is supported on $\{\boldsymbol{r}\,|\,|\boldsymbol{r}| \le 1\}\cup\{\boldsymbol{r}\,|\,|\boldsymbol{r}|_\calC \ge d_N |\boldsymbol{r}|\}$.  \cite{CFKS} call this a Ruelle-Simon partition of unity since Ruelle constructed such partitions in his work on QFT scattering theory.  I showed easily that if $f$ is a continuous function of compact support, $\left[f(H)-f(H(\calC))\right]j_\calC$ is compact.  This is because $I(\calC) j_\calC$ decays in all directions.  One then writes
\begin{equation}\label{6.25}
  f(H) = \sum_{\calC\in\bbP} \left[f(H)-f(H(\calC))\right]j_\calC + \sum_{\calC\in\bbP} f(H(\calC))j_\calC
\end{equation}
to conclude that if $f$ is supported on $(-\infty,\Sigma)$, then $f(H)$ is compact which implies Theorem \ref{T6.2}.

Many years later, in 2006, Last and I \cite{LSHVZ} returned to this subject in a much more general context.  I'll state our result for Schr\"{o}dinger operators, $-\Delta+V$, on $L^2(\bbR^{\gamma})$ when $V$ is uniformly continuous (which is true for $N$-body systems if all $V_{ij}$ are continuous and go to zero at infinity). We need the notion of limit at infinity. By the Arzel\`{a}-Ascoli Theorem \cite[Theorem 2.3.14]{RA}, $V(\cdot+y)$ restricted to large balls lies in a compact set as $y$ varies through $\bbR^\gamma$.  It follows that for any $y_m$ going to infinity, there is a subsequence $y_{m_j}$ so that $V(\cdot+y_{m_j})$ converges to some $W$ uniformly on compact subsets of $\bbR^\gamma$.  If $y_{m_j}/|y_{m_j}| \to e\in S^{\gamma-1}$, we say that $W$ is a limit of $V$ at infinity in direction $e$.  We let $\calL_e$ be the set of such $W$.  By the compactness noted above, each $\calL_e$ is non-empty.  Here is what Last and I \cite{LSHVZ} proved.

\begin{theorem} \lb{T6.3} For any Schr\"{o}dinger operators, $-\Delta+V$, on $L^2(\bbR^{\gamma})$ with $V$ that is uniformly continuous, one has that
\begin{equation}\label{6.26}
  \sigma_{ess}(-\Delta+V) = \bigcup_{e\in S^{\gamma-1}}\bigcup_{W\in\calL_e} \sigma(-\Delta+W)
\end{equation}
\end{theorem}

\begin{remarks} 1. \cite{LSHVZ} also has results for Schr\"{o}dinger operators where $V$ is allowed to have local singularities (stated in terms of uniformly local Kato class) and for Jacobi matrices, CMV matrices and, as I'll mention in the next section, for Schr\"{o}dinger operators with magnetic field.

2. The proof is really quite simple based on Weyl sequences \cite[Problem 3.14.5]{OT}, i.e. if $A$ is self-adjoint, then $\lambda\in\sigma_{ess}(A)$ if and only if there exists a sequence of unit vectors, $\{\varphi_n\}$ going weakly to zero with $\norm{(A-\lambda)\varphi_n}\to 0$.  We used localization ideas going back to Sigal \cite{IMSS} and G\aa rding \cite{Garding} to show one could pick the Weyl sequence to live in a large ball (of $n$ independent size) and then compactness to get a trial sequence for a limit at infinity.

3. Earlier in \cite{LSEigen}, Last and I had introduced the notion of right limit for Schr\"{o}dinger operators on the half line and proven that in that case the right side of \eqref{6.26} is a subset of the left side.  Limits at infinity generalize the notion of right limit.  \cite{LSEigen} also have results relating right limits and a.c. spectrum which were generalized in a beautiful and spectacular way by Remling \cite{RemlingRL}.  The work of Last-Simon \cite{LSHVZ} and Remling \cite{RemlingRL} is presented in \cite[Chapter 7]{SimonSz}.  By exploiting analogy, Breuer and I \cite{BrSiNB} used Remling's idea as an organizing tool in understanding an issue in classical complex analysis: which power series lead to natural boundaries on their disk of convergence.

4. Forms of \eqref{6.26} seem to have been in the air after 2000.  As discussed in Last-Simon \cite{LSHVZ} and \cite[Chapter 7]{SimonSz} (where references can be found) several other groups from very different communities found variants of \eqref{6.26}.  Their proofs used much more machinery than \cite{LSHVZ}.  In particular, \cite{LSHVZ} required the closure of the set of the right side of \eqref{6.26} but using ideas of Georgescu-Iftimovici \cite{GeorgeI}, one can show that the set is closed.

5. \eqref{6.26} implies the HVZ theorem (if the $V_{ij}$ are continuous and going to zero; using the extension mentioned in Remark 1, one can get the full HVZ result).  Given $e\in S^{\nu N-1}$, one defines $\calC(e)$ by putting $i$ and $j$ in the same cluster if and only if $e_i=e_j$.  It is immediate that the only right limit in $\calL_e$ is $H(\calC(e))$.  \cite{LSHVZ} also have an interesting result on approach to a periodic isospectral torus.
\end{remarks}

Next, I turn to my contributions to the questions of self-adjointness of Schr\"{o}dinger operators and the more general issue of the proper definition of self-adjoint quantum Hamiltonians.  In this regard, I should mention my work on defining these operators by the method of quadratic forms beginning with my PhD. thesis which was published as a book \cite{SimonThesis}.  This thesis studied $-\Delta+V$ on $L^2(\bbR^3)$ for $V$'s obeying
\begin{equation}\label{6.27}
  (4\pi)^{-2}\int\int \frac{|V(x)||V(y)|}{|x-y|^2} < \infty
\end{equation}
a class that I called the \emph{Rollnik class}, $R$, after \cite{Rollnik}.  Since the left side \eqref{6.27} is the square of the Hilbert-Schmidt norm of $|V|^{1/2}(-\Delta)^{-1}|V|^{1/2}$, it was rediscovered earlier than my work by many others.  In particular, Birman \cite{BirmanBS} and Schwinger \cite{SchwBS} used it in their work (mentioned in Section \ref{s8}) on bounds on the number of bound states and Grossman-Wu \cite{GW} used it in a study of two body scattering theory.

The thesis had an interesting source. Wightman was on leave in my third year of graduate school (1968-69).  When he left I didn't have a definite thesis problem although it seemed possible I'd do a thesis on the work I was doing on the anharmonic oscillator. George Tiktopoulos was a High Energy Theory Postdoc at Princeton (later a Professor in Athens) and gave a topics course in potential scattering which, while not mathematically precise, was more mathematically careful than many of the other High Energy Theorists.  He developed things for $H \equiv -\Delta+V$ for $V\in R\cap L^1$.  Such $V$'s were not necessarily locally $L^2$ so Kato's theorem didn't apply and you couldn't define $H$ as an operator sum.  I kept complaining, sometimes being a little obnoxious as smart graduate students can be, that he needed to add the condition $V\in L_{loc}^2$ to be able to use Kato's theorem (Grossman-Wu had done exactly this).  He was insistent that because he could define a Green's function for $(H-E)^{-1}$ for $E$ very negative via a convergent perturbation series, there must not be a problem.  Moreover the physics should work for potentials with a $|x|^{-\alpha}$ local singularities so long as $\alpha<2$.  $L^2_{loc}$ though requires $\alpha < 3/2$ while (for local singularities), $R$ works up to $2$.

I eventually realized that Tiktopoulos was right and one could do everything using quadratic forms and I wrote a long thesis where, among other things, I rigorously discussed scattering theory through the proof of dispersion relations and the HVZ theorem.  Wightman liked it so much that he proposed making it a volume in the book series he edited for Princeton Press.  I started an instructorship in September, 1969 with the thesis largely written, but Wightman asked me to hold off submission of the formal thesis until he had a chance to carefully read it and make suggestions.  Since I had a job (in those days, Universities weren't as picky about postdocs without being actual docs yet) and he (and I) were busy, submission kept being postponed.  The math and physics departments proposed promoting me to Assistant Professor and the Dean was very unhappy when he learned that I didn't officially have a degree and refused to process the appointment until I did.  Bob Dicke, the chair in Physics made it clear to Wightman he'd better deal rapidly with the roadblock and suddenly within a weekend, he'd read my entire thesis!

I wasn't the first one to use quadratic form methods to define quantum Hamiltonians.  From the earliest days of quantum mechanics, mathematicians had the idea to use the Friedrichs' extension which is essentially a quadratic form construction \cite[Section 7.5]{OT}.  The perturbation approach that I used had been used in a related context already in Kato's book \cite{KatoBk} and by Nelson \cite{NelsonQF} from whom I learned it.  Forms were also used by several of Nelson's and Wightman's students slightly before me.  What I did was show that large parts of the then existing theory could be carried over to a quadratic form point of view.  Afterwards and, in part, because of my work, forms became a tool more widely used by mathematical physics studying NRQM.

I returned to form ideas, essentially as a simplifying tool, many other times later in my career.  On the purely mathematical side, I wrote two papers on the subject of monotone convergence theorems for forms, an area where the first results appeared in the first edition of Kato's book \cite{KatoBk}.  Both papers \cite{SimonLSCForms, SimonMonotoneForms} resolved an issue left open by Kato in the case of monotone increasing forms; independently, Davies \cite{DaviesMonotoneForms}, Kato (in the revised second edition of his book) and Robinson \cite{Robinson} had also settled this issue.  In \cite{SimonMonotoneForms}, I found a decomposition analogous to the Lebesgue decomposition of measures which allowed a significant improvement of the  result for monotone decreasing forms.  Also on the mathematical side, Alonso and I \cite{AlonSi} wrote a paper that systematized the theory of self--adjoint extensions of semi-bounded operators in terms of quadratic forms.

As noted in Section \ref{s2}, I (some of it joint with Reed) used quadratic form techniques to simplify some of the technical issues around the complex scaling results of Balslev-Combes.  I developed a quadratic form version of the Cook method in scattering theory \cite{SimonScatteringQF}; Kato \cite{KatoScatteringQF}, Kuroda \cite{KurodaScatteringQF} and Schechter \cite{SchechterScatteringQF} also had results on that question.  As I'll discuss soon, in \cite{SimonMaxMinForm}, I discussed the form analog of Kato's famous $L^2_{loc}$ result.

Turning to self-adjointness proper, my most impactful paper was \cite{SimonPositiveSA}.  Following Kato's work \cite{KatoHisThm}, a number of authors studied what you needed for essential self-adjointness, aka esa, (on $C_0^\infty(\bbT^\nu)$) of the operator $-\Delta+V$ on $L^2(\bbR^\nu)$ in terms of $L^p(\bbR^\nu)$ conditions of $V$.  Call $p$, \emph{$\nu$--canonical} if $p=2$ for $\nu \le 3$, $p > 2$ if $\nu=4$ and $p=\nu/2$ if $p \ge 5$. Then the optimal $L^p$ extension of Kato's theorem is
\begin{theorem} \lb{T6.4}  Let $p$ be $\nu$--canonical.  Then if $V \in L^p(\bbR^\nu)+L^\infty(\bbR^\nu)$, then $-\Delta+V$ is esa on $C_0^\infty(\bbR^\nu)$.
\end{theorem}
Except for the improvement that one can have $p=\nu/2$ (he assumed that $p>\nu/2$) if $\nu\ge 5$, this is a result of Brownell \cite{Brownell}.  For later purposes, we note that rather than $L^p$ conditions, Stummel \cite{Stummel} stated conditions on $V$ in terms of norms like (when $\nu\ge 5$)
\begin{equation}\label{6.28}
  \lim_{\alpha\downarrow 0}\left[ \sup_x \int_{|x-y|<\alpha} |x-y|^{4-\nu} |V(y)|^2\,dy\right]=0
\end{equation}
See the discussion in \cite[Section 1.2]{CFKS}.  $L^p$ conditions imply Stummel conditions but Stummel conditions are more flexible.

That Theorem \ref{T6.4} is optimal can be seen when $\nu\ge 5$, then for $C$ large $-\Delta-C|x|^{-2}$ is \emph{not} esa on $C_0^\infty(\bbT^\nu)$. This implies that Theorem \ref{T6.4} fails for $p$ larger than the canonical value!  In particular, when $\nu \ge 5$, there are $L^2$ potentials for which $-\Delta+V$ is \emph{not} esa on $C_0^\infty(\bbT^\nu)$.  What I realized in \cite{SimonPositiveSA} is that there is an asymmetry between conditions on the positive and negative parts of $V$.  In particular, I proved that
\begin{theorem}  \lb{T6.5} If $V \ge 0$ and
\begin{equation}\label{6.29}
    V \in L^2(\bbR^\nu,e^{-cx^2}\,d^\nu x) \text{ for some } c > 0
\end{equation}
then $-\Delta+V$ is esa on $C_0^\infty(\bbR^\nu)$.
\end{theorem}
My discovery (and proof) of this result show the advantage of working in multiple fields, because this was an outgrowth of my work in CQFT!  As I discussed in Section \ref{s3}, for abstract hypercontractive semigroups (i.e. obeying \eqref{3.1} and \eqref{3.2}), Segal's method \cite{SegalSA1} showed that $H_0+V$ is esa on $D(H_0)\cap D(V)$ if $V$ obeys \eqref{3.3}-\eqref{3.4}.  One of the things that H{\o}egh-Krohn and I \cite{SiHK} realized is that if \eqref{3.3} is replaced by the stronger condition $V\ge 0$, one could replace \eqref{3.4} by the weaker condition that $V\in L^2$ (we did this to handle the spatially cutoff two dimensional $:\exp(\varphi(x)):$ field theory, a favorite model of H{\o}egh-Krohn).  I realized that this implied that when \eqref{6.29} holds, then for a suitable $d$, $-\Delta+dx^2+V$ is esa on $C_0^\infty(\bbR^\nu)$.  An additional trick allowed me to subtract the $dx^2$ and obtain Theorem \ref{T6.5}. Given this new result, I made the natural conjecture for $V\in L^2_{loc}(\bbR^\nu,d^\nu x)$.

When I finished writing the preprint of \cite{SimonPositiveSA}, I mailed a copy to Kato (in those days, papers were typed and, given that Xeroxing was costly, a very few Xerox copies were sent by snail mail - this was years before \TeX\, and email).  About six weeks later (counting the time for ground mail from Princeton to Berkeley and back!), I got Kato's paper \cite{KatoKI1} in which he proved my conjecture by showing that

\begin{theorem} \lb{T6.6} If $V \ge 0$ and $V\in L^2_{loc}(\bbR^\nu,d^\nu x)$, then $-\Delta+V$ is esa on $C_0^\infty(\bbR^\nu)$.
\end{theorem}

Kato's method was totally different from mine.  He first proved what I eventually called \emph{Kato's inequality}, that
\begin{equation}\label{6.30}
  \Delta |u| \ge \text{Re}(\text{sgn}\Delta u)
\end{equation}
(here $u$ is complex valued and $\text{sgn}(u) \equiv \lim_{c\downarrow 0} \bar{u}(x)/(|u(x)|^2+c^2)^{1/2}$).  A novel feature of \eqref{6.30} is that Kato proved it as a distributional inequality under the conditions that $u \in L^1(\bbR^\nu)$ and $\Delta u \in L^1(\bbR^\nu)$.  This inequality came from left field - I'm not aware of anything earlier that was close to it in the work of Kato or anyone else.  Moreover, given the inequality, the proof of Theorem \ref{T6.6} is a few lines (see, for example, \cite[pg 622]{OT}).

Kato's paper did much more than just prove \eqref{6.30} and show how to use that to prove Theorem \ref{T6.6}.  He also proved a version of \eqref{6.30} for magnetic fields, namely for smooth $\overrightarrow{a}$
\begin{equation}\label{6.31}
  \Delta |u| \ge \text{Re}(\text{sgn}(\overrightarrow{\nabla}-i\overrightarrow{a})^2 u)
\end{equation}
again as a distributional inequality.  He used this to show that
\begin{equation}\label{6.32}
  H(\overrightarrow{a},V) = -(\overrightarrow{\nabla}-i\overrightarrow{a})^2+V
\end{equation}
is esa on $C_0^\infty(\bbR^\nu)$ if $V\in L^2_{loc}(\bbR^\nu), V\ge 0$ and $\overrightarrow{a}$ is $C^1$.  I then improved this \cite{SimonSingMag} to only require that $\text{div}(\overrightarrow{a})\in L^2_{loc}(\bbR^\nu)$ and $\overrightarrow{a}\in L^p_{loc}(\bbT^\nu)$ where $p$ had a slightly stronger condition than $2p$ being $\nu$ canonical.  Finally, Leinfelder-Simader \cite{LS} proved the optimal result requiring that $\text{div}(\overrightarrow{a})\in L^2_{loc}(\bbR^\nu)$ and $\overrightarrow{a}\in L^4_{loc}(\bbR^\nu)$.

Kato also had results where $V$ had a negative part.  In that context, he introduced what I later called the \emph{Kato class}, $K_\nu$.  $V\in K_\nu\iff$
\begin{equation}\label{6.33}
 \left\{
               \begin{array}{ll}
                 \lim_{\alpha \downarrow 0} \left[\sup_x \int_{|x-y| \le \alpha} |x-y|^{2-\nu} |V(y)|\, d^\nu y\right] =0, & \hbox{ if } \nu >2\\
                 \lim_{\alpha \downarrow 0} \left[\sup_x \int_{|x-y| \le \alpha}  \log(|x-y|^{-1}) |V(y)|\, d^\nu y\right] =0, & \hbox{ if } \nu =2\\
                      \sup_x \int_{|x-y| \le 1} |V(y)| \, dy < \infty, & \hbox{ if } \nu=1
                \end{array}
                   \right.
\end{equation}
Ironically, this is not optimal (but it is close to optimal) for self-adjointness, but it is optimal for various $L^p$ semigroup condition as shown by Aizenman-Simon \cite[Theorem 1.3]{AizSimon}, for example $V\in K_\nu\iff e^{-t(-\Delta-|V|)}$ is bounded from $L^\infty$ to $L^\infty$ with $\lim_{t\downarrow 0} \norm{e^{-t(-\Delta-|V|)}}_{\infty,\infty}=1$.  The Kato class will appear several times below.

Before leaving the subject of Kato's inequality and self-adjointness, I note the following form analog of Theorem \ref{T6.6} that I proved using connections of $L^p$ semigroup bounds and Kato's inequality that I'll discuss shortly.

\begin{theorem} [Simon \cite{SimonMaxMinForm}] \lb{T6.7} Let $V \ge 0$ be in $L^1_{loc}(\bbR^\nu,d^\nu x)$ and let $\overrightarrow{a} \in L^2_{loc}(\bbR^\nu,d^\nu x)$ be an $\bbR^\nu$ valued function.  Let $Q(D_j^2) = \{\varphi \in L^2(\bbR^\nu,d^\nu x)\,|\, (\nabla_j-ia_j)\varphi \in L^2(\bbR^\nu,d^\nu x)\}$ with quadratic form $\jap{\varphi,-D_j^2\varphi}= \norm{(\nabla_j-ia_j)\varphi}^2$.  Let $h$ be the closed form sum $\sum_{j=1}^{\nu} -D_j^2+V$.  Then $C_0^\infty(\bbR^\nu)$ is a form core for $h$.
\end{theorem}

This paper also has a proof of Theorem \ref{T6.6} using $L^p$ semigroup bounds.  It doesn't explicitly use Kato's inequality but, by then, I knew that his inequality was a expression of the positivity preserving behavior of semigroups.

Both Kato and I were taken with his inequality and each of us wrote additional papers on the subject (Kato \cite{KI2, KI3, KIComp1, KI4,KIComp2,KIComp3, KILp}).  I focused on what the analog is in a much more general context.  In \cite{SimonKI1}, I proved

\begin{theorem}[Simon \cite{SimonKI1}]  \lb{T6.8} Let $A$ be a positive self--adjoint operator on $L^2(M,d\mu)$ for a $\sigma$--finite, separable measure space $(M,\Sigma,d\mu)$.  Let $Q(A)=D(|A|^{1/2}$ be the form domain of $A$ and $q_A(u)=\norm{|A|^{1/2}u}^2$, the quadratic form of $A$. Then the following are equivalent:

(a) ($e^{-tA}$ is positivity preserving)
\begin{equation*}
         \forall u \in L^2,\, u\ge 0, t \ge 0 \Rightarrow e^{-tA}u \ge 0
\end{equation*}

(b) (Beurling--Deny criterion) $u \in Q(A) \Rightarrow |u| \in Q(A)$ and
\begin{equation}\label{6.34}
  q_A(|u|) \le q_A(u)
\end{equation}

(c) (Abstract Kato Inequality) $u \in D(A) \Rightarrow |u| \in Q(A)$ and for all $\varphi \in Q(A)$ with $\varphi \ge 0$, one has that
\begin{equation}\label{6.35}
  \jap{A^{1/2}\varphi,A^{1/2}|u|} \ge \Real\jap{\varphi,\mathrm{sgn}(u) Au}
\end{equation}
\end{theorem}

The equivalence of (a) and (b) for $M$ a finite set (so $A$ is a matrix) is due to Beurling--Deny \cite{BD}.  For a proof of the full theorem (which is not hard), see Simon \cite{SimonKI1} or \cite[Theorem 7.6.4]{OT}.  In that paper, I also conjectured the analog of \eqref{6.31} in a similar general context, a result then proven independently by me \cite{SimonKI2} and by a group of three others \cite{HessKI}

\begin{theorem}[Hess--Schrader--Uhlenbrock \cite{HessKI}, Simon \cite{SimonKI2}]  \lb{T6.9} Let $A$ and $B$ be two positive self--adjoint operators on $L^2(M,d\mu)$ where $(M,\Sigma,d\mu)$ is a $\sigma$--finite, separable measure space.  Suppose that $\varphi \ge 0 \Rightarrow e^{-tA}\varphi \ge 0$.  Then the following are equivalent:

(a) For all $\varphi \in L^2$ and all $t \ge 0$, we have that
\begin{equation*}
  |e^{-tB}\varphi| \le e^{-tA}|\varphi|
\end{equation*}

(b) $\psi \in D(B) \Rightarrow |\psi| \in Q(A)$ and for all $\varphi \in Q(A)$ with $\varphi \ge 0$ and all $\psi \in D(B)$ we have that
\begin{equation}\label{6.36}
  \jap{A^{1/2}\varphi,A^{1/2}|\psi|} \le \Real\jap{\varphi,\mathrm{sgn(\psi) B\psi}}
\end{equation}
\end{theorem}

For a proof, see the original papers or \cite[Theorem 7.6.7]{OT}.  There is further discussion of this result in the context of diamagnetic inequalities in quantum mechanics in Section \ref{s7}.

These semigroup ideas are intimately related to properties of eigenfunctions of Schr\"{o}dinger operators, a subject I often looked at in the 1970's.  One issue that particularly attracted me was that of exponential decay.  In 1969, the only results on decay of discrete eigenfunctions of $N$-body quantum Hamiltonians with $N>2$ had very severe restrictions like $N=3$ or only Coulomb potentials.  I gave looking at $N$--body systems to Tony O'Connor, my first graduate student (who began working with me when I was a first year instructor).  He had the idea of looking at analyticity of the Fourier transform and obtained results in the $L^2$ sense (i.e. $e^{a|x|}\psi \in L^2$) that were optimal in that you couldn't do better in terms of isotropic decay.  Here $|x|$ is a mass weighted measure of the spread of the $N$ particles, explicitly, in terms of the inner product \eqref{6.5} and the center of mass $\rho_N$ of \eqref{6.6}, one has that
\begin{equation}\label{6.37}
  M|x|^2 = \jap{r-\rho_N,r-\rho_N}
\end{equation}
O'Connor found one had the $L^2$ bound if $|a|^2<2M(\Sigma-E)$.

His paper \cite{TO} motivated Combes--Thomas \cite{CT} to an approach that has now become standard of using boost analyticity.
It is widely applicable although in the $N$-body case it exactly recovered O'Connor's bound.  Over the years, I had a six paper series on the subject of exponential decay \cite{SimonExp1, SimonExp2, SimonExp3, SimonExp4, SimonExp5, SimonExp6}.  In the first three papers, I looked at getting pointwise bounds.  In the first paper, I obtained optimal pointwise isotropic bounds for $N$--body systems.  In the second paper, I considered the case where $V$ goes to infinity at infinity and proved pointwise exponential decay by every exponential (Sch'nol \cite{Schnol} earlier had a related result).  In the third paper, I assumed $|x|^{2m}$ lower bounds on $V$ and got $\exp(-|x|^{m+1})$ pointwise upper bounds on the eigenfunctions.  When one has an upper bound on $V$ of this form, one gets lower bounds of the same form on the ground state.  Papers 1-2 were written during my fall 1972 visit to IHES, one of my most productive times when Lieb and I did most of the Thomas--Fermi work and I developed new aspects of correlation inequalities and Lee--Yang for EQFT.

The fourth paper \cite{SimonExp4} (joint with Deift, Hunziker and Vock; Deift had been my student and we continued working on this while he was a postdoc.  I learned that Hunziker was looking at similar questions so we joined forces -- Vock was his master's student) explored non--isotropic bound for $N$-body systems.  We found a critical differential inequality that if $f$ obeys it, then $e^f \psi \in L^\infty$ and in some cases were able to find explicit formula for the optimal $f$ (but only in a few simple situations).  Later, Agmon \cite{Agmon} found the optimal solution of the differential inequality as a geodesic distance in a suitable Riemann metric (discontinuous in the case of $N$--body systems) -- this is now called the \emph{Agmon metric}, a name that appeared first in the fifth paper of this series by Carmona--Simon \cite{SimonExp5}, which also proved lower bounds for the ground state complementary to Agmon's upper bounds.  We proved that if $\psi(x)$ is the ground state and $\rho(x)$ the Agmon metric distance from $x$ to $0$, then $\lim_{|x| \to \infty}  - \log |\psi(x)| / \rho(x) = 1$.  In some ways, the fourth paper is made obsolete by \cite{Agmon,SimonExp5} although the explicit closed form for $\rho$ in some cases remains of significance.  The sixth paper with Lieb \cite{SimonExp6} studied $N$-body system in the special region where subclusters remained bound but were distant from each other.

Carmona--Simon \cite{SimonExp5} used path integral techniques in NRQM so I pause to say something about that subject which due to pioneering work of Lieb and Nelson was a kind secret weapon around Princeton which I also used so extensively that I wrote what became a standard reference \cite{SimonFI}.  It was based on lectures I gave in Switzerland in the summer of 1977.  I was on leave in 1976-77 and also gave lectures at the University of Texas which also turned into a book \cite{SimonTI} on my other secret functional analytic weapon, the theory of trace ideals.  It has had a rebirth of use since it is a tool in quantum information theory.

One thing that I used path integral methods for is to study more general issues of properties of eigenfunctions and integral kernels (for the semigroup and resolvent) than exponential decay,  although they also allowed stronger results and simpler proofs also for exponential decay.  I did this in the Functional Integration book just mentioned but even more in two articles, one with Aizenman \cite{AizSimon} and one that was billed as a review article \cite{SimonSchSmgp}.  The article with Aizenman, which won the Stampacchia prize, proved Harnack inequalities and subsolution estimates on eigenfunctions of Schr\"{o}dinger operators under only $K_\nu^{loc}$ conditions on $V$ (see also \cite{CR}).  The 80 page review article is my fifth most cited publication (the only more cited items are three books and the Berry's phase paper) and proves many results, for example, on continuum eigenfunction expansions, under greater generality than previously.  I should mention that this work was influenced by a beautiful paper of Ren\'{e} Carmona \cite{Car1} (and later \cite{Car2}) that emphasized a simple way to get $L^\infty$ bounds.  When I learned of this work, I invited Ren\'{e} to visit Princeton leading to \cite{SimonExp5}.  Later, he and I teamed up with an Irvine graduate student of his \cite{CMS} to discuss analogs of these Schr\"{o}dinger operators results when $-\Delta$ is replaced by other generators of positivity preserving semigroups, most notably the one, $\sqrt{-\Delta+m^2}$, associated to relativistic quantum theory.

A little more on subsolution estimates (which had been discussed for more general elliptic operators but with greater restrictions on the regularity of coefficients in the PDE literature, especially by Trudinger).  These imply that if $Hu=Eu$, then
\begin{equation}\label{6.38}
  |u(x)| \le C\int_{|y-x|\le 1} |u(y)| \, d^\nu y
\end{equation}
where $C$ only depends on $K_\nu$ norms of $V$ restricted to the ball of radius $1$ about x.  These immediately imply that $L^2$ estimates, for example those found by O'Connor \cite{TO}, imply pointwise estimates, for example, those proven by me in \cite{SimonExp1}.

Eigenfunction properties and expansions recurred in my later work many connected with 1D problems, including the discrete case, especially with applications to a.c. spectrum and/or almost periodic problems (see Section \ref{s9}).  I mention four: \cite{SimonACBdd} has a simple proof that, for 1D discrete and continuum Schr\"{o}dinger operators, if all eigensolutions are bounded for energies in an set, $S$, then the spectrum is purely a.c. on $S$, a result of Gilbert-Pearson \cite{GP}. It also uses this theorem to analyze 1D Schr\"{o}dinger operators with potentials of bounded variation, recovering results of Weidmann \cite{WeidmannBV} and extending them to the Jacobi case.  \cite{SimonSC} gave a simple proof, using rank one perturbation theory (specifically Theorem \ref{T11.2}(b)), that for such operators on $L^2(\bbR)$, the singular spectrum is always simple, a result proven earlier by Kac \cite{KacSC} and Gilbert \cite{Gilbert} in a more complicated way.  Last-Simon \cite{LSEigen} has many results on the connection of eigenfunctions to spectral behavior depending on the growth of transfer matrices for ODEs and Kiselev-Last-Simon \cite{KLSi} has additional results on growth of transfer matrices and spectral properties, including the subtle borderline $x^{-1/2}$ decaying random potential.

Before leaving the subject of eigensolutions, I should mention a paper with Schechter \cite{SchSi} also written the 1975-76 year that I was on leave (Schechter was at Yeshiva University where I spent two days a week that year).  Carleman \cite{Carleman} had studied the issue of unique continuation (if a solution of $Hu=Eu$ vanishes on an open set, it vanishes identically), a subject for which almost all work since has used what have come to be called Carleman estimates after that paper.  In 1959, Kato \cite{KatoPE} understood that unique continuation was an element of a proof of the non-existence of positive eigenvalues.  What Schechter and I realized is that Carleman estimates were limited to bounded potentials and it was natural to consider the problem for some unbounded $V$'s.  We proved the first such results although we stated that we believed our conditions were far from optimal.  We hoped that we'd motivate the harmonic analysis community and there were a number of papers that our work stimulated.  Most notable were Jerison-Kenig \cite{JerKen} and Koch-Tataru \cite{KochTat}.  An optimal result (from \cite{JerKen}) says that one has unique continuation for $-\Delta+V$ is $V\in L_{loc}^{\nu/2}(\bbR^\nu)$ (with $\nu>2$).  In terms of local $L^p$ conditions this is optimal but it has been realized recently (Garrigue \cite{Garrigue}) that if $\nu=N\mu$ and $V$ has an $N$ body form it is not optimal.  One would hope that there is a result for $L^p_{loc}$ when $p > \mu/2$ rather than $p > \nu/2$ (which is bad  for $N$ very large.  Ironically, the result of Schechter-Simon \cite{SchSi} that $p>\mu$ (if $\mu>4$) suffices is among the strongest results for this general $N$-body case.  In any event, there is work remaining to be done.

I'd first heard of unique continuation theorems as a graduate student in the context of Kato's result \cite{KatoPE} that if $V$ is a continuous function on $\bbR^\nu$ so that $|x| |V(x)| \to 0$ at infinity, then $-\Delta+V$ has no eigenvalues in $(0,\infty)$. In one of my first serious papers \cite{SimonNPE}, I found a result on no positive eigenvalues that allows $V$ to be a sum of two pieces, $V_1$ that obeys Kato's condition and a piece, $V_2$, that obeys $V_2\to 0$ and $|x|\partial V_2/\partial x\to 0$ at infinity. For more on positive eigenvalues, see my review of Kato's work in NRQM \cite[Section 12]{SimonKato}.

From Wightman, I'd learned of the paper of Wigner-von Neumann \cite{vNW} that constructed an example of $V$ going to $0$ at $\infty$ so that $-\Delta+V$ has  a positive eigenvalue.  The example they actually write down has $V(x) = \text{O}(|x|^{-2})$ at infinity and violates Kato's theorem!  I discovered that they had clearly used $\cos x/\sin x=\tan x$ which caused a miraculous cancellation of the $ \text{O}(|x|^{-1})$ terms!  My paper seems to have been the first to note the error and write down the correct explicit form they should have.  At one point, I had to ask Wigner a question about something else and I asked him about  if he knew that this paper, written 40 years, before had this error.  He thought for a moment and then replied ``No, I didn't know'' and, after a pause ``Johnny did that calculation.''  I note that the construction of the analog of this example in dimension $\nu\ne 1,3$ is not so straight-forward.  It can be found in a paper I wrote many years later with Frank \cite{FSWvN}.

Kato's result says that on $(0,\infty)$, if $\lim_{x\to\infty} x|V(x)|=0$, then $h=-\tfrac{d^2}{dx^2}+V(x)$ has no eigenvalues in $(0,\infty)$ and many years later Kiselev, Last and I \cite{KLSi} proved that if $\lim_{x\to\infty} x|V(x)|<\infty$, the set of positive eigenvalues is discrete with only $0$ as a possible limit point (indeed, the sum of the positive eigenvalues, if any, must be finite).  In \cite{SimonDPS}, I constructed $V$'s where $x|V(x)|$ had arbitrarily slow growth at infinity (in particular, some for which it was known that the a.c. spectrum was $[0,\infty)$ by Kiselev \cite{KisSlowAC}) with any desired uncountable set of positive energy eigenvalues, even dense sets.  I was motivated by an earlier paper of Naboko \cite{NabDPS} who was able to construct such examples so long as the set of positive eigenvalues had the form $E_n=\kappa_n^2$ with the $\kappa_n$ rationally independent.  Hsu et al. \cite{BSC} have a recent review of physically relevant examples with bound states in the continuum.

\bigskip

Next, I turn to scattering theory, in particular the question of N-body asymptotic completeness (big problem 4) where my most significant result involves the Deift-Simon wave operators \cite{DS2}.  To put it in context, I begin with a lightning summary of the high points of $2$ and $N$ body scattering.  One needs to bear in mind that big problem 3 (absence of s.c. spectrum) is often intimately related to big problem 4 in that sometimes the techniques to solve them (namely detailed analysis of the boundary values of the resolvent) are close; indeed Reed-Simon \cite{RS3} calls the combination of the two, \emph{strong asymptotic completeness}.

In abstract scattering theory, one defines wave operators for a pair of self-adjoint operators by (as above, the funny $\pm$ convention is taken from the physics literature and often the opposite to the convention in the mathematics literature).
\begin{equation}\label{6.39}
  \Omega^\pm(A,B) = \lim_{t\to\mp\infty} e^{itA}e^{-itB}P_{ac}(B)
\end{equation}
where $P_{ac}(B)$ is the projection onto those the subspace of those vectors whose spectral measure for $B$ is purely absolutely continuous.  The insertion of the $P_{ac}(B)$ (which is redundant in the usual case of two body physics where $B=-\Delta,\,A=-\Delta+V$ with $V$ short range) is a wonderful realization of Kato \cite{KatoTrace1}, who understood that if we call $\Omega^\pm(A,B)$ \emph{complete} if and only if
\begin{equation}\label{6.40}
  \ran\,\Omega^\pm(A,B) = \calH_{ac}(A)\equiv \ran\,P_{ac}(A)
\end{equation}
then one has that

\begin{theorem} \lb{T6.10} Suppose that $\Omega^\pm(A,B)$ exist.  Then, they are complete if and only if $\Omega^\pm(B,A)$ exist.
\end{theorem}

The first mathematical results on existence of wave operators was a simple argument of Cook \cite{Cook}, improved by Hack \cite{Hack1} and Kuroda \cite{Kuroda1}.  The latter two got existence on $\bbR^\nu$ for $V$'s decaying as $|x|^{-1-\varepsilon}$.  The first completeness results were obtained by Kato \cite{KatoTrace1, KatoTrace2} and Rosenblum \cite{RosenTrace} whose best result says that if $A-B$ is trace class, then $\Omega^\pm(A,B)$ exist so, by symmetry and Theorem \ref{T6.10}, they are complete (see \cite[Section 13]{SimonKato} for a lot more on the history and extension of the Kato-Rosenblum theorem).  By shifting from trace class to differences of resolvent being trace class, these results imply completeness on $\bbR^3$ if $V$ decays like $|x|^{-3-\varepsilon}$.  It then took about ten years, to get to solving big problems 3 and 4 for $N=2$ and $V$ bounded by $|x|^{-1-\varepsilon}$.  This was first accomplished by Agmon \cite{AgmonAC} (Agmon announced this at the 1970 ICM; at the same conference, Kato \cite{KatoAC} announced the solution of big problem 4 for this class, extending some ideas of Kato-Kuroda.  Then Kuroda \cite{KurAC1, KurAC2} realized that by borrowing one technical device from Agmon, their method also solved big problem 3 for this class.  For more on the history and details of this work see \cite[Section 15]{SimonKato}).

Already in 1963, Faddeev \cite{Fadd} obtained asymptotic completeness for certain $3$ body equations.  Because his basic condition were written on the Fourier transform, it is difficult to write them in terms of the $V_{ij}$ but his assumptions required decay faster than $(1+|x|)^{-2-\varepsilon}$.  He also supposed the two body subsystems didn't have zero energy resonances.  In any event, his work had limited impact on the mathematical physics literature and his methods were never extended to $N\ge 4$ nor were they a major factor in the eventual successful resolution of problems 3 and 4.

For decay faster than Coulomb, two body problems were well understood by 1972.  It took another 15 years for the complete solution of big problem 4 for general $N$-body systems. During that period, it was a major open question and several people, formally or informally, announced solutions which turned out to have errors.  At one point, Agmon wryly remarked to me ``those whom the gods would drive mad, they teach of the problem of $N$-body asymptotic completeness.''  While I was certainly aware of the problem and several times did work related to it, I never tried to systematically approach it because I didn't see a fruitful approach.  Two high points of the fifteen year intermediate period were work of Enss and Mourre, each of which played important roles in the eventual resolutions.  Because Mourre's work is more connected with big problem 3, I'll postpone its discussion of it.

Enss \cite{EnssAC} revolutionized scattering, especially two body scattering.  At a heuristic level, scattering is a time-dependent phenomenon but prior to Enss, the most powerful results in quantum scattering used time independent methods (i.e. focus on resolvents rather than the unitary groups) - Faddeev's work and the Agmon-Kato-Kuroda work mentioned above.  Enss used purely time-dependent methods without any resolvents anywhere.  He combined Cook's method with two extra ingredients.  The first was geometric, motivated, in part, my work with Deift \cite{DS2}, discussed below, and the geometric approach to the HVZ theorem by Enss \cite{EnssHVZ} and me \cite{SimonHVZ}, discussed above.  The other was to localize in phase space. He did this while respecting the uncertainty principle, by, in essence, projecting on spectral subspaces for the dilation operator, $A = \tfrac{1}{2}[\mathbf{x}\cdot\mathbf{p}+\mathbf{p}\cdot\mathbf{x}]$.  This suggested that a natural way of approaching his work was to use an eigenfunction expansion for $A$, i.e. the Mellin transform, which is precisely the approach used by Perry \cite{PerryEnss} in a thesis done under my direction.

I was taken with this work of Enss and talked it up using my then considerable influence.  I wrote a long (50 page) article \cite{SimonEnss} showing how to apply it in a large number of scattering theory situations (Reed and I had just finished our scattering theory volume \cite{RS3}, so I knew of lots of scattering problems beyond $2$-body NRQM).  When Enss visited the Institute, we used some of these ideas to study total cross-sections \cite{EnssSimon1, EnssSimon2}.

Fifteen years after Enss' work, his techniques was critical to an analysis by me and others of some intriguing examples of Neumann Laplacians.  Recall that Dirichlet and Neumann Laplacians are described most naturally in terms of quadratic forms \cite[Section 7.5]{OT}.  Given an open set, $\Omega\subset\bbR^\nu$, one defines $Q(-\Delta_N^\Omega)$ to be the set of functions, $\psi\in L^2(\Omega,d^\nu x)$ so that $\nabla\psi\in L^2(\Omega,d^\nu x)$.  The sesquilinear form $\psi\mapsto\int |\nabla\psi|^2\,d^\nu x$, where $\nabla\psi$ is the distributional gradient, defines a self adjoint operator, $-\Delta_N^\Omega$, the Neumann BC Laplacian for $\Omega$.  If we restrict the form to the closure of $C_0^\infty(\Omega)$ in form norm, the corresponding operator is $-\Delta_D^\Omega$, the Dirichlet BC Laplacian for $\Omega$.  If $\Omega$ has a smooth boundary, the functions $C^\infty$ with Dirichlet or Neumann boundary conditions are an operator core for $-\Delta_D^\Omega$ and $-\Delta_N^\Omega$ respectively.

As I'll explain in Section \ref{s8}, I had considered the case $\Omega=\{(x,y)\,|\, |xy|<1\}\subset\bbR^2$ and had shown that despite the fact that $\Omega$ has infinite volume, $-\Delta_D^\Omega$ has discrete spectrum.  The intuition is that the horns got narrow so the lowest Dirichlet eigenvalue associated with cross sections went to infinity.  When chatting at a conference in Gregynog, Wales, Brian Davies and I stated discussing $-\Delta_N^\Omega$.  With Neumann BC, the lowest eigenvalue is zero so one expects one ac mode in each of the four horns.  It was easy to construct wave operators to get existence and we realized that using Enss theory, we could prove \cite{DavSiHorns} that $-\Delta_N^\Omega$ had ac spectrum of multiplicity exactly four (one for each horn).

The opposite phenomenon to Dirichlet Laplacians of infinite volume but discrete spectrum is Neumann Laplacians with finite volume and some essential spectrum. Already Courant-Hilbert \cite{CH2} had found bounded regions with $0\in \sigma_{ess}(-\Delta_N^\Omega)$ and Hempel, Seco and I \cite{HSSRooms} had shown that for any closed set $S\subset [0,\infty)$, there was a bounded $\Omega$ so that $\sigma_{ess}(-\Delta_N^\Omega)=S$; these examples had empty ac spectrum. Davies and I found that for $\Omega = \{(x,y)\,|\,x>1, |y| < f(x)\}$, one has that $-\Delta_N^\Omega$ has ac spectrum of multiplicity one on all of $[0,\infty)$ so long as all of $k_1(x) = |f'(x)|, k_2(x)=|f'(x)|^2 f(x)^{-1}$ and
\begin{equation} \lb{6.40A}
  V(x) = \frac{1}{4}\left(\frac{f'}{f}\right)^2+\frac{1}{2}\left(\frac{f'}{f}\right)'
\end{equation}
are $\text{O}(|x|^{-1-\varepsilon})$ as $x\to\infty$.  (In Section \ref{s8}, I'll discuss results of \cite{JMS} on eigenvalue asymptotics when $V(x)\to\infty$). For example if $f(x)=x^{-\alpha},\alpha>0$, one gets ac spectrum even though if $\alpha>1$, then $\Omega$ has finite volume. I realized \cite{SimonJellyRoll} that one could wrap such a finite volume horn up and construct a bounded region whose $-\Delta_N^\Omega$ had ac spectrum on $[0,\infty)$!

Returning to the theme of general $N$-body asymptotic completeness, the big breakthrough and first solution was by Sigal-Soffer \cite{SigSof1}.  The paper uses in impressive combination of Mourre estimates, Deift-Simon wave operators and phase space estimates motivated by Enss theory.  Unfortunately, as the MathSciNet review says: \emph{It is disappointing that this important result has not received the exposition it deserves. The paper contains numerous misprints, points of unclarity, obscure notation, and minor technical errors. Because the details of the proof are inaccessible to all but the most dedicated specialist, there has been considerable speculation about the validity of the result. Fortunately, two experts who have studied the paper thoroughly have assured the reviewer (open letter from W. Hunziker and B. Simon, dated September 1, 1987) that essentials of the proof are correct.}  I should say a little more about my role in this.  The 70+ page paper was clearly very important but also not ideally written. Walter Hunziker and I felt a duty as leading figures (and also, because, if I recall, one or both of us were referees for the Annals) to determine if the results were correct.  I visited ETH in the summer of 1986 and for three weeks, I arrived about 10 in the morning, worked with Walter until 3 in the afternoon (with a break for lunch) painfully plowing through the paper line by line.  We found lots of little errors - typically a lemma was wrong but when we figured out how it was used, the lemma and/or its proof could be slightly modified to work.  The authors had apparently decided to state each lemma in the most general form they could, often so general, it was now wrong.  We made numerous suggestions for changes, decided the paper was basically correct and recommended publication even though we agreed that even after changes it was not a model of exposition.  When the Math Reviews reviewer contacted me, Walter and I produced a public document vouching for the result.

Later proofs sharing some elements with \cite{SigSof1}, are due to Graf \cite{GrafAC}, Derezi\'{n}ski \cite{DereAC} and Yafaev \cite{YafAC}.  Coulomb and long range potentials are treated in \cite{SigSof2, DereAC}.

Having put it in context with this summary, I turn to exactly what Deift and I did in \cite{DS2}, a paper used in all the proofs mentioned above \cite{SigSof1, SigSof2, GrafAC, DereAC, YafAC}.  We proved a kind of $N$-body analog of Theorem \ref{T6.10}.  For any partition, $\calC$, we defined $|\boldsymbol{r}|_\calC$ in \eqref{6.24} as the minimal distance between components.  We also can define
\begin{equation}\label{6.41}
  \rho_\calC(\boldsymbol{r}) = \max_{(jq)\subset \calC}|\boldsymbol{r_j}-\boldsymbol{r_q}|
\end{equation}
which describes how far particles within the components are from each other.  One defines functions $\tilde{J}_\calC$ for each $\calC\in\bbP$ which are $1$ in the region where $|\boldsymbol{r}|>1$ and
\begin{equation}\label{6.42}
  \rho_\calC(\boldsymbol{r}) \le \left[|\boldsymbol{r}|_\calC\right]^{1/2}
\end{equation}
and supported in the union of the set where $|\boldsymbol{r}|\le 1$ and the set where \eqref{6.42} holds with the right side multiplied by $2$.  Thus when $|\boldsymbol{r}|$ is large, points in the support have distances within the clusters much less than the average distance between particles.  Given two partitions $\calC$ and $\calD$, we write $\calD\rhd\calC$ if $\calD$ is a refinement of $\calC$, i.e. if every subset in $\calC$ is a union of sets in $\calD$.  Deift and Simon define
\begin{equation}\label{6.43}
  J_\calC = \tilde{J}_\calC - \sum_{\calD\rhd\calC,\,\calD\ne\calC} \tilde{J}_\calD
\end{equation}
which eliminates configurations where the particles in subsets of $\calD$ are very close to each other while particles in different subsets of $\calD$, but in the same subset of $\calC$, are quite far but still small compared to the maximum intercluster distance.  The \emph{Deift-Simon wave operators} are defined by
\begin{equation}\label{6.44}
 \Omega^\pm_\calC = \text{s-lim}_{t\to\mp\infty} e^{itH(\calC)}J_\calC e^{-itH}P_{ac}(H)
\end{equation}
If this limit exists and, say $\psi=\Omega^+_\calC\varphi$, for $\varphi\in\calH_{ac}(H(\calC))$ then as $t\to -\infty$, $e^{-itH}\psi$ is asymptotic to $J_\calC e^{-itH}(\calC)\varphi$.  If all the component $H(C_j)$'s have no sc spectrum and obey asymptotic completeness, then $J_\calC e^{-itH}(\calC)\varphi$ will look exactly like a sum of $e^{-itH}\eta_t$ for $t$ a threshold associated to $\calC$ and suitable $\eta_t\in\ran\Omega^+_t$.  Thus one expects, and Deift-Simon prove, an analog of Theorem \ref{T6.10}

\begin{theorem} [Deift-Simon \cite{DS2}] \lb{T6.11} Suppose $H$ is an $N$ body quantum Hamiltonian so that all the proper subset Hamiltonians $H(C)$ have no singular spectrum and obey asymptotic completeness. Then $H$ is asymptotically complete if and only if all the wave operators $\Omega^\pm_\calC$ exist.
\end{theorem}

This clearly suggests an approach to proving asymptotic completeness inductively.

I should say something about the origin of this work with Percy Deift.  Percy, who is actually a year older than me, grew up in South Africa and, encouraged to study practical subjects, got a master's degree there in Chemical Engineering.  When he got interested in more theoretical things, he decided to go Princeton, applying to a strange program in applied math.  At the time Princeton had no Applied Math Dept.  Martin Kruskal, because of his famous work on the Schwarzschild solution, had an appointment in Astronomy, but his true love was mathematics (later, once his soliton work became famous, I persuaded my colleagues in mathematics to give him a joint appointment, but that is another story).  He convinced the administration to allow him to run a PhD program where students had to have a home department although they could do their thesis work in any area and their preliminary exams were in applied math with various faculty in other departments on their exam committee.  It took students who often wouldn't have been able to get into the departments in which they did their thesis work by the front door.  It's students were a mixed bag.  Some of the most painful qualifying exams I ever served on were in that program, students who were woefully prepared and didn't come close to passing.  On the other hand, it had some spectacular successes.  Ed Witten, who would never have been admitted to the math or physics departments because he'd had almost no courses in those subjects when he applied, did get into Kruskal's program (one of the decisions I made as director of graduate studies in Physics was, after getting rave reviews from some professors who had him in courses, to allow him to transfer to the regular physics track).  Percy Deift, who has been my most successful student, was in Kruskal's program, originally with Chemical Engineering as a home department.

I was on leave in Percy's first year in Princeton and he came to me in the middle of his second year saying that he wanted to do a thesis under my direction.  I was skeptical but after consulting Lon Rosen, with whom he'd taken a math methods course, I said I'd give him a try.  We discussed various open problems in the then several areas I was working on.  He came back and told me that he'd like to work on $N$-body asymptotic completeness.  I told him that was a totally crazy problem for a graduate student to work on without some wonderful idea that looked almost sure to lead to success -- it was too risky that it would lead to total failure after several years.  He went away and came back and said that he understood it was too hard but could I suggest some problems that would lead in the direction of eventually solving $N$-body asymptotic completeness. His approach made me decide that maybe this student might get somewhere after all.  In the end, his nice thesis was in another direction \cite{DeiftThesis} but we agreed to discuss scattering theory on the side.  This was before the work that Enss and I did on HVZ or Enss' work and we wound up discovering the usefulness of using geometric ideas (indeed \cite{DS2} was motivation for my geometric approach to HVZ \cite{SimonHVZ}).  We first wrote a cute paper \cite{DS1} that used trace class methods, Dirichlet decoupling and path integral techniques to prove that (positive) local singularities were irrelevant to questions of existence and completeness of two body Schr\"{o}dinger operator wave operators and our second paper was \cite{DS2}.  (We wrote several other papers together later on in his career).

\bigskip

That completes what I want to say about big problem 4 so I turn to big problem 3 and the work I did with Perry and Sigal on extending Mourre theory to $N$-body systems \cite{PSS}.  Around 1977, Eric Mourre wrote a preprint and submitted it to one of the then editors of Communications in Mathematical Physics handling Schr\"{o}dinger operators (starting two years after that, I served as the main editor for that subject for more than 35 years).  Then, as now, an editor has the freedom to look over a manuscript and reject it without any refereeing and the editor decided that Mourre's paper wasn't important enough for CMP and rejected it.  Mourre placed the manuscript in his desk drawer rather than submit it elsewhere.  I also got the preprint, thought it might be interesting, but wasn't sure since it was hard to follow.  I was short on time, so I passed copies on to two of my former students asking them to take a careful look and let me know if there was anything interesting there.  They each eventually reported there didn't seem to be anything worth spending a lot of time on.  Despite these initial opinions, the paper was one of the most significant in the study of $N$-body NRQM spectral theory!

Part of Mourre's motivation was work by Lavine \cite{Lav1,Lav2,Lav3,Lav4} on $N$ body systems with repulsive potentials who, in turn, was extending work of Putnam \cite{PutnamPK}, Kato \cite{KatoPK}, Weidmann \cite{WeidVirial} and Kalf \cite{KalfVirial}.  Lavine and Mourre centrally use the generator of dilations (which is also central, in a different way to the work of Enss and Perry as I've discussed):
\begin{equation}\label{6.45}
  A = \frac{1}{2i}\left(\boldsymbol{x}\cdot\boldsymbol{\nabla}+\boldsymbol{\nabla}\cdot\boldsymbol{x}\right)
\end{equation}
What makes repulsive potentials special is that one has a positive commutator
\begin{equation}\label{6.46}
  [H,iA] > 0
\end{equation}
where $H$ is the $N$-body Hamiltonian.  Putnam (with a later slick proof of Kato) proved that for \emph{bounded} operators, positive commutators implies both have purely a.c. spectrum (related to this and useful in Mourre theory is the Virial theorem which is where the work of Weidmann and Kalf comes in).  Lavine figured out how to modify things to apply to the unbounded operators that enter in \eqref{6.46} and also how to sometimes get scattering theory results using an extension of Kato's smoothness theory \cite{KatoSmooth, KatoPK}.

In his paper, what Mourre realized is that one could get a lot from a local version of \eqref{6.46}.  Namely, he considered what have come to be called \emph{Mourre estimates}
\begin{equation}\label{6.47}
  P_\Delta(H)[H,iA]P_\Delta(H) \ge \alpha P_\Delta(H)+K
\end{equation}
where $\Delta=(a,b)$ is an open interval, $P_\Delta$ a spectral projection, $\alpha>0$ and $K$ a compact operator.  First of all, Mourre showed that when \eqref{6.47} holds, $H$ has no s.c. spectrum on $\Delta$ and that $H$ has only finitely many embedded eigenvalues in any $[a+\delta,b-\delta],\,\delta>0$, each of finite multiplicity.  Secondly, he showed that \eqref{6.47} held for fairly general $3$-body Schr\"{o}dinger Hamiltonians with decaying potentials for $\Delta$ any interval avoiding $0$ and the two body thresholds (which means eigenvalues of two body subsystems).

In one sense, it is surprising that Mourre's path breaking work wasn't rapidly recognized but there are several reasons.  First the paper was in French (the CMP editor was French so that wasn't an issue on the rapid initial rejection).  It was partly the originality of many of the ideas.  The estimate \eqref{6.47} with the compact error and its proof via differential inequalities were so new it was a little difficult to grasp what was going on.  But most of all, the paper was terribly written.  It was often unclear what the author was doing and what his steps meant.

Fortunately, the paper didn't merely wind up in his desk drawer because Mourre gave talks on it at conferences and, at one, Israel Sigal realized there might be something important here (and indeed Mourre estimates were a critical step in Sigal's 15 year successful quest to prove $N$-body asymptotic completeness).  So Sigal decided to try and extend Mourre's work to $N>3$.  The first problem he faced is that the preprint was in French and he didn't speak the language.  But he learned that Peter Perry, who was then my graduate student, was fluent in French, so they started working together and when trying to overcome some technical issues, they decided to ask me to join them.  By exploiting some rather complicated arguments, we were able to prove

\begin{theorem} [Perry-Sigal-Simon \cite{PSS}] \lb{T6.12} Let $H$ be a reduced $N$-body Hamiltonian with two body potentials obeying:
\begin{equation}\label{6.48}
  |V_{ij}(x)| \le C(1+|x|)^{-1-\varepsilon}
\end{equation}
for some $C,\varepsilon>0$.  Then the closure of the set of thresholds is countable, and a Mourre estimate of the form \eqref{6.47} with $A$ given by \eqref{6.45} holds for any closed interval in the complement of the closure of the thresholds.  This implies that $H$ has no singular continuous spectrum and that any such closed interval has at most finitely many eigenvalues, each of finite multiplicity.
\end{theorem}

\begin{remarks} 1. \eqref{6.48} is stated for simplicity of exposition; local singularities are allowed and one can even have slower than $|x|^{-1}$ decay if the first or second derivative decays as in \eqref{6.48}.

2. We first of all showed (following Mourre) that Mourre estimates implies absence of sc spectrum and finiteness of embedded eigenvalues.  One then gets the result inductively on thresholds so the key is the proof of the Mourre estimates which in our paper is quite involved.

3. Prior to our work, with one exception all results on absence of sc spectrum were single channel, requiring either repulsive potentials \cite{Lav1,Lav2,Lav3,Lav4} or weak coupling \cite{IO}.  The exception is the results of Balslev-Combes \cite{BC} using dilation analyticity that required analytic potentials.  Our work was the first that, for example, handled $C^\infty_0$ potentials.
\end{remarks}

This solved big problem 3 in great generality and provided tools of use in scattering theory.  After we obtained our results, I contacted Mourre to find out where his paper had been published.  When I learned it was sitting in his drawer, with his permission I contacted the original rejecting editor and the editor-in-chief of CMP (where I was, by then, an editor) to get their OK to have the paper reconsidered.  Peter Perry (with Mourre's permission) translated the paper into English and it appeared as \cite{Mourre}.  Before leaving the subject of Mourre theory, I should mention two lovely papers of Froese-Herbst and one paper for which I was a coauthor.  In \cite{FH1}, Froese-Herbst found a considerably streamlined proof of $N$-body Mourre estimates and, in \cite{FHExp}, they used Mourre estimates to obtain some remarkable results on exponential decay. In \cite{SimonEtAlStark}, Bentosela-Carmona-Duclos-Simon-Souillard-Weder used Mourre theory (but with $A=i\partial/\partial x_1$ which works well because of the $Fx_1$ term) to prove that a Stark Hamiltonian, $-\Delta+V(x)+Fx_1$, with $V$ fairly smooth and $F_1\ne 0$, has no sc spectrum and only isolated point spectrum of finite multiplicity (in 1D, a separate argument proved no eigenvalues).

\bigskip

That completes what I will say about few-body quantum systems.  Since this is the main section that includes a discussion of scattering theory, I will use the rest of this section to discuss some other work of mine connected to scattering starting with inverse scattering.  My most important inverse result involves an alternative to the Gel'fand-Levitan \cite{GL} approach to determining the potential, $V$, for $h=-\tfrac{d^2}{dx^2}+V$ on $[0,\infty)$ with Dirichlet boundary conditions at $0$ from its spectral measure (in the physics literature this is usually discussed in terms of determining $V$ from a reflection coefficient, bound state energies and norming constants but they determine the spectral measure, i.e. the measure $d\rho$ with
\begin{equation}\label{6.49}
  \int f(x) \,d\rho(x) = \pi^{-1}\lim_{\varepsilon\downarrow 0} \int f(x) \jap{\delta',(h-x-i\varepsilon)^{-1}\delta'}\,dx
\end{equation}
and it is the process $\rho\mapsto V$ that concerned Gel'fand-Levitan).  I wrote three papers on this alternate approach (the second and third with, respectively, Gesztesy and Ramm) \cite{SimonInverse1, SimonInverse2, SimonInverse3} and a fourth with Gesztesy \cite{GSConn} applying it.

To understand my motivation, one needs to understand the discrete analog of this question - going from a probability measure of bounded support on $\bbR$ to the Jacobi parameters, $\{a_j,b_j\}_{j=1}^\infty$ where each $a_j$ is strictly positive and each $b_j$ is real (see \cite[Section 1.2]{SimonSz} for discussion of Jacobi parameters).  The simplest way is as recursion coefficients for orthogonal polynomials on the real line (aka OPRL): given, $d\mu$, one forms the orthonormal polynomials and finds the recursion parameters \cite[(1.2.15)]{SimonSz}.  But there is another method associated with $19^{th}$ century work of Jacobi, Markov and Stieltjes.  If $d\mu$ is a measure on $\bbR$ of compact support, one defines the $m$-function by:
\begin{equation}\label{6.50}
  m(z) = \int \frac{d\mu(x)}{x-z}
\end{equation}

The operator of multiplication by $x$ in $L^2(\bbR,d\mu)$ represented in the OPRL basis yields a tridiagonal matrix with $b_n$ on diagonal and $a_n$ off diagonal called a \emph{Jacobi matrix}.  Mark Kac once gave talks around the topic he described as ``be wise, discretize''.  Indeed, around 1980, there was a notable shift in my work and the work of many mathematical physicists (and even earlier in some of the condensed matter theoretical physics literature) toward difference rather than differential operators which are often technically simpler.  On the half line, this was often \emph{discrete Schr\"{o}dinger operators} which are Jacobi matrices with $a_n$ all equal to $1$.  One also studies its analogs on all of $\bbZ$ and on $\bbZ^\nu$ and the more general Jacobi case.  Around 2000, when I added orthogonal polynomials on unit circle (aka OPUC) to my repertoire (see \cite{OPUC1, OPUC2}) the spectral theory of the associated operator theory and its matrix representation (the CMV matrix discussed in \cite[Chapter 4]{OPUC1}) also became an interest.

If $d\mu$ has Jacobi parameters $\{a_j,b_j\}_{j=1}^\infty$, one let's $d\mu_1$ be the measure with Jacobi parameters $\{a_{j+1},b_{j+1}\}_{j=1}^\infty$ where we drop the first two parameters and knock indices down.  One can prove that the $m$-function, call it $m_1$ of $d\mu_1$ is related to $m$ by (see \cite[Theorem 3.2.4]{SimonSz})
\begin{equation}\label{6.51}
  m(z) = \frac{1}{-z+b_1-a_1^2m_1(z)}
\end{equation}
This suggests another way to recover the Jacobi parameters from $d\mu$.  From $d\mu$, compute $1/m(z)$ using \eqref{6.50}.  By \eqref{6.51}, the leading Laurent series at infinity is $-z+b_1-a_1^2z^{-1}+\text{O}(z^{-2})$ so one finds the first two Jacobi parameters.  Then use \eqref{6.51} to compute $m_1$ and iterate.  The iteration of \eqref{6.51} gives the continued fraction:
\begin{equation}\lb{6.52}
  m(z) = \cfrac{1}{-z+b_1-
           \cfrac{a_1^2}{-z+b_2 -
             \cfrac{a_2^2}{-z+b_3  \cdots}}}
\end{equation}
a representation that goes back to Jacobi, Markov and Stieltjes.  In particular, Stieltjes proved that this expansion converges on the complement in $\bbC$ of the convex hull of the support of $d\mu$; see \cite[Section 7.7]{OT}.  For this reason, the solution of the inverse problem (of going from the measure back to the Jacobi parameters from the spectral measure) is called the continued fraction approach as opposed to the first approach called the OP approach.

As Gel'fand-Levitan remark in their paper, their approach to the inverse problem for Schr\"{o}dinger operators is an analog of the OP approach to the inverse Jacobi matrix problem.  About 1985, I began to wonder what the Schr\"{o}dinger operator analog was to the continued fraction approach.  \cite{SimonInverse1} is one of the papers I am proudest of for the following reason.  I've felt one of my weaknesses is a lack of persistence.  If I couldn't solve some problem fairly quickly, I'd drop it and, while I might not totally ignore it, I didn't usually return to it without some really good idea in advance.  But this question is one I'd spend a few weeks thinking about every few years until I finally solved it 1996-97!  Early on I realized that there was an analog of \eqref{6.51} for the continuum case, namely the celebrated Riccati equation for the Weyl $m$-function
\begin{equation}\label{6.53}
  \frac{dm}{dx}(z,x) = V(x)-z-m(z,x)^2
\end{equation}
where
\begin{equation}\label{6.54}
  m(z,x) = \frac{u'(x,z)}{u(x,z)}
\end{equation}
(when $\text{Im}(z)>0$) with $u$ the solution of $-u''+Vu-zu=0$ that goes to zero as $x\to\infty$.  The issue was deciding what might be the continuum analog of a continued fraction and it turned out to be a Laplace transform!

These papers also have results for operators on $[0,a]$ with $a<\infty$, but for simplicity I will (except for some remarks) discuss the case where $V$ is bounded, in $L^1([0,\infty))$ and continuous.  In that case, I proved the following

\begin{theorem} [\cite{SimonInverse1}] \lb{T6.13}  Let $V$ be bounded, in $L^1([0,\infty])$ and continuous and let $m(z,x)$ be the Weyl $m$-function \eqref{6.54}.  Then there exists a jointly continuous function $A(\alpha,x)$ on $(0,\infty)\times [0,\infty)$ so that one has that
\begin{equation}\label{6.55}
  m(-\kappa^2, x) = -\kappa-\int_{0}^{\infty} A(\alpha,x) e^{-2\alpha\kappa}\,d\alpha
\end{equation}
whenever $\kappa>\tfrac{1}{2}\int_{0}^{\infty} |V(x)|\,dx$. Moreover, $A(\alpha,x)$ depends only on $\{V(y)|x\le y\le x+\alpha\}$ and
\begin{equation}\label{6.56}
  \lim_{\alpha\downarrow 0} A(\alpha,x) = V(x)
\end{equation}
and one has that
\begin{equation}\label{6.57}
  \frac{\partial A(\alpha,x)}{\partial x} = \frac{\partial A(\alpha,x)}{\partial\alpha} + \int_{0}^{\alpha} A(\beta,x)A(\alpha-\beta,x)\,d\beta
\end{equation}
where if $V$ is $C^1$, then $A$ is jointly $C^1$ and this equation holds in classical sense and in general it holds in a suitable weak sense.
\end{theorem}

\begin{remarks}  1. The solution of the inverse problem should be clear.  To go from $m(z,x=0)$ to $V$, one uses \eqref{6.55} to obtain $A(\alpha,x=0)$ using uniqueness of the inverse of Laplace transform ($A$ is determined by asymptotics of $m$ so, one only needs \eqref{6.58} below and not the stronger \eqref{6.55}).  Then using the integrodifferential equation \eqref{6.57}, one finds $A(\alpha,x)$ for all $x$ (indeed, $A(\alpha,x)$ is determined by $A(\beta,x=0)$ for $\beta\in [0,\alpha+x]$).  Then one finds $V$ using \eqref{6.56}.  The result is that $A(\alpha,x=0)$ for $\alpha<B$ determines $V(y)$ for $y<B$ and vice-versa.

2. Continuity of $V$ is not critical.  One shows for discontinuous $V$, one has that $A(\alpha,x)=V(x+\alpha)+E(\alpha,x)$ where is $E$ is continuous and $\lim_{\alpha\downarrow 0} E(\alpha,x) = 0$ so, if $V$ is only locally $L^1$, one has that \eqref{6.56} holds in the sense of local $L^1$ convergence rather than pointwise in $x$.  Moreover, $E$ is smoother than $V$, so, for example if $V$ is locally $C^1$ with kinks, $A$ has kinks precisely along the lines where $x+\alpha$ is a kink point.

3. Simon \cite{SimonInverse1} discusses $V$'s that are locally $L^1$ and (more or less) bounded from below.  Gesztesy-Simon \cite{SimonInverse2} extends the theory to arbitrary locally $L^1$ potentials, even those which are limit circle at $\infty$.  In place of \eqref{6.55}, for general $V$, one has the formula
\begin{equation}\label{6.58}
  m(-\kappa^2, x) = -\kappa-\int_{0}^{a} A(\alpha,x) e^{-2\alpha\kappa}\,d\alpha+\text{O}(e^{-(2a-\varepsilon)\kappa})
\end{equation}
for all $\varepsilon>0, a>0$.  This formula, for each $x$, determines $A(\alpha,x)$ from $m(-\kappa^2,x)$.  This formula alone doesn't go directly from $A(\cdot,x)$ to $m(\cdot,x)$ but one can do that by going through the inverse construction discussed in remark 1. \cite{SimonInverse2} also has an explicit way to go directly from the spectral measure, $d\rho$ to $A(\cdot,x=0)$, namely
\begin{equation}\label{6.59}
  A(\alpha,x=0) = -2\int_{-\infty}^{\infty} \lambda^{-1/2} \sin(2\alpha\sqrt{\lambda})\,d\rho(\lambda)
\end{equation}
(the integral diverges so the formula requires a proper interpretation!). Ramm-Simon \cite{SimonInverse3} discuss asymptotics of $A(\alpha,x=0)$ as $\alpha\to\infty$ when $V(x)$ has very nice behavior at infinity.
For further developments, see Gesztesy-Sakhnovich \cite{FritzA} and Remling \cite{RemlingA}.
\end{remarks}

One important consequence of this work is a local version of the following theorem of Borg \cite{BorgBM} and Marchenko \cite{MarchBM}:

\begin{theorem} [Borg-Marchenko Theorem] \lb{T6.14} Let $V_1$ and $V_2$ be two locally $L^1$ functions on $[0,\infty)$ and let $m_1, m_2$ be the $m$-functions associated to $-\tfrac{1}{2}\tfrac{d^2}{dx^2}+V_j$.  Then $V_1(x)=V_2(x)$ for all $x\in [0,\infty)\iff m_1(z)=m_2(z)$ for all $z \in (-\infty,0)$ with $|z|$ large.
\end{theorem}

The local version, which originally appeared in \cite{SimonInverse1}:

\begin{theorem} [Local Borg-Marchenko Theorem] \lb{T6.15} Let $V_1$ and $V_2$ be two locally $L^1$ functions on $[0,\infty)$ and let $m_1, m_2$ be the $m$-functions associated to $-\tfrac{1}{2}\tfrac{d^2}{dx^2}+V_j$.  Then $V_1(x)=V_2(x)$ for all $x\in [0,a)\iff$ asymptotically for $z \in (-\infty,0)$ one has that:
\begin{equation}\label{6.60}
  \forall_{\varepsilon>0}\,\exists_{C,R>0}\,\forall_{z \in (-\infty,R)}\, |m_1(z)-m_2(z)| \le C e^{-|z|(2a-2\varepsilon)}
\end{equation}
\end{theorem}

This follows easily from Theorem \ref{T6.13} because \eqref{6.60} is equivalent to $A_1(\alpha,x=0)=A_2(\alpha,x=0)$ for all $x\in[0,a]$ and by the differential equation, that happens if and only if $V_1(x)=V_2(x)$ for all $x\in[0,a]$.  While this proof is quite illuminating, it does require one to develop an elaborate machinery.  In \cite{GeszSiBM}, Gesztesy-Simon found a simple direct proof of Theorem \ref{T6.15} and then Bennewitz \cite{BeneBM} showed that an argument based on Borg's method in \cite{BorgBM} provides a really short proof.

One memorable aspect of this work was a talk I gave about it at Rutgers while the work was being written up.  I was excited to give it because Gel'fand (of Gel'fand Levitan and other fame), who had moved there after he left Russia was in the audience.  He asked some pointed questions, made some positive comments and, in particular, pointed out that one positive element was that it was easy to extend to matrix valued potentials. That was in the fall of 1997 when I was 51 and Gel'fand was 84.  I remember thinking to myself ``Gee, I hope I'm that sharp when I'm 84.''  I mentally paused and then thought ``No, you wish you were that sharp when you were 48!''.

Using \cite{SimonInverse1, SimonInverse2}, Gesztesy and I proved \cite{GSConn} that

\begin{theorem} \lb{T6.15A} Let $V_0$ and $V_1$ be two potentials on $[0,\infty)$ so that the two operators $-\tfrac{1}{2}\tfrac{d^2}{dx^2}+V_j,\,j=0,1$ both have discrete spectrum which are identical.  Then there is a smooth family of potentials $V_t;\, 0\le t\le 1$ interpolating between them which each have the same spectrum.
\end{theorem}

\begin{remarks}  1.  The proof shows that if $V_j$ are both $C^k$, then the $V_t$'s we construct are also $C^k$.

2.  Using $A$'s, the proof is almost trivial.  One takes $A_t=tA_1+(1-t)A_0$!

3. This result is interesting because the analogous result on the whole line is open.  For example, it is not even known if the $C^\infty$ potentials going to $\infty$ at $\pm\infty$ whose spectrum is the same as the harmonic oscillator is an connected set!
\end{remarks}

\bigskip

Useful tools in inverse theory are so called trace formula that provide the potential as an integral of some kind of spectral or scattering object.  Gesztesy and I \cite{GSxi} found a fairly general trace formula with further developments in papers we wrote with others \cite{GHSxi, GHSZxi1, GHSZxi2,GSxi2}.  If $V$ is a continuous function on $\bbR$ bounded from below on defines $H=-\tfrac{1}{2}\tfrac{d^2}{dx^2}+V$ and $H_{D;x}$ is the same operator with a Dirichlet boundary condition forced at $x$.  $\xi(x,\lambda)$ is then the Krein spectral shift (see, e.g. \cite[Sections 5.7-5.8]{OT}) going from $H_{D:x}$ to $H$.  Thus
\begin{equation}\label{6.61}
  \tr(e^{-tH}-e^{-tH_{x;D}}) = t\int_{0}^{\infty} e^{-t\lambda}\xi(x,\lambda)\,d\lambda
\end{equation}
and the general trace formula in \cite{GSxi} is that
\begin{equation}\label{6.62}
  V(x) = \lim_{\alpha\downarrow 0}\left[E_0+\int_{E_0}^{\infty} e^{-\alpha\lambda}[1-2\xi(x,\lambda)]\,d\lambda\right]
\end{equation}
for an $E_0$ below the bottom of the spectrum of $H$.

\bigskip

Another set of inverse type problems that I studied in five papers with Gesztesy (one also jointly with Del Rio who first told me of Theorem \ref{T6.16} below) \cite{GSPartial1, GSPartial2, GSPartial3, GSPartial4, GSPartial5} was motivated in part by a remarkable theorem of Hochstadt-Lieberman \cite{HochL} (we state this with Dirichlet boundary conditions and continuous $V$ although they hold more generally):

\begin{theorem} \lb{T6.16} Let $V$ be a continuous on $[0,1]$ and let $H$ be $-\tfrac{1}{2}\tfrac{d^2}{dx^2}+V$ on $L^2(0,1;dx)$ with $u(0)=u(1)=0$ boundary conditions.  Then the set of eigenvalues of $H$ and $V$ on $[0,1/2]$ determine $V$.
\end{theorem}

Typical of our results is that if one considers the $H$ operator with boundary conditions $u(a)=u(b)=0$ with $0\le a< b\le 1$ on $L^2(a,b;dx)$, then, for $a\in (0,1)$, the spectra of the three operators on $[0,a]$, $[a,1]$ and $[0,1]$ determine $V$ \cite{GSPartial5}! The point of our papers is that eigenvalues are zeros of suitable $m$-functions and factorization theorems for analytic functions together with known asymptotics (and some partial information on V) determine $m$ and so by a version of Borg-Marchenko for bounded intervals they determine $V$.  For some results, we also use Phragm\'{e}n-Lindel\"{o}f theorems to allow us to only need information on some of the eigenvalues.

\bigskip

I close this subsection on scattering theory ideas by mentioning two of my papers that have somewhat unusual applications of the trace class completeness theory (the Kato-Rosenblum theorem mentioned earlier in this section).  One of these exploits an extension of the Kato-Rosenblum trace class scattering theorem due to Pearson \cite{PearTrace}:

\begin{theorem} \lb{T6.17} Let $A$ and $B$ be two self-adjoint operators and $J$ a bounded operator so that $AJ-JB$ is trace class.  Then
\begin{equation}\label{6.62A}
  \Omega^\pm(A,B;J) = \lim_{t\to\mp\infty} e^{itA}Je^{-itB}P_{ac}(B)
\end{equation}
exist.
\end{theorem}

\begin{remarks} 1. For a proof and a discussion of implications, see Reed-Simon \cite[Theorem XI.7]{RS3}.

2.  If $A$ and $B$ are unbounded, one has to worry about the meaning of ``$AJ-JB$ is trace class''.  The more precise hypothesis is that there is a trace class operator $C$ so that for all $\varphi\in D(A)$ and $\psi\in D(B)$, one has that
\begin{equation*}
  \jap{\varphi,C\psi} = \jap{A\varphi,J\psi} - \jap{\varphi,JB\psi}
\end{equation*}

3. In applications, it is a useful (and easy to prove) fact that if $J$ is compact, the limit in \eqref{6.62A} exists and is $0$.

4. One example that shows the power of Pearson's extension is that it implies that if $(A+i)^{-1}-(B+i)^{-1}$ is trace class, then the ordinary wave operators, \eqref{6.39}, exist and are complete (a result sometimes called the Birman-Kuroda theorem).  For this assumption implies the one of Pearson's theorem with $J= (A+i)^{-1}(B+i)^{-1}$.  Then using remark 3, we get existence for $J=(B+i)^{-2}$.  Applying that limit to vectors of the form $(B+i)^2\psi$ proves existence of the ordinary wave operators and then Theorem \ref{T6.10} implies completeness.

5. The following extension (and corollary of) of Pearson's theorem is useful: if
\begin{equation}\label{6.63}
    (A+i)^{-1}[AJ-JB](B+i)^{-1}
\end{equation}
is trace class, then Pearson's theorem applies with $J_1=(A+i)^{-1}J(B+i)^{-1}$ and, by mimicking the argument in remark 4, one sees that the limits in \eqref{6.62A} exist.
\end{remarks}

Davies and I \cite{DavSiLR} studied very general 1D Schr\"{o}dinger operators, $H=-\tfrac{d^2}{dx^2}+V$.  Let $J_r$ be a smooth function which is $0$ on $(-\infty,-1)$ and $1$ on $(1,\infty)$ and $J_\ell = 1-J_r$.  Then under great generality on $V$ (e.g. any bounded $V$; see below), one sees that Pearson's theorem implies that $\Omega^\pm(H,H;J_r)$ and $\Omega^\pm(H,H;J_\ell)=1-\Omega^\pm(H,H;J_r)$ exist and it is not difficult to show that they are projections $P^\pm_r$ and $P^\pm_\ell$.  Letting $\calH^\pm_r=\ran\,P^\pm_r$ (and similarly for $\ell$), one sees that:

\begin{theorem} [Davies-Simon \cite{DavSiLR}] \lb{T6.18} Let $V$ be a potential on $\bbR$ whose positive part is in $L^1_{loc}$ and negative part is a form bounded perturbation of $-\tfrac{d^2}{dx^2}$ with relative bound less than $1$.  Let $\calH_{ac}$ be the absolutely continuous subspace for $H$.  Then
\begin{equation}\label{6.64}
  \calH_{ac} = \calH^-_r\oplus\calH^-_\ell = \calH^+_r\oplus\calH^+_\ell
\end{equation}
Moreover if $\varphi\in\calH^-_r$, then as $t\to\infty$, one has that for any $R$ the probability that $e^{-itH}\varphi$ lies in $\{x\,|\,x<R\}$ goes to zero and similarly for the other four subspaces.
\end{theorem}

\begin{remarks} 1. For discussion of form bounded perturbations, see, for example, \cite[Section 7.5]{OT}.  Under the hypotheses, one can define $H$ as a closed form sum on $Q(-\tfrac{d^2}{dx^2})\cap Q(V_+)$.

2.  What this theorem says is that every $\varphi\in\calH_{ac}$ is the sum of a piece that goes in $x$ to plus infinity as $t\to\infty$ and a piece that goes to minus infinity.  Similarly, there is a decomposition as $t\to -\infty$.

3.  There are also results in \cite{DavSiLR} on operators on $\bbR^\nu$ which are periodic in all directions but one.

4.  \cite{DavSiLR} also has a powerful method for sometimes eliminating s.c. spectrum called the twisting trick.  I will not say more about it except to note that subsequently, Davies \cite{DavTwist} used a variant for a penetrating analysis of double well Hamiltonians, a subject I will discuss further in Section \ref{s8}.
\end{remarks}

Once you have this set up there is a natural notion of reflectionless: we say that a potential is \emph{reflectionless} if and only if $\calH^+_\ell=\calH^-_r$, that is all states that come in from the left go out entirely on the right with no reflection. Deift and I \cite{DSAP} conjectured in 1983 that a 1D almost periodic Schr\"{o}dinger operator is reflectionless in this sense.  A different notion of reflectionless arose in the theory of solitons (see, for example, \cite[Chapter II]{Soliton}) and there has arisen a huge literature on this notion capped by Remling's characterization \cite{RemlingRL} of right limits of a.c. spectrum mentioned in the remarks after Theorem \ref{T6.3}.  Namely, a 1D Schr\"{o}dinger operator, $H$, is called \emph{spectrally reflectionless} if and only if for all $x$ and Lebesgue a.e. $E$ in the a.c. spectrum of $H$, one has that (with $G$ the Green's function, i.e. integral kernel of the resolvent of $H$)
\begin{equation}\label{6.65}
  \lim_{\varepsilon\downarrow 0} \text{Re}\, G(x,x;E+i\varepsilon)=0
\end{equation}

In 2010, more than 25 years after the conjecture of Deift-Simon, Breuer, Ryckman and Simon \cite{BRS} proved that conjecture by proving the much more general (they also have this result for Jacobi matrices and two sided CMV matrices):

\begin{theorem} [\cite{BRS}] \lb{T6.19} A one dimensional Schr\"{o}dinger operator is spectrally reflectionless if and only if it is reflectionless in the sense of Davies-Simon.
\end{theorem}

The other trace class paper that I should mention is by Simon and Spencer \cite{SiSp}.  Typical of our results is

\begin{theorem} [\cite{SiSp}] \lb{T6.21} Let $h$ be a discrete Schr\"{o}dinger operator on $\bbR$ so that $\limsup_{n\to\infty} |b_n|=\limsup_{n\to -\infty} |b_n|=\infty$.  Then $h$ has no a.c. spectrum.
\end{theorem}

\begin{remarks} 1.  I want to emphasize this involves $\limsup$ rather than $\lim$.  If it were $\lim$, the spectrum would be discrete but it is easy to construct examples with the $\limsup$ condition but spectrum all of $\bbR$.  This result says that tunnelling through high barriers destroys ac spectrum which is an intuitive result.

2.  The proof is an easy application of the Birman-Kuroda theorem.  One picks a two sided subsequence, $\{n_j\}_{j=-\infty}^\infty$ going to $\pm\infty$ as $j\to\pm\infty$ and so that $\sum_{j=-\infty}^{\infty} |b_{n_j}|^{-1} < \infty$ and by simple estimates shows that if $h_\infty$ is $h$ with the sites at all $n_j$ removed (what you get if $b_{n_j}$ were taken to $\pm\infty$), then $(h+i)^{-1}-(h_\infty+i)^{-1}$ is trace class.  So since $h_\infty$, as a direct sum of finite matrices, has no ac spectrum, neither does $h$.

3. Simon-Spencer \cite{SiSp} use this idea in many other ways.  Not only the obvious ones like continuum Schr\"{o}dinger operators where the barriers not only have to be high but also not too narrow but also some limited results in higher dimensions.  There is even a proof that certain one dimensional random discrete Schr\"{o}dinger have no ac spectrum (as we'll recall in Section \ref{s11}, more is true).

4. Once I started working in the theory of orthogonal polynomials, I learned that Dombrowski \cite{Domb} had the same idea eleven years prior to us in a different but closely related (and also simpler!) context.  Namely, she considered Jacobi matrices, $J$, and proved that if $\liminf_{n\to\infty} a_n = 0$, then $J$ has no ac spectrum.  For one picks a subsequence $n_j$ so that $\sum_{j=1}^{\infty} a_{n_j} < \infty$.  Dropping those $a_{n_j}$'s gives a trace class perturbation which turns the matrix into a direct sum of finite matrices and one uses the Kato-Birman theorem!
\end{remarks}

\bigskip

I close this section on N-Body Quantum Mechanics by mentioning that my books by Reed-Simon \cite{RS2, RS3, RS4} and Cycon et al. \cite{CFKS} have been a useful introduction to many researchers trying to learn about the subject.

\section{Magnetic Fields in NRQM} \lb{s7}

One of the things that made the 70's so fruitful for my research is that I kept finding subareas where mathematical physicists hadn't looked, so I could wander around an orchard and pick the low hanging fruit and write papers which, because they set the framework, could be widely used and quoted later.  One of these areas was physics of NRQM with magnetic fields where Avron Herbst and I wrote a series of papers \cite{AHS1, AHS2, AHS3, AHS4} with lots of intriguing results, a major part of this section.  The basic objects studied are magnetic Hamiltonians
\begin{equation}\label{7.1}
  H(\boldsymbol{a},V) = -(\boldsymbol{\nabla}-i\boldsymbol{a}(x))^2+V(x)
\end{equation}

It may seem strange to imply that there wasn't earlier work on this subarea given that, for example, fifteen years before, Ikebe-Kato \cite{IK} had proven a fairly general result on self-adjointness in magnetic field, at least when the vector potential, $a_j(x)$, is $C^2$.  Kato, as discussed in the last Section, included magnetic fields in his Kato inequality paper as did some of my followup work.  But this earlier work focused almost entirely on self-adjointness or issues where one treated the magnetic vector potential as just a coefficient in the PDE but didn't focus on the physics underlying the magnetic field nor the special roles of gauge invariance nor the role of the non-commutativity of the components of $-i\nabla_j-a_j$.  We were really the first researchers to look at these problems as mathematical physicists rather than as mathematicians.

Avron and Herbst were following up on their earlier beautiful work on constant electric field \cite{AvronHerbstStark}.  They may have asked me to join them for magnetic fields because of my earlier work about a year before related to Theorem \ref{T6.9}. To begin with, I proved

\begin{theorem} [\cite{SiDiam}] \lb{T7.1}  Let
\begin{equation}\label{7.2}
  E(\boldsymbol{a},V) = \inf_{\varphi\in L^2(\bbR^\nu)} \jap{\varphi,H(\boldsymbol{a},V)\varphi}
\end{equation}
Then for any $\boldsymbol{a},V$, we have that
\begin{equation}\label{7.3}
  E(\boldsymbol{a},V) \ge E(\boldsymbol{a}=0,V)
\end{equation}
\end{theorem}

\begin{remarks} 1. ``any $\boldsymbol{a},V$'' of course means for which one can reasonably define the operators and for which they are bounded below.

2. In other words, energies go up if one turns on any magnetic field.

3. Formally, the proof is easy: if $\varphi=|\varphi|e^{i\psi}$, then $(\boldsymbol{\nabla}-i\boldsymbol{a})\varphi =(\boldsymbol{\nabla}|\varphi|-i\boldsymbol{a}|\varphi|+i\boldsymbol{\nabla}\psi|\varphi|)e^{i\psi}$ so $|\boldsymbol{\nabla}-i\boldsymbol{a})\varphi|\ge \boldsymbol{\nabla}|\varphi|$.  Squaring and integrating, one gets
\begin{equation}\label{7.4}
  \jap{\varphi,H(\boldsymbol{a},V))\varphi} \ge \jap{|\varphi|,H(\boldsymbol{a}=0,V))|\varphi|}
\end{equation}
from which \eqref{7.3} results.  A rigorous proof isn't much harder.
\end{remarks}

Since \eqref{7.4} involves $|\varphi|$ and if $\varphi$ is antisymmetric in particle coordinates, $|\varphi|$ is symmetric, this argument works for bosons but not fermions.  Moreover, if a particle has spin, one adds a $\boldsymbol{\sigma}\cdot\boldsymbol{B}$ term which destroys \eqref{7.3}, and the theorem only holds for spinless particles (in fact, Lieb (his result and proof were published as an appendix to \cite{AHS1}) proved in constant magnetic field for the Pauli equation of spin $\tfrac{1}{2}$ electrons, energies went down when magnetic fields are turned on; for a while there was a conjecture that for the Pauli equation Theorem \ref{T7.2} below held with the direction of the inequality reversed for general magnetic field but Avron and I \cite{ASPara} found a counterexample) so I wrote a paper entitled ``Universal Diamagnetism for Spinless Bosons'' and submitted it to Phys. Rev. Lett.  The referees report was memorable.  Essentially it said \emph{Since there are no stable spinless bosons in nature, the result of this paper is of limited physical applicability. But it is nice to see something nontrivial proven in just a few lines, so this paper should be accepted as an example to others.}  So the paper was accepted \cite{SiDiam}!

The next step illustrates the dynamics of the brown bag lunch at Princeton. At one, I described Theorem \ref{T7.1} and mentioned that I conjectured that this was a zero temperature result and that there should be a finite temperature result that was an inequality between integral kernels of semigroups, and I was working on it. Almost immediately, Ed Nelson interjected: ``You know that follows from the stochastic integral magnetic field version of the Feynman Kac formula.'' Stirred by this, I found a direct proof from Kato's inequality which, in typical fashion, Ed refused to be a coauthor of. These inequalities which I dubbed \emph{diamagnetic inequalities} are used often in the study of quantum mechanics in magnetic fields (some tried to call them Nelson-Simon inequalities but my name was catchier).  They say:

\begin{theorem} [Diamagnetic Inequalities] \lb{T7.2} Let $\boldsymbol{a}\in L^2_{loc}$, $V_+\in K_\nu^{loc}, V_-\in K_\nu$.  Then $C^\infty_0(\bbR^\nu)$ is a form core for $H(\boldsymbol{a},V)$ and pointwise
\begin{equation}\label{7.5}
  |\exp[-tH(\boldsymbol{a},V)]\varphi| \le \exp[-tH(\boldsymbol{a}=0,V)] |\varphi|
\end{equation}
\end{theorem}
The reader will recognize that this follows from Theorem \ref{T6.9} and \eqref{6.32}.  Of course, this is only if $\boldsymbol{a}$ is smooth.  The full result I proved in Simon \cite{SimonMaxMinForm} and is obtained by using the Trotter product formula and gauge transformations.  Many years later, Hundertmark and I \cite{HundSiMag} wrote a paper that proved diamagnetic inequalities for differences of semigroups with Neumann and Dirichlet boundary conditions.  This allowed a quick proof of invariance under change of boundary condition of the density of states in magnetic field, something whose prior proofs had been complicated.  This paper also had a new quadratic forms proof of diamagnetic inequalities.

\bigskip

I turn now to the results in the papers with Avron and Herbst.  A fraction of them specifically discuss constant magnetic field.  Classically, 2D electrons in such a field go in circles in two dimensions while is 3D they can move out to infinity in the direction of the field.  Quantum mechanically, one has the celebrated Landau levels - in 2D pure point spectrum (although of infinite multiplicity).  As we'll see this produces various enhanced binding scenarios.  Here are some of the main results.

1. \emph{Center of Mass Reduction}.  The physics of $N$ particles in \textbf{constant} magnetic field is invariant under translations of all of the particle coordinates but the mathematics of the reduction can be very different from what is described in Theorem \ref{T6.1}.  If the total charge is not $0$, the unitaries associated with translations in different directions perpendicular to the field no longer commute but only commute up to a phase.  Put differently, the components of the conserved generators of translation (called quasimomentum) obey canonical commutation relations.  It is surprising that this subject wasn't worked out in the physics literature before us, but it wasn't.  Many of the citations of this paper are in the physics literature as seen by the fact that it has 355 citations on Google Scholar but only 26 on MathSciNet!

2. \emph{Borel summability and dilation analyticity of atoms in constant magnetic field}. In paper III \cite{AHS3}, Borel summability of the perturbation series in magnetic field strength for constant field is proven for eigenvalues of multielectron atoms which are discrete and simple on the space where $L_z$, the total azimuthal angular momentum about the axis of the field, is fixed.  Paper I \cite{AHS1} discusses dilation analyticity for hydrogen in a constant magnetic field and paper III \cite{AHS3} discusses this for multielectron atoms.  Stability is a key technical issues.  These results (and more for hydrogen in constant field) were announced in \cite{AHSAnon1}.

3. \emph{Large B}. Paper IV \cite{AHS4} discusses asymptotic behavior for hydrogen in a large constant field $B$ as $|B|\to\infty$.  Because terrestrial magnetic fields in natural units are tiny, this is of interest only in astrophysical contexts.  If
\begin{align}
  H_0(B) &= (-i\boldsymbol(\nabla)-\tfrac{1}{2}B\hat{z}\times\boldsymbol{r})^2 \lb{7.6} \\
  H(B,\lambda) &= H_0(B)-\lambda|\boldsymbol{r}|^{-1} \lb{7.7}
\end{align}
then, by scaling $H(B,z)$ is unitarily equivalent to $B(H(1,zB^{-1/2}))$, so large B with $z=1$ is equivalent to studying fixed field and small $\lambda$.

Since the magnetic field binds in the two dimensions perpendicular to $\boldsymbol{B}$, this is effectively a 1D weak coupling problem where I'd discovered small coupling asymptotic series (see the discussion in Section \ref{s8}) but the decay is slower than in my earlier work.  However, one can modify that work and obtain the first few terms which are very complicated (the leading order is $\log(B)$ as first suggested by Ruderman \cite{Rud}) but there are $\log(\log(B))$ and $\log(B)^{-1}$ terms!

4. \emph{The Hydrogen Zeeman Ground State has $L_z=0$}. In paper III \cite{AHS3}, we proved that for hydrogen in magnetic field, the ground state has $L_z=0$.  For zero magnetic field, the ground state is positive (see, for example, \cite[Section XIII.12]{RS3}) so in an azimuthally symmetric potential, the ground state has $L_z=0$.  But paper I \cite{AHS1} constructs examples of azimuthally symmetric potentials so that for small field the lowest part of the spectrum is discrete and the lowest energy is not $L_z=0$ (Lavine-O'Carroll \cite{LOC} also have such examples).  The deep result that for attractive Coulomb potentials the ground state has $L_z=0$ uses the monotonicity of $V(r)$ in $r$ and correlation inequalities as extended to quantum systems as discussed in Section \ref{s3}.  For a proof of a more general result which doesn't use correlation inequalities, see Grosse-Stubbe \cite{GSt}.

5. \emph{Enhanced Binding}. As discussed in Section \ref{s8}, if $V\le 0$ is in $C_0^\infty(\bbR^3)$, then for small $\lambda$, the operator $-\Delta+\lambda V$ has no eigenvalues in $(-\infty,0)$ but for $C_0^\infty(\bbR)$, $-\tfrac{d^2}{dx^2}+\lambda V$ always has at least one negative eigenvalue.  Things are different if $-\Delta$ is replaced by $H_0(B)$ (given by \eqref{7.6}) with $B\ne 0$.  For
\begin{equation}\label{7.8}
  H_0(B) = \wti{H}_0(B)\otimes\bdone+\bdone\otimes\left(\tfrac{-d^2}{dx^2}\right)
\end{equation}
on $L^2(\bbR^3)=L^2(\bbR^2)\otimes L^2(\bbR)$ where $\wti{H}_0(B)$ has point spectrum ${\{(2n+1)B\}_{n=0}^\infty}$, each of infinite multiplicity.

Further analysis (see \cite[Section 3]{AHS1}) shows that if one restricts to $L_z=m$, one has point spectrum ${\{2\max(-m,0)+n+1\}_{n=0}^\infty} \equiv\{E_m(n)\}_{n=0}^\infty$, each of multiplicity $1$.  Thus $H_0(B)\restriction L_z=m$ is a direct sum of copies of $-\tfrac{d^2}{dx^2}+E_m(n)$.  If $V\le 0$ is azimuthal (so commuting in $L_z$) each space with $L_z=m$ and $m\ge 0$ has at least one eigenvalue of $H_0(B)+V$ below its essential spectrum, so we conclude that

\begin{theorem} [Avron, Herbst and Simon\cite{AHS1}] \lb{T7.3} If $V\in C_0^\infty(\bbR^3)$ is a non-negative (not identically $0$) function of only $z$ and $\rho=\sqrt{x^2+y^2}$, then $H_0(B)+V$ for any $B\ne 0$ has essentially spectrum $[|B|,\infty)$ and infinitely many eigenvalues in $(-\infty,|B|)$.
\end{theorem}

\begin{remark} By diamagnetic inequalities, if $\lambda$ is so small that $-\Delta+\lambda V$ has no negative spectrum, then all the eigenvalues are in $[0,|B|)$ for all $B\ne 0$.  Moreover, for $B$ fixed, for all small $\lambda$ there is exactly one eigenvalue for each of $H_0(B)+V\restriction(L_z=m)$ for each $m\ge 0$ (if $B>0$).
\end{remark}
6. \emph{Negative Ions}. One form of enhanced binding discovered by Avron-Herbst-Simon involves negative ions in magnetic field.  In nature, most neutral atoms do not bind any extra electrons although there is no rigorous proof that this is even true for $He^-$.  By using ideas discussed in point 5 above, AHS in an announcement \cite{AHSAnon2} in Phys. Rev. Lett. and in paper III \cite{AHS3} prove that every neutral atomic Hamiltonian in non-zero constant magnetic field will bond at least one additional electron.

7. \emph{Magnetic Bottles}. In paper I \cite{AHS1}, we looked at the question of whether one can produce Hamiltonians with compact resolvent with just magnetic field alone (i.e. $[H(a,V=0)+1]^{-1}$ compact).  It is easy to see how to do this in even dimension, e.g. $\bbR^2$ by making $B_z\to\infty$ as $x^2+y^2\to\infty$ (even though there is no $z$ direction, $B_z \equiv \partial_xa_y-\partial_ya_x$ is defined), but, apriori, it is not clear how do this in odd dimension.  However, we found lots of examples, e.g. in 3D, $\boldsymbol{B}=(x,y,-2z)$ (since $\boldsymbol{\nabla}\cdot\boldsymbol{B}=0$, there is $\boldsymbol{a}$ with $\boldsymbol{B}=\boldsymbol{\nabla}\times\boldsymbol{a}$).

8. \emph{General Spectral and Scattering Theory in constant magnetic field}. In papers I, II and III \cite{AHS1, AHS2, AHS3}, Avron, Herbst and I tried to extend much of the theory of $2-$ and $N-$body systems when $-\Delta$ is replaced by $H_0(B)$ (given by \eqref{7.6}).  Much of it is straightforward but there are interesting twists.  For example, for perturbations, $V$ of $-\Delta$, the Agmon-Kato-Kuroda theory needs $(1+|x|)^{-1-\varepsilon}$ decay but the (2-body) analog for $H_0(B)$ developed in \cite{AHS1} only needs $(1+|z|)^{-1-\varepsilon}$ decay and any kind of decay in the orthogonal directions.  Cook's method as developed in \cite{AHS1} for existence of $\Omega^{\pm}(H_0(B)+V,H_0(B))$ needs $(1+|z|)^{-1-\varepsilon}$ decay in $V$ but allows growth (!) in the $x$ and $y$ directions by less than an inverse Gaussian since Landau levels decay in a Gaussian manner!  Because of the unusual form of reduction of the center of mass discussed in \cite{AHS2}, the HVZ theorem proved there is more involved than in the zero field case.

\bigskip

I turn now to some later work on NRQM in magnetic field.  AHS, because of its focus on the constant field case, left an important issue on the table.  The standard analysis using Weyl's criterion \cite[Problem 3.14.5]{OT} and localized test functions proves that $\sigma_{ess}(H(\boldsymbol{a},V))=[0,\infty)$ if $\boldsymbol{a}(x)\to 0$ and $V(x)\to 0$ as $x\to\infty$.  But this gets the physics wrong.  If say $\nu=2$ and
\begin{equation}\label{7.9}
  B_z(x,y) = C(1+\rho)^{-\alpha}
\end{equation}
then in any fixed gauge, $\boldsymbol{a}\to 0$ if and only if $\alpha >1$ while one expects that $\sigma_{ess}(H(\boldsymbol{a},V))=[0,\infty)$ so long as $\alpha>0$, i.e. rather than requiring $\boldsymbol{a}\to 0$, we should only need $\boldsymbol{B}\to 0$ at $\infty$.  The first theorem of this type was proven in the PhD. thesis of my student Keith Miller \cite{Miller} which he never published since he decided not to take an academic job.  He used test functions and Weyl's criterion but functions with an $x$-dependent phase factor that implements a change of gauge in which the new $\boldsymbol{a}$ is small on the support of the test function.  The standard reference for this is a joint paper that he and I wrote \cite{MS} which I will turn to shortly.  There is now a huge literature on the this issue which is summarized in Last-Simon \cite{LSHVZ}, a paper that has an HVZ type result in terms of limits at infinity of magnetic fields (we limited ourselves to bounded, uniformly H\"{o}lder continuous magnetic fields).

Miller and I \cite{MS} found the following remarkable fact:

\begin{theorem} [Miller-Simon \cite{MS}] \lb{T7.4} Let $H(\alpha)$ be the quantum Hamiltonian of a $2D$ particle with $V=0$ and magnetic field given by \eqref{7.9}.  For all $\alpha>0$, $\sigma(H(\alpha))=[0,\infty)$.  If $0<\alpha<1$, $H(\alpha)$ has dense pure point spectrum in all of $[0,\infty)$.  If $\alpha=1$, there is $E_0$ (depending on $C$) so that the spectrum is dense pure point on $[0,E_0]$ and purely a.c. on $[E_0,\infty)$.  If $\alpha>1$, the spectrum is purely a.c.
\end{theorem}

\begin{remark} There is an arithmetic mistake in the calculation of $E_0$ in \cite{MS} that was recently noted and corrected by Avramska-Lukarska et al \cite{ALHK}.
\end{remark}

The proof is easy.  That $\sigma(H(\alpha))=[0,\infty)$ follows from Miller's argument about $\boldsymbol{B}\to 0$.  On the other hand since $B$ is azimuthal symmetric, one can pick a gauge in which $H(\boldsymbol{a},V)$ commutes with rotations and look at fixed $L_z=m$ where the operator is unitarily equivalent to $-\tfrac{d^2}{dx^2}+V_{\alpha,m}(x)$ on $L^2([0,\infty))$.  If $0<\alpha<1$, then $V_{\alpha,m}\to\infty$ so each $H(\alpha)\restriction L_z=m$ has purely discrete spectrum (although they have to fit together to give dense point spectrum)!  If $\alpha=1$, each $V_{\alpha,m}\to E_0$ and if $\alpha>1$, each $V_{\alpha,m}\to 0$ at a power rate.

There is a interesting classical physics underlying this.  If $\alpha>1$, the classical orbits are all unbounded, if $0<\alpha<1$, all orbits are bounded while if $\alpha=1$ the orbits are either bounded or unbounded depending on whether $E<E_0$ or $E>E_0$.

While the physics is not related, there is an intriguing result of Hempel et. al \cite{HHHK} that has similar mathematics.  $H=-\Delta+\cos |x|$ on $L^2(\bbR^3)$ has alternating bands of a.c. and dense point spectrum!  The argument is similar to that of \cite{MS}: near $\boldsymbol{x}=(x,0,0)$ with $x$ large, $H$ looks like $-\tfrac{d^2}{dx^2}+\cos x-\tfrac{d^2}{dy^2}-\tfrac{d^2}{dz^2}$ which lets one show that $\sigma(H)=[E_0,\infty)$ for suitable $E_0$.  On the other hand, if $S$ is the (band) spectrum of $-\tfrac{d^2}{dx^2}+ \cos x$ on $L^2(\bbR)$, the restriction of $H$ to each fixed angular momentum space is $-\tfrac{d^2}{dr^2} + \tfrac{\ell(\ell+1)}{r^2}+\cos r$ which has a.c. spectrum on $S$ and eigenvalues in the gaps.

\bigskip

Finally, I have two papers \cite{ASCont, SiCont} that looked at continuity in $B$ of various objects associated to $H_0(B)+V$ when $V$ is periodic.  My interest was originally sparked by a lack of continuity in frequency, $\alpha$, of the density of states (and the spectrum) of the almost periodic Jacobi matrix (see section \ref{s9}), ${Hu(n)=u(n+1)+u(n-1)+\lambda \cos(\pi\alpha n+\theta)u(n)}$ which is supposed to be a strong coupling approximation of a $2D$ periodic operator in magnetic field (Thouless explained the apparent puzzle of the continuity of the density of states shown in \cite{SiCont} and this lack of continuity of the above $H$ in $\alpha$ for fixed $\theta$. The correct analog is not the density of states associated to the operator $H$ for fixed $\theta$.  Rather the correct analog is the integral over $\theta$ and this is known to be continuous in $\alpha$ \cite[Theorem 3.3]{ASAP2}). The first paper \cite{SiCont} proves continuity of the density of states in $B$.  The key is to note that while $\boldsymbol{a}$ in any fixed gauge is misbehaved at infinity the diagonal heat kernel $e^{-t(H_0(B)+V)}(x,x)$ is  gauge invariant and so periodic in $x$.  This yields continuity of the Laplace transform of the density of states.  \cite{ASCont} deals with the more subtle issue of continuity of the spectrum in $B$.

\section{Quasi-classical and Non--quasi-classical limits} \lb{s8}

The structures of quantum and classical mechanics are very different, so it is remarkable that a world we believe is described by quantum theory is consistent in the right realm with classical mechanics.  Thus their connection has been a compelling subject for both physicists and mathematicians.  I have a lot of work of work that explores their connection and when the naive connection needs modification.

One of the simplest connections goes back to Weyl \cite{WeylQC} in 1912, a number of years before the new quantum mechanics and so, obviously, done in different context, namely classical electromagnetism.  Looking at $-\hbar^2\Delta$ below a fixed energy $E$ at small $\hbar$ is the same as looking at $-\Delta$ below a very large energy.  Weyl fixed a compact set $\Omega\subset\bbR^\nu$ with smooth boundary and looked at the number, $N_\Omega(E)$, of eigenvalues below $E$ for $-\Delta_D^\Omega$, the Laplacian in $\Omega$ with Dirichlet boundary conditions and proved that
\begin{equation}\label{8.1}
  \lim_{E\to\infty} \frac{N_\Omega(E)}{E^{\nu/2}} = \frac{\tau_\nu}{(2\pi)^\nu}|\Omega|
\end{equation}
where $\tau_\nu$ is the volume of the unit ball in $\bbR^\nu$ and $|\Omega|$ is the volume of $\Omega$ (an exposition of the proof of \eqref{8.1}, close to Weyl's, can be found, for example, in \cite[Section 7.5]{OT} which also explains the interesting history of how Weyl came to consider this problem).  The right side of \eqref{8.1} has two volumes in $\bbR^\nu$ and can be interpreted as a volume in phase space.  $-\Delta_D^\Omega$ is the quantum Hamiltonian when the units are $\hbar=1=2m$, so $E=p^2$ and the volume in phase space of $\{(x,p)\,|\, x\in\Omega,p^2\le E\}$ is $\tau_\nu E^{\nu/2}$.  Thus \eqref{8.1} says that for large $E$, $N_\Omega(E)$ looks like the volume in phase space where the energy is less than $E$ times $(2\pi)^{-\nu}$.  $\hbar=1\Rightarrow h=2\pi$ so this says each state takes a volume of $h^\nu$.

We now shift to a particle with interaction.  As above, $-\hbar^2\Delta+V$ in the small limit is related to $-\Delta+\lambda V$ in the large $\lambda$ limit.  If we are interested in the number of negative energy states, the relevant volume is $\{(x,p)\,|\, p^2+\lambda V(x)\le 0\}$, so the semiclassical number of states is
\begin{equation}\label{8.2}
  N_{cl,V}(\lambda) = \frac{\tau_\nu \lambda^{\nu/2}}{(2\pi)^\nu}\int_{\bbR^\nu} |V(x)|^{\nu/2}\,d^\nu x
\end{equation}

From early on, while a graduate student, I had an interest in bounds on the number of bound states of quantum systems, although I didn't initially think about quasi-classical limits.  I found in the literature two well-known results.  Bargmann \cite{Barg} proved that for $-\tfrac{d^2}{dx^2}+V(x)$ on $L^2(0,\infty)$ with $u(0)=0$ boundary conditions, the number of negative eigenvalues, $n(V)$, obeys
\begin{equation}\label{8.3}
  n(V) \le \int_{0}^{\infty} r|V(r)| \, dr
\end{equation}
(Bargmann, who viewed this as a bound on $s$-waves for a $3D$ problem with a central potential, also had results for higher angular momentum).  Schwinger \cite{SchwBS} (see below for the work of Birman) proved on $L^2(\bbR^3)$, that one has that $N(V)$, the number of negative eigenvalues (counting multiplicity) of $-\Delta+V$, is bounded by
\begin{equation}\label{8.4}
  N(V) \le \frac{1}{(4\pi)^2}\int \frac{|V(x)||V(y|}{|x-y|^2}\,d^3xd^3y
\end{equation}
With this in mind, as a graduate student, I realized that the growth of $N(\lambda V)$ as $\lambda\to\infty$ was interesting and not what one might expect naively from \eqref{8.4} when $\nu=3$.  Rather I found \cite{SimonLargeCouplingGrad} when $V$ is very nice there are $\lambda^{\nu/2}$ upper and lower bounded (but I did not prove strict $\lambda^{\nu/2}$ asymptotic behavior nor did I realize at the time the connection to quasi-classical behavior).

Quoting my paper, several years later, Martin \cite{Martin} proved the much stronger quasi-classical result
\begin{equation}\label{8.5}
  \lim_{\lambda\to\infty} \frac{N(\lambda V)}{N_{cl,V}(\lambda)} = 1
\end{equation}
on $\bbR^\nu$ when $V$ is H\"{o}lder continuous (as noted earlier, Birman-Borzov \cite{BirmanBorzov}, Robinson \cite{Robinson} and Tamura \cite{Tamura} proved the same result (some with somewhat weaker hypotheses on $V$) in a similar time frame).  These authors all used a variant of Weyl's argument.  Interestingly enough, the result without proof (essentially doing a quasi-classical computation and asserting its correctness) appeared in 1948 as a solved problem in the quantum mechanics book of Landau-Lifshitz \cite[Section 48, Problem 1]{LL} (I could only check this in the second edition of the English translation; 1948 is the date of the first edition).

I pause in the discussion of \eqref{8.5} for a side trip to the tool behind the next steps.  In 1976, Valya Bargmann reached the age of 68 and had to retire and the remaining senior joint appointments in mathematical physics at Princeton edited a festschrift in his honor \cite{LSWBarg}.  I wrote two reviews for that book on subjects where Bargmann had been a pioneer, one \cite{SiBargQDyn} on how to go from time automorphisms to Hamiltonians in Quantum Mechanics (where the foundational work was done by Bargmann and Wigner) and one on bounds on the number of bound states \cite{SiBargBdSt}. While the second was there because of Bargmann's bound, an especial point concerned the bound then universally associated in the West with Schwinger.  I had found that in the same year as Schwinger's paper, Birman had published \cite{BirmanBS} a long paper that included the same bound as Schwinger proven by the same method.

This method considers eigenvalues $E<0$ of $H_0+V$ where $\sigma(H_0)\subset [0,\infty)$ and $V$ is relatively form compact and self-adjoint so, if $V^{1/2}\equiv|V|^{1/2}\sgn(V)$, then
\begin{equation}\label{8.6}
  K_E = -|V|^{1/2}(H_0-E)^{-1}V^{1/2}
\end{equation}
is compact.  One shows that the dimension of the solutions of $K_E\varphi=\varphi$ is the multiplicity of $E$ as an eigenvalue of $H_0+V$ (essentially because if $\varphi=|V|^{1/2}\psi$, then $K_E\varphi=\varphi\iff(H_0+V)\psi=E\psi$).

Birman and Schwinger had the further idea of adding a coupling constant, $\lambda$, and looking at eigenvalues, $E_j(\lambda)$, of $H_0+\lambda V$.  Using the fact that $E_j(\lambda)$ is a strictly monotone function of $\lambda$, one proves that

\begin{theorem} \lb{T8.1} The number of eigenvalues of $H_0+V$ less than $E<0$ (counting multiplicity) is the number of eigenvalues (counting multiplicity), $\mu>1$ of $K_E$.
\end{theorem}

\begin{remarks} 1. Birman and Schwinger only considered the case where $V\le 0$ so $K_E$ is self-adjoint and the compact operator, $K_E$ is self-adjoint so its eigenvalues are real.  But one can prove in general that the eigenvalues are real even though $K_E$ may not be self-adjoint.

2. $\mu=\lambda^{-1}$.

3. By taking limits, one can show that if $K_E$ has a limit, $K_0$, as $-E\downarrow 0$, then the number of eigenvalues $E<0$ is bounded by the number of $\mu > 1$ for $K_0$.  When $V\le 0$ so $K_E \ge 0$, this in turn is bounded by $\tr(K_0)$ and $\tr(K_0^2)$.  This allowed Birman and Schwinger to provide an alternative proof of \eqref{8.3} and the first proof of \eqref{8.4}.
\end{remarks}

I felt it important to get Birman some credit for what was known as Schwinger's bound, so in \cite{SiBargBdSt}, I dubbed Theorem \ref{T8.1} the \emph{Birman-Schwinger principle} and thereafter $K_E$ became known as the \emph{Birman-Schwinger operator} (or when written as an integral operator, \emph{Birman-Schwinger kernel}) and \ref{8.4} became the \emph{Birman-Schwinger bound}.  I am very glad I succeeded in this.  There is an interesting postscript: several years afterwards I got a letter from Birman thanking me several times for mentioning his work but then essentially asking ``but why did you include Schwinger -- my paper was dated almost a year earlier''.  While Birman was correct about submission dates, the result already widely had Schwinger's name and there were rumors this was one of many things that Schwinger had written in the notebooks he kept while working on radar during the Second World War and doing real physics in his spare time.

I return to my analysis of \eqref{8.5}.  All prior results that I knew of required $V$ to at least be continuous so I wondered about $V$'s with singularities and, more generally only $L^p$ conditions.  In \cite{SiWkTrace}, I proved several theorems and conjectures about this situation.  In particular, I showed that

\begin{theorem} [\cite{SiWkTrace}] \lb{T8.2} Let $\calB$ be a Banach space of functions on $\bbR^\nu$ in which $C_0^\infty(\bbR^\nu)$ is dense and with $\norm{\cdot}_{\nu/2} \le C\norm{\cdot}_\calB$ for some $C$.  Suppose one has a bound of the form
\begin{equation}\label{8.7}
  N(V) \le c_1\norm{V}_\calB^{\nu/2}
\end{equation}
Then \eqref{8.5} holds for all $V\in\calB$.
\end{theorem}

The proof is by an approximation argument using the fact that if $A$ and $B$ are self-adjoint operators and $N(\cdot)$ the number of negative eigenvalues (counting multiplicity), then $N(A+B)\le N(A)+N(B)$.  I also noted that because of Theorem \ref{T8.1}, by looking at $\lambda V$ as $\lambda\to\infty$, a bound like \eqref{8.7} is equivalent to
\begin{equation}\label{8.8}
  \mu_n((-\Delta)^{-1/2}|V|^{1/2}) \le c_2 n^{-1/\nu}\norm{V}_\calB^{1/2}
\end{equation}
I developed a version of weak trace ideals analogous to weak $L^p$ spaces (related ideas were already in Goh'berg-Krein \cite{GK}).  Thinking of $(-\Delta)^{-1/2}$ as a function of the Fourier transform variable, \eqref{8.8} with $\calB=L^{\nu/2}(\bbR^\nu)$ is implied by (called Conjecture 2 below)
\begin{equation}\label{8.9}
   \norm{f(p)g(x)}_{\nu,w} \le c_3 \norm{f}_{\nu,w}\norm{g}_\nu
\end{equation}
where $\norm{\cdot}_{p,w}$ on the left side is a weak trace ideal norm and on the right a weak $L^p$ norm
(\cite[Section 2.2]{HA}).  In \cite{SiWkTrace}, I conjectured (a slightly weaker version of) \eqref{8.9} for $2<p<\infty$ and noted that it implied for $\nu\ge 3$ (called Conjecture 1 below; while Conjecture 2 $\Rightarrow$ Conjecture 1, the latter was of interest even if proven by other means)
\begin{equation}\label{8.10}
  N(V) \le c_\nu \norm{V}_{\nu/2}^{\nu/2}
\end{equation}
which I separately conjectured.  And I note that by Theorem \ref{T8.2}, this implies \eqref{8.5} for the maximal set where $V\in L^{\nu/2}(\bbR^\nu)$.

I was already familiar with bounds a little like \eqref{8.9}.  In \cite{SeilerSi3}, Erhard Seiler and I had proven that for $2\le q<\infty$ one has that
\begin{equation}\label{8.11}
  \norm{f(p)g(x)}_{q} \le C  \norm{f}_{q}\norm{g}_q
\end{equation}
By using interpolation ideas and this bound, I succeeded in \cite{SiWkTrace} in proving \eqref{8.7} for $\norm{\cdot}_\calB = \norm{\cdot}_{\nu/2+\varepsilon}+\norm{\cdot}_{\nu/2-\varepsilon}$ for all small $\varepsilon>0$ (with a constant that might diverge as $\varepsilon\downarrow 0$) and so \eqref{8.5} for $V\in L^{\nu/2+\varepsilon}\cap L^{\nu/2-\varepsilon}$ but I couldn't prove \eqref{8.9}.

I did this work in the spring of 1975.  In the fall, Charlie Fefferman, who I'd told about my conjecture 2, introduced me to Michael Cwikel, a visitor at IAS, whom he described as an expert on interpolation theory and might be the one to solve my conjecture.  I took Cwikel aside and described the problem to him.

A couple of months later, I was leaving my physics office planning to check my math mailbox on my way home.  As I passed his office, Elliott Lieb beckoned to me saying something like ``You know your conjecture.  I think I've solved it.'',  He proceeded to describe to me his beautiful proof of Conjecture 1 \cite{LiebCLR1, LiebCLR2} using path integrals.  I then went to my math mailbox and found a note from Cwikel saying that he had proven Conjecture 2 and thereby Conjecture 1 enclosing a sketch of his proof \cite{CwikelCLR}.  I couldn't imagine my finding Lieb's tour de force but found it ironic that I failed to find Cwikel's proof because I only thought of using interpolation theory while Cwikel, who was an expert on interpolation was smart enough to instead use in a clever way a  standard harmonic analysis trick of breaking a function into the sets where it lies between $2^k$ and $2^{k+1}$ that I'd seen Stein use many times in grad courses I'd taken not long before. I should mention that while I was unaware of it when I wrote \cite{SeilerSi3, SiWkTrace}, Birman and Solomjak had written a number of papers on trace ideal properties of $f(p)g(x)$.  While they didn't have Cwikel's result, motivated by his result they proved some additional estimates on such operators.  Much of this work is summarized in Birman-Solomjak \cite{BirSolSV}.  I should also mention here two papers with improved versions of Cwikel's estimates: Frank \cite{FrankCLR} and Hundertmark et al. \cite{HundCLR}.

In July of 1976, I went to a conference in the Soviet Union (one of only two trips I made there) which was ideal in terms of location and my interest.  The conference was in a small town outside Leningrad but organized by the Moscow based group of Dobrushin and Sinai so I could talk to them about the work on phase transitions described in Section \ref{s5}.  And because it was near Leningrad, Birman and his group could come out to meet me (I only learned later, it was not easy for them to get permission to do so).  They began by saying that my paper \cite{SiWkTrace} was very interesting but, while they didn't quite have a counterexample, they were fairly sure that my conjecture was wrong at which point I told them about Cwikel and Lieb.  There was confusion because the conjecture they meant was Conjecture 2 but I thought they meant Conjecture 1!  In fact, they handed me a reprint in Russian of a paper of Rozenblum \cite{RozCLR1} who had announced a result equivalent to my Conjecture 1 in 1972 (a detailed exposition only appeared after my visit \cite{RozCLR2}).  Eventually, I gave the bound the name \emph{CLR bound}, a name which stuck.  I believe that Rozenblum feels this is unfair but given the methods are totally different and the work independent, I think it appropriate.  Before leaving this subject, I should mention a later different proofs of Conlon \cite{ConlonCLR}, Fefferman \cite{FeffNQC}, Li-Yau \cite{LiYauCLR} and a non-path integral variant of Lieb's proof by Rozenblum-Solomyak \cite{RozSolCLR}.

In the same time frame as my conjecture, Lieb and Thirring \cite{LTSM}, as part of their brilliant proof of the stability of matter, exploited another quasi-classical bound, namely, if $E_j(V)$ are the negative eigenvalues (counting multiplicities) of $-\Delta+V$ on $L^2(\bbR^3)$, then
\begin{equation}\label{8.12}
  \sum_j (-E_j(V)) \le c_{1,3} \int |V(x)|^{5/2}\,d^3x
\end{equation}
Interestingly, their proof only relied on the Birman-Schwinger bound \eqref{8.4} even though \eqref{8.12} has the right large coupling behavior and \eqref{8.4} does not. Of course, if $V_-(x)=\max(0,-V(x)$, it is easy to see that $-E_j(V)\le -E_j(V_-)$, so once one has \eqref{8.12}, one immediately has the stronger result where $|V|$ on the right is replaced by $V_-$.  For simplicity of exposition, we'll continue to use $|V|$ but the reader should bear in mind that this implies the stronger result.

In the same Bargmann Festschrift referenced above, they \cite{LTonLT} exploited the same proof to show what are now called \emph{Lieb-Thirring bounds}
\begin{equation}\label{8.13}
  \sum_j (-E_j(V))^p \le c_{p,\nu} \int |V(x)|^{p+\nu/2}\,d^3x
\end{equation}
for $-\Delta+V$ on $L^2(\bbR^\nu)$ so long as
\begin{equation}\label{8.14a}
  p>0 \text{ if } \nu\ge 2; \quad p>1/2 \text{ if } \nu=1
\end{equation}
The CLR bounds are not included but are at the borderline.  As we'll see below, there cannot be a bound at $\nu=2, p=0$ nor for $\nu=1, p<1/2$.  That left the case $\nu=1, p=1/2$ which was open for 20 years until a bound was proven by Weidl \cite{Weidl}.  Then Hundertmark et. al \cite{HLT} using a different method proved bounds with optimal constants for that case.

Speaking of optimal constants, Lieb-Thirring in their first paper on the general inequality \cite{LTonLT} already raised the question of the optimal constant in \eqref{8.13}, not surprising given Lieb's then recent work on the best constants in Young, Hausdorff-Young and Sobolev inequalities.  There are two obvious lower bounds on the constant.  By Weyl type arguments, $\lim_{\lambda\to\infty} \lambda^{-(p+\nu/2)}\sum_j (-E_j(\lambda V))^p$ is a universal, computable number $c^{c\ell}_{p,\nu}$, so clearly, $c^{c\ell}_{p,\nu} \le c_{p,\nu}$.  It is also not hard to see that the bound on $-E_1$ implied by \eqref{8.13} implies a Sobolev inequality, so the known best constants in the Sobolev inequalities implies another lower bound, $c^{Sob}_{p,\nu}$.  Lieb-Thirring \cite{LTonLT} conjectured that $c_{p,\nu}=\max(c_{p,\nu}^{c\ell},c_{p,\nu}^{Sob})$ and using KdV sum rules (see \eqref{13.1}) proved this for $p=3/2,\nu=1$ (where it happens that $c_{p,\nu}^{c\ell}=c_{p,\nu}^{Sob}$) and Laptev-Weidl \cite{LapW} extended $c_{p,\nu}=c^{c\ell}_{p,\nu}$ to $p=3/2$ and all $\nu$.  Aizenman-Lieb \cite{AL} proved that if $c_{p_0,\nu}=c_{p_0,\nu}^{c\ell}$ for some $p_0$ than the same is true for all $p\ge p_0$.  In particular, the Lieb-Thirring conjecture holds for all $\nu$ when $p\ge 3/2$ and (by \cite{HLT}) for $\nu=1, p=1/2$.

Shortly after their conjecture, Glaser et al \cite{GGM} found it was false for $p=0, \nu \ge 8$ and recently Frank et al \cite{FGL} found it false for any $\nu \ge 2$ and some $p$.  That leaves $\nu=1$, where $\tfrac{1}{2}<p<\tfrac{3}{2}$ is open (and where $c_{p,\nu}^{Sob}>c_{p,\nu}^{c\ell}$.)  This interested me so much that it is an entry in my 2000 open problems list \cite[Problem 15]{Simon15} of 2000.  See Frank et. al. \cite{FLW} and Frank \cite{FrankLTRev} for more on Lieb-Thirring inequalities.

Around 2000 and for several years afterwards, I returned to issues connected with the critical $1D$ (i.e. $\nu=1, p=\tfrac{1}{2}$) Lieb-Thirring inequality.  This is connected to the issue of \emph{Szeg\H{o} asymptotics} of OPRL \cite{SimonSz} (see the discussion in Section 6 around \eqref{6.50} for the definitions of OPRL and OPUC).  This asymptotics says that for certain classes, $\{p_n\}_{n=0}^\infty$, of OPRL whose spectral measure has essential support $[-2,2]$, one has that for all $z\in\bbD\setminus\{0\}$ that
\begin{equation}\label{8.14}
  \lim_{n\to\infty} z^n p_n(z+\tfrac{1}{z})
\end{equation}
exists and is not identically zero (which determines asymptotics of $p_n(x)$ for $x\notin [-2,2]$).  The name comes from work of Szeg\H{o} \cite{SzAsym} on asymptotics of OPUC \cite{OPUC1}.

For OPRL, from a measure theory point of view, a natural family of measures analogous to OPUC is measures on $[-2,2]$, typically with pure a.c. measure, i.e.
\begin{equation}\label{8.15}
  d\rho(x)=f(x)\,dx, \quad x\in [-2,2]
\end{equation}
(see \cite[Section 13.1]{OPUC2}).  For such measures, early on it was realized the critical condition on such measures is
\begin{equation}\label{8.16}
  \int_{-2}^{2} \log\,f(x) (4-x^2)^{-1/2} dx > -\infty
\end{equation}
a condition known as the Szeg\H{o} condition after an analog for OPUC used by Szeg\H{o} in \cite{SzAsym}. If one thinks about Jacobi parameters rather than measures, it is natural to allow pure points outside $[-2,2]$, possibly even countably many so long as the only limit points are $\pm 2$.  In this regard, the best possible result when Killip and I began our work (discussed below) was

\begin{theorem} [Peherstorfer-Yuditskii \cite{PY}] \lb{T8.3} Let $d\rho$ be a probability measure on $\bbR$ which has a pure a.c. part on $[-2,2]$ of the form \eqref{8.15} and additional pure points $\{E_j^\pm\}_{j=1}^{N_\pm}$ on $\pm (2,\infty)$ so that $f$ obeys the Szeg\H{o} condition \eqref{8.16} and, in addition,
\begin{equation}\label{8.17}
  \sum_{j,\pm} \sqrt{|E_j^\pm| - 2} < \infty
\end{equation}
Then the associated OPRL, $\{p_n\}_{n=0}^\infty$, obey Szeg\H{o} asymptotics \eqref{8.14}.
\end{theorem}

Taking into account that $\pm 2$ are the edges of the spectrum of the free Jacobi matrix (i.e. $b_n\equiv 0, a_n\equiv 1$), one sees that \eqref{8.17} is a kind of critical Lieb-Thirring sum.  Two other relevant facts: First, Nevai \cite{Nevai} made a conjecture about Jacobi parameters:
\begin{equation}\label{8.18}
  \sum_{n=1}^{\infty} |a_n-1|+|b_n| < \infty \Rightarrow \eqref{8.16} \qquad \text{[Nevai]}
\end{equation}
In my work with Killip \cite{KS}, we proved that
\begin{equation}\label{8.18A}
  \eqref{8.17}+\text{LHS of }\eqref{8.18}\Rightarrow \eqref{8.16}\qquad \text{[Killip-Simon]}
\end{equation}

It was then clear that a suitable critical Lieb-Thirring inequality for Jacobi matrices would prove Nevai's conjecture.  Fortunately, Dirk Hundertmark, one of the coauthors of \cite{HLT} had just come to Caltech as a postdoc and we proved

\begin{theorem} [Hundertmark-Simon \cite{HundSi}] \lb{T8.3A} One has that
\begin{equation}\label{8.19}
  \sum_{j,\pm} \sqrt{(E_j^\pm)^2 - 4} \le \sum_{n=1}^{\infty} |b_n|+ 4 \sum_{n=1}^{\infty}  |a_n-1|
\end{equation}
\end{theorem}

Indeed, mimicking the proof of \cite{HLT} fairly easily proves \eqref{8.19} when $a_n\equiv 1$ and we found a method to go from that case to the full \eqref{8.19}.  Of course, \eqref{8.19} says that LHS of \eqref{8.18}$\Rightarrow$\eqref{8.17} and so \eqref{8.18A}$\Rightarrow$\eqref{8.18} proving Nevai's conjecture.  \cite{HundSi} also proved general $p>\tfrac{1}{2}, \nu=1$ Lieb-Thirring and Bargmann bounds for Jacobi matrices.

Over the next few years, with graduate students and postodcs, I explored various extensions.  Zlato\v{s} and I \cite{SZ} proved a variant of \cite{KS} for oscillatory $b_n$ and $a_n-1$ and Damanik-Hundertmark-Simon \cite{DHS} proved \eqref{8.17} and a Szeg\H{o} condition for some oscillatory Jacobi matrices where the sum on the right side of \eqref{8.19} is infinite, for example, $a_n=1+\tfrac{(-1)^n\alpha}{n}$ or $b_n=\tfrac{(-1)^n\beta}{n}$.

It was widely believed in the OP community that if Szeg\H{o} asymptotics holds, one must have a Szeg\H{o} condition so it came as a surprise when Damanik-Simon \cite{DSInv} found necessary and sufficient conditions for Szeg\H{o} asymptotics to hold that allowed many examples where one has Szeg\H{o} asymptotics even though the Szeg\H{o} condition and finiteness of the LHS of  \eqref{8.17} fail; indeed, \cite{DSInv} has examples where the sum of $(|E_j^\pm|-2)^\alpha$ is infinite for all $\alpha<\tfrac{3}{2}$.

I was also involved in several projects leading to Lieb-Thirring bounds for perturbations of non-free Schr\"{o}dinger operators and Jacobi matrices.  Frank, Weidl and I \cite{FSW}, to quote our result for Schr\"{o}dinger operators, proved that if there is a solution, $u_0$, of $(-\Delta+V_0)u_0=0$ and $c_1, c_2\in\bbR$ so that $0<c_1\le u_0(x)\le c_2<\infty$ for all $x$, then Lieb-Thirring inequalities for perturbations of $-\Delta$ imply them for perturbations of $-\Delta+V_0$ (with adjusted constants).  In particular, this implies perturbations of periodic Schr\"{o}dinger obey a Lieb-Thirring bound at the bottom of the spectrum.  A similar analysis gives such bounds for perturbations of periodic Jacobi matrices at the top and bottom of the spectrum and also for perturbations of almost periodic finite gap Jacobi matrices \cite{SY, CSZFG2}.  These results depend on a ground state representation for Schr\"{o}dinger operators that goes back to Jacobi, discussed above before \eqref{3.1}, and which was heavily used in work in constructive quantum field theory.  Somewhat surprisingly, \cite{FSW} seems to be the first place that this representation was worked out for Jacobi matrices.  This ground state representation was used earlier to compare operators by Kirsch and me \cite{KSGap2} in a paper, that, in particular, got interesting bounds on effective masses in solid state Hamiltonians.  This representation has also been used recently by Christiansen, Zinchenko and me \cite{CSZPer} in the study of periodic Jacobi matrices on trees.

It was natural to ask about Lieb-Thirring bounds and the analog of the Nevai conjecture for eigenvalues in gaps of perturbations of periodic and finite gap almost periodic Jacobi matrices.  For the periodic problem with all gaps open, this was accomplished by Damanik, Killip and me \cite{DKS} and for finite gap problems by Frank and me \cite{FSFG} after partial results by Birman \cite{BirmanFG} and Hundertmark-Simon \cite{HundSiFG}.

\bigskip

We saw that \eqref{8.10} was only proven for $\nu\ge 3$.  There is a good reason for this.  A bound like \eqref{8.10} implies that for $\lambda$ small, $-\Delta+\lambda V$ has no bound states (for say, all $V\in C^\infty_0$) but it was known that for a negative square well, i.e. $V$ the negative of the characteristic function of $(-a,a)$ in $\bbR$ or of a disk of radius $a$ in $\bbR^2$, $-\Delta+\lambda V$ has a negative eigenvalue for all $\lambda$.  I asked what happens for general $V$ and proved \cite{SiWkCp} that

\begin{theorem} [Simon \cite{SiWkCp}] \lb{T8.4} Let $V$ be a real-valued function on $\bbR$, not identically $0$, obeying
\begin{equation}\label{8.20}
  \int (1+|x|)^2 |V(x)|\,dx < \infty
\end{equation}
Then $-\tfrac{d^2}{dx^2}+\lambda V(x)$ has a negative eigenvalue, $E(\lambda)$ for all small, positive $\lambda$ if and only if
\begin{equation}\label{8.21}
  \int V(x)\, dx \le 0
\end{equation}
and if that is the case, one has that
\begin{equation}\label{8.22}
  \alpha(\lambda)\equiv (-E(\lambda))^{1/2} = -\tfrac{\lambda}{2}\int V(x)\, dx - \tfrac{\lambda^2}{4}\int V(x)|x-y| V(y)\, dx dy + \text{o}(\lambda^2)
\end{equation}
\end{theorem}

\begin{remarks} 1. This work was motivated by Murph Goldberger, who at the time was department chair of physics at Princeton (I was Director of Graduate Students) and who later was the President of Caltech at the point when I was recruited.  He had organized a group of particle theoretical physicists (Jason) who worked on DoD projects for a few weeks each summer.  While studying some problems on sound waves in water (not quantum mechanics!), Murph with Henry Abarbanel and Curt Callen got interested in negative eigenvalues in one dimension and found \eqref{8.22} as a formal series. Murph wanted to know if I could prove something.

2. For $\alpha$ to always be positive when \eqref{8.21} holds, one must have that $\int V(x)\, dx = 0 \Rightarrow \int V(x)|x-y| V(y)\, dx dy \le 0$, a fact summarized by saying the $|x-y|$ is a conditionally negative definite kernel; this is indeed true.

3. Blankenbecler, Goldberger and I \cite{BGS}, and independently Klaus \cite{KlausWkCp}, showed that \eqref{8.20} could be replaced by the weaker
\begin{equation}\label{8.23}
   \int (1+|x|) |V(x)|\,dx < \infty
\end{equation}

4. If $V(x)\sim -ax^{-\beta}$ as $x\to\infty$ with $1<\beta<2$, something interesting happens.  There are now infinitely many eigenvalues but most are $\text{O}(\lambda^{2/(2-\beta)})$ while the lowest eigenvalue is $\text{O}(\lambda^{2/(3-2\beta)})$.  This is proven in \cite{BGS} which also has results when $\beta=2$.

5. Not only does \eqref{8.16} fail if $\nu=1,2$, but there is a result \cite[Remark 3 on pg 315]{SiBargBdSt} that if $\norm{\cdot}$ is any translation invariant norm on a vector space of functions that includes some non-zero, everywhere non-positive continuous functions of compact support, then for any $N$ and any $\varepsilon$, there is a $V$ with $\norm{V}<\varepsilon$ and so that $-\Delta+V$ has at least $N$ negative eigenvalues!

6.  We think of this result which violates a putative quasi-classical bound as an example of non-quasi-classical eigenvalue behavior.

7. \cite{SiWkCp} also proves that if $V$ decays exponentially, then $(-E(\lambda))^{1/2}$ is analytic in $\lambda$ at $\lambda=0$.

8. \cite{SiWkCp} also has results when $\nu=2$.  In that case, if $\int V(x)\, d^2x <0$, we have a negative eigenvalue for $-\Delta+\lambda V$ but for small $\lambda$ it is $\text{O}(\exp(-d/\lambda))$!
\end{remarks}

The proof of the theorem is not hard.  The one-dimensional Birman-Schwinger kernel for $E=-\alpha^2$ has the form
\begin{equation}\label{8.24}
  K_\alpha(x,y) = |V|^{1/2}(x)\exp(-\alpha|x-y|)V^{1/2}(y)
\end{equation}
The Birman-Schwinger principle says that $E$ is an eigenvalue of $-\tfrac{d^2}{dx^2}+\lambda V$ if an only if $-\lambda^{-1}$ is an eigenvalue of $K_\alpha$.  In more than two dimensions, $\norm{K_\alpha}$ is bounded as $\alpha\downarrow 0$ so if $\lambda < \sup \norm{K_\alpha}$, then $-\Delta+\lambda V$ has no negative eigenvalues, but in $1$ (or $2$) dimensions, $\norm{K_\alpha}$ diverges  as $\alpha\downarrow 0$.  The reason that there is only one negative eigenvalue when $\lambda$ is small (at least when \eqref{8.20} or \eqref{8.23} holds) is that the divergent piece is rank one.  To see this, \cite{SiWkCp} writes
\begin{equation}\label{8.25}
  K_\alpha=L_\alpha+M_\alpha
\end{equation}
\begin{equation}\label{8.26}
  L_\alpha(x,y) = |V|^{1/2}(x) V^{1/2}(y)/2\alpha
\end{equation}
\begin{equation}\label{8.27}
  M_\alpha(x,y) = (2\alpha)^{-1} |V|^{1/2}(x)\left[e^{\alpha|x-y|}-1\right] V^{1/2}(y)
\end{equation}
so $\lim_{\alpha\downarrow 0}M_\alpha \equiv M_0$ exists where
\begin{equation}\label{8.28}
  M_0(x,y) =  |V|^{1/2}(x)|x-y|V^{1/2}(y)
\end{equation}
Thus, the Birman-Schwinger principle is equivalent to
\begin{equation}\label{8.29}
  \alpha = -\tfrac{\lambda}{2} \jap{V^{1/2},(1+\lambda M_\alpha)^{-1}|V|^{1/2}}
\end{equation}
The leading term on the right is $-\tfrac{\lambda}{2} \int V(x)\,dx$ and the next is $\tfrac{\lambda^2}{4}\jap{V^{1/2},M_0|V|^{1/2}}$.

This work lead to several threads of later work.  In \cite{SiLinear}, I asked when a new eigenvalue, $E(\lambda)$, issuing from $E=0$ at $\lambda=\lambda_0$ is $\text{O}(\lambda-\lambda_0)$ and proved that

\begin{theorem} \lb{T8.5} Suppose that $B$ is a relatively compact symmetric perturbation of a self-adjoint operator, $A$, that $\sigma_{ess}(A)$ includes $[0,\varepsilon]$ for some $\varepsilon>0$ and that $\lambda_0>0$ is such that $A+\lambda B$ has exactly one more negative eigenvalue (counting multiplicity) for $\lambda\in (\lambda_0,\lambda_0+\delta)$ than for $\lambda\in (\lambda_0-\delta,\lambda_0)$.  Then there is one eigenvalue, $E(\lambda)$, near $0$ for $\lambda\in (\lambda_0,\lambda_0+\delta)$ and
\begin{equation}\label{8.30}
  \lim_{\lambda\downarrow\lambda_0} \frac{E(\lambda)}{\lambda-\lambda_0} = \alpha
\end{equation}
exists.  Moreover, $\alpha\ne 0$ if and only if $A+\lambda_0 B$ has eigenvalue $0$, and in that case, the eigenvalue is simple and $\alpha=\jap{\varphi,B\varphi}$ where $(A+\lambda_0 B)\varphi=0$ with $\norm{\varphi}=1$.
\end{theorem}

In \cite{SiCritical,SiKlCrit1}, I made what turned out to be an important definition.

\begin{definition} Let $\nu \ge 3$ and $V\in L^p(\bbR^\nu)\cap L^q(\bbR^\nu)$ for some $p<\tfrac{\nu}{2}<q$.  $V$ is called

\emph{supercritical} $\iff\,\, \inf\spec(-\Delta+V)<0$

\emph{subcritical} $\iff\,\, \inf\spec(-\Delta+(1+\varepsilon)V)=0 \text{ for some } \varepsilon>0$

\emph{critical} $\iff\,\,-\Delta+(1+\varepsilon)V \text{ is supercritical for all } \varepsilon>0 \text{ and } \inf\spec(-\Delta+V)=0$
\end{definition}

Thus critical $V$'s are ones that like $-\Delta$ in $1$ and $2$ dimensions are about to give birth to bound states.  A main result of \cite{SiCritical} is that $V$ is subscritical $\iff \sup_t\norm{e^{-(-\Delta+V)t}}_{\infty,\infty}<\infty$ (the norm from $L^\infty$ to $L^\infty$).  Both \cite{SiCritical} and \cite{SiKlCrit1} show that if $V$ and $W$ are both critical in three dimensions, then $-\Delta+V(x)+W(x-R)$ has a bound state of energy, $E(R)<0$ for large $R$ with $E(R)\sim -\beta R^{-2}$.  \cite{SiKlCrit1} even computes the universal value of $\beta$.  This result is connected to the Effimov effect.  A considerable literature has developed on the study of critical operators; \cite{PinchRev} reviewed the literature as of 2005.  I note that while \cite{SiKlCrit1} focused on $\bbR^3$, it has a remark about what happens in $\bbR^\nu; \nu\ge 4$ which is wrong; for the correct results, see \cite[Theroems 8.5-8.6]{PinchRev}.

In \cite{SiKlCrit2, SiKlCrit3}, Klaus and I consider the variety of coupling constant threshold behavior that can occur for $-\Delta+V+\lambda W$ (with $V$ and $W$ short range) when as $\lambda\downarrow\lambda_0$, some eigenvalue is absorbed in the continuous spectrum (the first paper deals with the two body problem and the second with a limited set of $N$ body systems).  The results are quite complicated and supplement/illuminate some work of Jensen-Kato \cite{JenKato}.

\bigskip

Our next topic concerns the situation where $V$ goes to infinity at infinity (or, at least, is bounded away from zero there) but $\min(V(x))=0$ and we are interested in the lowest eigenvalues of $-\Delta+\lambda^2 V$ as $\lambda\to\infty$.  In \cite{SiSC1} I proved

\begin{theorem} [\cite{SiSC1}] \lb{T8.6} Let $V, W$ be two $C^\infty$ functions on $\bbR^\nu$ so that

(1) For some $A, R>0$, one has $V(x)\ge A$ if $|x|\ge R$

(2) $V(x) \ge 0$ for all $x$

(3) $V(x)$ vanishes only at $\{x^{(j)}\}_{j=1}^M$ ($M\ge 1$) and at each such minimum, the matrix $A^{(j)}_{k\ell}=\frac{\partial V}{\partial x_k \partial x_\ell}(x^{(j)})$ is strictly positive.

(4) $W$ is bounded from below

\noindent Let $H^{(j)} = -\Delta+\sum_{k,\ell} A^{(j)}_{k\ell} x_k x_\ell + W(x^{(j)})$.  Let $\{e_\alpha\}_{\alpha=1}^\infty$ be an ordering of the union over $j$ of the eigenvalues of $H^{(j)}$ (counting multiplicity) so that $e_1 \le e_2 \le \dots$.  Then
\begin{equation}\label{8.30a}
  H(\lambda) \equiv -\Delta+\lambda^2 V+\lambda W
\end{equation}
has eigenvalues, $\{E_\alpha(\lambda)\}_{\alpha=1}^\infty$, with $E_1(\lambda) \le E_2(\lambda) \le \dots$ at the bottom of its spectrum and for any $\alpha=1,2,\dots$ we have that
\begin{equation}\label{8.31}
  \lim_{\lambda\to\infty} \frac{E_\alpha(\lambda)}{\lambda} = e_\alpha
\end{equation}
Moreover, each $E_\alpha(\lambda)$ has an asymptotic series in $\lambda^{-1}$ to all orders
\begin{equation}\label{8.32}
  E_\alpha(\lambda) = e_\alpha\lambda + a_\alpha^{(0)}+a_\alpha^{(1)}\lambda^{-1}+\dots
\end{equation}
\end{theorem}

I always thought of the paper in which this theorem appeared as Ed Witten's homework assignment because one motivation for this work was his wonderful paper on the supersymmetric proof of Morse inequalities and the Morse index theorem \cite{Witten}.  In it, he used this theorem (or rather its generalization when functions on $\bbR^\nu$ are replaced by the tangent bundle on a compact manifold and $-\Delta$ by a Laplace-Beltrami operator (also discussed in \cite{SiSC1})).  When using this result, Witten says \emph{Although the rigorous theory has apparently not been
developed for operators acting on vector bundles on manifolds, the method used in Reed and Simon \cite{RS3}, pp. 34--38, to treat the double well potential should suffice with some elaboration for this case.}  In fact, the argument in \cite{RS3} he refers to is one-dimensional and uses some other properties of the simple double well.  The general one-dimensional case had been done by Combes et al \cite{CDS}, but their arguments also depended on one dimension and are somewhat involved, so I wrote my paper in part to get a proof that works in multiple dimensions and also one that is fairly simple.

\cite{SiSC1} was the first paper of a series.  The other papers dealt with eigenvalue splitting in the situation where multiple wells have the same eigenvalue.  It is simpler to discuss the case of the lowest eigenvalue assuming a double degeneracy.  A basic role is played by the Agmon metric, mentioned in Section \ref{s6} (in the page after \eqref{6.37}), which was known to determine the rate of decay of eigenfunctions.  In this situation it is defined as the distance in the Riemann metric $V(x)(dx)^2$, i.e.
\begin{equation}\label{8.33}
  \rho(x,y) = \inf \left\{\int_{0}^{1} \sqrt{V(\boldsymbol{\gamma}(s))}|\dot{\boldsymbol{\gamma}}(s)|\,ds \,\middle|\, \boldsymbol{\gamma}(0)=x,\boldsymbol{\gamma}(1)=y\right\}
\end{equation}
over all smooth paths, $\boldsymbol{\gamma}(s),\,0\le s \le 1$, between $x$ and $y$.  The main result of \cite{SiSCAnon, SiSC2} is

\begin{theorem} [\cite{SiSC2}] \lb{T8.7} Let $V$ be a $C^\infty$ function on $\bbR^\nu$ (and $W=0$) obeying conditions (1)-(3) of Theorem \ref{T8.6}.  Suppose there are two points $a\ne b$ where $V$ vanishes and that $e_1=e_2< e_3$ so that $e_1$ and $e_2$ are eigenvalues associated to the operators, $H^{(j)}$, at the points $a$ and $b$.  Let $j_a$ (resp $j_b$) be characteristic functions of small balls about $a$ (resp $b$), balls that are so small that they are disjoint.  Let $\Omega_\lambda$ be the normalized ground state of the operator, $H(\lambda)$, of \eqref{8.30a}. Suppose that
\begin{equation}\label{8.34}
  \liminf \norm{j_a\Omega_\lambda}\norm{j_b\Omega_\lambda} > 0
\end{equation}
Then
\begin{equation}\label{8.35}
  \lim_{\lambda\to\infty} -\lambda^{-1} \log\left[E_2(\lambda)-E_1(\lambda)\right]=\rho(a,b)
\end{equation}
\end{theorem}

\begin{remarks} 1.  \eqref{8.34} says that the ground state lives near both minima.  One condition that guarantees this is if there is a Euclidean rotation or reflection of order $2$ that leaves $V$ invariant with $Ra=b$.  In that case the limit is exactly $1/4$.  That holds for the famous $1D$ double well where $V(x)=x^2(x-1)^2$.  In that case $\rho$ is given by a WKB integral and this results was proven by various authors in the ten years before my result (see \cite{SiSC2} for references) but my work was the first rigorous result in more than one dimension.

2. The proof controls eigenfunction decay by writing the eigenfunctions in terms of path integrals and using the method of large deviations to single out a minimum action solution.  It is a basic fact of classical mechanics that minimum action is equivalent to minimum distance in a suitable metric.

3. The importance of minimum action paths to leading order tunnelling in multi-dimensions was noted in the theoretic physics literature several years before my work.  These solutions were called \emph{instantons}; see \cite{SiSC2} for references to that literature.

4. Shortly after my work, Helffer-Sj\"{o}strand \cite{HelS1} developed a powerful microlocal analysis approach to these problems and recovered the results of Theorems \ref{T8.6} and \ref{T8.7} that got higher order terms, established some of Witten's conjectures and worked in a more general setting as they discussed in a number of later papers.

5. I wrote two later papers \cite{SiSC3, SiSC4} on some specialized situations related to Theorem \ref{T8.7}.
\end{remarks}

Kirsch and I \cite{KSGap1} proved an interesting universal tunnelling bound

\begin{theorem} [Kirsch-Simon \cite{KSGap1}] \lb{T8.7A} Let $V$ be a continuous function on $\bbR$ so that $-\tfrac{d^2}{dx^2}+V(x)$ has discrete eigenvalues $\{E_j\}_{j=1}^N$ below any essential spectrum.  Let $n<N+1$ and suppose for some $\alpha>0$, we have that $V(x)\ge E_n+\alpha^2$ on $\bbR\setminus [a,b]$.  Let $\lambda=\max_{E\in [E_{n-1},E_n]; x\in (a,b)} \sqrt{|E-V(x)|}$. Then
\begin{equation}\label{8.35A}
  E_n-E_{n-1} \ge \pi \lambda^2 \alpha (\lambda+\alpha)^{-1} e^{-\lambda(b-a)}
\end{equation}
\end{theorem}

\bigskip

The last topic that I discuss in this section concerns another non-quasi-classical situation.  Just as Theorem \ref{T8.5} was motivated by a question posed to me by Goldberger and Theorem \ref{T8.6} by Witten, this work was motivated by a query from some theoretical, non-mathematical, physicists.  In this case, I was asked by Jeffrey Goldstone and Roman Jackiw if the two dimensional Schr\"{o}dinger operator
\begin{equation}\label{8.36}
  H_1= -\frac{\partial^2}{\partial x^2}-\frac{\partial^2}{\partial x^2}+x^2y^2
\end{equation}
has purely discrete spectrum or not.  They noted that one was used to the condition for purely discrete spectrum of $-\Delta+V$ being that $V(x)\to\infty$ as $x\to\infty$.  While this failed for $V(x,y)=x^2 y^2$ since $V$ vanished on the axes, they suspected the spectrum was discrete since it went to infinity in all but four directions. In fact, the natural quasi-classical condition of finite phase space volume, i.e. $|\{(x,p)\,|\, p^2+ V(x)\le E\}| < \infty$ for all $E$ also fails in this case.  A closely related question involves the operator
\begin{equation}\label{8.37}
  H_2= -\Delta_D^\Omega; \qquad \Omega=\{(x,y)\in\bbR^2\,\mid\,|xy| \le 1\}
\end{equation}
Motivated by their question, in \cite{SiNQC1}, I proved that

\begin{theorem} [\cite{SiNQC1}] \lb{T8.8} The operators $H_1$ of \eqref{8.36} and $H_2$ of \eqref{8.37} both have purely discrete spectrum.
\end{theorem}

\begin{remarks} 1. \cite{SiNQC1} gives six proofs that $H_2$ has purely discrete spectrum.  The simplest proves that $H_1$ has purely discrete spectrum (and that easily implies that so does $H_2$) as follows:  It follows by scaling and the fact that $-\tfrac{d^2}{dq^2}+q^2$ has smallest eigenvalue $1$ that $-\tfrac{d^2}{dq^2}+\omega^2 q^2 \ge |\omega|$, which implies that $-\tfrac{\partial^2}{\partial x^2}+x^2 y^2 \ge |y|$.  Interchanging $x$ and $y$, adding the two and multiplying by $1/2$ shows that
\begin{equation}\label{8.38}
  H_1 \ge \tfrac{1}{2}(-\Delta+|x|+|y|) \equiv H_3
\end{equation}
Since $H_3$ has purely discrete spectrum, by the min-max principle \cite[Theorem 3.14.5]{OT}, so does $H_1$.  The ``defect'' in this proof is that it turns out that $H_1$ for large energies is much larger than $H_3$.  The number of eigenvalues of $H_3$ larger than $E$ grows like $E^3$ (by a quasi-classical estimates) while for $H_1$ only like $E^{3/2}\ln E$ (as discussed below) which is much smaller.

2. The operator $H_1$ that Goldstone and Jackiw asked me to look at was a toy model for a more involved model they were really interested in.  Let $\bbA$ be a semi-simple Lie algebra and let $-\Delta_A$ be the Laplacian in the inner product on $\bbA$ given by the negative of the Killing form.  For $\nu\ge 2$, let $\bbA^\nu$ be the set of $\nu$-tuples, $(A_1,\dots,A_\nu)$, of elements of $\bbA$.  Then they were interested in the operator on $L^2(\bbA^\nu)$
\begin{equation}\label{8.39}
  H_4 = -\sum_{i}\Delta_{A_i} - \sum_{i<j} \tr\left(\left[A_i,A_j\right]^2\right)
\end{equation}
This had been proposed as a model of zero momentum Yang-Mills fields.  In \cite{SiNQC1}, I also found one proof that this more involved operator with a potential that stays zero on unbounded narrow sets  has purely discrete spectrum.  Another model that I considered had been proposed by Feynman - namely take three particles in three dimensions and let the interaction be the area of the triangle whose vertices are positions of the three particles.  This potential also stays zero on an unbounded narrow set and I proved it has purely discrete spectrum.

3. There was much work earlier on Dirichlet Laplacians like $H_2$ where $\Omega$ can have infinite volume but still have purely discrete spectrum.  In 1948, Rellich \cite{RellichNQC} considered a class of Dirichlet operators that includes $H_2$ and proved that they have purely discrete spectrum.  In 1953, Mol\v{c}anov \cite{MolchanovNQC} found necessary and sufficient conditions on $\Omega$ for $-\Delta_D^\Omega$ to have purely discrete spectrum in terms of fairly involved conditions involving capacity, with an optimal version of Mol\v{c}anov's result in Maz'ya-Shubin \cite{MaSh}.  Maz'ya \cite{Mazya} has a review of the uses of these kinds of capacity conditions.  For results of this type with magnetic field, see Helffer-Mohamed \cite{HelMo}

4. Fefferman and Phong, using ideas described and summarized in \cite{FeffNQC}, have an illuminating picture of when the naive phase space picture of eigenvalue counting fails.  Their ideas are used in the most general result in \cite{SiNQC1} and, in particular, in the proof that $H_4$ has purely discrete spectrum.

5. I revisited the issues connected to Theorem \ref{T8.8} many years later \cite{SiNQC3}.
\end{remarks}

Robert \cite{RobertNQC}, Solomyak \cite{SolomNQC}, Tamura \cite{TamuraNQC} and I \cite{SiNQC2} also obtained results on the eigenvalue counting asymptotics of operators like $H_1$ and $H_2$.  In one sense, one can say this is consistent with the quasi-classical expectation in that, for example, \eqref{8.1} holds for $H_2$ since both sides are infinite.  But the growth in this case is not the $\text{O}(E)$ that \eqref{8.1} gives when $|\Omega|<\infty$ but at a different rate which we think of as non-quasi-classical. In particular, \cite{SiNQC2} proves that
\begin{equation}\label{8.40}
  \lim_{E\to\infty} E^{-1}(\log E)^{-1}   N_{H_2}(E)  = 1/\pi
\end{equation}
and
\begin{equation}\label{8.41}
  \lim_{E\to\infty} E^{-3/2}(\log E)^{-1}   N_{H_1}(E)  = 1/\pi
\end{equation}

There is also discussion in the literature of the analog of $H_2$ when Dirichlet boundary conditions are replaced by Neumann boundary conditions.  As I described in Section \ref{s6} around equation \eqref{6.40A}, if one looks at the Neumann Laplacian of
\begin{equation}\label{8.42}
  \Omega=\{(x,y)\in\bbR^2\,\mid\,x>1, |y|<f(x)\}
\end{equation}
then if the $V$ of \eqref{6.40} goes to zero slowly, $-\Delta_N^\Omega$ has some a.c. spectrum.  Evans and Harris \cite{EH} found necessary conditions on when this operator has purely discrete spectrum. For many such operators, Jak\v{s}i\'{c}, Mol\v{c}anov and I \cite{JMS} found the leading asymptotics for the number of eigenvalues, $N_N^\Omega(E)$, of asymptotics as $E\to\infty$. In particular, we found for the interesting case $f(x)=\exp(-x^\alpha)$ that
\begin{equation}\label{8.43}
  N_N^\Omega(E)\sim \left\{
                      \begin{array}{ll}
                        \frac{1}{2}|\Omega|E, & \hbox{ if } \alpha>2 \\
                        \frac{1}{2}\left(|\Omega|+\frac{1}{2}\right)E, & \hbox{ if }\alpha=2 \\
                        C_\alpha E^{1/2+1/(2(\alpha-1))}, & \hbox{ if } 1<\alpha<2
                      \end{array}
                    \right.
\end{equation}
where
\begin{equation*}
  C_\alpha = \frac{1}{4(\alpha-1)\sqrt{\pi}}\left(\frac{\alpha}{2}\right)^{1/(1-\alpha)} \frac{\Gamma(1/(2(\alpha-1)))}{\Gamma(3/2+1/(2(\alpha-1)))}
\end{equation*}
For $\alpha>2$, we have a quasi-classical Weyl behavior, but for other alpha, we have non-quasi-classical behavior.

A final remark on non-quasi-classical eigenvalue behavior.  Kirsch and I \cite{KSNQC} (motivated again by a question from a non-mathematical physicist - in this case, Michael Cross) found such behavior for the growth of the number of eigenvalues below $E$ as $E\uparrow 0$ for ${-\Delta+c(1+|x|)^{-1}}$.

\section{Almost Periodic and Ergodic Schr\"{o}dinger Operators} \lb{s9}

In AY 1980-81, I visited Caltech as a Fairchild Distinguished Scholar (I got an offer during the year and stayed).  I was looking forward to a year with no teaching, only one postdoc (Yosi Avron, who also had a leave from Princeton) and only one grad student (Peter Perry), a year where I expected to be able to focus on research with few distractions.  I had the impression that many of the areas I had focused on were winding down, at least as far as my involvement.  CQFT was mainly using involved expansions and estimates, not my forte, and the leap to four dimensional space-time which required going beyond superrenormalizable theories seemed daunting (and still hasn't happened!).  The hottest open question in $N$-body NRQM was asymptotic completeness and, while I was hopeful the $N$-body Mourre estimates that I'd recently proven with Perry and Sigal (see Section \ref{s6}) would be useful, I had no plan for how to proceed.  So I suggested to Yosi that we look at moving into a new area.  There seemed to be two to consider: Schr\"{o}dinger operators with almost periodic potentials (where I was aware of some interesting non-rigorous work of Aubry \cite{Aubry, AA}) and quasi-classical eigenvalue counting (where I was aware of a then recent preprint of Helffer and Robert \cite{HR} - a kind of multidimensional version of Bohr-Sommerfeld quantization rules).  They both seemed interesting and promising.  After thinking about it, I said to Yosi: ``Let's try to do both.  We'll do almost periodic first - it doesn't look very complicated or involved.  We'll finish it up in six months and then we can turn to quasi-classical''.  Little did I realize!  Almost periodic Schr\"{o}dinger operators was a major focus of my work for more than 15 years and, now, 40 years later, while there has been remarkable progress, it is still an active area with its own separate conferences.  While, in the few years after that, I did some quasi-classical research (see Section \ref{s8}), I never worked on detailed eigenvalue locations and related issues although it has become an active area (see, for example, Zelditch \cite{Zelditch} for a recent review of some aspects).

Before turning to the details of this subject, I should point out that it is intimately connected to the subject of Section \ref{s11} (random Schr\"{o}dinger operators) and to the subject of Section \ref{s12} (singular spectrum) so some papers may only be mentioned here and discussed in more detail in later sections.  Moreover, our formal discussion below will start with the general framework of Schr\"{o}dinger operators and Jacobi matrices with ergodic potentials which encompasses random and almost periodic potentials as special cases.  We will be very brief in this presentation referring the reader to the relevant sections of my book with Cycon et al. \cite{CFKS}, the lovely review of Jitomirskaya \cite{JitoOneFoot} or, for more comprehensive discussion, the books of Aizenman-Warzel \cite{AW}, Carmona-Lacroix \cite{CL}, Damanik-Filman \cite{DamFil}, Pastur-Figotin \cite{PF}, or Stollmann \cite{StollBk}.

My initial work, much of it with Avron \cite{ASTR, ASAP1, ASAP2,BSCantor, CrSi1, CrSi2, SiKotani,DSAP} was a big part of my research during the three year period 1981-1983 (which was extremely fruitful including also my early work on TKNN integers and Berry's phase and their geometric significance \cite{ASS1, SiBerry} (see Section \ref{s10}), my work on ultracontractivity \cite{DavSi} (see Section \ref{s3}), my work on multiwell problems \cite{SiSC1, SiSC2} and on nonclassical eigenvalue asymptotics \cite{SiNQC1, SiNQC2} (see Section \ref{s8}), my discovery of localization for slowly decaying random potentials \cite{SiRPt} (see Section \ref{s11}), the completion and publication of my influential review article on Schr\"{o}dinger semigroups \cite{SimonSchSmgp} (with over 1400 citations on Google Scholar), several miscellaneous papers on NRQM with Coulomb potentials \cite{SimonEtAlStark, LSST} as well as the preparation of my 45 hour, Bayreuth lecture course in the summer of 1982 which turned into \cite{CFKS}).

I gave a review talk on the early work on almost periodic Schr\"{o}dinger operators at the 1981 Berlin ICMP \cite{SiAPFlu} and the paper based on the talk became known as the almost periodic flu paper because I started by remarking on the fact that there seemed to be a worldwide explosion of work in this new area that I dubbed the almost periodic flu.  Indeed, besides my work in California with Avron, there was work by Bellissard and collaborators in France (reviewed with lots of references in \cite{BellAP}; notable was his use of $C^*$-algebra methods), Chulaevsky  \cite{ChulAP} in Moscow and by Moser \cite{MoserAP}, Johnson-Moser \cite{JohnMAP} and Sarnak \cite{SarnAP} in New York (notable was the Johnson-Moser invention of rotation number and the resulting gap labelling).

The basic framework is a probability measure space $(\Omega,\Sigma,\mu)$ with expectation, $\bbE$, a distinguished bounded function, $f:\Omega\mapsto\bbR$, and a distinguished group $g\mapsto T_g$ of ergodic measure preserving maps indexed by the reals or the integers (see \cite[Sections 2.6-2.9]{HA} for more on the ergodic and subadditive ergodic theorems).  In the continuous case, one considers a potential $V_\omega(x) = f(T_x(\omega))$ and ergodic Schr\"{o}dinger operator $H_\omega=-\tfrac{d^2}{dx^2}+V_\omega(x)$ acting on $L^2(\bbR)$ and in the discrete case, one takes diagonal elements $b_n(\omega)=f(T_n(\omega))$ and ergodic discrete Schr\"{o}dinger operator
\begin{equation}\label{9.1}
  (H_\omega u)_n = u_{n+1}+u_{n-1}+b_n(\omega)u_n
\end{equation}
acting on $\ell^2(\bbZ)$.  While this is the simplest example, one often generalizes (and we will occasionally below) in three ways: one can allow suitable unbounded $f's$ (often bounded from below), one can replace $\bbR$ or $\bbZ$ by $\bbR^\nu$ or $\bbZ^\nu$ with the multidimensional Laplacians, and, finally, one can consider ergodic Jacobi matrices rather than only discrete Schr\"{o}dinger operators (i.e. allow ergodic $a_n$'s).

Two special cases are the random (discussed mainly in Section \ref{s11}) and almost periodic cases (the latter is the subject of this section after a general discussion of some common objects).  For the discrete random case, $\Omega=[a,b]^\bbZ$, $f(\{\omega_j\}_{j\in\bbZ}) = \omega_0, T_1(\{\omega_j\}_{j\in\bbZ})=\{\omega_{j+1}\}_{j\in\bbZ},$ and
\begin{equation}\label{9.1A}
   d\mu(\{\omega_j\}_{j\in\bbZ}) = \otimes_{j\in\bbZ} d\kappa(\omega_j)
\end{equation}
so $b_j(\omega)$ is a sequence of independent identically distributed random variables (aka, iidrv).  The special case where $d\kappa$ is uniform distribution on an interval is usually called the \emph{Anderson model} (after \cite{Anderson} for which Anderson got the Nobel prize for claiming the model had localized states as we'll discuss in Section \ref{s11}). I'll sometimes call the general iidrv case the \emph{generalized Anderson model}. Sometimes what I called the generalized Anderson model is called just the Anderson model but it pays to have separate names.  One sometimes studies unbounded distributions or even non-independence but demands the $b_j$ be ``really random'', at least defined by a Markov process.  Since I never worked directly on continuum random operators, I'll leave the description of those models to the books mentioned above, especially Aizenman-Warzel \cite{AW} and Stollmann \cite{StollBk}.  I do however note that to accommodate models with say a fixed potential localized about lattice points in $\bbR^\nu$ with iidrv coupling constants, one needs to modify the set up to only require ergodicity under a discrete group even for $\bbR^\nu$ models.

The other case is almost periodic functions.  (For more on the general theory of almost periodic functions, see, e.g. \cite[Section 6.6]{OT}).  In this case, $\Omega$ is a separable, compact, abelian group, called the \emph{hull}, there is a homomorphism $S:G\to\Omega$ (with $G=\bbR$ or $\bbZ$) so that $T_g(\omega)=\omega S(g)$ and $f:\Omega\to\bbR$.  Two important special cases are where $S(G)$, the image, is a winding line on a finite dimensional torus, $\Omega$, viewed as a product of copies of $\partial\bbD$ with complex product as the group product, in which case the potential is called \emph{quasiperiodic}, and the case where the potential is a uniform limit of periodic functions of longer and longer commensurate periods (e.g. $V(x) = \sum_{n=1}^{\infty} 2^{-n} \cos(x/2^n)$), in which case the potential is called \emph{limit periodic}.  The most famous example is
\begin{equation}\label{9.2}
  (H_{\alpha,\lambda,\theta}u)_n = u_{n+1}+u_{n-1}+\lambda \cos(\pi \alpha n+\theta)u_n
\end{equation}
the \emph{almost Mathieu operator} (henceforth AMO). In much recent literature, what I call $\lambda$ is called $2\lambda$ (so the self dual point is $\lambda=1$) but I'll follow the convention of the older literature.  I like to joke that there have been more papers in the \emph{Annals of Mathematics} about the AMO than about any other single mathematical object.  In the physics literature, this is called Harper's equation when $\lambda=2$ and arose as a tight binding approximation to a 2D electron in magnetic field ($\alpha$ is then the magnetic flux per unit cell).  The name almost Mathieu equation is one I introduced in \cite{ASAP2} and \cite{SiAPFlu}.  I took it from the fact that the differential equation
\begin{equation}\label{9.3}
  -\frac{d^2u}{dx^2}+ \lambda \cos(x) u(x) = E u(x)
\end{equation}
is called the Mathieu equation (with Avron \cite{ASMath}, I had then recently studied the asymptotics of its gap widths as $E\to\infty$).  My name is a joke based on the fact that \eqref{9.2} is almost \eqref{9.3} and is also only almost periodic if $\alpha$ is irrational (while \eqref{9.3} is periodic).

Two basic objects associated to one dimensional ergodic operators are the density of states (DOS) and the Lyapunov exponent.  The DOS, unlike the Lyapunov exponent, makes sense in higher dimensions, but, for simplicity, let us mainly focus on the one dimensional discrete case.  For each $\omega$, $H_\omega$ defines a self-adjoint operator on $\ell^2(\bbZ)$, and so defines a spectral measure, $d\mu_\omega(E)$, defined by
\begin{equation}\label{9.4}
  \int f(E) d\mu_\omega(E) = \jap{\delta_0, f(H) \delta_0}
\end{equation}
The DOS measure $dk(E)$ is defined by
\begin{equation}\label{9.5}
  dk = \bbE(d\mu_\omega)
\end{equation}
The integrated density of states (\emph{IDS}) is then defined by
\begin{equation}\label{9.6}
  k(E) = dk((-\infty,E))
\end{equation}
If $\chi_L$ is the characteristic function of $\{n\,\mid\,-L\le n\le L\}$, then translation covariance shows that if $P_B(H_\omega)$ is the spectral projection for a set, $B$, then
\begin{equation}\label{9.7}
  (2L+1)^{-1} \bbE(\tr(\chi_L P_B(H_\omega) \chi_L)) = \int_B dk(E)
\end{equation}
This together with translation covariance and the Birkhoff ergodic theorem \cite[Theorem 2.6.9]{HA} imply that for a.e. $\omega$ and all continuous functions $f$ on $\bbR$ of compact support one has that
\begin{equation}\label{9.8}
 \lim_{L\to\infty} (2L+1)^{-1} \tr(\chi_L f(H_\omega) \chi_L)) = \int f(E) dk(E)
\end{equation}
A simple argument (e.g. restricting to moments) then shows that $dk$ is also the limit of the eigenvalue density of $H_\omega$ restricted to large boxes with either periodic or Dirichlet boundary conditions.

The earliest mathematical work on the DOS was by Benderski\u{i}-Pastur \cite{BPDOS} who defined it in the random case as a limit of box eigenvalue counting.  See \cite[pg. 175]{CFKS} for additional references on work on the random case prior to Avron-Simon \cite{ASAP2} whose principal theme was the DOS for the almost periodic case (and, as we'll discuss below, the Thouless formula, Aubry duality and spectral properties of the AMO).  We introduced the definition via \eqref{9.5} and the formula \eqref{9.8} which we proved held for every $\omega$ in the hull, rather than just almost every $\omega$.  We also proved the equality to the definition via \eqref{9.5} to the definition via periodic or Dirichlet boundary conditions and also (up to a factor of $\pi$) to the then recently defined \emph{rotation number} of Johnson-Moser \cite{JohnMAP}.  This equality under boundary conditions was natural given the statistical mechanical analogy and the proof was not hard.  Twenty-five years later, its analog turned out to be very useful to prove a result \cite{SiOPBC} in the theory of orthogonal polynomials that surprised many experts in that subject.

We also proved that for any $\omega$, the spectrum of $H_\omega$ is equal to the support of the measure $dk$ and, in particular, $\spec(H_\omega)$ is the same for all $\omega$ in the hull. We also proved that $k$ was continuous in $E$ (and in $\alpha$ at irrational values of $\alpha$ - it is definitely \emph{not} continuous at rational $\alpha$!) in 1D and noted the importance of continuity since that implies that for every $E$ one has that (here $H_\omega^{L,D}$ is the restriction of the Hamiltonian, $H_\omega$, to the box $\{n\,\mid\,-L\le n\le L\}$ with Dirichlet boundary conditions which one could replace by periodic boundary conditions).
\begin{equation}\label{9.9}
  k(E) = \lim_{L\to\infty} (2L+1)^{-1} \# \{\text{eigenvalues of } H_\omega^{L,D} \le E\}
\end{equation}

Our proof of the continuity of $k$ in 1D depended on the fact that in 1D eigenvalues have multiplicity at most $1$ (all that mattered was the finiteness) so we suggested, but couldn't prove, that $k$ was continuous in general dimension.  Moreover, the proof only showed that {$\lim_{\varepsilon\downarrow 0}[k(E+\varepsilon)-k(E-\varepsilon)]=0$} (which implies continuity for monotone functions) with nothing about how small differences are.  Later, Craig and I \cite{CrSi1} proved log H\"{o}lder continuity (using subharmonic function methods introduced in the subject by Herman \cite{Herman}) and then extended this to any dimension \cite{CrSi2} to get the continuity in any dimension that \cite{ASAP2} had conjectured.  Since the proofs of \cite{CrSi1, CrSi2} use the Thouless formula, I'll discuss it below.  I note that shortly after us, Delyon-Souillard \cite{DSDOS} found a distinct, really short, proof of the multidimensional continuity (but not log H\"{o}lder continuity).

Before leaving the DOS, I should mention gap labelling, not because I contributed to it, but because it will be relevant to the discussion of Cantor spectrum and the ten martini problem.  If $H$ is a discrete Schr\"{o}dinger operator of period $L$, the usual Bloch wave analysis shows the spectrum can have up to $L-1$ gaps and that if there is such a gap, the IDS value in it (which has to be constant) is an integral multiple of $1/L$.  Johnson-Moser \cite{JohnMAP} (once one has the equality of their rotation number and the IDS) found an analog for the 1D almost periodic continuum case (their method was extended to the discrete case by Delyon-Souillard \cite{DSGap}).  Independently, Bellissard \cite{BellGL1}, proved the same result using $C^*$-algebra methods, eventually using the same idea for certain higher dimensional operators \cite{BellGL2}.  Bellisard's work caused Johnson \cite{JohnSAC} to find another approach to gap labelling. For any almost periodic function, $f$, its average, $\bbA(f)= \lim_{R\to\infty} (2R)^{-1} \int_{-R}^{R} f(x)\,dx$ is easily shown to exist.  Given a real, $\omega$, $f$ is said to have a non-zero Fourier coefficient at $\omega$ if and only if $\bbA(e^{-2\pi ix\omega}f)\ne 0$. Since in the case that $x\in\bbZ$, these Fourier coefficients only depend on the fractional part of $\omega$, we view $\omega$ as an element of $\bbR/\bbZ$ which we write as $[0,1)$ and add mod $\bbZ$. One can prove that the set of $\omega$ with non-zero coefficient is a countable set.  The set of reals which are finite sums and differences of those $\omega$ with non-zero coefficients is called the \emph{frequency module of $f$}.  It is easy to see that $f$ is quasiperiodic if and only if the frequency module is finitely generated.  Moreover, unless the potentials are periodic, the frequency module is dense in $[0,1)$ or $[0,\infty)$.

\emph{Gap labelling} is the assertion that in any gap of the spectrum, the value of the IDS is a number in the frequency module.  What this says in case $V$ or $a, b$ are periodic, where the frequency module is multiples of $1/p$ (with $p$ the ``true'' period), is that the constant value of the IDS in a gap is among $j/p, j=1,\dots$ (where in the discrete case $j$ runs through $p-1$).  We speak of a gap being \emph{open} for a given $j$ if there is such an interval of constancy and \emph{closed} if there is a single energy with $k(E)=j/p$. Earlier in 1976, I had proven in \cite{SiGenericGaps} that in the continuum case for a dense $G_\delta$ of $V$'s of period one, all allowed gaps are open.  For the discrete case there is a much more precise analysis \cite{DMN, MvM} that shows the set of period $p$ Jacobi parameters with at least one gap closed is a finite union of closed varieties with codimension $2$ so the set with all gaps open is an open set that is much more than generic. A little thought shows that if all allowed gaps are open in the almost periodic case, the set of gaps is dense so that the spectrum of $H$ is a Cantor set (i.e. a closed, perfect, nowhere dense set).  Of course, it can be a Cantor set even if only many, rather than all, gaps are open.

Avron and I were struck by this Cantor spectrum.  Just as we were doing this work, pictures appeared from the Voyager flyby of Saturn which showed many more gaps in that planet's rings than previously known, so many that it almost appeared that the rings were nowhere dense!  We wrote a speculative paper \cite{ASSaturn} suggesting the structure might be due to an almost periodic Hill equation, although we pointed out that naive perturbation estimates of gap size were too small by several orders of magnitude so there would need to be some then not understood phenomena increasing this gap size.  Alas, there does not seem to be such a phenomenon and nature chose a different mechanism.

At this time, Avron and I \cite{ASAP1}, Chulaevski\u{i} \cite{ChulAP}, Moser \cite{MoserAP} and Pastur-Tkachenko \cite{PT} independently found classes of limit periodic Schr\"{o}dinger operators with Cantor spectrum.  We also proved the spectrum remained purely absolutely continuous so the spectrum was a positive measure Cantor set. (Much more is now known about limit-periodic Schr\"{o}dinger operators - a.c. spectrum is not typical; see Damanik-Gan \cite{DamGan}.)  We also discovered that such absolutely continuous spectrum still had all states with slow decay leading us to develop a refinement of a.c. spectrum \cite{ASTR} .

Mark Kac had moved to USC about the time I was visiting Caltech and at lunch one day in 1981, he and I discussed Cantor spectrum and the AMO.  We agreed that it was an interesting conjecture to prove that the operator $H_{\lambda,\alpha,\theta}$ of \eqref{9.2} had a Cantor spectrum for all irrational $\alpha$ and $\lambda\ne 0$ (if $\alpha$ is irrational, it is known (Avron-Simon \cite{ASAP2}) that the spectrum is $\theta$ independent). ``That's a grand conjecture'', said Mark, ``I'll offer ten martinis for its solution.'' He later repeated this offer at an AMS meeting and I popularized it as the ten martini problem. Added to the interest was the famous \emph{Hofstadter butterfly} \cite{Hofs1, Hofs2}, a picture (see Figure 2) showing the spectrum at the critical value $\lambda=2$ of the spectrum as a function of $\alpha$ (computed numerically for $\alpha=p/q$ with $q$ not too large) which looks like a fractal.  The ten martini problem was solved in full in 2004 by Avila-Jitomirskaya \cite{AJ10} (mentioned in Avila's Fields Medal citation) after an important partial result by Puig \cite{Puig}.  This is weaker than the result that all gaps are open, something known as the dry form of the ten martini problem (still partially open).

\begin{figure}[H]
\includegraphics[scale=0.8,clip=true]{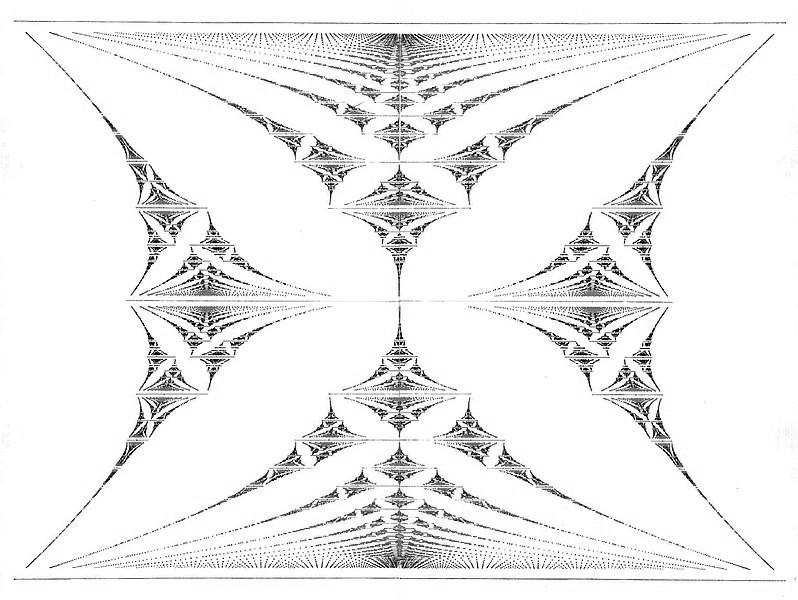}
\caption{The Hofstadter Butterfly}
\end{figure}

A year after my lunch with Kac, Bellissard and I \cite{BSCantor} used the strategy of my periodic result \cite{SiGenericGaps}.  We first proved that if $\alpha=p/q$ is rational and $q\theta$ is not a multiple of $\pi$, then all gaps were open (i.e. the spectrum has $q-1$ gaps).  This non-trivial analytic fact was proven using ideas motivated by the classical result of Ince \cite{Ince} that the continuum Mathieu operator \eqref{9.3} has all gaps open.   Once we knew that, with further analysis and some continuity results on $k(E)$ from Avron-Simon \cite{ASAP2}, the magic of the Baire category theorem showed that for a Baire generic set of $(\alpha,\lambda)$ the spectrum is a Cantor set! It is remarkable that with one's Baire hands one can learn something about the irrational case (Cantor spectrum) by knowing something about the rational case even though, of course, in the rational case, the spectrum is never Cantor.

When I told Mark about this on the phone admitting it wasn't the full result, he remarked ``But it is still interesting!  I'll give you three martinis for it.'' So I always think of this as the three Martini result.  Alas, before we met again, Mark was dead of abdominal cancer (the same disease that felled the other half of the Feynman-Kac formula not too long afterwards).

Returning to my basic series with Avron, I need to define the Lyapunov exponent.  I'll do it first for the discrete case.  Given a pair of potential Jacobi parameters, $a>0,b\in\bbR$ and $z\in\bbC$, one defines the single step transfer matrix:
\begin{equation}\label{9.10}
   A(a,b;z) = \frac{1}{a} \left(
                \begin{array}{cc}
                  z-b & -1 \\
                  a^2 & 0 \\
                \end{array}
              \right)
\end{equation}
so the difference equation
\begin{equation}\label{9.11}
  a_n u_{n+1} + b_n u_n + a_{n-1} u_{n-1} = z u_n
\end{equation}
is equivalent to $\left(
                   \begin{array}{c}
                     u_{n+1} \\
                     a_n u_n \\
                   \end{array}
                 \right)
  = A(a_n,b_n;z)\left(
                   \begin{array}{c}
                     u_{n} \\
                     a_{n-1} u_{n-1} \\
                   \end{array}
                 \right)$.  I learned the trick of putting a factor of $a$ in the lower component which yields an $A$ with $\det(A)=1$ from Killip in about 2000 and it didn't appear in the earlier papers.

One defines the \emph{transfer matrix}
\begin{equation}\label{9.12}
  T_n(\{a_j,b_j\}_{j=1}^n;z) = A(a_n,b_n;z)A(a_{n-1},b_{n-1};z)\ldots A(a_1,b_1;z)
\end{equation}
We use $T_n(z;\omega)$ for the transfer matrix with $a_n(\omega),b_n(\omega)$. The Furstenberg-Kesten theorem \cite[Theorem 2.9.1 ]{HA} then implies that for each fixed $z$, for a.e. $\omega$, one has that the Lyapunov exponent
\begin{equation}\label{9.13}
  \gamma(z) = \lim_{n\to\infty} n^{-1}\log\left(\norm{T_n(z;\omega)}\right)
\end{equation}
exists and is a.e. $\omega$ independent. More can be proven: The multiplicative ergodic theorem \cite[Theorem 2.9.10]{HA} says that for each $z$ and a.e. $\omega$, not only does \eqref{9.13} hold but there is a one dimensional subspace $V_{z;\omega}\subset\bbC^2$ so that
\begin{equation}\label{9.14}
  \lim_{n\to\infty} n^{-1}\log\left(\norm{T_n(z;\omega)v}\right) = \left\{
                                                    \begin{array}{ll}
                                                      -\gamma(z), & \hbox{ if } v\in V\setminus\{0\} \\
                                                      \gamma(z), & \hbox{ if } v\in\bbC^2\setminus V
                                                    \end{array}
                                                  \right.
\end{equation}

Thus for such $z,\omega$, if $\gamma>0$, then all solutions of \eqref{9.11} on a half line either grow or decay exponentially.  We emphasize that the need for a.e. $\omega$ rather than every is not a mere technicality but, as we will see (in the discussion two paragraphs prior to \eqref{9.23}), can have dramatic consequences.

An important role is played by what is called the \emph{Thouless formula}:
\begin{equation}\label{9.15}
  \gamma(E)=\int \log|E-E'| dk(E)
\end{equation}
which relates the Lyapunov exponent, $\gamma$, to the IDS, $k(E)$ in the discrete case. This is the form for the discrete Schr\"{o}dinger case where $a_n\equiv 1$; in general, one has an extra term $\bbE(-\log(a(\omega))$ with the added condition that this expectation is finite. It has the name because of the 1974 work of Thouless \cite{ThouHisForm} although it appeared earlier in the physics literature in a paper of Herbert and Jones \cite{HJThou}.  In fact, closely related ideas, although not the exact formula, go back to Szeg\H{o} in 1924 \cite{SzPot} who realized an important connection to two dimensional potential theory (for discussion of the basics of potential theory, see \cite{Helms, MFPot, Ransford} or \cite[Sections 3.6-3.7]{HA}) since the right side of \eqref{9.11} is the (negative of the) logarithmic potential of $dk$.  I was not aware of this related work from the OP community in the 1980's but only many years later, at which point I wrote a summary article \cite{SiPotTh} that explained the use of potential theory ideas to spectral theorists and the opposite direction to the OP community as well as some new insights.

Thouless' basic idea is that the elements of the transfer matrix when all $a_n\equiv 1$ are monic polynomials, $P_n(z)$, whose zeros are the eigenvalues of the Hamiltonian in a box. Since $\log(P_n(z))=\sum_{j=1}^{n}\log|z-E_j|$, where the sum is over the eigenvalues, \eqref{9.15} then follows from the fact that $dk$ is the limit of the density of eigenvalues in a box.  Avron and I realized this argument worked flawlessly when $z$ lay outside the (convex hull) of the spectrum of $H$, but because of infinities in the log was problematic for $z$ on the real axis.  Indeed, we noted in the almost periodic case for $z$ off the real axis, it held for all $\omega$ rather than just a.e. $\omega$.  In \cite{ASAP2}, we were able to use the fact that the integral on the right side of \eqref{9.15} is the Hilbert transform of $k(E)$ and the $L^2$ continuity of Hilbert transform to prove that \eqref{9.15} holds for Lebesgue a.e. $E$ in $\bbR$ and this suffices for some applications we made.

Slightly later, Craig and I \cite{CrSi1} were able to prove \eqref{9.15} for all $E\in\bbR$.  The key was the observation of Herman \cite{Herman} that the limit in \eqref{9.13} was subharmonic.  Since the integral on the right side of \eqref{9.15} is also subharmonic and, by Thouless' argument, the equation holds for $z$ non-real, it holds for all $z$ by a regularity result on subharmonic functions.  Craig and I realized that in general for fixed $\omega$, the quantity $ \limsup_{n\to\infty} n^{-1}\norm{T_n(z;\omega)}$ might not be upper semicontinuous which implied that this $\limsup$ might only have the right side of \eqref{9.15} as an upper bound.  We also realized that since $\gamma(E)\ge 0$, the measure $dk$ in \eqref{9.15} can't give too great weight to small sets which implied the log H\"{o}lder continuity.  By looking at averages of positive Lyapunov exponents on strips, we could even extend the continuity result to higher dimension.

Avron-Simon \cite{ASAP2} also began the study of a fascinating subject, the possible $\omega$ dependence of spectral components.  Recall \cite[Theorem 5.1.12]{OT} that one can refine the spectrum into pure point, a.c. and s.c. pieces.  It is a theorem of Kunz-Souillard \cite{KunzS} that these spectral pieces are a.e. constant in the general ergodic case (one might think this is obvious by the ergodic theorem but the subtlety is proving the measurability in $\omega$ of the projections onto various spectral pieces). As mentioned above, Avron and I proved a.e. could be replaced by all in the almost periodic case for the spectrum but as we'll see shortly, that is not true for two of the three spectral components.

For the AMO, \eqref{9.2}, there are interesting dependencies of spectral types on the coupling constant and frequency.  A key aspect is what is called Aubry duality \cite{Aubry, AA}.  Formally, the Fourier transform maps $H_{\alpha,\lambda}$ to $\tfrac{\lambda}{2} H_{\alpha,4/\lambda}$ since it maps the finite difference operator to multiplication by $\cos$ and turns multiplication by $\cos$ into a finite difference operator.  Of course, this can only be formal since Fourier transform in $\ell^2(\bbZ)$ maps not to itself but to $L^2(\partial\bbD,d\theta/2\pi)$!  One version of \emph{Aubry duality} says that the IDS, $k(\alpha,\lambda;E)$, of $H_{\alpha,\lambda}$ obeys
\begin{equation}\label{9.16}
  k(\alpha,\lambda;E)= k(\alpha,4/\lambda;2E/\lambda)
\end{equation}
One way of understanding the dual relation is to view the direct integral of $H_{\alpha,\lambda,\theta}$ over $\theta$ as an operator on $L^2(\bbR)$ and apply the appropriate Fourier transform. Alternatively, following \cite{AA}, one looks at $\alpha=p/q$ with $\theta=0$ on a set with $q$ points on which finite Fourier transform maps $\ell^2(\bbZ_q)$ to itself.  One obtains \eqref{9.16} by approximating irrational $\alpha$ by rationals.  Aubry-Andr\'{e}'s argument \cite{AA} for the limit was formal; Avron and I \cite{ASAP2} proved the necessary continuity (which only holds at $\alpha$ irrational!) to get the first rigorous proof of \eqref{9.16}.  Two immediate consequences of Aubry duality are (here $\alpha$ is irrational)
\begin{align}
  \spec(H_{\alpha,\lambda}) &= \frac{\lambda}{2} \spec(H_{\alpha,4/\lambda}) \label{9.17} \\
   \gamma(\lambda,\alpha;E) &= \gamma(4/\lambda,\alpha;2E/\lambda) + \log(\lambda/2) \label{9.18}
\end{align}

Aubry-Andr\'{e} \cite{AA} have a number of conjectures about the almost Mathieu equation which I (and others) made some progress on in my work.  The first involves the conjecture about the Lebesgue measure, $|\spec(H_{\alpha,\lambda})|$, of the spectrum
\begin{equation}\label{9.19}
  |\spec(H_{\alpha,\lambda})| = 2|2-|\lambda||
\end{equation}
based on numeric calculations.  We note this implies zero Lebesgue measure when $\lambda=2$ (which has led to a lot of literature on what the Hausdorff measure is of the set in that case; \cite{JK} has a summary of some of that literature as well as new results), an earlier conjecture of Hofstadter \cite{Hofs1, Hofs2}.

In \cite{ASvM}, Avron, von Mouche and I attempted to prove \eqref{9.19} and proved the equality for rational $\alpha$ if the left side is the Lebesgue measure of the intersection over $\theta$ of $\spec(H_{\alpha,\lambda,\theta})$ and proved the convergence of the Lebesgue measure of the union over $\theta$ as any sequence of rationals approaches an irrational $\alpha$.  This implies that the left side of \eqref{9.19} is $\ge$ the right side for $\alpha$ irrational.  The techniques of \cite{ASvM} have been used in many later works studying this problem.  For $\alpha$ whose continued fraction expansions are not bounded, Last \cite{LastMeas1, LastMeas2} proved the complete \eqref{9.19} for all $\lambda$.  The set of $\alpha$ with bounded integers in their continued fraction expansion is easily seen to have Lebesgue measure zero and to be a nowhere dense $F_\sigma$ so Last's result covers ``most'' irrationals but not all and, in particular, it does not cover the golden mean which has been used in many numeric calculations.  The result is now known for all irrational $\alpha$ through a series of papers.  The history is reviewed in Jitomirskaya-Krasovsky \cite{JK} which has a simple proof of the general result.

The other conjecture in \cite{AA} concerns spectral types.  One starts with \eqref{9.18} and the fact that always $\gamma \ge 0$ which implies that for $\lambda>2$
\begin{equation}\label{9.20}
  \gamma(E) \ge \log(\lambda/2)
\end{equation}
(proven rigorously by Avron-Simon \cite{ASAP2} and Herman \cite{Herman}) which later was proven to hold with equality on the spectrum by Bourgain-Jitomirskaya \cite{BJ}.  By a result of Pastur \cite{PasturAP} and Ishii \cite{IshiiAP}, strict positivity of the Lyapunov exponent implies that the spectrum there has no a.c. component so \cite{AA} suggested that when $\lambda>2$, the spectrum is pure point.  The Fourier transform of a rapidly decaying eigenfunction looks like a plane wave so their conjecture on spectral type was a.c. spectrum when $0<\lambda<2$ and pure point spectrum when $\lambda>2$.  They realized that $\lambda=2$ would be subtle and suggested perhaps there would be eigenfunctions with power law decay.

When Avron and I started thinking about the issue of spectral type for the AMO, Peter Sarnak, then a graduate student, suggested to me that spectral properties might depend on the Diophantine properties of the irrational frequencies (see also \cite{SarnAP}), that is how well those irrationals are approximated by rationals.  The \emph{Liouville numbers} are those irrationals $\alpha$ for which there exist rationals $p_k/q_k$ with
\begin{equation}\label{9.21}
  \left|\alpha-\frac{p_k}{q_k}\right| \le k^{-q_k}
\end{equation}
while we say that $(\alpha_1,\dots,\alpha_\ell)$ have \emph{typical Diophantine properties} if there exist $C$ and $k$ so that for all integers, not all zero, we have that
\begin{equation}\label{9.22}
  \min_{m\in\bbZ}\left|m-\sum_{j=1}^{\ell} n_j\alpha_j\right| \ge C(n_1^2+\dots+n_\ell^2)^{-k}
\end{equation}
We say that $\alpha$ is \emph{Diophantine} if \eqref{9.22} holds with $\ell=1$ and $\alpha_1=\alpha$.  It is well known that the Diophantine rationals in $[0,1]$ have full Lebesgue measure while the disjoint set of Liouville numbers is a dense $G_\delta$ (providing an interesting demonstration that the two notions of generic are distinct).  Avron and I decided that the picture of Aubry-Andr\'{e} was likely wrong when $\alpha$ was a Liouville number.  I visited Moscow in the spring of 1981 and explained our expectation and Molchanov came up to me with a young mathematician, Sasha Gordon, who had shown \cite{GordonHisLemma} that if a potential, $V$, is well enough approximated by periodic potentials, then $-\tfrac{d^2}{dx^2}+V$ on $L^2(\bbR,dx)$ has no square integrable eigenfunctions.  Avron and I found the easy extension to the discrete case and used it and the Pastur-Ishii result to prove \cite{ASAPAnon, ASAP2} that when $\alpha$ is a Liouville number and $\lambda>2$, then the AMO, $H_{\alpha,\lambda,\theta}$, has purely singular continuous spectrum for all $\theta$.  This was only one of the times that Gordon had a significant impact on my work - the other most significant one was his impact on my work on generic singular continuous spectrum (see Section \ref{s12}).  We also had two joint papers \cite{GJMS, GJLS}. I think Gordon, who was a very inventive mathematician, never got the recognition that he deserved and I felt guilty that I might have benefitted from his brilliance as much or more than he did!

The result with Avron on examples with purely singular spectrum and $\gamma>0$ on the spectrum shows the subtlety of tracking measure zero sets.  By Fubini's theorem and the multiplicative ergodic theorem, one concludes that for a.e. $\theta$, one has that for a.e. $E\in\bbR$, every solution of the difference equation $H_{\alpha,\lambda,\theta}u=Eu$ either decays or grows exponentially at both $\pm\infty$.  Gordon's lemma implies that in the Liouville case, the solutions decaying in one direction can't in the other so the spectral measures of $H_{\alpha,\lambda,\theta}$ live on the sets where Lyapunov behavior fails to hold.  The moral is that the a.e. set of $\theta$ where \eqref{9.13} might not hold, not only exists but can be where all the important stuff is happening!

For Diophantine $\alpha$, with $\lambda>2$, the Aubry-Andr\'{e} conjecture was proven by Jitomirskaya \cite{JitoLocalAMO} who proved for such values of the parameters, one has dense point spectrum of $H_{\alpha,\lambda,\theta}$ for Lebesgue a.e. $\theta$.  It is not a limitation that the proof is only for a.e. $\theta$ since the spectrum is purely singular continuous for a dense $G_\delta$ of $\theta$ (see the discussion of my work with Jitomirskaya \cite{SiSingC3} in the next paragraph).  That leaves $\alpha$ which is neither a Liouville number nor Diophantine, a non-empty, uncountable, set of irrationals that is both of Lebesgue measure zero and a subset of a nowhere dense $F_\sigma$ (so, in a sense, rare).  For such $\alpha$ one looks at the continued fraction approximations $p_n/q_n$ which are, the best rational approximations \cite[Section 7.5]{CAA}, and defines
\begin{equation}\label{9.23}
  \beta(\alpha) \equiv \limsup_{n\to\infty} \left(\frac{\log q_{n+1}}{q_n}\right)
\end{equation}
(This measure of approximation in the context of almost periodic Schr\"{o}dinger operators goes back to my paper on the Maryland model \cite{SiMaryland} in an equivalent form $\beta(\alpha) \equiv \limsup_{n\to\infty} -n^{-1} \log(|\sin(\pi\alpha n)|)$.  Diophantine $\alpha$ have $\beta=0$ and Liouville $\alpha$ have $\beta=\infty$. Avila-You-Zhou \cite{AYZ} (see also Jitomirskaya-Liu \cite{JitLiu2}) proved a conjecture from \cite{JitoEvery} that if $2<\lambda<2e^{\beta(\alpha)}$ the spectrum of $H_{\alpha,\lambda,\theta}$ is purely singular continuous for all $\theta$ and if $\lambda>2e^{\beta(\alpha)}$, then the spectrum of $H_{\alpha,\lambda,\theta}$ is dense pure point for a.e. $\theta$.  See Jitomirskaya-Liu \cite{JitLiu} for more on this case including a review of the literature and a detailed analysis of the eigenfunctions.

One of the results of the singular continuous revolution that I'll discuss in Section \ref{s12} is that Jitomirskaya and I \cite{SiSingC3} proved that if $a_n(\omega)=1$ and $b_n(\omega)$ is an even almost periodic function, then for a dense $G_\delta$ of $\omega$ in the hull, $h_\omega$ has no eigenvalues.  If it is a model where $\gamma>0$ on the spectrum so that the Pastur-Ishii theorem implies no a.c. spectrum, this yields purely s.c. spectrum for a Baire generic set of $\omega$.  Since there are models (like AMO for $\lambda>2$ and $\alpha$ Diophantine) where it is known there is dense point spectrum for a.e. $\omega$, we see there are examples where neither the point spectrum nor the s.c. spectrum are $\omega$ independent even though we know that they are a.e. $\omega$ independent. However, independently, Kotani \cite{KotaniAC} and Last and I \cite{LSEigen} showed that a.c. spectrum is always $\omega$ independent.

I later wrote a paper with Hof and Knill \cite{HKS} in which we proved, using a relative of the ideas in \cite{SiSingC3} that certain weakly almost periodic potentials taking only finitely many values (which are known to have no a.c. spectrum \cite{KotaniFinite}) have purely s.c. spectrum for a dense $G_\delta$ set in their hull.

Next, I turn to AMO when $0<\lambda<2$.  The results on point spectrum with exponentially decaying eigenfunctions for $\alpha$ Diophantine and $\lambda>2$ plus Aubry duality immediately imply lots of a.c. spectrum when $\alpha$ is Diophantine and $\lambda<2.$  For several years after \cite{ASAP2}, it was assumed that the dual of singular continuous spectrum must be purely singular continuous, so it came as a big surprise when Last \cite{LastMeas1} proved that for \emph{all} irrational $\alpha$ one has that $|\sigma_{ac}(H_{\alpha,\lambda,\theta})|\ge 4-2|\lambda|$.  At the time, Last was a graduate student of Avron at Technion and Avron told me of Last's result.  I assured Avron that I was sure Last was wrong.  Since I would be coming soon to Israel, rather than plow through the paper and figure out the error, I suggested we meet so I could determine where the error was and tell him! To my surprise, Last convinced me that his proof was correct and that he could prove a kind of lower semicontinuity on $|\sigma_{ac}(H)|$ and then use the fact that Avron, van Mouche and I \cite{ASvM} had proven the inequality for rational $\alpha$.  Once the blinders were removed, I realized that my work then in progress with Gesztesy \cite{GSxi} provided a new proof of Last's result!  Indeed, we could slightly improve his result since where he had $|\sigma_{ac}|$, we could obtain the potentially smaller $|\Sigma_{ac}|$ ($\Sigma_{ac}$ is the essential support of the ac spectrum, that is the minimal class of sets mod sets of Lebesgue measure $0$ that supports all the a.c. spectral measures).  We proved that
\begin{theorem} [Gesztesy-Simon \cite{GSxi}] \lb{T9.1} Let $H^{[n]}$ be a sequence of periodic discrete Schr\"{o}dinger operators so that for each fixed $m$, $b^{[n]}_m$ has limit $b_m$ and let $H$ be the discrete Schr\"{o}dinger operator with potential $b$.  Then for any open interval $(\alpha,\beta)\subset\bbR$, we have that
\begin{equation}\label{9.24}
  |(\alpha,\beta)\cap\Sigma_{ac}(H)| \ge \limsup_{n\to\infty} |(\alpha,\beta)\cap\Sigma_{ac}(H^{[n]})|
\end{equation}
\end{theorem}
For AMO with $|\lambda|<2$, this leaves the question of the point and singular continuous spectra which were expected to be empty and proving that first for a.e. $\theta$ and then for all $\theta$ was open for many years; indeed, proving this for non-Diophantine $\alpha$ (Jitomirskaya \cite{JitoLocalAMO} had handled the Diophantine case for a.e. $\theta$) was one of the list of problems I sent to the 2000 ICMP \cite{Simon15}.  The full result was settled by Avila \cite{AvilaProb6} (It is most unfortunate that this paper has never appeared.  It seems the blame is shared by the top journal that rejected it and by the author who then refused to send it elsewhere).

Finally, in our discussion of AMO, I mention the self-dual point $|\lambda|=2$ which is often quite subtle.  In \cite{GJLS}, Gordon, Jitomirskaya, Last and I claimed that if the spectrum of $H_{\alpha,\lambda,\theta}$ has zero measure for $\lambda=2$ and some irrational $\alpha$ (the spectrum is $\theta$ independent), then for a.e. $\theta$ the spectrum is purely singular continuous.  At the time the zero measure result was known for most, but not all, irrational $\alpha$ but, as just mentioned, it is now known for all irrational $\alpha$. As explained in the discussion under (2) below, \cite{GJLS} was fooled by a sloppily stated result in \cite{DSAP} so there was a gap in the proof and the result was only established in Avila et al. \cite{AJM}. That left the question of whether there might be an exceptional set with some (or even all) point spectrum.  Very recently, Jitomirskaya \cite{JitoCritical} proved there are no point eigenvalues for any $\alpha$ and any $\theta$.  This paper includes a discussion of earlier work between the 1997 paper of Gordon et al \cite{GJLS} and her 2020 breakthrough.

That concludes our survey of the refined spectral analysis of AMO and I conclude this section with a summary of some of my other papers on almost periodic operators.

(1) \emph{Kotani Theory}. In 1982, I received a brilliant paper by S. Kotani \cite{KotaniTheory} which dealt with ergodic continuum 1D Schr\"{o}dinger operators. Since it dealt with a.c. spectrum, it was mainly of interest for the almost periodic case.  It had three main results

(a) A kind of converse of the Pastur-Ishii theorem, namely, if $\gamma(E)=0$ on a Borel subset $A\subset\bbR$ of positive Lebesgue measure, then for a.e. $\omega$, $H_\omega$ has a.c. spectrum on $A$.

(b) If $\gamma(E)=0$ on an open interval $I\subset\bbR$, then the spectrum is purely a.c. on $I$.

(c) If $\gamma(E)=0$ on a Borel subset $A\subset\bbR$ of positive Lebesgue measure, then, the process $\x\mapsto V_\omega(x)$ is deterministic which means there is no a.c. spectrum in ``truly random'' cases.

I discovered it was not so straightforward to extend this to the case of discrete Schr\"{o}dinger operators but I succeeded in \cite{SiKotani} which was used by many later authors.  In \cite{KoSiStrip}, Kotani and I joined forces to extend these results to discrete strips.

(2) \emph{Deift-Simon Theory}. I wrote a paper with Deift \cite{DSAP} that focused on aspects of a.c. spectrum motivated by Kotani \cite{KotaniTheory} and some results of Moser \cite{MoserAP} who had proven that the rotation number, $\alpha(E)=\pi k(E)$, obeys
\begin{equation}\label{9.25}
  d\alpha^2(E)/dE \ge 1
\end{equation}
on the spectrum of periodic continuum Schr\"{o}dinger operators and also for the particular limit periodic potentials he studied in \cite{MoserAP}.  In \cite{DSAP}, we noted that \eqref{9.25} could not hold in general for all ergodic Schr\"{o}dinger operators, essentially because of the phenomena of Lifshitz tails (see Section \ref{s11}) but we proved in the continuum case, it holds on the set where $\gamma(E)=0$ and for the discrete Schr\"{o}dinger operator (i.e. Jacobi matrix with $a\equiv 1$) one has the stronger
\begin{equation}\label{9.26}
  2 \sin(\alpha) d\alpha(E)/dE \ge 1
\end{equation}
on $A\equiv \{E\,|\,\gamma(E)=0\}$ which implies that $|A|\le 4$. We also proved that the a.c. spectrum has multiplicity 2. While we regarded these as the most important results in the paper (as seen by our abstract), this paper is probably best known for two more technical aspects.  First we construct $L^2$ (in $\omega$) eigenfunctions for energies in the a.c. spectrum, which, for example, plays a critical role in the work in (4) below. Secondly, there is a claim in \cite{GJLS} that our results imply mutual singularity of the singular parts of the spectral measures for a.e. pair $(\omega,\omega')$ based on a theorem in \cite{DSAP} that for every real $E$, a certain set of $\omega$ associated to the singular spectrum has measure zero.  Unfortunately, the theorem in \cite{DSAP} is sloppily stated in that in the section it appears, there is an implicit condition that $\gamma>0$ which the authors of \cite{GJLS} forgot (despite the fact that I was a coauthor of both - blush!).  Since \cite{GJLS} apply this at the self-dual $\lambda$ where $\gamma=0$, that paper has one incorrect claim.  I note that the claimed mutually singularity when $\gamma=0$ is still open although many expect that it is true.

(3) \emph{The Maryland Model} Dick Prange and two of his postdocs at Maryland found and studied \cite{Maryland1, Maryland2, Maryland3, Maryland4} a fascinating exactly solvable almost periodic model which I dubbed ``the Maryland model'', a name that has stuck in later literature.  I wrote two papers \cite{SiLorentz, SiMaryland} on rigorous aspects of the model which had some overlap with another independent rigorous analysis by Figotin-Pastur \cite{FPMaryland}.  The model is just like AMO but $\cos$ is replaced by $\tan$, i.e.
\begin{equation}\label{9.27}
   (H_{\alpha,\lambda,\theta}u)_n = u_{n+1}+u_{n-1}+\lambda \tan(\pi \alpha n+\theta)u_n
\end{equation}
Since $\tan$ is unbounded, one needs to eliminate the countable set of $\theta$'s where for some $n$, $\pi \alpha n+\theta$ is of the form $(k+\tfrac{1}{2})\pi;\, k\in\bbZ$.  Then one has a well defined but unbounded operator.  When $\alpha$ is Diophantine, the Maryland group found an explicit set of eigenfunctions but didn't prove it complete and their computation of the density of states had other formal elements.

One surprise is that the DOS was the same as for the Lloyd model which is the random model whose single site distribution is $\pi^{-1}\tfrac{\lambda}{\lambda^2+x^2}\,dx$ which physicists call Lorentzian and mathematicians the Cauchy distribution. $\lambda$ is called the \emph{half-width} of the distribution. This distribution has the following weird property.  If $X_1, X_2$ are two independent random variables each with a Cauchy distribution of the same half-width, then for any $t\in [0,1]$, $tX_1+(1-t)X_2$ is also a Cauchy random variable with half width $\lambda$.  \cite{SiLorentz} noted that this implies that the Lloyd model has the same DOS as the free Hamiltonian with a random Cauchy constant added to it and so does the Maryland model with the same $\lambda$ (note that if $\theta$ is uniformly distributed on $[0,\pi]$, then $\lambda \tan(\theta)$ is Cauchy distributed with half width $\lambda$).

In addition to a proof of completeness of those eigenfunctions that had been found by Prange's group, \cite{SiMaryland} studies other properties of this model including the fact that for $\alpha$ a Liouville number the spectral measures are purely singular continuous and the structure of the (non-normalizable) eigenfunctions for such $\alpha$.  I note that this model was one of the first times that non-mathematical physicists had to face singular continuous spectrum; they gave the corresponding eigenfunctions which decay in an average sense but are not normalizable, the name ``exotic states.''

(4) \emph{Clock Spacing of Zeros for Ergodic Jacobi Matrices with Absolutely Continuous Spectrum} We saw above that the DOS describes the bulk properties of the eigenvalue distribution of Schr\"{o}dinger operators in large boxes in that the eigenvalue counting distribution converges to the DOS.  But there is the issue of the fine structure, in particular the spacing between nearby eigenvalues.  The earliest results on this problem are in the random Anderson type case where Molchanov \cite{MolchanovPoisson} in $1D$ and Minami \cite{MinamiPoisson} in higher dimensions proved the distribution is asymptotically Poisson.  For random matrices, the fine spacing in the bulk is governed by the Wigner surmise for GOE/GUE (see Mehta \cite{MehtaGUE} or Deift \cite{DeiftGUE} for discussion, proof and history).   Basically, because of strong localization in the Anderson case, nearby eigenvalues don't impact each other and the placement of such eigenvalues are close to independent of each other. In the random matrix case, eigenfunctions are more spread out, there is some eigenvalue repulsion so eigenvalues are less likely to be too close to each other.

I gave the problem of extending the Poisson results to OPUC to my then graduate student, Mihai Stoiciu which he solved \cite{StoiciuPoisson}.  Along the way, I suggested that he do some numeric calculations and, for comparison, suggested he look at some rapidly decaying Verblunsky coefficients.  Here is the striking result that he found for the zeros of $\Phi_{n=20}$ when $\alpha_j=\left(\tfrac{1}{2}\right)^{j+1}$:

\begin{figure}[H]
\includegraphics[scale=0.6,clip=true]{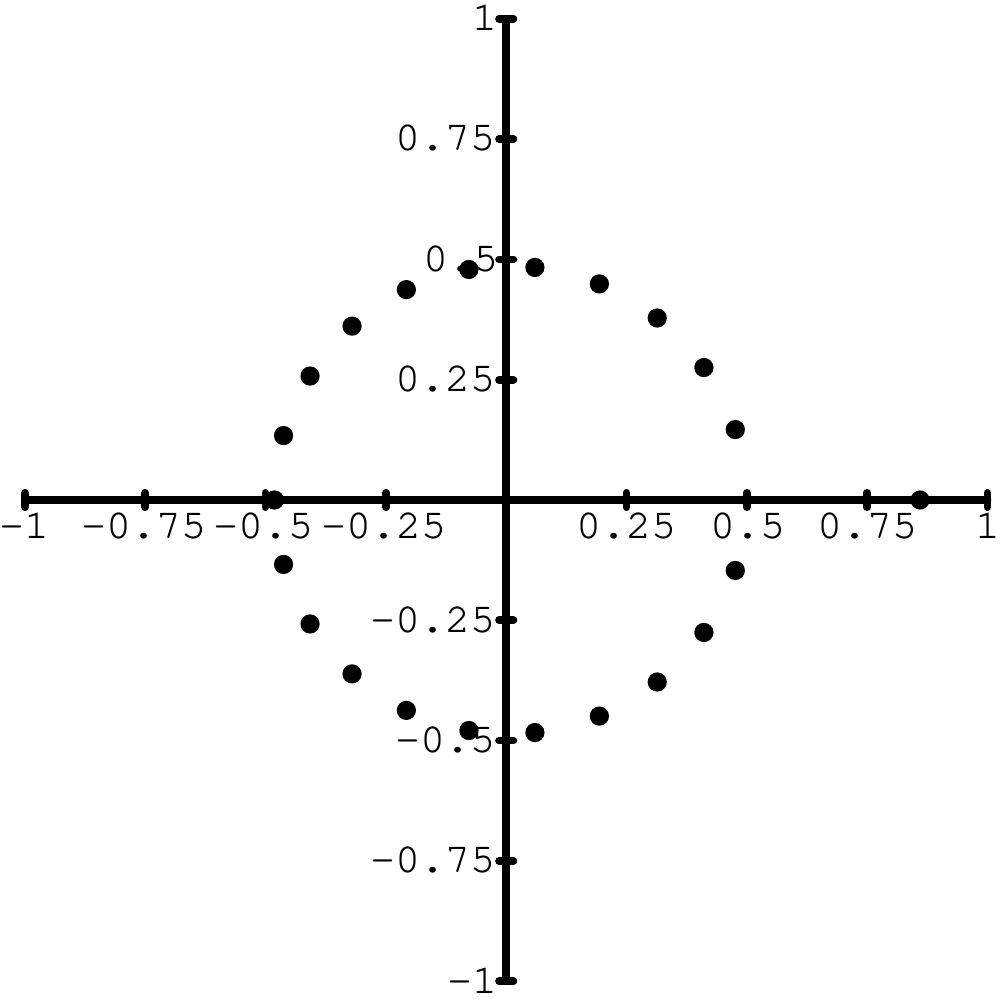}
\caption{Zeros of an OPUC}
\end{figure}

By a Theorem of Mhaskar-Saff \cite{MhSa}, it was known that the counting measure for the zeros in this case converges to a uniform distribution on the circle of radius $\tfrac{1}{2}$ but I was amazed when I saw that the eigenvalue repulsion was so strong they seemed to be spaced like the numerals on a clock.  I called this ``clock spacing'', a name which stuck even when eventually applied to OPRL when the spacing was locally rigid but globally not equally spaced because the limiting DOS didn't have a uniform density.  I wrote a series (the last with Last) \cite{SiClock1, SiClock2, SiClock3, SiClock4} for situations there Verblunsky or Jacobi parameters converged to a constant or periodic sequence.

Motivated by this, Lubinsky \cite{LubClock1, LubClock2}, found a new approach to clock behavior for OPRL with a.c. spectrum $[-1,1]$ based on proving a universality result for the Christoffel-Darboux kernel
\begin{equation}\label{9.28}
  K_n(x,y) = \sum_{j=0}^{n}p_j(x)p_j(y)
\end{equation}
namely that, one says that \emph{bulk universality} holds at $x_0$ if and only if
\begin{equation}\label{9.29}
 \frac{K(x_0+a/n,x_0+b/n)}{K_n(x_0,x_0)}\rightarrow \frac{\sin(\pi\rho(x_0)(b-a))}{\pi\rho(x_0)(b-a))}
\end{equation}
uniformly in bounded $a,b$ where $\rho(x)$ is the weight in the DOS which is assumed to be absolutely continuous.  Earlier, although Lubinsky and I didn't realize it until later, Freud \cite{FreudOP} had studied the fine structure of zeros, had proven \eqref{9.29} in a less general setting and realized that it implied clock spacing in the sense that if $\ldots x^{[n]}_{-k}<\ldots<x^{[n]}_{-1}<x^{[n]}_{0}\le x_0<x^{[n]}_{1}<\ldots$ are the zeros near $x_0$, then
\begin{equation}\label{9.30}
  \lim_{n\to\infty} n(x^{[n]}_{j+1}-x^{[n]}_{j})=1/\rho(x_0)
\end{equation}
for all $j\in\bbZ$.  Levin \cite{LLClock} independently rediscovered this connection, so the fact that bulk universality implies clock behavior is sometimes called the Freud-Levin theorem.

Lubinsky proved bulk universality for a large class of measures supported on $[-1,1]$ with a.e. non-vanishing a.c. weight there.  Totik \cite{TotikClock} and I \cite{SiClock5} were able to replace $[-1,1]$ by fairly general sets $\fre\subset\bbR$ but we required that $\fre$ have a large interior, so large it was dense in $\fre$.  Last and I realized that Lubinsky's second approach \cite{LubClock2} might allow one to handle various almost periodic Jacobi matrices with a.c. spectrum even though that spectrum is nowhere dense (e.g. the AMO with $|\lambda|<2$) but we ran into a couple of hard technical problems.  Fortunately, we were able to convince Avila to attack these issues and the three of us \cite{ALSClock} were able to prove bulk universality and clock behavior for a.e. $x_0$ in $\sigma_{ac}$ for all ergodic Jacobi matrices.

One of the goals when I wrote my two OPUC books \cite{OPUC1, OPUC2} was to extend the spectral analysis of Jacobi matrices to OPUC (i.e. replacing Jacobi parameters by Verblunsky coefficients).  Included were two sections concerning ergodic Verblunsky coefficients, one on random \cite[Section 12.6]{OPUC2} and one on subshifts \cite[Section 12.8]{OPUC2}, a class of weakly almost periodic functions.

I end this section on almost periodic Jacobi matrices and Schr\"{o}dinger operators by emphasizing that because it has focused on my own work, there is no discussion of some issues and limited discussion of later work on the issues we do discuss.  In particular, I have not said anything on Hausdorff dimension of the spectrum in the case where $\sigma(H)$ has Lebesgue measure zero nor about long time behavior of powers of the position (although I do have two papers on the latter \cite{SiBall1, SiBall2}). Nor have I discussed subshifts and substitution models except for the OPUC work just mentioned and \cite{HKS}. For more on these subjects, the reader can consult a number of books and review articles \cite{DamOpen, CL, DamRevSub, DamFil, JitoEvery, JitoOneFoot, JitoMarx, LastEvery, PF} (that said, we really do need a more recent comprehensive review of the AMO).

\section{Topological Methods in Condensed Matter Physics} \lb{s10}

I was a pioneer \cite{Nature} in the use of topology and geometry (mathematicians sometimes use ``geometry'' when there is an underlying distance and ``topology'' for those geometric object that don't rely on a distance) in NRQM.  In particular, Avron, Seiler and I \cite{ASS1} realized that the approach of Thouless  et al. \cite{TKNN} to the quantum Hall effect (for which Thouless got the Nobel prize) was basically an expression of the homotopic invariants (aka Chern integers) of a natural line bundle that arises in certain eigenvalue perturbation situations, and I realized \cite{SiBerry} that the phase that Berry \cite{BerryHisPhase} found in the quantum adiabatic theorem is holonomy in this bundle and that the quantity Berry \cite{BerryHisPhase} used to compute this phase (and which independently had been found by Avron et al. \cite{ASS1}), now called the \emph{Berry curvature}, is just the curvature in this line bundle.  I emphasize that Thouless  et al. \cite{TKNN} never mention ``topology'' and that Thouless learned they'd found a topological invariant, essentially the Chern class, from me.  And the only mention of curvature or holonomy in Berry \cite{BerryHisPhase} is where he remarks that \emph{Barry Simon, commenting on the original version of this paper, points out that the geometrical phase factor has a mathematical interpretation in terms of holonomy, with the phase two-form emerging naturally (in the form (7 b)) as the curvature (first Chern class) of a Hermitian line bundle}.

As a mathematician, I am mainly an analyst and most of my training and expertise is analytic so, as background, I should explain something about how I came to know enough toplogy/geometry to realize its significance in NRQM.  As a freshman at Harvard, I took the celebrated Math 55 Advanced Calculus course whose first half did differential calculus in Banach spaces and second half integral calculus on manifolds.  This was a dip into the sea of geometry but from an analytic point of view without any discussion of Riemannian metrics or curvature.  I did some self study of general relativity but the true topology/geometry was hidden since my study was in physics books (and before the era of those that emphasized the geometry).  A key part of my education was a course on Algebraic Topology given my senior year by Valentin Po\'{e}naru, then a recent refugee from Romania, who was visiting Harvard.  It was a wonderful course and I really got into the subject, so much that Po\'{e}naru took me aside and tried to convince me to give up mathematical physics and switch to topology.  I was particularly taken with the homotopy group long exact sequence of a fibration. (See Hatcher \cite{Hatcher} for background on this and other topological issues).

Let me mention one of the simplest examples of fibrations of interest in physics, namely, the Hopf fibration, which is a natural map of $S^3$ to $S^2$.  Let $\mathbf{\sigma}_j; j=1,2,3$ be the usual Pauli $\sigma$ matrices.  If $\mathbf{a}=(a_0,a_1,a_2,a_3)=(a_0,\mathbf{\overrightarrow{a}})$ is a unit vector in $\bbR^4$, then $U(\mathbf{a})\equiv a_0\bdone+ i \mathbf{\overrightarrow{a}}\cdot\mathbf{\sigma}$ is a unitary matrix with determinant $1$ if and only if $\mathbf{a}\in S^3$. This sets up a homeomorphism between $S^3$ and the group $SU(2)$ of $2\times 2$ unitary matrices of determinant $1$. In that case there is a rotation $R(\mathbf{a})$ on $S^2$ defined by $U(\mathbf{a})(\mathbf{b}\cdot\mathbf{\sigma})U(\mathbf{a})^{-1} =R(\mathbf{a})(\mathbf{b})\cdot\mathbf{\sigma}$.  This is the Cayley-Klein parametrization of rotations, a map of $SU(2)$ onto $SO(3)$.  If $e_3$ is a unit vector in the $z$ direction, then $\mathbf{a}\mapsto R(\mathbf{a})e_3$ defines the Hopf fibration, $H$, which maps $S^3$ to $S^2$.  The point is that it is easy to see (for example, by looking at the inverse images of the north and south poles) that inverse images of distinct points under $H$ are circles which are linked so the map is homotopically non-trivial proving that $\pi_3(S^2)$ is non-zero (in fact, this homotopy group is generated by $H$ and is just $\bbZ$).

Of course, geometry in the naive sense was present, even central, to some of my work in 1970's, for example the work on phase space methods in N-body NRQM (see Section \ref{s6}) and I had even mentioned that the Agmon metric was the geodesic distance in a suitable Riemann metric but if one thinks of ``real'' geometry needing curvature and ``real'' topology needing homology or homotopy invariants, I'd not used them in my research in the '70's.  In Section \ref{s8}, I mention work that was motivated by Witten's seminal paper \cite{Witten} on the supersymmetry proof of the Morse inequalities and index theorem.  This paper has been celebrated not only for the results itself but because of the bridge it opened up between high energy theorists studying gauge (and later string) theories and topologists but it also impacted me in leading me to consider certain geometric ideas that I needed in the work I'll describe in this Section.  This is not so much in those of my papers directly motivated by Witten \cite{SiSC1, SiSC2} but through other mathematics motivated by it.  For Witten motivated several reworkings of the proof of the Atiyah-Singer Index theorem, in particular, a preprint of Getzler \cite{Getz} (see \cite[Chapter 12]{CFKS} for additional references) which caught my attention in the period just after I gave the Bayreuth lectures which eventually appeared as \cite{CFKS}.  I had lectured on Witten's proof of the Morse inequalities there and decided to add a chapter on this further extension (the chapter, chapter 12, was actually the only chapter that I wrote in \cite{CFKS} - the other chapters were written by my coauthors based on and usually expanding the lectures I'd given).

For pedagogical reasons, I decided to give details only in the special, indeed, classical case of the Gauss-Bonnet theorem where it turns out that Getzler's proof is essentially one found in 1971 by Patodi \cite{Pat} who didn't know that he was speaking supersymmetry!  While I'd heard of the Gauss-Bonnet theorem, I hadn't known exactly what it said until following up on Witten taught me all about it.  Since it will explain some of my later work, let me say a little about this theorem (and also holonomy) in the case of $S^2$, the sphere of radius $R$ embedded in $\bbR^3$. At each point, the Gaussian curvature is $1/R^2$ so, if $K$ is the curvature and $d^2\omega$ the surface area, we have that
\begin{equation}\label{10.1}
  \frac{1}{2\pi}\int K\,d\omega = \frac{1}{2\pi} \frac{1}{R^2} 4\pi R^2 = 2
\end{equation}
The remarkable fact is that if you deform the sphere to another surface, say, an ellipsoid, then the curvature is no longer constant but the integral in \eqref{10.1} is still $2$.  But this is not true for the torus.  The integral is still independent of the underlying metric needed to define $K$, but it is $0$, as can be seen by looking at the flat torus $\bbR^2/\bbZ^2$ with the Euclidean metric on $\bbR^2$ (which cannot be isometrically embedded into $\bbR^3$ but can in $\bbR^4$).  In fact, for any surface in $\bbR^3$ (and for hypersurfaces in general dimension) the integral is the Euler characteristic of the surface (Euler-Poincar\'{e} characteristic in higher dimension).  This is the Gauss-Bonnet theorem.  It says that the integral of a natural geometric quantity lies in a discrete set and is determined by topological invariants.

To explain holonomy, consider someone carrying a spear around the earth trying at all times to keep the spear tangent to the sphere and parallel to the direction it was pointing (which may or may not be parallel to the direction the person is walking).  Imagine, going along the equator through one quarter of the earth, turning left, going to the north pole, turning left and going back to the original point.  Suppose the spear is parallel to the equator at the start.  The person turns to move along a line of longitude, but being careful not to turn the spear, it will point directly to his right.  After the next turn, the spear will point backwards.  So despite having tried to keep it parallel , upon return, it has rotated by $90^\circ$, i.e. $\pi/2$ radians.  This rotation after parallel transport is \emph{holonomy}.  The path encloses one eighth of the earth, a area of $4\pi R^2/8 =\pi R^2/2$ so the integral of the curvature over the enclosed area is the holonomy!

Perhaps relevant to my work is the following amusing story.  Avron and I were talking in my office about our work on almost periodic things and Dick Feynman burst in and exclaimed ``how do you compute the homotopy groups of spheres?''  There had been several papers in the high energy literature that mentioned those and he was puzzled why the higher homotopy groups were not trivial.  I told him about the Hopf fibration which had always struck me and then retrieved from my memory the exact sequence of a fibration.  The conclusion of our discussion continues to amuse me.  When I finished Avron looked at me and said: ``Barry, I didn't realize you knew anything about that''.  Before I could answer, Dick with a huge grin on his face turned around waved his hands at my rather full bookshelves and exclaimed in his trademark New York accent: ``Whadya mean?  He's a Professor, of course he knows it!''  I might have recalled all that when I needed homotopy and the exact sequence of a fibration several months later in my work with Avron and Seiler, but it helped that I'd had this interaction.

In early 1983, Yosi Avron told me about the paper of Thouless et al. paper \cite{TKNN} which gave a novel explanation of the quantum Hall effect, a subject that had fascinated Yosi.  The striking aspect of that effect is that a resistance was quantized.  In the TKNN approach (we quickly came up with that abbreviation, sometimes TKN$^2$, especially TKNN integers, a name which has stuck), this arose because, using the Kubo formula, they got the resistance (in a certain idealized situation) was given by an integral over a torus that turned out to be an integer (in suitable units).

We quickly realized that their integers were associated to a single band which was assumed non-degenerate (i.e. at every point in the Brillouin zone, the eigenstate for that band is simple) and their integrand involved the change of eigenfunction.  We also realized that since the integrand was an integer it had to be invariant under continuous change and so an indication of a homotopy invariant of maps from the two dimension torus $T^2$ to unit vectors in Hilbert space mod phases (equivalently a continuous assignment of a one dimensional subspace in the Hilbert space to each point in $T^2$).  After more thought and study, we learned that the homotopy class of maps from $T^2$ could be classified by maps from $S^1$ and $S^2$ and so the underlying homotopy groups of $\bbP(\infty)$, the one dimensional subspaces of a Hilbert space.  We also considered that there might be non-trivial homotopy invariants depending on several bands so what we wanted to consider was the homotopy groups of the set, $\calN$, of compact operators with non-degenerate eigenvalues.  We got excited since if, for example, we found a non-trivial $\pi_3$, there would be new topological invariants for the physically relevant three-dimensional torus.

By a continuous deformation, we could consider maps to a fixed set of simple eigenvalues but variable eigenspaces.  Given the phase change this was the same as the quotient of all unitary maps by the diagonal unitary maps $\calU(\calH)/D\calU(\calH)$.  So these homotopy groups might be computable via the exact sequence of the fibration that my talk with Feynman had reminded me about!  Indeed, since it was known that the set of all unitaries $\calU(\calH)$ is contractible, it has no homotopy, i.e. all homotopy groups are trivial and, thus, by the exact sequence of the fibration, we knew that $\pi_j(\calN)=\pi_{j-1}(D\calU(\calH))$.  Since the diagonal unitaries is just an infinite product of circles, $T^1$, and $\pi_j(T^1)$ is $\bbZ$ for $j=1$ and $0$ for all other $j$, we had discovered that the \emph{only} non-trivial homotopy group of $\calN$ was $\pi_2$, that the same was true for $\bbP(\infty)$ and that $\pi_2(\calN)$ was just an infinite product of $\pi_2(\bbP(\infty))$'s.  In other words, the only homotopy invariants were the TKNN integers.

We added Reudi Seiler, whom Yosi had been consulting, to the authors and published this negative result in \emph{Physical Review Letters} \cite{ASS1}. We made a big deal of our new result that if two non-degenerate bands with TKNN integers $n_1$ and $n_2$ went through a degeneracy as parameters were varied so that afterwards they were again non-degenerate with TKNN integers $n_3$ and $n_4$, then $n_1+n_2=n_3+n_4$.  But there were results that were more important although only noted in passing.  Most basic was the new one that the TKNN integers were homotopy invariants, something that would be clarified by my work on Berry's phase which I turn to shortly.  We also found two compact formulae for the integrand that eventually became commonly used in further work.  First if that $\psi_j$ is the eigenstate of band $j$, then the corresponding TKNN integer, $n_j$, is given by
\begin{equation}\label{10.2}
  n_j = \frac{1}{2\pi}\int_{T^2} K_j;  \qquad K_j = i\jap{d\psi_j,d\psi_j}
\end{equation}
We were especially fond of a second formula, that if $P_j$ is the projection onto $\psi_j$, then
\begin{equation}\label{10.3}
  K_j = i \tr(dP_j P_j dP_j)
\end{equation}
We liked this because while \eqref{10.2} requires a choice of phase in each space, \eqref{10.3} is manifestly phase invariant.  The operator $d$ in the last two expressions is the exterior derivative and there is an implicit wedge product.  The reader might worry that because $df\wedge df=0$, if there were no trace and $P_j$ in \eqref{10.3} was a function, the quantity \emph{would} be 0.  But because $P_j$ is operator valued, it is not $0$.  Indeed,
\begin{align}
  K &= i \sum_{k, \ell} \tr\left(\frac{\partial P_j}{\partial x_k}P_j\frac{\partial P_j}{\partial x_\ell}\right) dx_k\wedge dx_\ell \nonumber  \\
    &= \frac{i}{2}\sum_{k, \ell} \tr\left(P_j\left[\frac{\partial P_j}{\partial x_k},\frac{\partial P_j}{\partial x_\ell}\right]\right) dx_k\wedge dx_\ell \nonumber \\
    &= \frac{i}{2}\sum_{k, \ell} \bigjap{\psi_j,\left[\frac{\partial P_j}{\partial x_k},\frac{\partial P_j}{\partial x_\ell}\right]\psi_j} dx_k\wedge dx_\ell \label{10.4}
\end{align}
where $[\cdot,\cdot]$ is commutator and we used the antisymmetry of  $dx_k\wedge dx_\ell$.

The next part of this story took place in Australia, so I should mention that trip in the summer of 1983 (well, the winter in Australia!) almost didn't happen.  My fourth child, Aryeh, was born in Dec., 1982 and given the time to get his birth certificate and passport, it was only the end of April that I was able to contact the Australian consul in Los Angeles to get visas for all of us including a work visa for me.  He sent a long medical form for me requiring a new general exam from a doctor and xray.  I'd had them 3 months before at Kaiser but was told by the consul that I had to do them over.  I've been raised to avoid unnecessary xrays and I wasn't sure Kaiser would agree to a second exam.  As far as I could tell, this was a restriction put in place to make it difficult for Asians to come and work and I tried to use my invitation from the Australian Academy of Sciences to get a waiver.  The consul was uncooperative, almost nasty.  This was not only pre-Skype but email was almost non-existent and intercontinental phone calls were very expensive, so I sent a telex to my host, Derek Robinson, explaining that, because of visa issues, I would probably have to cancel my trip.  The next day, he called me, which impressed me given the cost of international calls, telling me to stay calm and he'd fix it.  I didn't know that Derek was the secretary of the Australian Academy of Sciences.  But three days later, I get a call from the consul saying ``Sir, I am anxious to issue your visas, but I need you to return the forms I sent you.''  I replied ``But what about the medical form.''  ``Oh, you don't need that, sir.''  According to the current vogue, I should feel guilty for having used my white privilege, but given how important this visit turned out to be, I am glad.

Derek was actually away for the first two weeks of my visit but Brian Davies had also just arrived so we collaborated together on the work on ultracontractivity that is mentioned in Section \ref{s3}.  Midway through my visit, I heard that Michael Berry, whom I'd meet several years before at Joel Lebowitz' seminar, was visiting physics at Australian National University where Derek was in mathematics and where I was visiting.  He'd given a seminar, but before I'd learned he was there, so I called and asked him for a private version which he kindly agreed to.  He explained to me his work on an extra phase he'd found in the adiabatic theorem (see below) and gave me a copy of the manuscript \cite{BerryHisPhase} that he'd recently submitted to Proc. Roy. Soc.  He mentioned that Bernard Souillard, when he heard about Berry's work, told Berry that he thought it might have something to do with the paper of Thouless on TKNN integers but then Berry added that when he asked Thouless about it, Thouless said that he doubted there was any connection (no pun intended).  I replied I thought there probably was and that night, I figured out all the main points that appeared in my paper \cite{SiBerry}!

Berry's paper dealt with the quantum adiabatic theorem. This theorem deals with a time dependent Hamiltonian $H(s); 0\le s \le 1$ and considers $T$ large and $H(s/T)$ so one is looking at very slow changes.  $\varphi_T(s)\equiv\wti{U}_T(s)\varphi; \, 0\le s \le T$  solves $\dot{\varphi}_T(s)=-iH(s/T)\varphi_T(s);\varphi_T(0)=\varphi$.  Let $E(s)$ be an isolated, simple eigenvalue of $H(s)$ and let $P(s)$ be the projection onto the corresponding eigenspace.  The adiabatic theorem says that if $P(0)\varphi=\varphi$, then $\lim_{T\to\infty}(\bdone-P(s/T))\varphi_T(s/T)=0$, i.e. if you start in an eigenspace you stay in it adiabatically.  Berry asked and answered the question, what happens if $H(1)=H(0)$ so you end where you start.  What is the limiting phase of $\varphi_T(T)$.  The surprise he found (it turned out that in 1956 Pancharatnam \cite{Panch} had done the same thing, but it had been forgotten) is that the naive guess that $\varphi_T(T)\sim e^{-iT\int_{0}^{1} E(s) ds}\varphi$ is wrong but that there is an additional phase, $e^{i\Gamma}$. In my paper, I gave $\Gamma$ the name it is now known by - \emph{Berry's phase}.

Berry originally wrote $\Gamma$ as a line integral but, then, assuming that the family $H(s)$ was a closed curve in a parameter space, he used Stokes theorem to write $\Gamma$ as the integral over a surface, $S$, in parameter space whose boundary was the closed curve in the form

\begin{equation}\label{10.5}
  \Gamma = \int_S K(\omega)\,d\omega
\end{equation}
\begin{equation}\label{10.6}
   K = \Ima \sum_{m\ne 0} \frac{\jap{\varphi_m(\omega),\nabla H(\omega)\varphi_0(\omega)}\times\jap{\varphi_0(\omega),\nabla H(\omega)\varphi_m(\omega)}} {(E_m(\omega)-E_0(\omega))^2}
\end{equation}
where he supposed the interpolating Hamiltonian $H(\omega)$ had a complete set $\{\varphi_m\}_{m}$ of simple eigenfunctions with $H(\omega)\varphi_m(\omega) = E_m(\omega)\varphi_m(\omega)$ and $P(\omega)\varphi_0(\omega)=\varphi_0(\omega);\,E(\omega)=E_0(\omega)$.

What I did in my paper \cite{SiBerry} is realize that what Berry was doing was simple and standard geometry in the exact same setting as TKNN. I'd learned in the meantime that the TKNN integers were called the Chern invariant and the curvature $K$ was called the Chern class and used those names for the first time in this context.  The adiabatic theorem defines a connection, i.e. a way of doing parallel transport and Berry's phase was nothing but the holonomy in this connection.  Berry had used \eqref{10.2} as an intermediate formula in his paper but didn't have the phase invariant formula of Avron-Seiler-Simon.  Despite the fact that our independent work was earlier (dates of submission for our paper is May 31, 1983 and his June 13, 1983) and that the geometric ideas were in our paper (and more explicitly with the name curvature in \cite{SiBerry}), $K$ is universally known as the \emph{Berry curvature}.

Berry also realized that in situations where the parameter space could be interpolated into higher dimensions, that eigenvalue degeneracies were sources of curvature, a theme I developed in \cite{SiBerry}.

In the vast literature related to these issues, I should mention two especially illuminating points.  The first involves the fact that the first mathematically precise and, in many ways, still the best proof of the quantum adiabatic theorem is Kato's 1950 proof \cite{KatoAdi} (see \cite[Section 17]{SimonKato} for an exposition).  Without loss, one can suppose E(s)=0 (otherwise replace $H(s)$ by $H(s)-E(s)\bdone$). Kato constructs a comparison dynamics solving
\begin{equation}\label{10.7}
  \frac{d}{ds}W(s) = iA(s)W(s),\, 0 \le s \le 1; \qquad W(0)=\bdone
\end{equation}
\begin{equation}\label{10.8}
  iA(s) \equiv [P'(s),P(s)]
\end{equation}
for which
\begin{equation}\label{10.9}
  W(s)^{-1}P(s)W(s) = P(0)
\end{equation}
by an explicit calculation and he proves that
\begin{equation}\label{10.10}
  \norm{W(s)P(0)-U_T(s)P(0)} = \textrm{O}(1/T)
\end{equation}
The relevant point here is that $W(s)$ defines a connection whose differential, by \eqref{10.8}, is $[P,dP]$ so that its differential, the curvature, is given by \eqref{10.4}.  Thus the Avron-Simon-Seiler formula for the Berry curvature was almost in Kato's paper nearly 35 years before!

Secondly, as noted in \cite{SiBerry}, when the Hilbert space is $\bbC^n$, this connection appeared a 1965 paper of Bott-Chern \cite{BottChern}.  As noted later by Aharonov-Anadan \cite{AhAn}, this connection is induced by a Riemannian metric going back to Fubini \cite{Fub} and Study \cite{Study} at the start of the twentieth century.

I returned to the subject of the quantum Hall effect and Berry's phase twice. As background, I note that from Berry's paper onwards, a key observation was that Berry's phase is zero if all the $H(s)$ can be taken simultaneously real (indeed, Berry tells the story that prior to this work, he noted a curiosity in eigenvalue perturbation theory; if one has real matrices depending on two parameters with an eigenvalue degeneracy only at $0\in\bbR^2$, then going around the degeneracy causes a sign flip in the eigenvector.  In this case, because eigenvectors are chosen real, there is only a $\pm$ degeneracy and so a unique way of continuing.  He talked about this result and someone asked him what happened in the complex case and he replied, there was no difference.  But after the talk, he realized that in the complex case, phase ambiguity meant there was no unique way to continue under just perturbation of parameters and then, that the adiabatic theorem did give a way of continuing which in the complex case could lead to a non-trivial phase).  Since the curvature must be real, the $i$ in \eqref{10.4} (or the $\Ima$ in \eqref{10.6}) show if all the $P$'s are real then $K=0$ and there is no Berry phase.  For spinless particles, time reversal just complex conjugates the wave function so the mantra became ``time reversal invariance kills Berry's phase''.  Magnetic fields destroy reality of the operators (and are not time reversal invariant).  Indeed, the basic example is to take a constant magnetic field, $\mathbf{B}\in\bbR^3$ and $H(\mathbf{B})=\mathbf{B}\cdot\mathbf{\sigma}$ where $\mathbf{\sigma}$ is a spin $s$ spin.  The curvature is then $(2s+1)\mathbf{B}/B^3$.

In work with Avron and two then Caltech postdocs Sadun and Seigert \cite{ASSSAnon, ASSS}, I discovered that for fermions you could have a non-zero Berry phase even with time reversal invariance and that there was a remarkable underlying quaternionic structure relevant to their study.  The underlying issue goes back to a 1932 paper of Wigner \cite{Wigner} on time reversal invariance, $T$, in quantum mechanics.  He first proved his famous theorem that symmetries in quantum mechanics are given by either unitary or anti-unitary operators and then argued that $T$ was always antiunitary with $T^2=\bdone$ for bosons and $T^2=-\bdone$ for fermions.  In the Bose case, that means $T$ acts like a complex conjugate and so the argument of no Berry's phase applies but not in the fermion case.  Instead $J\equiv T$ and, $I$, the map of multiplication by $i$ are two anticommuting operators whose squares are each $-\bdone$, so  they and $K=IJ$ turn the underlying vector space into one over the quaternions! It was Dyson \cite{DysonSym} who first realized that fermions under time reversal have a quaternionic structure (although he first notes the relevance of sympletic groups and that the connection between such groups and quaternions is well known in the mathematical literature on group representations).

We worked out the details, especially for half integral spin systems.  Just as the simplest example of Berry's phase is a spin $1/2$ magnetic dipole, our simple example is a spin $3/2$ electric quadrupole. An interesting feature concerns the fact that eigenspaces are never simple but always even complex dimension.  This degeneracy is known as Kramers degeneracy (after \cite{Kramers}) - one point of Wigner's paper \cite{Wigner} is to explain this degeneracy as a result of time reversal symmetry for fermions.  Thus one looks for simple holonomy in systems with quaternionicly simple eigenvalues, i.e. eigenvalues of fixed complex multiplicity $2$.  The Berry phase is thus a $2\times 2$ unitary matrix.

One noteworthy aspect of \cite{ASSS} is its abstract which reads in full: \emph{Yes, but some parts are reasonably concrete}. While I had introduced topological ideas, I was somewhat dismayed about all the terribly fancy stuff that appeared in the math physics literature, especially throwing around the term ``fiber bundle''.  Yosi and I used to joke that some people seemed to suffer from bundle fibrosis.  So we were concerned about some of the abstruse language in \cite{ASSS} and decided to work out several examples in full as a counterweight.  We liked our abstract, but getting it into the journal was not easy, an interesting story that I'll not include (but see http://www.math.caltech.edu/SimonFest/stories.html\#barry).  Almost twenty five years after our paper, the abstract earned a fan blog post entitled \emph{Abstract Snark} \cite{MRag} that declared our abstract and one other ``almost Zen in their simplicity and perfection''.

My other work in this area is three related papers that I wrote with Avron and Seiler \cite{ASS2, ASS3, ASS4} that followed up on an alternate approach to the quantum Hall effect due to Bellissard \cite{BellHall} in which topology entered as an index in $C^*$-algebraic K-theory.  We developed an index theory for the simpler case where certain subsidiary operators were Fredholm. To me, some of the mathematics we developed was most fascinating.  In particular we proved

\begin{theorem} \label{T10.1} Let $P$ and $Q$ be two orthogonal projections so that $P-Q$ is trace class.  Then $\tr(P-Q)$ is an integer.
\end{theorem}

\begin{remarks} 1. For discussion of trace class, see \cite[Section 3.6]{OT}.

2. This is a result that begs to be proven by Goldberger's method \footnote{Murph Goldberger was one of my professors at Princeton (see the remark after \eqref{8.22}) and, in his day, a famous theoretical physicist who used to joke about things that just had to be true: \emph{oh, you just use Goldberger's method which is a proof by reductio ad absurdum. Suppose it's false; why that's absurd!} Murph was the President of Caltech when I moved and played a role in my recruitment.}.

3. Our proof relied on two operators used extensively by Kato \cite{KatoBk}, $A=P-Q$ and $B=1-P-Q$ which he showed obeyed $A^2+B^2=\bdone$.  We noted \cite{ASS4} that one also had the supersymmetry relation $AB+BA=0$. Since $A$ is trace class and self-adjoint, using a basis of eigenfunctions and the Hilbert-Schmidt Theorem \cite[Theorem 3.2.1]{OT} shows that
\begin{equation}\label{10.11}
  \tr(A) = \sum_\lambda \lambda d_\lambda
\end{equation}
where we sum over eigenvalues and $d_\lambda=\dim(\calH_\lambda)$ with $\calH_\lambda=\{\varphi\,\mid\,A\varphi=\lambda\varphi\}$.  The supersymmetry implies that $\psi\in\calH_\lambda\Rightarrow B\psi\in\calH_{-\lambda}$.  Since $B^2\restriction\calH_\lambda=(1-\lambda^2)\bdone$, we see if $\lambda\ne\pm 1$, then $B$ is a bijection of $\calH_\lambda$ and $\calH_{-\lambda}$ so, for such $\lambda$, we have that $d_\lambda=d_{-\lambda}$.  Thus \eqref{10.11} implies that $\tr(A)=d_1-d_{-1}\in\bbZ$.

4. Slightly earlier, this result was proven by different methods by Effros \cite{Effros}.  His proof is sketched in \cite[Problem 3.15.20]{OT} and our proof can also be found \cite[Example 3.15.19]{OT}.  I found another proof using the Krein spectral shift which is sketched in \cite[Problem 5.9.1]{OT}.  Amrein-Sinha \cite{AmSi} has a fourth proof.

5. For a review of some of the literature on pairs of projections, see \cite[Section 5]{SimonKato}.  I have several more recent papers on pairs of projections \cite{SiPairs, classify1}.
\end{remarks}

\section{Anderson Localization: The Simon-Wolff Criterion} \lb{s11}

I have some contributions to random Schr\"{o}dinger operators, especially in one dimension.  While the first of my papers in the area predates slightly the work of the last section, I've placed this here because my two most significant contributions were finished near the end of 1984, so close to each other that there was a joint announcement \cite{SiTaWoAnon}!  One is the work with Tom Wolff \cite{SiWo} on a necessary and sufficient condition for point spectrum which appears in the title of this section and the other is work with Michael Taylor \cite{SiTa} on regularity of the density of states in the Anderson model.  While the Simon-Taylor work was done first (indeed, I talked about it at the conference where I learned of Kotani's work that motivated Simon-Wolff), I begin with \cite{SiWo}.

The generalized Anderson model is described in Section \ref{s9}. Suppose that the single site distribution, $d\kappa$, is acwrt Lebesgue measure. If $H_\omega$ has dense point spectrum for a.e. $\omega$, then, by independence at distinct sites, if we fix all sites but one, we will have dense point spectrum for Lebesgue a.e. choice of the potential at the remaining point in the a.c. support of the single site distribution.  So it is natural to discuss a family of rank one perturbations,
\begin{equation}\label{11.1}
  A_\lambda = A+\lambda Q,\qquad Q=\jap{\varphi,\cdot}\varphi
\end{equation}
where is A is a self-adjoint operator with simple spectrum (I discuss in remark $1$ why assuming simplicity is no loss), $Q$ the projection onto a unit vector $\varphi$ and $\lambda\in\bbR$. If $d\mu_\lambda$ is the spectral measure for $A_\lambda$ in vector, $\varphi$, a key role is played by the function
\begin{equation}\label{11.2}
   K(E) = \int \frac{d\mu_0(E')}{(E-E')^2}
\end{equation}
which is well-defined as a function with values in $(0,\infty]$ including the possible value of $\infty$.  This function will play a crucial role in the next section also. One main theorem of Simon-Wolff \cite{SiWo} is
\begin{theorem} [Simon-Wolff criterion \cite{SiWo}] \lb{T11.1} Suppose $A$  is a self adjoint operator with cyclic unit vector $\varphi$. Fix an open interval $(\alpha,\beta)$ in $\spec(A)$.  The following are equivalent:

(a) For (Lebesgue) a.e. real $\lambda$, $A_\lambda$ has dense point spectrum in $(\alpha,\beta)$

(b) For (Lebesgue) a.e. real $E\in (\alpha,\beta)$, we have that $K(E)<\infty$.
\end{theorem}

\begin{remarks} 1. We supposed that $A$ has simple spectrum, with $\varphi$ a cyclic vector.  For general $A$ and $\varphi$, we can restrict $A$ to the cyclic subspace generated by $\varphi$ and that restriction obeys the simplicity and cyclicity assumptions, so we can conclude something about the spectral measure $d\mu_\lambda$.  For an Anderson type model, if we know each $\delta_\beta, \beta\in\bbZ^\nu$ has a spectral measure with dense pure point spectrum, we get the result for the full operator.  We also note that it was later shown that any single $\delta_\beta$ is cyclic in the localization region; see the end of this section.

2. Simon-Wolff \cite{SiWo} further note that if $G(\beta,\gamma;z)={\jap{\delta_\beta,(H-z)^{-1}\delta_\gamma}}$ is the Green's function and $d\mu$ is the spectral measure for $H$ and $\delta_0$, then
\begin{equation}\label{11.3}
  \int |E-E'|^{-2} d\mu(E') = \lim_{\varepsilon\downarrow 0} \sum_{\beta\in\bbZ^\nu} |G(\beta,0;E+i\epsilon)|^2
\end{equation}
The two main approaches to the spectral analysis of multidimensional generalized Anderson models are the multiscale analysis of Fr\"{o}hlich-Spencer (see Fr\"{o}hlich-Spencer \cite{FSLoc1, FSLoc2} for the original work and \cite{StollBk} for a pedagogical presentation) and the method of fractional moments of Aizenmann-Molchanov (see Aizenmann-Molchanov \cite{AMLoc} for the original work and \cite{AW} for a pedagogical presentation).  Both most directly prove exponential decay of Green's functions with some kind of uniformity as one approaches the real axis and prove the finiteness of the right side of \eqref{11.3} for a.e. $\omega$ and a.e. $E$ in some interval so, by Theorem \ref{T11.1}, they imply dense point spectrum.
\end{remarks}

The proof of Theorem \ref{T11.1} relies on two elements - a general analysis of the spectral type under rank one perturbations due to Aronszajn \cite{AronAD} and Donoghue  \cite{DonAD} (Aronszajn discussed the special case of variation of boundary condition for ODEs and Donoghue extended to general rank one perturbations; some elements appeared earlier in their joint work \cite{AronDon}).  We need the Stieltjes (aka Borel, aka Cauchy) transforms
\begin{equation}\label{11.4}
  F_\lambda(z) = \int \frac{d\mu_\lambda(E')}{E'-z}
\end{equation}
and we define various subsets of $\bbR$ using $F_{\lambda=0}$ and $K$:
\begin{equation}\label{11.5}
  \begin{split}
  &S_\lambda = \{x\,\mid\,\lim_{\varepsilon\downarrow 0} F_0(x+i\varepsilon) = -\lambda^{-1}\} \text{ for }\lambda\ne 0; \\
  &S_0 = \{x\,\mid\,\lim_{\varepsilon\downarrow 0} \Ima F_0(x+i\varepsilon) = \infty\}
   \end{split}
\end{equation}
\begin{equation}\label{11.6}
  \begin{split}
  &P = \{x\,\mid\,K(x)<\infty\}; \qquad P_\lambda=S_\lambda\cap P \text{ for }\lambda\ne 0; \\
  &P_0 = \{x\,\mid\,\limsup_{\varepsilon\downarrow 0} \varepsilon\Ima F_0(x+i\varepsilon) >0 \}
  \end{split}
\end{equation}
\begin{equation}\label{11.7}
  \begin{split}
    &L = \{x\,\mid\,\lim_{\varepsilon\downarrow 0} \Ima F_0(x+i\varepsilon) \in (0,\infty) \}; \\
    &B=\bbR\setminus\left(\bigcup_{\lambda\in\bbR} S_\lambda \cup L\right)
  \end{split}
 \end{equation}
where, when we write a $\lim$ is equal to some value, it includes the statement that the limit exists.

As preliminaries, we note first that, by the dominated convergence theorem, if $K(x)<\infty$, we have that $\lim_{\varepsilon\downarrow 0} F_0(x+i\varepsilon)$ exists and lies in $\bbR$ so $P$ is $\bigcup_{\{\lambda\ne 0\}}P_\lambda$ plus the set where the limit is $0$.  Secondly, the general theory of Stieltjes transforms implies that each $S_\lambda$ has measure zero.  Note also that the sets $P_\lambda$ are disjoint from each other and from $L$.  We say that $Z\subset\bbR$ \emph{supports a measure}, $\nu$, if and only if
\begin{equation}\label{11.8}
  \nu(\bbR\setminus Z)=0
\end{equation}
Then the work of Aronszajn-Donoghue implies

\begin{theorem} [Aronszajn-Donoghue \cite{AronAD, DonAD}] \lb{T11.2} Let $A_\lambda$ be a family of rank one perturbations.  Then

(a) The a.c. parts of the measures $d\mu_{\lambda,ac}$ are mutually absolutely continuous for all $\lambda\in\bbR$ and are supported on $L$.

(b) The singular parts of the measures $d\mu_{\lambda,sing}$ are mutually singular and for distinct $\lambda\in\bbR$ and each is supported on $S_\lambda$.

(c) For all $\lambda\in\bbR$, the pure point part of the measure, $d\mu_{\lambda,pp}$ is supported on $P_\lambda$ and the singular continuous part of the measure is supported on $S_\lambda\setminus P_\lambda$.

(d) The set $B$ has Lebesgue measure zero and, for all $\lambda\in\bbR$, we have that $\mu_\lambda(B)=0$.
\end{theorem}

\begin{remarks}  1.  For proofs, see \cite[Section 5.8]{OT} or \cite[Section 12.2]{SimonTI}.

2.  One can say much more about $P_\lambda$ and $L$.  First, $L$ is the essential support of the all the $d\mu_{\lambda,a.c.}$.  Secondly, for $\lambda\ne 0$, each point, $x_0$ in $P_\lambda$ is a pure point with $d\mu_{\lambda,pp}(x_0) = (\lambda^2 K(x_0))^{-1}$.

3. After my introduction to rank one theory in the course of this work, I was motivated to do a lot more in the subject.  First, the work on Baire generic singular continuous components \cite{SiSingC2, SiSingC4} discussed in Section \ref{s12}.  Secondly, I worked on the natural meaning of $A_\lambda$ when $\lambda=\infty$ \cite{GSiInf}  and the extension of the theory when $A$ is unbounded and $\varphi$ is very singular \cite{KSiInf}.  Finally, I extended the theory to multiplicative rank one perturbations of unitary operators, a subject useful in OPUC \cite{OPUC1, SiPOPUC}.

4. After those works, I wrote some lecture notes on rank one perturbations \cite{SiVanc}.  When the AMS decided to reprint my trace ideals book, which had gone out of print, it made sense to include those notes as some extra chapters in the second edition \cite{SimonTI}.
\end{remarks}

The second element of the Simon-Wolff analysis was our result that has come to be called \emph{spectral averaging}:

\begin{theorem} [Spectral Averaging \cite{SiWo}] \lb{T11.3} For general rank one perturbations, one has that
\begin{equation}\label{11.9}
  \int \left[d\mu_\lambda(x)\right]\, d\lambda = dx
\end{equation}
in the sense that
\begin{equation}\label{11.10}
  \int \left[f(x)d\mu_\lambda(x)\right]\, d\lambda = \int f(x)\, dx
\end{equation}
for any continuous function, $f$, of compact support on $\bbR$.
\end{theorem}

\begin{remarks}  1.  Theorems \ref{T11.2} and \ref{T11.3} immediately imply Theorem \ref{T11.1} because, by Theorem \ref{T11.2}, (a) is equivalent to $d\mu_\lambda((\alpha,\beta)\setminus P)=0$ for a.e. $\lambda$ and by Theorem \ref{T11.3}, $\int \left[d\mu_\lambda((\alpha,\beta)\setminus P)\right] = |(\alpha,\beta)\setminus P|$.

2. There are variants of spectral averaging that predate \cite{SiWo}.  In 1971, Javrjan \cite{Jav} proved equivalent formulae for the special case of boundary condition variation of Sturm-Liouville operators on $[0,\infty)$.  For some applications all that is needed is the consequence of spectral averaging that if a set $Q\subset\bbR$ has Lebesgue measure zero, then for a.e. $\lambda$ one has that $d\mu_\lambda(Q)=0$ for a.e. $\lambda$.  This fact (or the stronger one that some average of $d\mu_\lambda$ with an a.e. positive weight is dominated by an a.c. measure) appears in the literature in several place prior to \cite{SiWo}: for example Carmona \cite{Car3} and Kunz-Souillard \cite{KunzS}.
\end{remarks}

I conclude the discussion of Simon-Wolff \cite{SiWo} with a bit about the history of its genesis.  In the summer of 1984, Kotani reported on some interesting work at a conference in Maine.  I didn't hear about this work until he and I attend a conference in Bremen in November although he eventually published his work in the Proceeding of the conference in Maine \cite{KotaniPoint}.  While Kotani focused, as he often did, on continuum Schr\"{o}dinger operators, I'll discuss the discrete case which he mentioned in passing.  He looked at an ergodic discrete Schr\"{o}dinger operators on a half line ($n\ge 1$) (i.e. $a_n\equiv 1$, $b_n(\omega)$ samples of an ergodic process) in an energy region, $(\alpha,\beta)$ where one knew the Lyapunov exponent was positive.  He considered operators $h_\omega^\theta$ where the eigenfunction equation $hu(n)=u(n+1)+u(n-1)+b_n u(n)=Eu(n),\, n\ge 1$ was supplemented by the boundary condition $\cos(\theta) u(1)+\sin(\theta) u(0) = 0$.  This is equivalent to truncating the doubly infinite matrix but replacing $b_1$ by $b_1-\cot(\theta)$.  As explained in Section \ref{s9}, one has exponentially growing or decaying solutions except for an $\omega$-dependent set of energies in $(\alpha,\beta)$.  By making explicit an argument of Carmona \cite{Car3}, he showed that for Lebesgue a.e. $\theta$, the spectral measures were supported on the set where one had this exponential dichotomy.  Thus he had the shocking result that in cases like the one where Avron and I proved there was purely singular spectrum (AMO with large coupling and Liouville frequencies), the half line problem had pure point spectrum for a.e. boundary condition $\theta$.

It was immediately clear to me that these ideas might say something about dense point spectrum for the higher dimensional Anderson model where Fr\"{o}hlich-Spencer had recently announced results on exponential decay of Green's functions.  I asked Kotani if he'd thought about such applications and when he said no, I asked if he minded if I thought about it and he said fine.  I returned to Caltech and quickly realized the relevance of the Aronszajn-Donoghue theory and understood the key was finding some abstract version of the Carmona argument.  I decided it was a question connected with Hilbert transforms and so consulted Wolff and we came up with spectral averaging.

Bernard Souillard was also at the conference in Bremen and he also realized the possible applicability of Kotani's scheme to multidimensional localization and he, together with Delyon and L\'{e}vy also developed an approach \cite{DLSou1, DLSou2, DLSou3} to these problems.  They did not phrase it in terms in general rank one perturbations and required exponential decay (rather than only $\ell^2$ decay) and didn't have a necessary and sufficient theorem so, Simon-Wolff has been much more generally quoted. But their ideas worked more easily in some non-rank one situations and, indeed, Delyon, Souillard and I \cite{DSiS2} used their approach to prove some results about random operators with so-called off-diagonal disorder (which are rank $2$)!

After my work with Wolff, I wrote two papers, one with Kotani \cite{SiLoc1, KoSiLoc2} applying these ideas to discrete Schr\"{o}dinger operators in strips.

Before leaving the subject of spectral averaging, I should mention a later work of mine \cite{SimonSpecAv} that extends it to trace class perturbations and averages over finite intervals and relates it to a wonderful formula of Birman-Solomjak \cite{BirSolTrace}.  It involves the Krein spectral shift (see \cite[Section 11.4]{SimonTI} for references and the theory), $\xi_{A,B}(x)$ which, whenever $B-A$ is trace class, can  be defined by
\begin{equation}\label{11.11}
  \tr(f(B)-f(A)) = \int f'(x) \xi_{A,B}(x) dx
\end{equation}
Javrjan \cite{Jav} actually had a local version of \eqref{11.9} which generalized to arbitrary rank one perturbations says that
\begin{equation}\label{11.12}
  \int_{\lambda_0}^{\lambda_1} \left[d\mu_\lambda(x)\right]\, d\lambda = \xi_{A+\lambda_0 Q,A+\lambda_1 Q}(x)  dx
\end{equation}
from which \eqref{11.9} follows because $\lim_{\lambda\to\infty}\xi_{A-\lambda C,A+\lambda C}(x) = \rank(C)\bdone$ (for all $x$) if $C$ is finite rank.  The main result in \cite{SimonSpecAv} considers general families, $A(s); s_0\le s \le s_1$, of self-adjoint operators with a weak derivative $C(s)$ which is trace class, positive and continuous in trace norm.  I defined $d\mu_s(x) = \tr(C(s)^{1/2} dE_s(x) C(s)^{1/2})$ (with $A(s)= \int x dE_s(x)$ in the spectral resolution form of the spectral theorem \cite[Section 5.1]{OT}) and proved that
\begin{equation}\label{11.13}
  \int_{s_0}^{s_1} \left[d\mu_s(x)\right]\, ds = \xi_{A(s_0),A(s_1)}(x)  dx
\end{equation}
I showed that this was equivalent to the formula of Birman-Solomjak that
\begin{equation}\label{11.14}
  \frac{d}{ds}\tr(f(A(s))) = \tr(C(s)f'(A(s)))
\end{equation}
and provided a half page proof of \eqref{11.14}.

I turn next to my work with Taylor \cite{SiTa} which concerns the issue of regularity of the IDS, $k(E)$.  The most heavily quoted and used regularity result is the estimate of Wegner \cite{Wegner} that for the generalized Anderson model on $\bbZ^\nu$, if the $b_n$ are iidrv with distribution \eqref{9.1A} where
\begin{equation}\label{11.15}
   d\kappa(x) = g(x) dx
\end{equation}
with $g\in L^\infty$, then one has the \emph{Wegner estimate}
\begin{equation}\label{11.16}
  |k(E)-k(E')| \le 2\norm{g}_\infty |E-E'|
\end{equation}
This estimate is easy to prove and can be deduced from spectral averaging (although it predates it!).  It (or rather its finite volume analog) is the starting point for most variants of multiscale analysis. This estimate and others that are known in general dimension are of the form that $k$ is as regular as $E\mapsto\kappa(-\infty,E)$.  What Taylor and I proved was the possibility of significantly greater smoothness in one dimension (at least once $\kappa$ has some minimal smoothness).

Why did I think this might be true?  For the free case,
\begin{equation}\label{11.17}
  k(E) = \left\{
           \begin{array}{ll}
             0, & \hbox{ if } E \le -2\\
             \tfrac{1}{\pi}\arccos\left(\tfrac{-E}{2}\right), & \hbox{ if }-2\le E \le 2 \\
             1, & \hbox{ if } E\ge 2
           \end{array}
         \right.
\end{equation}
which implies that $k(E)$ is $C^\infty$ on $(-2,2)$ with $dk/dE=\left[2\pi\sqrt{4-E^2}\right]^{-1}$ so there is a singularity in $dk/dE$ at $E=\pm 2$.  In general, one would expect there are singularities at the edges of the spectrum.  Indeed, this $k(E)$ is globally H\"{o}lder continuous of order $1/2$ and no higher order.  For the Anderson model, I knew there were Lifshitz tails (see below) which implies that $k(E)$ went to zero as $E\downarrow\Sigma_-$, the bottom of the spectrum faster than the inverse of any power of $(E-\Sigma_-)^{-1}$ consistent with $k(E)$ being $C^\infty$ as $E$ shifts from the spectrum to below the spectrum.  I discussed this with Tom Spencer who was dubious that $k$ was $C^\infty$ for the original Anderson model, so we made a 25 cent bet on whether it was true (I would win if someone, not necessarily me, proved it true and he would win if someone, not necessarily him, proved it false).

Of course one expected this not merely for the Anderson model (where $g$ is the characteristic function of an interval), but for at least some generalized Anderson models.  I found that one needed some minimal regularity on $\kappa$ because Bert Halperin \cite{Halp} had shown that there were examples where $d\kappa$ was a two point measure, where $k$ was not even $C^1$.  While Halperin went on to become a distinguished condensed matter theorist, he wrote this paper as a junior undergrad at Harvard!  While the argument was solid, there were missing points of mathematical clarity, so that Taylor and I, who wanted to advertise the result, included details in an appendix \cite{SiTa}.  The model has
\begin{equation}\label{11.18}
  d\kappa(x) = (1-\theta) \delta(x) + \theta \delta(x-\lambda)
\end{equation}
with $0<\theta<1/2$.  This model came to be called the \emph{Bernoulli-Anderson model}.  We showed that $k(E)$ was not H\"{o}lder continuous of any order larger than $\alpha_0=2|\log(1-\theta)|/\arccosh(1+\tfrac{1}{2}|\lambda|)$ so by taking $\theta$ small and/or $\lambda$ large, one can assure lack of H\"{o}lder continuous of any prescribed order.  We also gave heuristics and conjectured that for those extreme values $dk$ should have a singular component (we recall that the Cantor function is H\"{o}lder continuous of order $\log (2)/\log(3)$, so $dk$ can be singular continuous even though $k$ is H\"{o}lder continuous).  Motivated by this, Carmona et. al. \cite{CKM} proved this conjecture (and more importantly proved localization in Bernoulli Anderson models) and Martinelli-Micheli \cite{MM} even proved for any fixed $\theta$, $dk$ was purely singular continuous for all large $\lambda$.

The main result of Simon-Taylor is

\begin{theorem} \lb{T11.4} Let $k$ be the IDS for a generalized Anderson model in $\nu=1$ dimension with $d\kappa$ of the form \eqref{11.15} where $g$ has compact support and for some $\alpha>0$, one has that $(1+k^2)^{\alpha/2} \hat{g}(k)$ is the Fourier transform of an $L^1$ function.  Then $k$ is $C^\infty$.
\end{theorem}

When $g$ is the characteristic function of an interval, the hypothesis holds for any $\alpha<1$, so this won my 25 cent bet with Spencer!  Let me say something about the strategy and genesis of this result.  Most of the early proofs of localization in the 1D Anderson model relied on a theorem of Furstenberg \cite{Furst}, who proved that, under certain circumstances, products of iidrv $\bbS\bbL(2,\bbR)$ matrices had positive Lyapunov exponent.  His proof relied on the action of $\bbS\bbL(2,\bbR)$ on $\bbR\bbP(1)$, real projective space (by $(A,[\varphi])\mapsto A[\varphi]$) and the induced natural convolution of measures on $\bbS\bbL(2,\bbR)$ with measures on $\bbR\bbP(1)$ to get measures on $\bbR\bbP(1)$.  If $\mu$ was the probability measure on $\bbS\bbL(2,\bbR)$ describing the distribution of individual matrices in the random product, Furstenberg showed and used that there was a unique measure $\nu$ on $\bbR\bbP(1)$ so that $\mu*\nu=\nu$.  In the Anderson case, for each real energy, $E$, there is a distribution of transfer matrix \eqref{9.10} and so an invariant measure, $\nu_E$ for each $E$.  I realized that by a discrete analog of the Sturm oscillation theorem, $k(E)$ was the weight that $\nu_E$ gave to those lines in $\bbR\bbP(1)$ with two coordinates of the same sign so that smoothness of $k$ should be implied by smoothness of $\nu_E$ in $E$.  Since $\nu_E$ was also invariant for multiple $\bbS\bbL(2,\bbR)$ convolutions of $\mu_E$, what one needed is that these multiple convolutions got smoother and smoother in $E$.  While I was interested in smoothness in $E$, I suspected (correctly it turns out) that what one really needed was that these high order convolutions of $\mu_E$ were a.c. wrt Haar measure on $\bbS\bbL(2,\bbR)$ with weights that were smoother and smoother in the group parameters.

This was a question in noncommutative harmonic analysis and I assumed the representation theory of $\bbS\bbL(2,\bbR)$ would play a major role, so I contacted Michael Taylor, who I'd heard was a big expert on the topic (shortly after this, he published two books on the subject \cite{TayBk1, TayBk2}) and suggested that we work on it.  At some point, I also spoke to Eli Stein who was also a big expert on the representation theory of $\bbS\bbL(2,\bbR)$ and he made the suggestion that it is often easier to control convolutions on $\bbS\bbL(2,\bbR)$ with one's ``bare
hands'' rather than by using the non-commutative Fourier transform which is what we did.  The underlying $\mu_E$ are certainly not a.c. wrt Haar measure on $\bbS\bbL(2,\bbR)$ since they are supported on a one dimensional subset of the three dimensional group but we proved that under the technical condition on $g$ in Theorem \ref{T11.4} the three fold convolution \emph{is} a.c. wrt Haar measure with a weight that has a tiny bit of smoothness so that in the standard way, the higher order convolutions of that will be smoother and smoother.  While conceptually the proof was straightforward, some of the technical details were formidable.  In particular, we strongly used the compact support hypothesis on $g$.

Our paper stimulated several others that obtained strengthening of our result - two by Klein and others \cite{KlReg1, KlReg2} and one by March-Snitman \cite{Snit}.  Their techniques were very different from ours and each other.  In particular, \cite{KlReg2} only needs the weak condition on $g$ that its Fourier transform is $C^\infty$ with all derivatives vanishing at $\infty$ (automatic if $g$ is of compact support and the analog is even true if $d\kappa$ is the Cantor measure!).

Besides these two major works on random potentials, I have papers on four other aspects ((1) and (3) only in one dimension).  Let me briefly discuss them.

(1)\emph{ localization for slowly decaying random potentials} I wrote a number of papers on the model (half or whole line) where $a_n\equiv 1$ and
\begin{equation}\label{11.19}
   b_n=(1+|n|)^{-\alpha}\omega_n
\end{equation}
where $\omega_n$ are iirdv (sometimes with restrictions on their common distribution) \cite{SiRPt, DSiSAnon, DSiS, KLSi} and \cite[Section 12.7]{OPUC2}.  The first and most basic paper \cite{SiRPt} showed that if $0<\alpha<1/2$, with minor assumptions on the distribution $d\kappa$, of $\omega$ (basically, it has the form \eqref{11.15} with $g$ bounded  and of compact support), then for a.e. $\omega$, $h_\omega$ has dense point spectrum in $[-2,2]$ with eigenfunctions decaying at least as fast as $e^{-C|n|^\beta};\, \beta=1-2\alpha$.  As noted there, the proof is an easy adaptation of the proof of localization in the one dimensional Anderson model by Kunz-Souillard \cite{KunzS}.  I pointed out that one knew (by the trace class theory) there was pure a.c. spectrum on $[-2,2]$ when $\alpha>1$ and that while $1/2\le \alpha \le 1$ was open, it was likely that localization required $\alpha<1/2$.

The transition region $\alpha=1/2$ was discussed in two papers that I wrote with Delyon and Souillard \cite{DSiSAnon, DSiS} which is especially interesting because the same ideas allow the analysis of a random Kronig-Penny model in non-zero electric field.  We fixed $g$ and added a coupling constant, $\lambda$ in $b_n=\lambda (1+|n|)^{-1/2} \omega_n$.  We showed for all sufficiently large $\lambda$, for a.e. $\omega$, the model has dense point spectrum with power decaying eigenfunctions and for all sufficiently small $\lambda$ no point spectrum.  Subsequently Delyon \cite{DelSC} proved purely singular continuous spectrum in these small $\lambda$ regions.

Kotani-Ushiroya \cite{KotUsh} studied a closely related set of models. They studied 1D random continuum Schr\"{o}dinger operators of the type studied by the Russian group \cite{GMP} but with the potential multiplied by $(1+|x|)^{-\alpha}$.  They proved purely a.c. spectrum on $[0,\infty)$ when $1/2<\alpha$ and sharpened the results of the last paragraph when $\alpha=1/2$.  Kiselev, Last and I \cite{KLSi} used discrete analogs of Pr\"{u}fer variables to recover and strengthen the results for the decaying discrete models.  In particular, if $b_n = \lambda n^{-1/2}\omega_n$ with $\omega_n$ as in the classical Anderson model, we proved that the spectrum is purely dense pure point if $\lambda^2\ge 12$ and if $\lambda^2 <12$, the spectrum is purely s.c. in the region $|E|\le\sqrt{4-\lambda^2/3}$ and dense pure point in the complementary part of $[-2,2]$.  We even found the local Hausdorff dimension of the spectral measures (see \cite[Section 8.2]{RA} for discussion of the local Hausdorff dimension of a singular measure) in the singular continuous region.

Random decaying operators, as we will see (Section \ref{s12}), play an important role in discussions of singular continuous spectrum for Baire generic decaying potentials.  Gordon et al \cite{GMJ} (I was a coauthor of an extension of this work to higher dimensions \cite{GJMS}) considered the random potentials of the form \eqref{11.19} with $\alpha<0$.  One might think that $|b_n|\to\infty$ so the spectrum is discrete but if $g$ has zero in its support, it might happen that although $\limsup |b_n| = \infty$, one has that $\liminf |b_n| = 0$!  Indeed, when $\omega$ is uniformly distributed in $[0,1]$, the spectrum is discrete if and only if $-\alpha > 1$.  When $0< -\alpha \le 1$, there is a semi-infinite interval of dense point spectrum.

(2)\emph{ Lifshitz tails} I made some contributions to the theory of Lifshitz tails \cite{LifSi1, LifKS, LifSi2} (I am embarrassed to say that I sometimes used the atypical spelling Lifschitz although I do note the original is Cyrillic).  There is a huge literature, so I'll only include my papers and the original one of Lifshitz referring the reader to the excellent review of Kirsch-Metzger \cite{KirschRev} from my 60th birthday festschrift for more references and more history.  Here's a rough heuristic argument close to Lifshitz original \cite{Lifs}.  Consider a model
\begin{align}
  H_\omega u(n) &= 2\nu u(n) - \sum_{|j|=1} u(n+j) + b_\omega(n)u(n) \nonumber \\
                &\equiv [(H_0+b_\omega)u](n)\label{11.20}
\end{align}
on $\bbZ^\nu$ with, say the $b_\omega(n)$ uniformly distributed in $[0,1]$.  The free term is written as a Laplacian so that $H_0$ is a positive operator whose spectrum is $[0,4\nu]$ and it is easy to prove that for a.e. $\omega$ the spectrum of $H_\omega$ is $[0,4\nu+1]$.  Imagine putting the system in a large box and looking for eigenvalues with energy less than $\varepsilon$ and normalized eigenfunction $\varphi$.  For $\jap{\varphi, H_0 \varphi}$ to be small, $\varphi$ must be spread out over a region of radius $R$ at least $\varepsilon^{-1/2}$ with $\text{O}(\varepsilon^{-\nu/2})$ sites.  For $\jap{\varphi, V \varphi}$ to also be small, we need $V$ to be small at (most of) these sites, certainly no less than $1/2$ so we are looking at probabilities of order $c^{\varepsilon^{-\nu/2}}$ with $0<c<1$.  (One could argue that $c$ should be $\varepsilon$ but that only introduces a log term in the exponent and would restrict the form of the single site probability).  In any event, the expectation is that at least %
\begin{equation}\label{11.21}
  \lim_{E\downarrow 0} \log(-\log(k(E)))/\log(E) = -\nu/2
\end{equation}
the weakest form of Lifshitz tails (and the only one that I proved).  The early rigorous results in this area used the method of large deviations.  My work was motivated by a breakthrough of Kirsch-Martinelli who found the first proof that used bare hands rather than some fancy probabilistic methods.  They only obtained results of the form \eqref{11.21} (which was weaker than some earlier work) but for more general models.  They relied on Dirichlet-Neumann bracketing \cite[Section 7.5]{OT} and treated continuum models.  I wrote \cite{LifSi1} mainly to advertise their work but also to extend it to the discrete case.  The most important contribution of that paper was the use of Temple's inequality which was often used in later works.  In \cite{LifKS}, Kirsch and I proved results like \eqref{11.21} for random perturbations of periodic problems near the bottom and top of the spectrum.  We could not handle the issue of showing there are also Lifshitz tails near the internal gap edges, a problem that, so far as I know, remains open, but I did handle the case of interior gaps in an Anderson model where there are gaps due to gaps in the support of $d\kappa$ \cite{LifSi2}.  I shouldn't leave this subject without mentioning there are interesting issues involving Lifshitz tails in random alloys with long range potentials and in magnetic fields which are discussed in \cite{KirschRev}.

(3) \emph{the notion of semi-uniform localization of eigenfunctions (SULE)} Del Rio, Jitomirskaya, Last and I \cite{SiSingCAnon2, SiSingC4} illuminated what exponential localization in random systems means (this work also discussed Hausdorff dimension of singular continuous spectrum, so I will return to it in Section \ref{s12}).  To use the title of our paper aimed to physicists we dealt with the question, ``What is localization?''.  At the time we wrote it, given the acceptance of Anderson's picture, many theoretical physicists would tell you that a system on $\bbZ^\nu$ is localized (at all energies) means there is a complete set of eigenfunctions, $\{\varphi_{\omega,m}\}_{m=1}^\infty$, each obeying
\begin{equation}\label{11.22}
  |\varphi_{\omega,m}(n)| \le C_{\omega,m}e^{-A|n-n_{\omega,m}|}
\end{equation}
where $n_{\omega,m}$ is the center of localization of the $m$th eigenfunction. Physically though, localization means that a function which at time zero lives on a finite set should remain not too spread out uniformly in time. The natural estimate is to expect that
\begin{equation}\label{11.23}
  \bbE\left(\sup_t|e^{-itH_\omega}(n,\ell)|\right) \le Ce^{-\tilde{A}|n-\ell|}
\end{equation}
Indeed, Delyon et al \cite{DKSou} proved this for 1D Anderson models and Aizenman \cite{AizAnd} proved this in high dimension for large coupling Anderson models.  One point of \cite{SiSingCAnon2, SiSingC4} is that there are (non-random) models where \eqref{11.22} holds but not only does \eqref{11.23} fail but in fact for any $\delta>0$ one has that $\jap{e^{-itH}\delta_0,n^2 e^{-itH}\delta_0}/t^{2-\delta} = \infty$! In fact, it is just a rank one perturbation (by $c\delta_0$) of the 1D AMO at coupling larger than $2$ with Liouville frequency.  Our point was that knowing the size of $C_{\omega,m}$ is critical for dynamic consequences of dense point spectrum.  One might guess that one can take $C$ independent of $m$ but that doesn't hold in large classes of models.  Instead, for a fixed $H$, we defined SULE to mean that for all $\delta>0$, there is $C_\delta$
\begin{equation}\label{11.24}
  |\varphi_{m}(n)| \le C_\delta e^{\delta |n_m|}e^{-A|n-n_{m}|}
\end{equation}
We proved that for operators $H$ with simple spectrum this is equivalent to (and, in general, it implies)
\begin{equation}\label{11.25}
  \sup_t|e^{-itH}(n,\ell)| \le C_\delta e^{\delta \ell}e^{-A|n-\ell|}
\end{equation}
We explicated the a.e. $\omega$ versions of this and noted that \eqref{11.23} implies \eqref{11.25} for a.e. $\omega$.

(4) \emph{simplicity of the spectrum in the localization regime} In \cite{SimonCyc}, I proved that for a generalized Anderson model in arbitrary dimension, if, for a.e. $\omega$, the spectrum is only dense pure point on an interval $[a,b]$, then for a.e. $\omega$ and every $n$, $\delta_n$ is cyclic for $H_\omega\restriction [a,b]$, i.e. finite linear combinations of $\{P_{[a,b]}(H_\omega)H_\omega^k\delta_n\}_{k=0}^\infty$ are dense in $\ran P_{[a,b]}(H_\omega)$.  In particular, this implies that the spectrum is simple on $[a,b]$.

Motivated in part by this, Jak\v{s}i\'{c}-Last \cite{JLCyc1, JLCyc2}, analyzed these questions more deeply.  In particular they proved the result if ``dense pure point on an interval $[a,b]$'' is replaced by ``has spectrum on all of $[a,b]$ with no a.c. part''.  They needed a result of Poltoratski on Hilbert transforms \cite{Polt} to control singular continuous spectra and in \cite{JLCyc3}, they provided a new proof of his result.  Poltoratski is a great expert on Hilbert transforms, so when, in our study of consequences of Remling's work, Zinchenko and I needed some facts about that transform, we joined forces with Poltoratski to prove what we needed \cite{PSZ}.

\section{Generic Singular Continuous Spectrum} \lb{s12}

I like to joke that I spent the first part of my career proving that singular continuous spectrum never occurs (following Wightman's ``no goo hypothesis'' dictum) and the second part showing that it is generic!  In 1978, Pearson \cite{PearsonSC} shocked most experts by showing that 1D continuum Schr\"{o}dinger operators with slowly decaying sparse potentials have purely singular continuous spectrum and, as discussed in Section \ref{s9}, Avron and I proved that for suitable coupling and frequency, the AMO also had purely singular continuous spectrum but the phenomenon was still regarded as exotic and highly atypical.  Starting in early 1993, I discovered that, at least in the sense of Baire, it was, in fact a generic phenomenon, a discovery sometimes called ``the singular continuous revolution''.  In the next few years I published eight papers \cite{SiSingC1, SiSingC2, SiSingC3, SiSingC4, SiSingC5, SiSingC6, SiSingC7, HKS} and two announcements \cite{SiSingCAnon1, SiSingCAnon2} on the subject. Later I studied the analog for OPUC \cite[Section 12.4]{OPUC2}.

I recall that the Baire category theorem \cite[Theorem 5.4.1]{RA} says that a countable intersection of dense open sets in a complete metric space is dense.  Thus countable unions of nowhere dense sets (called first category) are candidates for non-generic sets in that they are closed under countable unions and their complements (the supersets of the dense $G_\delta$'s) are dense.  So dense $G_\delta$ sets in complete metric spaces are call \emph{Baire generic}.  Subsets of $[0,1]$ of Lebesgue measure $1$ are called \emph{Lebesgue generic}.  The notions can be radically distinct in that one can find subsets $A$ and $B$ of $[0,1]$ which are disjoint with one Baire generic and the other Lebesgue generic (and we will see shortly lots of spectral theoretic cases where they are).  We have already seen after \eqref{9.22} that the Diophantine irrationals and the Liouville numbers provide such sets.  One application of the Baire category theorem \cite[Section 5.4]{RA} is for existence.  If a countable set of conditions each hold on a dense $G_\delta$, they all hold somewhere (indeed on a dense $G_\delta$).  The most famous example is an indirect proof of the existence of continuous, nowhere differentiable functions \cite[Problem 5.4.3]{RA}.

I should emphasize that the idea of s.c. spectrum being Baire generic under some conditions was discovered before me by Sasha Gordon.  In a paper submitted in 1991 \cite{GordonSC1}, he announced and in a paper \cite{GordonSC2} published about the same time as \cite{SiSingC2}, Gordon found the same result as in \cite{SiSingC2} (Theorem \ref{T12.2} below) with a different proof.  Our work was definitely later.  I also note that in 1981, Zamfirescu \cite{Zam} proved the suggestive result that among all measures on $[0,1]$ with a fixed bounded variation and no pure points, which is a complete metric space in the variation norm, a Baire generic measure is singular.  See \cite[Problem 5.4.8]{RA} for a proof of the related result that if the probability measures on $[0,1]$ is given the vague topology (in which it is a complete metric space) a Baire generic one is purely singular continuous. I also note that Choksi-Nadkarni in two papers \cite{CN1, CN2} (the first in 1990 predating my work but which the authors say appeared \emph{in a (somewhat inaccessible) conference proceedings}) proved results for unitary operators analogous to the results entitled \emph{Generic Self-Adjoint Operators} below.  That said, my presentation of the full panoply of situations with generic s.c. spectrum established the notion widely.

My original motivation for this work involved a visit to Caltech by Raphael del Rio who gave a seminar on a result \cite{delRioSC} related to the following theorem which appeared in this form in the paper of del Rio, Makarov and Simon \cite{SiSingC2}:

\begin{theorem} [\cite{SiSingC2}] \lb{T12.1} Consider a one parameter family of the form \eqref{11.1} where $A$ is a bounded self-adjoint operator.  Then there is a set, $B\subset\spec(A)$ which is a dense $G_\delta$ in $\spec(A)$ so that no $E\in B$ is an eigenvalue of any  $A_\lambda;\,\lambda\ne 0$.
\end{theorem}

del Rio's result was for boundary condition variation of Sturm-Liouville operators, only discussed a set being dense and uncountable. His proof was fairly involved but it had the key idea of studying the set we will define as $B$ below and applying the Aronszajn-Donoghue theory.  Namely, we let
\begin{equation}\label{13.1}
  B=\{E\,\mid\, K(E)=\infty\}
\end{equation}
and proved that it was dense and uncountable.  I was struck by this result which seemed surprising since, a priori, it certainly seemed possible that for the Anderson model, the eigenvalues filled the entire spectrum as $\lambda$ varied.  Within a couple of days, I realized that dense $G_\delta$'s were lurking and that there was a very short proof.  For suppose $E_0\in\spec(A)$ and there is an open interval, $C$, about $E_0$ which is disjoint from $B$ so that $K(E_1)<\infty$ for all $E_1\in C$. It is easy to see that this condition implies that $F(E_1) \equiv \lim_{\varepsilon\downarrow 0}F(E_1+i\varepsilon)$ exists and is real.  Since the imaginary part vanishes $E_1\notin L\cup S_0$ which, by Theorem \ref{T11.2}, is dense in $\spec(A)$.  It follows that $E_0\notin\spec(A)$ contradicting that $E_0\in\spec(A)$.  We conclude that $B$ is dense in $\spec(A)$.  Moreover, by a simple argument, $G$ is lower semicontinuous, so $\{E|K(E)>n\}$ is open and thus $B$ is a $G_\delta$.  By Theorem \ref{T11.2}(c), no $E\in B$ is an eigenvalue of some $A_\lambda;\,\lambda\ne 0$.

Much more interesting than the set of forbidden energies is the set of forbidden coupling constants and, in this regard, one has

\begin{theorem} [Gordon\cite{GordonSC2}, delRio-Makarov-Simon\cite{SiSingC2}] \lb{T12.2} Consider a one parameter family of the form \eqref{11.1} where $A$ is a bounded self-adjoint operator.  Then $\{\lambda\,\mid\, A_\lambda \text{ has no eigenvalues in } \spec(A)\}$ is a dense $G_\delta$ in $\bbR$.
\end{theorem}

The proof relies on the fact that on the set $P$ where $K(E)<\infty$, the boundary value, $F(E)$, exists and is real, and by Theorem \ref{T11.2}(c), the corresponding $\lambda$'s are given by $F(E)=-\lambda^{-1}$.  Since $\lambda\mapsto -\lambda^{-1}$ takes countable unions of closed nowhere dense subsets of $\pm (0,\infty)$ to such unions, it suffices to write $P$ as a union of such sets each of which is mapped by $F$ to such a set.  One does this by finding such a union so that on each such set, $F$ is Lipschitz.  One needs to go into the complex plane to do this.

This result implies the remarkable fact that in the Anderson model, that for a.e. choice of random potential at all sites but one, for a Lebesgue generic choice at the last point, the spectrum is entirely pure point while for a different Baire generic choice it is purely singular continuous. In particular, dense pure point spectrum will turn into singular continuous spectrum under some arbitrarily small perturbations.  But there is an asymmetry.  In the context of  general rank one perturbations, if there is dense pure point spectrum, there is always a dense set of couplings with purely singular continuous spectrum, but there are examples where for \emph{all} coupling, the spectrum is purely singular continuous.

\bigskip

This next step on forbidden coupling constant was quite natural and shortly after I figured out Theorem \ref{T12.1}, del Rio and I started working of what became Theorem \ref{T12.2} inviting Makarov to join us when we ran into difficulty.  As we were doing that, I asked myself if there might not be a general mechanism underlying this phenomenon of generic singular continuous spectrum and I realized that the key was some soft analysis.  I found the following

\begin{theorem} [\cite{SiSingC1}] \lb{T12.3} Let $\calA$ be a family of self-adjoint operators on a Hilbert space, $\calH$, which is given a metric topology in which convergence implies strong operator convergence of resolvents and in which $\calA$ is a complete metric space. Then the following sets are all $G_\delta$ sets:
\begin{SL}
  \item[\rm{(a)}] For each closed set, $C\subset\bbR$, the set of $A\in\calA$ with \emph{no} eigenvalues in $C$.
  \item[\rm{(b)}] For each open set, $U\subset\bbR$, and each fixed vector $\psi\in\calH$, the set of $A\in\calA$ so that the spectral measure obeys $\left(\mu_A^{(\psi)}\right)_{ac}[U]=0$.
  \item[\rm{(c)}] For each closed set, $K\subset\bbR$, the set of $A\in\calA$ with $K\subset\spec(A)$.
\end{SL}
\end{theorem}

\begin{remarks} 1. While my proof is not hard, it is a little awkward and unnatural.  Lenz-Stollmann \cite{LenzS} found a more natural and direct proof and also slight extensions where rather than putting a topology on $\calA$, one looked at continuous images of complete metric spaces and, in \cite{SimonLpDecomp}, I provided a different simplification of the proof.

2. I emphasize that this theorem says nothing about density.  That may or may not hold.
\end{remarks}

I singled out one consequence of this because it has such an Alice-in-Wonderland character:

\begin{theorem} [The Wonderland Theorem \cite{SiSingC1}] \lb{T12.4} Let $\calA$ be a family of self-adjoint operators on a Hilbert space, $\calH$, which is given a metric topology in which convergence implies strong operator convergence of resolvents and in which $\calA$ is a complete space.  Suppose that a dense set in $\calA$ has purely a.c. spectrum on an open interval $I\subset\bbR$ and another dense set has purely dense point spectrum on $I$.  Then for a dense $G_\delta$ of $A\in\calA$, $I$ lies in $\spec(A)$ and the spectrum is purely singular continuous there!
\end{theorem}

\begin{remarks} 1.  I stated the result this way for drama but the same conclusion holds under the weaker hypothesis that there is a dense subset with no eigenvalues in $I$ and another with no a.c. spectrum in $I$ and another with $I\subset\spec(A)$.

2. The proof is immediate from Theorem \ref{T12.3}.  If $I=(a,b)$ let $\calC_n$ be the set of $A\in\calA$ in Theorem \ref{T12.3}(a) when $K=[a+1/n,b-1/n]$.  If $\{\psi_m\}_{m=1}^\infty$ is a dense set in $\calH$, let $\calK_m$ be the set of $A\in\calA$ in Theorem \ref{T12.3}(b) when $U=I$ and $\psi=\psi_m$.  Finally let $\calP_k$ be the set of $A\in\calA$ with $[a+1/k,b-1/k] \subset\spec(A)$.  By the hypothesis, all these sets are dense and by Theorem \ref{T12.3} all are $G_\delta$ sets.  So, by the Baire category theorem, their intersection is a Baire generic set. If $A$ is in their intersection, it has $I\subset\spec(A)$ and there are no eigenvalues in $I$ and no a.c. spectrum.
\end{remarks}

Here are some of the applications of this set of ideas:

\begin{enumerate}
  \item \emph{Generic Self-Adjoint Operators} Let $\calA$ be the set of all self-adjoint operators, $A$ with $\norm{A}\le 1$ which is a complete metric space in the strong operator topology.  Then, a Baire generic $A\in\calA$ has spectrum all of $[-1,1]$ and purely singular continuous spectrum!  As mentioned above, Choksi-Nadkarni \cite{CN1} had proven the same result for unitary operators and later, they noted that using Cayley transforms, their results imply the result for self-adjoint operators.  They also noted that generically, the spectrum is simple (which is easy to prove with a simple ``this set is a $G_\delta$ argument'' and the fact that small perturbations of operators with dense point spectrum are operators with simple dense point spectrum).

\setlength{\leftskip}{0pt}

The proofs depend on a result of Weyl \cite{WvNW} (see \cite[Theorem 5.9.2]{OT}) that any self-adjoint operator is a norm limit of operators with point spectrum and then a short argument that one can norm approximate operators with dense point spectrum by ones with purely a.c. spectrum.  This result illuminates Weyl's.

\setlength{\leftskip}{0pt}

  \item \emph{A Generic Weyl-von Neumann Theorem}.  Let $\calI_p$ be the trace ideal with $\norm{\cdot}_p$ norm (\cite[Section 3.7]{OT} or \cite{SimonTI}).  von Neumann \cite{WvNvN} extended Weyl's result to allow small Hilbert-Schmidt ($\calI_2$) perturbations and Kuroda \cite{WvNKuroda} allowed arbitrarily small $\calI_p$ for any $p>1$ (it is false if $p=1$ by the Kato-Rosenbum Theorem \cite[Theorem XI.8]{RS3}).  Using this, I proved that for any self-adjoint operator $A$, and any $p>1$, for a $\norm{\cdot}_p$-topology Baire generic $B\in\calI_p$, the spectrum of $A+B$ on $\spec_{ess}(A)$ is purely singular continuous.
  \item \emph{Generic Discrete Schr\"{o}dinger Operators with Bounded Potential}.  Let $\Omega=\bigtimes_{n=-\infty}^\infty [\alpha,\beta]$ with the product topology and given $\omega\in\Omega$ let $A(\omega)$ by the Jacobi matrix with all $a_n=1$ and $b_n=\omega_n$. For a dense $G_\delta$, $A(\omega)$ has purely s.c. spectrum.  It is easy to see this: because of Anderson localization, the set of $\omega$ whose spectrum is $[-2+\alpha,2+\beta]$ and pure point is dense in $\Omega$ (and a $G_\delta$ by Theorem \ref{T12.3}).  Moreover, given any $\omega$, it is easy to see it is a weak limit of periodic $\omega$'s so the set $\omega$ for which $A(\omega)$ has only a.c. spectrum is dense.  One concludes that a Baire generic $A(\omega)$ has spectrum $[-2+\alpha,2+\beta]$ and purely singular continuous spectrum.
  \item \emph{Generic Schr\"{o}dinger Operators with Slowly Decaying Potential}.  In \cite{SiSingC1}, I proved if you look at $C_\infty(\bbR^\nu)$, the continuous functions on $\bbR^\nu$ vanishing at $\infty$, then, Baire generically, $-\Delta+V$ has purely singular continuous spectrum on $[0,\infty)$.  The continuous functions of compact support for which the spectrum of the associated Schr\"{o}dinger operator is pure a.c. on $[0,\infty)$ is dense.  Moreover, by the results of Deift-Simon \cite{DS1}, the a.c. spectrum is unchanged by modifying $V_0$ inside a finite ball, so if one finds a $V_0\in C_\infty(\bbR^\nu)$ with associated Schr\"{o}dinger operator having no a.c. spectrum, given an $W$, we find $V_n\in C_\infty(\bbR^\nu)$ equal to $W$ on the ball of radius $n$ and $V_0$ outside the ball of radius $n+1$, to show that the $V$'s whose associated Schr\"{o}dinger operator has no a.c. spectrum is dense.  One finds the required $V_0$ by taking a centrally symmetric decaying random potential and using \cite{SiRPt}.
  \item \emph{Generic Discrete $1D$ Schr\"{o}dinger Operators with Slowly Decaying Potential}. Using similar ideas, \cite{SiSingC1} fixes $\alpha\in (0,1/2)$ and looks at Jacobi matrices with $a_n\equiv 1$ and a Baire generic $b_n$ in the space of sequences so that $|n+1|^\alpha b_n\to 0$ in the complete metric space with norm $\sup_n (|n+1|^\alpha |b_n|)$ to get a Baire generic family of Jacobi matrices with purely singular continuous spectrum in $(-2,2)$. $\alpha<1/2$ enters so one can use Simon \cite{SiRPt} to get a dense set with no a.c. spectrum.
\end{enumerate}

\bigskip

In \cite{SiSingC3}, Jitomirskaya and I asked the analog for the hull in the almost periodic case of the same question that del Rio, Makarov and I \cite{SiSingC2} and Gordon \cite{GordonSC2} answered for coupling constant variation in the random case and we proved (we also had a result in the continuum case):

\begin{theorem} [Jitomirskaya-Simon \cite{SiSingC3}] \lb{T12.5} Let $H_\omega$ be a discrete Schr\"{o}dinger operator of the form \eqref{9.1}.  Suppose that for some $\omega_0\in\Omega$ $b_n(\omega_0)$ is even in $n$ (for example the AMO, \eqref{9.2}).  Then for a dense $G_\delta$, $U\subset\Omega$, $H_\omega$ has no eigenvalues if $\omega\in U$.
\end{theorem}

\begin{remarks} 1. In particular, if one knows that there is no a.c. spectrum (when \cite{SiSingC3} was written, the result of Kotani \cite{KotaniAC} and Last-Simon \cite{LSEigen} that the a.c. spectrum of $H_\omega$ was constant was not known, so instead, we noted that so long as it was known that there was at least one $\omega$ with no a.c. spectrum, one had it for a dense $G_\delta$), then for a Baire generic $\omega$, $H_\omega$ has purely s.c. spectrum.  In particular, this is true for the AMO when $\lambda>2$.  So when the frequency is Diophantine, we have a situation where for a Lebesgue generic $\theta$, the spectrum is dense pure point and for a Baire generic $theta$, the spectrum is purely singular continuous!

2.  We needed the condition that some function in the hull is even to use Gordon's lemma which is discussed after \eqref{9.22}.

3.  A year later, motivated by this results, Hof, Knill and I \cite{HKS} proved generic s.c. spectrum for a class of subshift potentials which while not strictly almost periodic or even are closely related.  See \cite{HKS} for details or Damanik \cite{DamRevSub} for more on subshifts.
\end{remarks}

\bigskip

Besides these four papers, I had several additional papers in the series \emph{Operators with singular continuous spectrum}.  The most substantial, indeed, by far the longest paper in the series is with del Rio, Jitomirskaya and Last \cite{SiSingC4} (announced in \cite{SiSingCAnon2}).  As already mentioned in Section \ref{s11}, that paper discussed localization for random quantum systems (see point (3) in the discussion including \eqref{11.22}) but its main focus involves the Hausdorff dimensions of the support of the singular continuous spectral measures for Anderson Hamiltonians and of the set of $\lambda$ in Theorem \ref{T12.2} (for more on Hausdorff dimension, see Falconer \cite{Falc} and for Hausdorff dimension of measure, see Rogers \cite{Rog} or Simon \cite[Section 8.2]{RA}).

In the context of general rank one perturbation theory, the set of $\lambda$ leading to singular continuous spectrum can be large, e.g. if the initial measure is pure point but the set $P$ of \eqref{11.6} is empty (which can happen for suitable initial measure $d\mu_0$), then, by Theorem \ref{11.2}, there is purely singular continuous spectrum for \emph{all} $\lambda\ne 0$.  A major point of \cite{SiSingC4} is that when one has SULE (see \eqref{11.22}), the complement of $P$, i.e. the set, $B$, of $E$ where $K(E)=\infty$ has Hausdorff dimension $0$.  Using this one finds:

\begin{theorem} [delRio et al. \cite{SiSingC4}] \lb{12.6} Consider a generalized Anderson model with SULE.  Then for a.e. choice of potential, if we vary the potential at a single point, say $b_0$, then the set of such $b_0$ which have any singular continuous spectrum has Hausdorff dimension zero.  Moreover, for any such value, the spectral measure is supported on a set of Hausdorff dimension zero and one has that for $t$ large, $\jap{\delta_0,x^2(t)\delta_0} \le C(\log |t|)^2$.
\end{theorem}

\bigskip

The last three papers in the series are addenda to the main themes.  Paper 5 with Stolz \cite{SiSingC5} has nothing to do with Baire genericity.  Rather it has criteria for sparse potentials to have no point spectrum, so, if one can also assure no a.c. spectrum, the spectrum is pure s.c.  For the examples of Pearson \cite{PearsonSC}, this provides an independent proof of the absence of point spectrum.  By getting no a.c. spectrum with the method of Simon-Spencer \cite{SiSp} (see Theorem \ref{T6.21}), one gets explicit examples with purely singular continuous spectrum.  Paper 6 \cite{SiSingC6} presented the first examples of graph Laplacians and Laplace Beltrami operators on manifolds with purely singular continuous spectra.  Paper 7 \cite{SiSingC7} constructed a multidimensional example with high barriers as in Simon-Spencer \cite{SiSp} but still having a.c. spectrum.  This might seem to have nothing to do with s.c. spectrum but the example is separable, built of two 1D operators which have s.c. spectrum, but with time decay one can compute and then use to prove the sum has a.c. spectrum (my work was in part motivated by an unpublished remark of Malozemov and Molchanov that because the convolution of two s.c. measures can be a.c., it might be possible to construct examples like this).

\section{Further Remarks} \lb{s13}

As I indicated earlier, while there have been references to some work after that cutoff, this paper mainly discusses research done before 1995.  My research since had its roots in the earlier work but went in a direction which seems to have less relevance to quantum physics.  As discussed in Section \ref{s6}, $H=-\Delta+V(x)$ has ``normal'' spectral behavior if $V(x)$ decays faster than $|x|^{-1}$, namely only discrete spectrum in $(-\infty,0)$ and purely a.c. spectrum on $(0,\infty)$.  On the hand as I discuss in Sections \ref{s11} and \ref{s12}, if the decay is slower than $|x|^{-1/2}$, one can sometimes have no a.c. spectrum.  This was realized by 1995.

I began to wonder what happens for decay like $|x|^{-\alpha}$ with $1/2<\alpha<1$.  If $V$ has a gradient decaying faster than $|x|^{-1}$, it was known for many years (see, e.g. \cite{WeidmannBV}) that one has ``normal'' spectral behavior and in the random case, one also has this (see, e.g. \cite{KotUsh}).  What could happen in the general case?  Fortunately, at this time, I had two very talented grad students - Sasha Kiselev from St. Petersburg and Rowan Killp from Auckland in New Zealand.  Interestingly enough, they both came to work with me upon the strong recommendation of Boris Pavlov \cite{PavlovBk} who moved from St. Petersburg to Auckland at a time to interact with each of them as undergraduates.  I gave Kiselev the problem of whether there was always a.c. spectrum for $|x|^{-\alpha}$ decay when $1/2<\alpha<1$ which I suspected was true.  For his thesis \cite{KisSlowAC}, he proved the occurrence for  $3/4<\alpha<1$ and not long after the general $1/2<\alpha<1$ case was done independently by Christ-Kiselev \cite{CKSlowAC} and Remling \cite{RemSlowAC} (work done while he was visiting my group).

The idea developed around Caltech, stated explicitly by Kiselev-Last-Simon \cite{KLSi}, that the exact borderline should be $V$ or $b$ in $L^2(\bbR,dx)$, a problem which caught Killip's fancy. Percy Deift visited Caltech and when Killip told him about the conjecture, given Percy's work on exactly integrable systems, he immediately thought about the sum rule of Gardner et al \cite{GGKM} that for any nice enough potential, $V(x)$, on $\bbR$, one has that
\begin{equation}\label{13.1}
  \int_{-\infty}^{\infty} V(x)^2\, dx = \frac{16}{3} \sum_j |E_j|^{3/2} + \frac{8}{\pi} \int_{-\infty}^{\infty} \log\left(\left|\frac{1}{T(k)}\right|\right)k^2\,dk
\end{equation}
where, say, $V$ is bounded with compact support, $\{E_j\}$ are the negative eigenvalues of $-\frac{d^2}{dx^2}+V(x)$ and $T(k)$ is the transmission coefficient at energy $E=k^2$.  Interestingly enough, more than 20 years before, in \cite{LTonLT}, Lieb and Thirring had dropped the last term in \eqref{13.1} to get a Lieb-Thirring inequality which they could prove was optimal, because for soliton potentials, the dropped term vanishes.  Deift and Killip \cite{DKill} dropped the middle term of \eqref{13.1} and got an inequality they could use to prove that if $V\in L^2(0,\infty)$ (resp $b\in\ell^2(\bbZ_+)$), then $H=-\tfrac{d^2}{dx^2}+V$ (resp. $hu(n)=u(n+1)+u(n-1)+b_nu(n)$) has a.c. spectrum $[0,\infty)$ (resp $[-2,2]$).  For the discrete result, they needed a Toda lattice analog of the KdV sum rule.

Given these results and the results of \cite{NabDPS, SimonDPS} on embedded dense point spectrum, I made one of the problems in my 2000 open problems list \cite{Simon15}: \emph{Do there exist potentials $V(x)$ on $[0,\infty)$ so that $|V(x)| \le |x|^{-1/2-\varepsilon}$ for some $\varepsilon > 0$ and so that $-d^2/dx^2+V(x)$ has some singular continuous spectrum}.  At the time, I didn't realize that an analogous problem had been solved in 1936 by Verblunsky \cite{Verb2}!  For the analog of the potential for OPUC are what are now called Verblunsky coefficients (a term I introduced in 2005), a sequence $\{\alpha_n\}_{n=0}^\infty$ of numbers in $\bbD$ associated to any probability measure, $d\nu$ on $\partial\bbD$.  What Verblunsky proved, extending a result of Szeg\H{o} \cite{SzHisThm, SimonSz}, is that this sequence lies in $\ell^2$ if and only if the a.c. part of $d\nu$ obeys a certain condition.  The singular part could be arbitrary so long as the total mass of $d\nu$ was 1.

These ideas were brought into the question of mixed spectrum for Schr\"{o}dinger operators by Sergey Denisov, then a graduate student in Moscow, in a preprint I first learned about in January 2001 (a version only appeared in print several years later \cite{Denisov}).  Nick Makarov and I were impressed enough that we invited him to be a postdoc at Caltech, of which more shortly.  Denisov used a continuum analog of OPUC called Krein systems and said that he could construct $V\in L^2((0,\infty),dx)$ so the corresponding $H$ had an arbitrary singular continuous part on some $[0,E_0]$.  Technically that didn't solve my problem since I had stated the result in terms of power decay, not $L^p$ (the power result was obtained shortly afterwards by Kiselev \cite{KisSC}) but to me morally it did.

Rowan Killip and I set out to understand Denisov's proof but, in part, because we had no prior experience with OPUC or Krein systems and could find little literature on the former and none on the later, we found the arguments opaque.  We did determine that the key to his proof seemed to be a sum rule.  We were interested in a result for Jacobi matrices, so we looked at some sum rules of Case \cite{Case}.  His sum rules were only formal so it wasn't clear when they hold although his arguments certainly could be made rigorous if $b_n$ and $a_n-1$ were of finite support.  We wanted them to always hold which presented two problems - to avoid possible cancellations of infinities, we needed both sides to be positive and one had to be able control the limits.

We found that none of Case's sum rules (which entered from successive terms in a Taylor series) had the necessary positivity but by fooling around, we found a linear combination of two of them that was positive.  It was mysterious why there was any such combination and the rather complicated functions that entered in the final sum rule were totally ad hoc.  Fifteen years later Gamboa, Nagel and Roault \cite{GNR} found a totally new proof using the method of large deviations on certain random matrix ensembles that explained why there was a positive quantity and the meaning of the previously ad hoc functions. For the OPUC analog, Breuer, Zeitouni and I found some other positive sum rules using large deviations \cite{BSZ2}.  Since the GNR paper wasn't very accessible to spectral theorists, we wrote a pedagogic exposition of their approach \cite{BSZ1}.

In going from the finite support case to the general, there was one tricky limit that stymied us for a while. In those days, jury duty in Los Angeles could mean coming in every day for two weeks waiting in the jury assembly room all day for assignment to a trial and in the summer of 2001, I had such a stint not even winding up on a jury!  Sitting around gave me lots of time to think about this holdup and I realized that since the object that we were having trouble with was a relative entropy, it had some semicontinuity properties that overcame our difficulty.  We were quite pleased by this discovery although we learned several years later that Verblunsky \cite{Verb2} had made the same discovery 65 years earlier in his related work! (Verblunsky didn't know he had an entropy but he had discovered and exploited the semicontinuity.)

One result of my work with Killip \cite{KS} was necessary and sufficient conditions on a spectral measure for associated Jacobi matrix to have $J-J_0$ a Hilbert Schmidt operator (where $J_0$ corresponds to $a_n=1, b_n=0$), a result now regarded as an OPRL analog of Szeg\H{o}'s theorem.  We also obtained results described in \eqref{8.18A} which led to a proof of Nevai's conjecture.

While we were writing this up, Denisov arrived at Caltech and gave a course on Krein systems which I sat in on.  What struck me was his initial few works where he described the theory of OPUC.  I was struck by the beauty and elegance of the subject although I found a much simpler proof of Szeg\H{o} recursion than he gave us (it turned out the proof that I'd found was in the literature but so little known that I surprised at least one expert with it).  It became clear to me that the similarities between OPUC and OPRL suggested one should be able to carry over (often with some gymnastics needed) much of the spectral theory of Jacobi matrices to OPUC.  Rather than lots of small papers on each subarea, I decided one long review article made sense.  I had to face the fact that there wasn't really any recent exposition of the basics of OPUC so I decided on a two long review articles which grow into a two volume set of books \cite{OPUC1, OPUC2} with over 1000 pages!

In many ways, the spectral theory of OPUC and its relation to OPRL became a major focus of my research for the time since.  Many of the major results involve Szeg\H{o}'s theorem and Szeg\H{o} asymptotics, among them the extension by Damanik-Killip-Simon \cite{DKS} of the work of Killip-Simon to perturbations of certain periodic Jacobi matrices, my study with Christiansen and Zinchenko on Szeg\H{o} behavior of finite gap operators \cite{CSZFG1, CSZFG2, CSZFG3} and the work with Damanik \cite{DSInv} on necessary and sufficient conditions for Szeg\H{o} asymptotics for OPRL.

OPRL can be viewed as the solutions of an $L^2$ minimization problem.  The analogous $L^\infty$ minimization problem define Chebyshev polynomials which depend on some compact subset, $\fre\subset\bbC$. In a brilliant paper 1969 paper, Widom \cite{WCheb} discussed how to modify Szeg\H{o} asymptotics for Chebyshev polynomials when $\fre$ is a finite union of disjoint sufficiently smooth Jordan curves.  He obtained partial results for finite gap sets in $\bbR$ and he made a conjecture about the expected asymptotics which we dubbed Szeg\H{o}-Widom asymptotics.  This conjecture remained open for over 45 years until proven by Christiansen, Zinchenko and me \cite{CSZCheb1}.  Our proof could be phrased in terms of discriminants of periodic Jacobi matrices which made the arguments natural.

It is appropriate to end with a story about the publication of that paper.  I felt the paper was important enough to warrant sending it to a top three journal but, for various reasons, one of my coauthors wanted to send it to a slightly less prestigious but still top journal.  Fairly quickly we got a reply that the person asked for a quick opinion thought it was nice that we had solved a 45+ year old conjecture but the paper wasn't up to the standard of this journal because the proofs were too simple! I was scandalized by this and insisted that we try a top three journal, which we did, where the paper was accepted.  Lest you think I disapprove of the system that top math journals use to decide which papers to publish, I feel that it is best described by Churchill's description of democracy - the worst possible method of evaluation except for all the others.

DATA AVAILABILITY

Data sharing is not applicable to this article as no new data were created or analyzed in this study.


\end{document}